\documentclass[a4paper,10pt,3p,onecolumn,compress]{elsarticle}

\usepackage[T1]{fontenc}
\usepackage{amsmath}
\usepackage{amssymb}
\usepackage{graphicx}
\usepackage{tcolorbox}
\usepackage{dcolumn}
\usepackage{bm}
\usepackage{xspace}
\usepackage{isotope}
\usepackage{units}
\usepackage{multirow}
\usepackage{booktabs}
\usepackage{natbib}
\usepackage{physics}
\usepackage{orcidlink}
\usepackage{rotating}
\usepackage[acronym,automake, sort=standard]{glossaries}
\makeglossaries

\newacronym{agn}{AGN}{active galactic nucleus}
\newacronym{ags}{AGS}{Alternating Gradient Synchrotron}
\newacronym{aladin}{ALADIN}{A Large Acceptance DIpole magNet}
\newacronym{bnl}{BNL}{Brookhaven National Laboratory}
\newacronym{bgo}{BGO}{bismuth germanium oxide}
\newacronym{bring}{BRing}{Booster synchrotron Ring}
\newacronym{bsm}{BSM}{Beyond Standard Model}
\newacronym{cbm}{CBM}{compressed baryonic matter}
\newacronym{cmb}{CMB}{cosmic microwave background}
\newacronym{cnao}{CNAO}{Centro Nazionale di Adroterapia Oncologica}
\newacronym{cr}{CR}{cosmic ray}
\newacronym{crdb}{CRDB}{cosmic-ray database}
\newacronym{dm}{DM}{dark matter}
\newacronym{ec}{EC}{electronic capture}
\newacronym{etf}{ETF}{External Target Facility}
\newacronym{fair}{FAIR}{Facility for Antiproton and Ion Research}
\newacronym{foot}{FOOT}{FragmentatiOn Of Target}
\newacronym{frs}{FRS}{FRagment Separator}
\newacronym{gntf}{GNTF}{Gamma-ray Neutron Test Facility}
\newacronym{gsfc}{GSFC}{Goddard Space Flight Center}
\newacronym{ganil}{GANIL}{Grand Accélérateur National d'Ions Lourd}
\newacronym{gce}{GCE}{Galactic centre excess}
\newacronym{gcr}{GCR}{Galactic cosmic ray}
\newacronym{glad}{GLAD}{GSI Large Acceptance Dipole}
\newacronym{gsi}{GSI}{GSI Helmholtz Centre for Heavy Ion Research}
\newacronym{hec}{HEC}{High Energy Cave}
\newacronym{hep}{HEP}{high-energy physics}
\newacronym{hfrs}{HFRS}{High-energy FRagment Separator}
\newacronym{hiaf}{HIAF}{Heavy-Ion Accelerator Facility}
\newacronym{himac}{HIMAC}{Heavy Ion Medical Accelerator in Chiba}
\newacronym{hirfl}{HIRFL}{Heavy Ion Research Facility in Lanzhou}
\newacronym{hpge}{HPGe}{high-purity Germanium detectors}
\newacronym{icru}{ICRU}{International Committee for Radiological Units}
\newacronym{ihep}{IHEP}{Institute of High Energy Physics}
\newacronym{is}{IS}{interstellar}
\newacronym{ism}{ISM}{interstellar medium}
\newacronym{imp}{IMP}{Institute of Modern Physics}
\newacronym{jenaa}{JENAA}{{J}oint {E}CFA-{N}uPECC-{A}PPEC {A}ctivities}
\newacronym{jinr}{JINR}{Joint Institute for Nuclear Research}
\newacronym{lartpc}{LArTPC}{liquid argon time-projection chamber}
\newacronym{lbnl}{LBNL}{Lawrence Berkeley National Laboratory}
\newacronym{ligo}{LIGO}{Laser Interferometer Gravitational-wave Observatory}
\newacronym{lyso}{LYSO}{lutetium–yttrium oxyorthosilicate}
\newacronym{mc}{MC}{Monte Carlo}
\newacronym{mdr}{MDR}{maximum detectable rigidity}
\newacronym{nasa}{NASA}{National Aeronautics and Space Administration}
\newacronym{nica}{NICA}{Nuclotron-based Ion Collider
Facility}
\newacronym{nist}{NIST}{National Institute of Standards and Technology}
\newacronym{nsrl}{NSRL}{NASA Space Radiation Laboratory}
\newacronym{pandora}{PANDORA}{Photo-Absorption of Nuclei and Decay Observation for Reactions in Astrophysics}
\newacronym{ppac}{PPAC}{Parallel Plate Avalanche Counter}
\newacronym{r3b}{R3B}{Reactions with Relativistic Radioaktiv Beams}
\newacronym{rhic}{RHIC}{Relativistic Heavy Ion Collider}
\newacronym{ribll}{RIBLL2}{Radioactive Ion Beam Line in Lanzhou}
\newacronym{ribs}{RIBs}{Radioactive Ion Beams}
\newacronym{rich}{RICH}{ring-imaging Cherenkov}
\newacronym{riken}{RIKEN}{Rikagaku Kenkyūsho, Institute of Physical and Chemical Research}
\newacronym{sm}{SM}{standard model}
\newacronym{ss}{SS}{Solar system}
\newacronym{toa}{TOA}{top of atmosphere}
\newacronym{tps}{TPS}{treatment planning systems}
\newacronym{tof}{TOF}{time-of-flight}
\newacronym{uhecr}{UHECR}{ultra-high-energy cosmic ray}
\newacronym{xscrc}{XSCRC}{Cross-Sections for Cosmic Rays at CERN}

\newacronym{aleph}{ALEPH}{Apparatus for LEP PHysics}
\newacronym{alice}{ALICE}{A Large Ion Collider Experiment}
\newacronym{amber}{AMBER}{Apparatus for Meson and Baryon Experimental Research}
\newacronym{argus}{ARGUS}{A Russian-German-United States-Swedish collaboration}
\newacronym{atlas}{ATLAS}{A Toroidal LHC Apparatus}
\newacronym{brahms}{BRAHMS}{Broad RAnge Hadron Magnetic Spectrometers}
\newacronym{cern}{CERN}{Conseil européen pour la recherche nucléaire (European Organization for Nuclear Research)}
\newacronym{cms}{CMS}{Compact Muon Solenoid}
\newacronym{compass}{COMPASS}{Common Muon and Proton Apparatus for Structure and Spectroscopy}
\newacronym{isr}{ISR}{Intersecting Storage Rings} 
\newacronym{kek}{KEK}{Kō Enerugī Kasokuki Kenkyū Kikō (High Energy Accelerator Research Organization)}
\newacronym{lear}{LEAR}{Low Energy Antiproton Ring}
\newacronym{lhc}{LHC}{Large Hadron Collider}
\newacronym{lhcb}{LHCb}{LHC beauty}
\newacronym{lhcf}{LHCf}{LHC forward}
\newacronym{ls}{LS}{Long Shutdown}
\newacronym{ps}{PS}{Proton Synchrotron}
\newacronym{shine}{SHINE}{a SPS Heavy Ion and Neutrino Experiment}
\newacronym{smog}{SMOG}{System for Measuring Overlap with Gas}
\newacronym{sps}{SPS}{Super Proton Synchrotron}
\newacronym{torch}{TORCH}{Time Of internally Reflected CHerenkov light}

\newacronym{acecris}{ACE-CRIS}{Advanced Composition Explorer Cosmic Ray Isotope Spectrometer}
\newacronym{aladino}{ALADInO}{Antimatter Large Acceptance Detector In Orbit}
\newacronym{ams}{AMS}{Alpha Magnetic Spectrometer}
\newacronym{bess}{BESS}{Balloon-borne Experiment with Superconducting Spectrometer} 
\newacronym{calet}{CALET}{CALorimetric Electron Telescope}
\newacronym{cream}{CREAM}{Cosmic Ray Energetics And Mass}
\newacronym{crn}{CRN}{Cosmic Ray Nuclei detector}
\newacronym{crres}{CRRES}{Combined Release and Radiation Effects Satellite}
\newacronym{ctao}{CTAO}{Cherenkov Telescope Array Observatory}
\newacronym{dampe}{DAMPE}{DArk Matter Particle Explorer}
\newacronym{gaps}{GAPS}{General AntiParticle Spectrometer}
\newacronym{grams}{GRAMS}{Gamma-Ray and AntiMatter Survey}
\newacronym{hawc}{HAWC}{High Altitude Water Cherenkov experiment}
\newacronym{heao}{HEAO3}{High Energy Astrophysical Observatory}
\newacronym{heaohne}{HEAO3-HNE}{Heavy Nuclei Experiment on HEAO3}
\newacronym{helix}{HELIX}{High Energy Light Isotope Experiment}
\newacronym{herd}{HERD}{High Energy cosmic-Radiation Detection}
\newacronym{hero}{HERO}{High-Energy Ray Observatory}
\newacronym{hess}{H.E.S.S.}{High Energy Stereoscopic System}
\newacronym{isee}{ISEE}{International Sun-Earth Explorer}
\newacronym{isomax}{ISOMAX}{ISOtope Magnet eXperiment}
\newacronym{iss}{ISS}{International Space Station}
\newacronym{lat}{LAT}{Large Area Telescope}
\newacronym{lhaaso}{LHAASO}{Large High Altitude Air Shower Observatory}
\newacronym{magic}{MAGIC}{Major Atmospheric Gamma Imaging Cherenkov telescope}
\newacronym{nicer}{NICER}{Neutron Star Interior Composition Explorer}
\newacronym{olimpiya}{OLIMPIYA}{OLIvined from Meteorites—Poisk {search for} heavy I {and} superheavy YAder {nuclei}}
\newacronym{pamela}{PAMELA}{Payload for Antimatter Matter Exploration and Light-nuclei Astrophysics}
\newacronym{phescami}{PHeSCAMI}{Pressurised Helium Scintillating Calorimeter for AntiMatter Investigation}
\newacronym{ta}{TA}{Telescope Array}
\newacronym{tiger}{TIGER}{Trans-Iron Galactic Element Recorder}
\newacronym{uhcreldef}{UHCRE-LDEF}{Ultra Heavy Cosmic Ray Experiment on the Long Duration Exposure Facility}

\newcommand{\pp}{pp\xspace}
\newcommand{\Hep}{{He}p\xspace}
\newcommand{\pHe}{p{He}\xspace}
\newcommand{\pC}{p{C}\xspace}
\newcommand{\HeHe}{{He}{He}\xspace}
\newcommand{\sqrtsnn}{\sqrt{s_\text{NN}}}
\newcommand{\neutron}{\ensuremath{\mathrm{n}}\xspace}
\newcommand{\proton}{\ensuremath{\mathrm{p}}\xspace}
\newcommand{\triton}{\ensuremath{\mathrm{t}}\xspace}
\newcommand{\deuteron}{\ensuremath{\mathrm{d}}\xspace}
\newcommand{\helium}{\ensuremath{\mathrm{He}}\xspace}
\newcommand{\antiproton}{\ensuremath{\mathrm{\overline{p}}}\xspace}
\newcommand{\antineutron}{\ensuremath{\mathrm{\overline{n}}}\xspace}
\newcommand{\antideuteron}{\ensuremath{\mathrm{\overline{d}}}\xspace}
\newcommand{\antitriton}{\ensuremath{\mathrm{\overline{t}}}\xspace}
\newcommand{\antihelium}{\ensuremath{\mathrm{\overline{He}}}\xspace}
\newcommand{\antiheliumthree}{\ensuremath{\mathrm{^3\overline{He}}}\xspace}
\newcommand{\antiheliumfour}{\ensuremath{\mathrm{^4\overline{He}}}\xspace}

\newcommand{\ijk}{{i+j\rightarrow k}}
\newcommand{\abc}{{a+b\rightarrow c}}
\newcommand{\C}{{\cal C}} 

\newcommand{\lt}{\ensuremath <}

\newcommand\quasiel{{\operatorname{quasi-el}}}

\usepackage{lineno}

\usepackage{hyperref} 
\hypersetup{
  colorlinks = true,
  citecolor  = magenta,
  linkcolor = black
}

\begin{document}

\title{
Precision cross-sections for advancing cosmic-ray physics and other applications: a comprehensive programme for the next decade
}

\author[1]{D.~Maurin\corref{cor1}\orcidlink{0000-0002-5331-0606}}
\ead{david.maurin@lpsc.in2p3.fr}
\cortext[cor1]{Corresponding author}
\author[2]{L.~Audouin\orcidlink{0000-0001-9899-6923}}
\author[3]{E.~Berti\orcidlink{0000-0002-5841-7760}}
\author[4]{P.~Coppin\orcidlink{0000-0001-6869-1280}}
\author[5]
{M.~Di Mauro\orcidlink{0000-0003-2759-5625}}
\author[6]{P.~von~Doetinchem\orcidlink{0000-0002-7801-3376}}
\author[5,7,8]{F.~Donato\orcidlink{0000-0002-3754-3960}}
\ead{fiorenza.donato@unito.it}
\author[9,10]{C.~Evoli\orcidlink{0000-0002-6023-5253}}
\author[11]{Y.~Génolini\orcidlink{0000-0002-7326-1282}}
\author[12]{P.~Ghosh\orcidlink{0000-0003-1293-3660}}
\author[13]{I.~Leya\orcidlink{0000-0002-3843-6681}}
\author[14,15,16]{M.~J.~Losekamm\orcidlink{0000-0001-7854-2334}}
\author[8]{S.~Mariani\orcidlink{0000-0002-7298-3101}}%
\ead{saverio.mariani@cern.ch}
\author[17]{J.~W.~Norbury\orcidlink{0009-0000-0740-0611}}
\author[18,19]{L.~Orusa\orcidlink{0000-0002-1879-457X}}
\author[4]{M.~Paniccia\orcidlink{0000-0001-8482-2703}}
\author[8]{T.~Poeschl\orcidlink{0000-0003-3754-7221}}
\author[11]{P.~D.~Serpico\orcidlink{0000-0002-8656-7942}}
\author[4]{A.~Tykhonov\orcidlink{0000-0003-2908-7915}}
\author[20]{M.~Unger\orcidlink{0000-0002-7651-0272}}
\author[21]{M.~Vanstalle\orcidlink{0000-0003-1193-7475}}
\author[22,23]{M.-J.~Zhao\orcidlink{0000-0001-6844-1409}}
\author[24,10]{D.~Boncioli\orcidlink{0000-0003-1186-9353}}
\author[5,7]{M.~Chiosso\orcidlink{0000-0001-6994-8551}}
\author[5]{D.~Giordano\orcidlink{0000-0003-0228-9226}}
\author[25]{D.~M.~Gomez~Coral\orcidlink{0000-0002-9200-6607}}
\author[3]{G.~Graziani\orcidlink{0000-0001-8212-846X}}
\author[8]{C.~Lucarelli\orcidlink{0000-0002-8196-1828}}
\author[26,27]{P.~Maestro\orcidlink{0000-0002-4193-1288}}
\author[14]{M.~Mahlein\orcidlink{0000-0003-4016-3982}}
\author[28]{L.~Morejon\orcidlink{0000-0003-1494-2624}}
\author[29]{J.~Ocampo-Peleteiro\orcidlink{0000-0002-7585-2641}}
\author[29]{A.~Oliva\orcidlink{0000-0002-6612-6170}}
\author[20]{T.~Pierog\orcidlink{0000-0002-7472-8710}}
\author[8]{L.~Šerkšnytė\orcidlink{0000-0002-5657-5351}}

\affiliation[1]{organization={LPSC, Université Grenoble-Alpes, CNRS/IN2P3}, postcode={38026}, city={Grenoble}, country={France}}
\affiliation[2]{organization={Université Paris-Saclay, IJC Lab, CNRS/IN2P3}, postcode={91405}, city={Orsay}, country={France}}
\affiliation[3]{organization={INFN, Sezione di Firenze},  postcode={I-50019 Sesto Fiorentino}, city={Florence}, country={Italy}}
\affiliation[4]{organization={DPNC, Université de Genève}, postcode={1211 Genève 4}, city={Geneva}, country={Switzerland}}
\affiliation[5]{organization={INFN, Sezione di Torino}, addressline={Via P. Giuria 1}, postcode={10125}, city={Torino}, country={Italy}} 
\affiliation[6]{organization={University of Hawaii at Manoa}, postcode={HI}, city={Honolulu}, country={USA}}
\affiliation[7]{organization={Università degli Studi di Torino}, city={Torino}, country={Italy}}
\affiliation[8]{organization={European Organization for Nuclear Research (CERN)}, city={Geneva}, country={Switzerland}}
\affiliation[9]{organization={Gran Sasso Science Institute (GSSI)}, addressline={Viale Francesco Crispi 7}, postcode={67100}, city={L’Aquila}, country={Italy}}
\affiliation[10]{organization={INFN-Laboratori Nazionali del Gran Sasso (LNGS)}, addressline={via G. Acitelli 22}, postcode={67100}, city={Assergi (AQ)}, country={Italy}}
\affiliation[11]{organization={LAPTh, CNRS, Université Savoie Mont Blanc}, postcode={F-74940}, city={Annecy}, country={France}}
\affiliation[12]{organization={NASA Goddard Space Flight Center}, postcode={Greenbelt, Maryland, 20771}, country={USA}} 
\affiliation[13]{organization={University of Bern, Space Sciences and Planetology}, postcode={CH-3012}, city={Bern}, country={Switzerland}}
\affiliation[14]{organization={Technical University of Munich, School of Natural Sciences}, city={Garching}, country={Germany}}
\affiliation[15]{organization={Excellence Cluster ORIGINS}, city={Garching}, country={Germany}}
\affiliation[16]{organization={Now at European Space Agency, ESTEC, Noordwijk, Netherlands}}
\affiliation[17]{organization={NASA Langley Research Center}, postcode={Hampton, Virginia, 23666}, country={USA}}
\affiliation[18]{organization={Department of Astrophysical Sciences, Princeton University}, postcode={NJ 08544}, city={Princeton}, country={USA}}
\affiliation[19]{organization={Department of Physics and Columbia Astrophysics Laboratory, Columbia University}, postcode={NY 10027}, city={New York}, country={USA}}
\affiliation[20]{organization={IAP, KIT}, city={Karlsruhe}, country={Germany}}
\affiliation[21]{organization={Université de Strasbourg, CNRS, IPHC-UMR7178}, postcode={F-67000}, city={Strasbourg}, country={France}}
\affiliation[22]{organization={Key Laboratory of Particle Astrophysics, Institute of High Energy Physics, Chinese Academy of Sciences}, postcode={100049}, city={Beijing}, country={China}}
\affiliation[23]{organization={China Center of Advanced Science and Technology}, postcode={100190}, city={Beijing}, country={China}}
\affiliation[24]{organization={Università degli Studi dell’Aquila, Dipartimento di Scienze Fisiche e Chimiche}, postcode={Via Vetoio, 67100}, city={L’Aquila}, country={Italy}}
\affiliation[25]{organization={Instituto de F{\'i}sica, Universidad Nacional Autónoma de México, Circuito de la Investigación Cient{\'i}fica}, city={Ciudad de México}, country={México}}
\affiliation[26]{organization={Department of Physical Sciences, Earth and Environment, University of Siena}, postcode={via Roma 56, 53100}, city={Siena}, country={Italy}}
\affiliation[27]{organization={INFN, Sezione di Pisa}, postcode={Polo Fibonacci, Largo B. Pontecorvo 3, 56127}, city={Pisa}, country={Italy}}
\affiliation[28]{organization={Bergische Universität Wuppertal}, postcode={Gausstrasse 20, 42117}, city={Wuppertal}, country={Germany}}
\affiliation[29]{organization={INFN, Sezione di Bologna}, postcode={40126}, city={Bologna}, country={Italy}}

\date{\today}

\begin{abstract}
Cosmic-ray physics in the GeV-to-TeV energy range has entered a precision era thanks to recent data from space-based experiments. However, the poor knowledge of nuclear reactions, in particular for the production of antimatter and secondary nuclei, limits the information that can be extracted from these data, such as source properties, transport in the Galaxy and indirect searches for particle dark matter. The {\em Cross-Section for Cosmic Rays at CERN} workshop series has addressed the challenges encountered in the interpretation of high-precision cosmic-ray data, with the goal of strengthening  emergent synergies and taking advantage of the complementarity and know-how in different communities, from theoretical and   experimental astroparticle physics to  high-energy and nuclear physics. 
In this paper, we present the outcomes of the third edition of the workshop that took place in 2024. We present the current state of cosmic-ray experiments and their perspectives, and provide a detailed road map to close the most urgent gaps in cross-section data, in order to efficiently progress on many open physics cases, which are motivated in the paper.   
Finally, with the aim of being as exhaustive as possible, this report touches several other fields -- such as cosmogenic studies, space radiation protection and hadrontherapy -- where overlapping and specific new cross-section measurements, as well as nuclear code improvement and benchmarking efforts, are also needed. We also briefly highlight further synergies between astroparticle and high-energy physics on the question of cross-sections.
\end{abstract}
\maketitle
\tableofcontents

\section{Introduction} 

\begin{figure}
\centering
\includegraphics[width=0.65\linewidth]{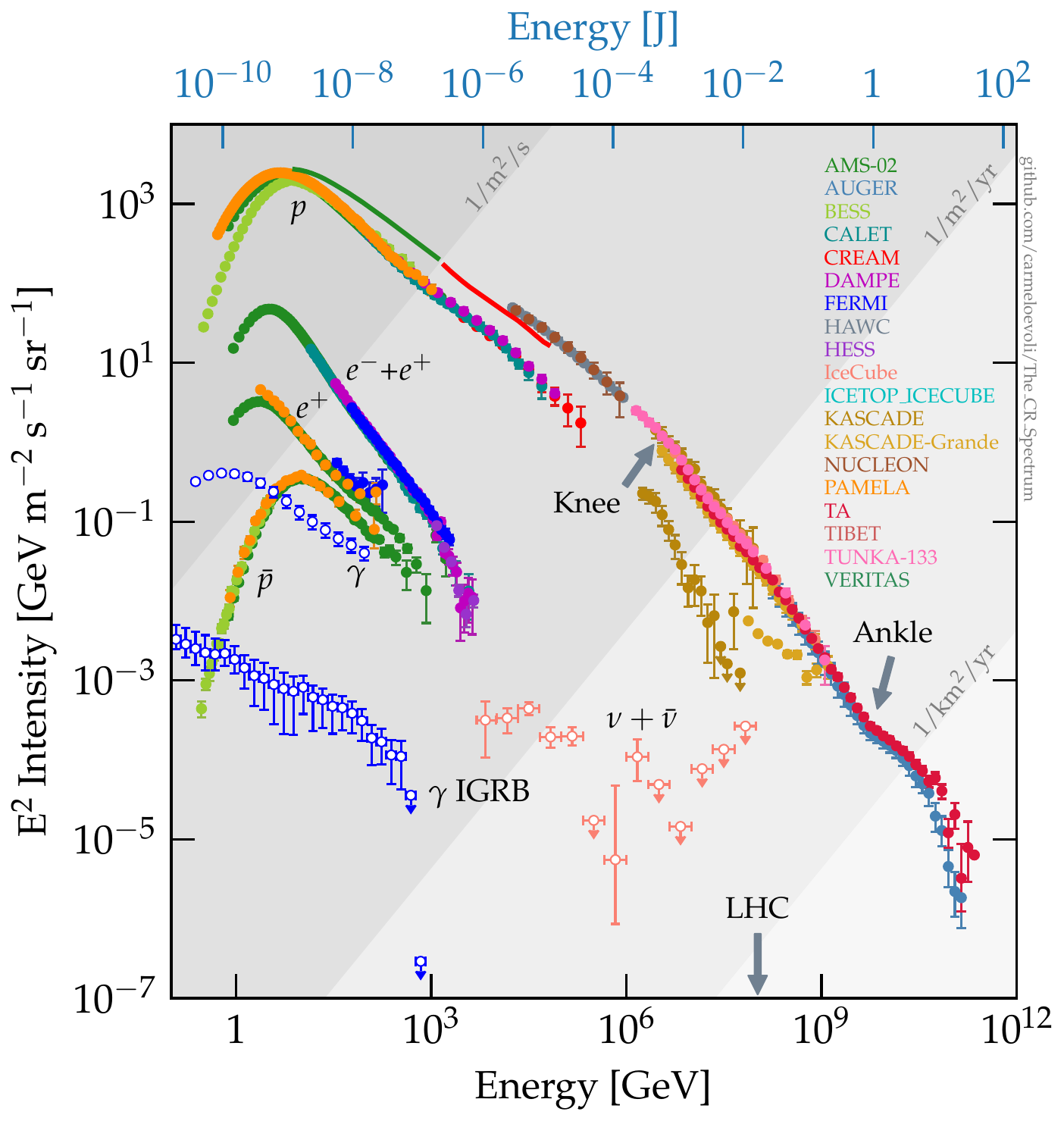}
\caption{Compilation of the \acrshort{cr} energy spectrum, scaled by $E^2$ to highlight spectral features (notably the \emph{Knee} at $\sim\!10^{6}$\,GeV and the \emph{Ankle} at $\sim\!10^{9}$\,GeV). Coloured markers and lines show measurements of the total (all-particle) spectrum and individual components (e.g., protons, electrons, positrons, antiprotons (=\antiproton)), while open symbols indicate neutral particles (diffuse $\gamma$ rays from the Galactic plane and from the isotropic $\gamma$-ray background, and diffuse neutrinos). Diagonal lines represent integral flux levels for reference. The charged-\acrshort{cr} data are taken from the \acrfull{crdb}~\cite{2014A&A...569A..32M, 2020Univ....6..102M, 2023EPJC...83..971M}, with additional $\gamma$-ray and $\nu$ data from Fermi-\acrshort{lat}~(\acrlong{lat})~\cite{Fermi-LAT:2012edv, Fermi-LAT:2015otn} and IceCube~\cite{IceCube:2013low}, respectively. 
The energy reached at the \acrshort{lhc} (\acrlong{lhc}) at \acrshort{cern} is also indicated.}
\label{fig:CRspectrum}
\end{figure}

Charged particles arriving at the Earth from space with energies\footnote{To avoid any ambiguity in the notations and units, throughout this review, the total and kinetic energy are denoted $E$ and $E_{\rm k}$, respectively, and are expressed in GeV (or other multiple of eV). The other energy variables and units employed are the momentum  $p$ in GeV/c, the rigidity $R=pc/Ze$ in GV and the kinetic energy per nucleon $E_{\rm k/n}$ in GeV/n. Different communities use different units in the literature for the latter, and the meaning of GeV/n is the same as, for instance, A\,GeV or GeV/u.} above 100\,MeV -- the so-called \acrfullpl{cr} -- are dominated by protons and helium nuclei (He in the following, mostly $^4$He), with some traces of heavier nuclei, electrons, and antiparticles.
Up to energies of about 100\,TeV, it is now feasible to detect these particles \emph{directly} using balloon-borne and space-based instruments, which can precisely determine their charge. At higher energies, the flux becomes exceedingly low, and \acrshort{cr}s can only be detected \emph{indirectly} through the extensive air showers they produce upon interacting with the atmosphere.
A striking and well-known characteristic of the \acrshort{cr} spectrum is its nearly featureless, power-law behaviour, spanning roughly 12~orders of magnitude in energy and 33~orders of magnitude in flux, as illustrated in Fig.~\ref{fig:CRspectrum}. Beyond the overall energy range and the typical power-law index $\gamma \simeq 2.5\text{--}3$, considerable attention has been devoted to explaining specific spectral features. These include the \emph{Knee} at a few PeV, where the spectral slope steepens by $\Delta\gamma \sim\,0.5$, the \emph{Ankle} at several EeV, where the spectrum flattens again, and a flux suppression at $E\sim10^{20}$\,eV. The \emph{Knee} and \emph{Ankle} are commonly interpreted as signatures of a transition from Galactic to extragalactic \acrshort{cr} components. Meanwhile, the flux suppression at ultra-high energies may indicate either a maximum acceleration limit in extragalactic sources and/or reflect significant energy losses of \acrshort{cr}s travelling through the cosmic background radiation field.

\acrshort{cr}s in the GeV--PeV range are typically referred to as Galactic Cosmic Rays (\acrshort{gcr}s), whereas those exceeding $10^{18}$\,eV, dubbed \acrfullpl{uhecr}, are generally considered to be of extragalactic origin. Within our Galaxy, $\mu$G-scale turbulent magnetic fields cause the trajectories of charged \acrshort{gcr}s to lose any directional memory, yielding a highly isotropic flux at the level of $10^{-4}\text{--}10^{-3}$.
Above $3 \times 10^{18}$\,eV, however,  the moderate strength and direction of the detected  anisotropy (pointing towards the Galactic anti-centre) strengthens the case for an extragalactic nature of \acrshort{uhecr}s, even further so if combined with the chemical composition indicators ~\cite{PierreAuger:2017pzq}. The Sun, for its part, can accelerate particles up to a few GeV during solar flares, and the heliosphere modulates \acrshort{gcr} fluxes up to several tens of GeV. This solar modulation is reflected in an increased suppression of the observed \acrshort{gcr} intensity with decreasing energy (at GeV energy, this suppression is $\gtrsim 10$).

The physics of \acrshort{gcr}s, which is the main focus of this paper, has entered a precision era. Thanks to satellite-borne and space-based experiments, data on many \acrshort{cr} species  reach a precision below 10\% in the GeV--TeV energy range (e.g.,~\cite{PAMELA:Galper2017, AMS:PhysRep2021}).
A major motivation to investigate messengers such as \acrshort{cr}s, as well as understanding their sources and our Galactic environment, is the indirect search for particle \acrfull{dm} signatures.
These relics from the early universe could  annihilate or decay in the halo of our Galaxy, leaving a feeble footprint in charged \acrshort{gcr}s and/or in photons. The disentanglement of a possible hint from these particles in the observed \acrshort{cr} fluxes must rely on a robust theoretical
framework for \acrshort{gcr}s as a whole.
However, the poor knowledge of nuclear reactions, in particular for the production of antimatter and secondary nuclei, limits the information that can be extracted from the \acrshort{cr} data about the source properties, the transport in the Galaxy, and the eventual presence of a signal from \acrshort{dm} annihilation or decay.

The huge array of cross-sections needed for a precise prediction of \acrshort{cr} data can actually be determined only by experiments at accelerators. While some of these data has already been collected in ongoing \acrshort{cern} experiments, significantly larger and more precise datasets are still required.
The issue of determining the cross-sections necessary for a precise modelling of \acrshort{gcr}s has been the topic of a series of workshops held at \acrshort{cern}~\cite{XSCRC2017, XSCRC2019, XSCRC2024}. This \emph{review} is one of the outcomes of the last edition held in October 2024, \acrshort{xscrc}\,2024 (\acrlong{xscrc})~\cite{XSCRC2024}. It aims at providing a road map to the most urgent cross-sections to (re-)evaluate, in order to progress on many physics cases, mostly related to \acrshort{gcr}s, but not only.

The paper is organised as follows. In Sec.~\ref{sec:physics_cases}, we review the most important physics cases related to \acrshort{gcr}s, highlighting specific situations where current cross-sections uncertainties prevent further progress. Improved cross-section measurements are also mandatory in several transverse topics, including cosmogenic studies, space exploration protection and hadron therapy. With the goal of building synergies between our communities (owing to overlapping or complementary needs), we also highlight their science cases.
In Sec.~\ref{sec:CRdata}, we review the current precision and energy reach of charged \acrshort{cr} data, their most striking features, and some recent advances they brought to the field. These data are the primary drive for improving cross-sections, so we also present the ongoing and future \acrshort{cr} experiments. The latter illustrate the existing long term and long-lasting programmes to get even more precise \acrshort{cr} data, which will further drive the need for even more precise nuclear data (than the ones already needed now).
In Sec.~\ref{sec:XSneeds}, we discuss the status and limitations of nuclear data, in the context of \acrshort{gcr} data interpretation. We provide an actionable list of reactions and energies to measure, and their required precision, sorting the highest priority ones. We cover nuclear fragmentation but also anti-particle and $\gamma$-ray production, illustrating how these new nuclear data would be a game changer for \acrshort{gcr}s. To be complete, we also discuss the status and impact of other cross-sections involved in the modelling of \acrshort{gcr}s (inelastic, annihilation, etc.).
In Sec.~\ref{sec:facilities}, we move to the facilities and experiments where some of these nuclear measurements are taking or could take place. For the highest energies, we detail how strong synergies between the \acrshort{gcr} and \acrfull{hep} communities have been recently built, highlighting recent successful results and ongoing measurements at \acrshort{cern} experiments. In the multi-GeV regime, we present ongoing experiments and forthcoming facilities (at the 2026--2027 horizon) where strong synergies with the nuclear physics community could be built. We also show that \acrshort{cr} experiments are themselves excellent apparatus for measuring important cross-sections, in an energy range not accessible elsewhere.
In Sec.~\ref{sec:Transverse}, we come back to the cross-sections needs from transverse communities. We present their specific requirements, in terms of reactions, energies and precision, along with their priority list. We also briefly touch some other topics, where promising \acrshort{hep} and astroparticle physics synergies take place.
Throughout this review, we provide summary tables and figures, to give an as clear and straightforward view as possible of: (i) the \acrshort{cr} experiment panorama and projects; (ii) the nuclear data status and needs; and (iii) the beam capabilities and opportunities at various facilities. Although we do not have a dedicated section on nuclear and \acrfull{mc} codes in this review, we highlight and discuss at many points the importance, uses, and associated benefits and difficulties of these tools.

\section{Physics cases in a high-precision era}
\label{sec:physics_cases}

This first section presents the physics cases and key questions in several topics where limitations appear because of cross-sections uncertainties. 
The main focus is on \acrshort{gcr} physics questions and related cross-sections (Sec.~\ref{sec:physics_case_GCR}), but we also present transverse fields where similar cross-sections are involved (Sec.~\ref{sec:physics_case_transverse}), and other astroparticle physics cases where different cross-sections, but similar needs for better measurements, arise (Sec.~\ref{sec:physics_case_other}).

\subsection{Astrophysics and beyond \texorpdfstring{\acrshort{sm}}{sm} physics cases for \texorpdfstring{\acrshort{gcr}s}{gcr}}
\label{sec:physics_case_GCR}

\begin{figure}
\centering
\includegraphics[width=0.5\linewidth]{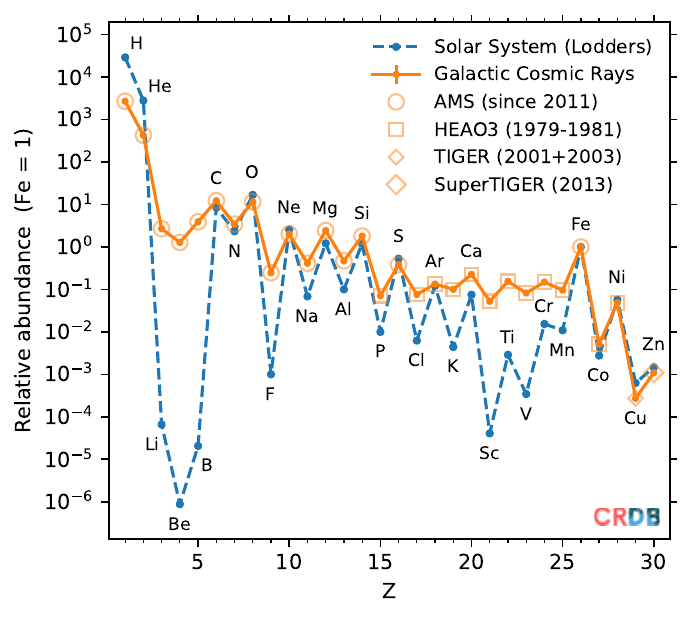}
\includegraphics[width=\linewidth]{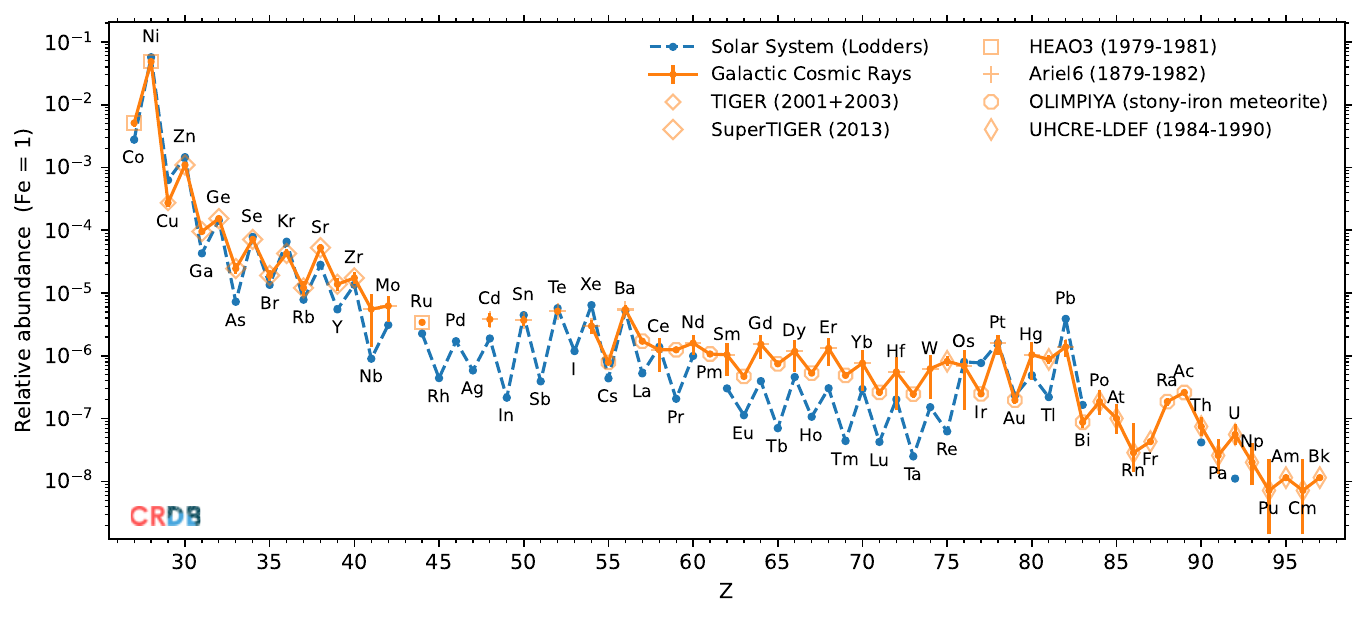}
\caption{Relative abundance of elements in the \acrshort{ss} (dashed blue line) \cite{2003ApJ...591.1220L} and in \acrshort{gcr}s (solid orange line), arbitrarily normalised to Fe$\;\equiv1$. The even-odd pattern in the abundances is related to the even-odd nuclear stability effect. In the top panel, beside H and He, elements with near-matching abundances (names above the orange line) are mostly of primary origin, while \acrshort{ss}/\acrshort{gcr}~$\ll 1$ (names below the blue line) are of secondary origin.   In the lower panel, the pattern of \acrshort{gcr}/\acrshort{ss} abundances is less clear, owing to more uncertain data. \acrshort{gcr} data were extracted and selected from
\acrshort{crdb}~\cite{2014A&A...569A..32M, 2020Univ....6..102M, 2023EPJC...83..971M}, at $R=5$\,GV for $Z\leq28$ and mostly at $E_{\rm k/n}\sim 1.5$\,GeV/n ($\sim\!3.82$\,GV) for the rest, rescaled to \acrshort{ams} flux ratio Fe(3.82\,GV)/Fe(5\,GV). Also, above $Z=30$, only ratio $x/y$ are measured with, for instance, $y$ the flux of Fe, of all elements $Z\geq55$, $Z\geq70$, or some other charge ranges. These ratios were multiplied by Fe (\acrshort{ams}) and combinations of (Yb+W+Pt+Pb)$/y$ flux data ratios to recover $x$. Data are \acrshort{ams} \cite{AMS:PhysRep2021, AMS:Fe-PRL2021, AMS:S-PRL2023}, \acrshort{heao} \cite{1990A&A...233...96E}, \acrshort{tiger} \cite{2009ApJ...697.2083R}, Super\acrshort{tiger}, \cite{2016ApJ...831..148M}, Ariel6 \cite{1987ApJ...314..739F}, \acrshort{olimpiya}~\cite{2022AdSpR..70.2674A} and \acrshort{uhcreldef} \cite{2012ApJ...747...40D}.}
\label{fig:CRabundances}
\end{figure}

In Fig.~\ref{fig:CRabundances}, the  composition in the \acrshort{gcr} flux is compared against the inferred one from  photospheric measurements and chondrites in the \acrfull{ss}~\cite{2003ApJ...591.1220L}, which is representative of the environment around a typical (population I) Galactic star. The overall similarity comforts the idea  that \acrshort{gcr}s are accelerated from an environment resembling the \acrfull{ism}. Yet, intriguing differences stand out: elements like lithium, beryllium, and boron are only one order of magnitude less abundant than carbon or oxygen in the \acrshort{gcr} flux, while merely present in traces in the \acrshort{ism}. Similar trends appear elsewhere, notably in fluorine and elements slightly lighter than iron (Sub-iron elements: Sc, Ti and V). These “over-represented” species are  interpreted as formed (almost) exclusively during propagation (hence they are dubbed secondary) from the fragmentation of heavier \acrshort{gcr} nuclei into lighter ones during interactions with the \acrshort{ism} targets. This secondary component provides relevant insights into the travel history of parent nuclei as they propagate through the \acrshort{ism} before reaching the Earth. For instance, they are one of the most solid evidences for some kind of diffusive propagation, since the amount of material crossed compatible with the measurements is orders of magnitude larger than what expected from a ballistic propagation.

\subsubsection{Can we reveal \texorpdfstring{\acrshort{dm}}{DM} with \texorpdfstring{\acrshort{cr}s}{CR}?}
Anomalous energy spectra, anisotropy, and composition patterns of the cosmic particle fluxes provide one of the main strategies (the so-called {\it indirect} one) in the searches for signals beyond the \acrfull{sm} of particle physics (see, e.g., \cite{Gaskins:2016cha, Cirelli:2024ssz} for reviews).
In particular, the rarest cosmic particles, such as antimatter (positrons, \antiproton and antinuclei) and stable neutral messengers, are privileged channels for indirect searches of byproducts of \acrshort{dm} particle decays or annihilations~\cite{Salati:2010rc}.
The current consensus is that the fluxes of all firmly detected energetic cosmic particles are dominated by established astrophysical processes, not all of them however precisely known. In particular, the bulk of \acrshort{cr} antimatter and diffuse $\gamma$ photons are related to interactions of \acrshort{gcr}s with the atoms of the \acrshort{ism} in the Milky Way.
This implies that any exotic contribution coming for instance from \acrshort{dm}  represents a subdominant contribution, with the possible exception of poorly probed energy ranges. To fully exploit the precision of current data, one should make sure that theoretical uncertainties are under control at least at a level comparable to the observational one, which is hardly the case today.
Which directions appear as the most promising ones to tackle this challenge?

The Galactic environment for \acrshort{gcr} propagation is typically modelled as a cylinder with a radius of about 20~kpc and a vertical size or thickness, labelled Galactic halo (half-)height $L$, whose value is only poorly constrained between 2 and 10~kpc~\cite{Evoli:2019iih, Weinrich:2020ftb, DiMauro:2023oqx}. The value of $L$ is especially relevant for the prediction of the flux of cosmic particles from \acrshort{dm} annihilation, since it controls how much of the injected flux is retained in the diffusive environment~\cite{Donato:2003xg,Evoli:2011id,Genolini:2021doh}. In fact, the flux of positrons and \antiproton from \acrshort{dm} annihilation is directly related to the uncertainty on $L$ (e.g., \cite{Genolini:2021doh}). This halo size is determined from ratios of radioactive to stable secondary \acrshort{gcr} species (e.g., \cite{2002A&A...381..539D}). The $\beta$-unstable species $^{10}$Be, $^{26}$Al, $^{36}$Cl and $^{54}$Mn, with half-live in the Myr range, of the order of the residence time in the Galaxy, are the considered species to constrain $L$~\cite{2002A&A...381..539D, 2003ApJ...586.1050M, 2010A&A...516A..66P}. Recent studies have focused on the $^{10}$Be/$^9$Be ratio~\cite{Evoli:2019iih, Weinrich:2020ftb, 2024PhRvD.109h3036Z}, and it was shown that nuclear cross-sections uncertainties on the production of Be isotopes were already limiting the interpretation of current data, and would similarly plague forthcoming \acrshort{cr} measurements.

\acrshort{gcr} \antiproton are expected to be produced mainly as secondaries. The \acrshort{ams} experiment has measured the flux of \antiproton between 0.5\,GeV and 500\,GeV, which is globally consistent with the expectations for a secondary origin~\cite{Boudaud:2019efq, DiMauro:2023jgg}. However, different groups have found a mild excess at around 10\,GeV that could be explained by \acrshort{dm} particles annihilating with a thermal cross-section into quarks~\cite{Cuoco:2016eej, Cui:2016ppb}. Nonetheless, the current uncertainties on the \antiproton production cross-sections, as well as propagation and the correlations in the experimental data~\cite{Cuoco_2019, Boudaud:2019efq}, make these claims shaky and dependent on the data analysis approximations adopted~\cite{Cuoco_2019, Calore:2022stf}. The NA49 experiment has released a preliminary measurement of antineutron (\antineutron) production from proton--proton (\pp) collisions that is about $20\%$ larger than the cross-section for \antiproton production, corroborating earlier indications~\cite{Fischer:2003xh}. This data suggests an isospin asymmetry between the \antiproton and \antineutron production that is indeed relevant for the interpretation of the \acrshort{ams} \antiproton flux, see e.g.,~\cite{Winkler:2017xor}.
The presence of an isospin asymmetry is expected considering the different production yields between the positive, negative, and neutral charged pions at low energies. However, regarding the \antineutron/\antiproton asymmetry, no published and precise measurements have been performed so far, and no theoretical predictions have been calculated from fundamental assumptions. Currently, the magnitude of the isospin effect is the main theoretical uncertainty for the \acrshort{ams} data interpretation.

The search for antideuteron (\antideuteron) and antihelium (\antihelium) nuclei in \acrshort{gcr}s is one of the most promising indirect search strategies for \acrshort{dm}~\cite{vonDoetinchem:2020vbj}. The reason is that at kinetic energies below a few GeV/n, the secondary production is expected to be at least one order of magnitude smaller than the most optimistic yields associated to viable thermal relic \acrshort{dm} models. However, theoretical predictions are not fully under control. Until recent years, \acrshort{cr}
antinuclei have been evaluated by considering simple coalescence models of antinucleons, whose free parameters are fitted to available antinuclei data from accelerator experiments. Currently, two data points from \acrshort{argus} (\acrlong{argus}) at the Upsilon mass resonances~\cite{ARGUS:1985cfz}, one data point from \acrshort{aleph}~(\acrlong{aleph})~\cite{ALEPH:2006qoi} at the $Z$ resonance, and data from \pp collisions with centre-of-mass energy $\sqrt{s}$ from 900\,GeV to 13\,TeV from \acrshort{alice} (\acrlong{alice})~\cite{ALICE:2015wav, ALICE:2017xrp, ALICE:2019dgz, ALICE:2020foi, ALICE:2021mfm, ALICE:2021ovi} are available.
In a recent analysis, Ref.~\cite{DiMauro:2024kml} reported that the main current theoretical uncertainties of the coalescence model resides in the \antideuteron production data from accelerator experiments. In fact, once different coalescence models are tuned using the same data point, e.g.,~the one from \acrshort{aleph}, the theoretical predictions vary at most by $10\text{--}15\%$ in the relevant energy range.
Tuning the coalescence parameters at the energies relevant for astroparticle physics requires very precise measurements of the antinuclei production, at centre-of-mass energies between 10 and 100\,GeV.
Alternatively, significant efforts have been made to develop a parameter-independent coalescence model, based on the Wigner function formalism. However, the nucleon yields and the nucleon emission source describing the relative distances at which nucleons are produced must be constrained by the data. This model has been applied successfully to predict \antideuteron spectra measured by \acrshort{alice} in \pp collisions at 13\,TeV~\cite{Mahlein2023} and other collision energies~\cite{Mahlein2024}.

On top of the rather uncertain production cross-sections of antinuclei, another uncertainty in their flux predictions is related to inelastic cross-sections, leading to antinuclei destruction during their propagation. Only recently a first measurement of such a quantity has been performed~\cite{ALICE2022II}, but further progress is required to validate currently used parametrisations \cite{DAngelo:2024vyn}.

Another indirect mean to probe particle \acrshort{dm}, albeit not strictly concerning charged \acrshort{gcr}s, is through $\gamma$ rays. The search for photons is not hampered by the diffusion on magnetic fields, and regions expected to exhibit high \acrshort{dm} densities can be targeted by telescopes. So far, the deepest and most sensitive searches have been conducted by Fermi-\acrshort{lat} \cite{Fermi-LAT:Atwood_2009}.
Strong bounds have been obtained by targeting a number of dwarf spheroidal galaxies, whose mass is largely  dominated by \acrshort{dm}~\cite{PhysRevLett.115.231301, McDaniel:2023bju}.
A \acrfull{gce} in Fermi-\acrshort{lat} data at GeV energies has been extensively investigated over the past two decades
\cite{Goodenough:2009gk, Ajello_2016}. Its angular and spectral features make it intriguingly similar to expectations from simple particle \acrshort{dm} models, although alternative explanations invoking an unresolved millisecond pulsar population in the bulge have gained momentum over the years \cite{bartels+16, Calore:2015bsx}. For a \acrshort{dm} origin of the \acrshort{gce}, a $\gamma$-ray signal is also expected from dwarf spheroidal galaxies: depending on the data analysis technique and the assumption on the \acrshort{dm} abundance in these systems, they put weak or strong constraints on the \acrshort{dm} hypothesis for the \acrshort{gce}, at a level comparable to \antiproton constraints (see, e.g., Fig.~3b in~\cite{Calore:2022stf}).
One major issue for \acrshort{dm} observations via $\gamma$ rays is the modelling of the diffuse Galactic emission, typically produced by \acrshort{gcr} nuclei scattering off the \acrshort{ism} and electrons interacting with the interstellar radiation fields. This emission, which unavoidably fills any line of sight, is dependent also on the nuclear production cross-sections.

\subsubsection{Where are \texorpdfstring{\acrshort{gcr}s}{GCR} synthesised, accelerated, and how?}

Both based on energetic considerations and multi-wavelength observations of their non-thermal spectra~\cite{BeckerTjus:2020xzg, Bykov:2018wrt}, the main sources of \acrshort{gcr}s have long been speculated to be supernova remnants (and, for leptons, also pulsar wind nebulae), with the role of star clusters, X-ray binaries or mergers of compact objects raised from time to time. The exact mechanisms that accelerate particles from these Galactic sources are not known, although variants of diffuse shock acceleration or magnetic reconnection are typically invoked (see, e.g., Ref.~\cite{Caprioli:2023orv, Amato:2024niq}). The observation of fine spectral features in the proton, electron, and photon spectra are required to identify signatures of subtle non-linear effects in the acceleration process, or the relative weight of hadronic and leptonic components in the accelerated yields~\cite{Winner:2020wxk, Corso:2023cmt}. The goal of identifying these signals is only meaningful, however, if uncertainties related to collisional physics, recently subject to re-evaluation~\cite{orusa2023new, Orusa_2022}, are well under control.

Isotopic anomalies, such as the measured  overabundance of the $^{22}$Ne isotope over $^{20}$Ne, may point to a prominent role of star clusters and superbubbles~\cite{Gabici:2023ihx}.  The detection of $^{60}$Fe in \acrshort{gcr}s, a short-lived radionuclide with a half-life of 2.6\,Myr synthesised in core-collapse supernovae, further constrains the history and source location of the \acrshort{gcr}s measured at Earth~\cite{Binns:2019mfq}.
These topics are entangled with the understanding of the site of the $r$-process for the nucleosynthesis of $^{60}$Fe, a transverse argument that links together the searches of ultra-heavy nuclei in \acrshort{gcr}s with geophysical methods, our theoretical and experimental understanding of cross-sections in nuclear astrophysics, and the astrophysics of binary neutron star mergers and core-collapse supernovae (as revealed by the electromagnetic counterpart \cite{2017ApJ...848L..19C, 2017ApJ...848L..18N} of the binary neutron star merger \acrshort{ligo}/Virgo GW170817 \cite{2017PhRvL.119p1101A}).

\subsubsection{\texorpdfstring{\acrshort{gcr}}{GCR} transport: refining the model or going beyond it?}

Galactic energetic particles propagate in the Milky Way environment in their journey from the sources to the Earth. As detailed in Sec.~\ref{sec:CRtransport}, these particles are affected by both non-collisional processes and energy loss mechanisms, the latter being particularly relevant for electrons and positrons~\cite{Evoli:2023kxd}. By a careful study of the energy spectra of these particles, notably the ratios of fluxes mostly produced by spallation in the \acrshort{ism} to fluxes mostly affected by source acceleration, one infers information on the rigidity and spatial dependence of the diffusion coefficient, which is ultimately informing on the physics of the Galactic magnetic turbulence~\cite{Amato:2017dbs, Evoli:2019wwu, DiMauro:2023jgg}.
The standard scenario assumes that \acrshort{gcr}s propagate onto externally assigned magnetic field turbulence, onto which they do not backreact.
Features of the \acrshort{cr} data uncovered in the past decade (see Ref.~\cite{Serpico:2015caa} for an early review), such as the different slope of proton and He \acrshort{gcr}s at energies below/above 100\,GeV, may hint at a change of regime~\cite{Genolini:2017dfb}, where lower-energy \acrshort{gcr}s would rather scatter onto the turbulence that they self-generate~\cite{Blasi:2012yr, Evoli:2018nmb}. However, non-factorisable spatial vs. power-spectrum properties of the turbulence may also accommodate the data~\cite{Tomassetti:2012ga}.
It is not yet clear if the transport is close to quasi-homogeneous, rather than being localised in small volumes of the \acrshort{ism}~\cite{Lemoine:2023sxw, 2024MNRAS.528.4245B, Kempski:2024pqe}. Disentangling these possibilities is an extremely challenging effort, which may be completely hampered by cross-section uncertainties.

\subsubsection{Going beyond the standard paradigm for the sources?}

The source terms are typically factorised in a continuum function of $(t, {\bf x})$\footnote{Usually, stationary conditions are considered and no time-dependence is assumed.} in the Galaxy, times an often universal power-law energy spectrum. With respect to the chemical composition, normalised abundances tracing the \acrshort{ism} composition are often assumed, though the data seem to point at a preferential acceleration of refractory elements contained in interstellar dust~\cite{2009ApJ...697.2083R, 2021MNRAS.508.1321T}. These various approximations can be questioned.

The stochastic distribution of astrophysical sources in space and time leads to spectral deviations from the average, see, e.g.,~\cite{Genolini:2016hte, Mertsch:2018bqd, Evoli:2021ugn}. Uncovering these effects at a statistical level requires however to keep errors associated to collisional effects at a percent-level or lower. This is especially true above 1--10\,TeV, where specific spectral features may hint at a predominant contribution of one or few local sources.

Some isotopes and elements, such as deuteron~\cite{AMS:Deuteron-PRL2024} and lithium~\cite{2020ApJ...889..167B}, seem to indicate some departure from the standard secondary origin~\cite{2024PhRvD.110l3030L, Dimiccoli:2025hrd}. The robustness of these hints is however plagued by the significant uncertainties affecting their predicted yields from spallation cross-sections, so that the predictions are currently still consistent with the expectations thanks to the large uncertainties~\cite{2024ApJ...974L..14Y, 2022A&A...668A...7M}.

One component that was supposed to be absent from primary sources until $\sim\!15$ years ago, but whose existence is now robustly accepted, is the positron flux~\cite{PAMELA:2008gwm, PAMELA:2013vxg, PhysRevLett.113.121101, AMS:positrons2019, 2012PhRvL.108a1103A}. Yet, its interpretation is still unclear.
While \acrshort{dm} could in principle contribute to the observed flux above 10\,GeV, which provoked a significant early excitement for this measurement, the needed annihilation intensity is nowadays in conflict with other bounds. The two most physically motivated interpretations are related to the acceleration of electron and positron pairs from pulsar wind nebulae~\cite{Hooper:2008kg, Blasi:2010de, Orusa:2021tts, Orusa:2024ewq, Evoli:2020szd}, or positrons produced and accelerated in supernova shocks sweeping the circumstellar medium~\cite{Blasi:2009bd, Mertsch:2014poa}.
In order to infer as precisely as possible a primary component in the positron flux, a very precise estimate of the positron production cross-sections, which are used for calculating the secondary production, is mandatory~\cite{DiMauro:2023oqx}.

To check the extent to which the \acrshort{gcr} injection spectrum and the propagation properties are homogeneous over the Galaxy, one may want to probe \acrshort{gcr} properties away from the location of Earth. This goal can be indirectly attained by studying the angular and spectral properties of the interstellar $\gamma$-ray emission (e.g., Ref.~\cite{2011ApJ...726...81A, Evoli:2012ha}), probed by Fermi-\acrshort{lat} and, to lesser extent, by imaging Cherenkov telescopes. Its main contribution is due to the fragmentation of \acrshort{gcr}s interacting with \acrshort{ism} atoms and their follow-up production of $\pi^0$, which subsequently decay into two photons ($\gamma$ rays).
Currently, data for the Lorentz invariant cross-section for the production of neutral pions are not sufficient to obtain a precise estimate of $\gamma$ ray spectrum originating from $\pi^0$ production~\cite{orusa2023new}.

\subsection{Transverse physics cases with overlapping nuclear cross-sections needs}
\label{sec:physics_case_transverse}
Beyond the purely astrophysical and cosmological topics of interest summarised above, nuclear cross-sections also play a key role in other topics related to \acrshort{cr}s. The key questions related to the time variation of \acrshort{gcr}s are introduced below, and then some links to societal aspects, namely space exploration and medicine, are given.

\subsubsection{Cosmogenic studies, impact on climate and life on Earth}
\label{sec:motiv_cosmogenic}
For most cosmogenic nuclide studies, constant \acrshort{gcr} fluences are assumed. There are, however, reasons why \acrshort{gcr} fluences in the \acrshort{ss} should vary over timescales of millions of years. \textit{First}, the periodic passage of the \acrshort{ss} through Galactic spiral arms might cause periodic \acrshort{gcr} variations. In spiral arms, star formation and supernova rates are higher, leading to an increased \acrshort{gcr} flux. \textit{Second}, the \acrshort{ss} periodically moves up and down the Galactic plane, which can affect the \acrshort{gcr} flux. There are some arguments that such periodic \acrshort{gcr} fluency variations can affect Earth's climate. At times of higher \acrshort{gcr} fluences, there are higher ionisation rates in the upper atmosphere, which can produce higher cloud coverage. This finally could produce a cooling effect and even start an ice age. This would most likely affect the origin and evolution of life. \textit{Third}, as the \acrshort{ss} occasionally moves through dense molecular clouds, the heliosphere can shrink and, as a consequence, \acrshort{gcr} modulation is diminished or disappears completely. Such an episode, in which the Earth might have been even outside the heliosphere and directly exposed to the \acrshort{ism} of the dense molecular cloud, would significantly affect the evolution of life on our planet. It has been proposed that the \acrshort{ss} might have passed such a dense molecular cloud $\sim\:$2\,Myr ago (e.g., \cite{Opheretal2024}). There are some arguments that cosmogenic nuclide studies in iron meteorites provide evidence for periodic \acrshort{gcr} fluency variations (e.g., \cite{Shaviv2002, Shaviv2003, Schereretal2006}). While some studies were supportive, many subsequent investigations have questioned the original hypothesis \cite{Wallmann2004, SloanWolfendale2013} and/or the interpretation of the database~\cite{Wieleretal2013}. In addition to periodic \acrshort{gcr} intensity variations, there are also indications for a onetime and sudden increase in the \acrshort{gcr} intensity with a higher flux in the past several million years, relative to the long-term average over the past
500--1000\,Myr~\cite{HampelSchaeffer1979, Aylmeretal1988, Lavielleetal1999, Wieleretal2013}.

To prove or reject \acrshort{gcr} fluency variations, cosmogenic nuclides stored in terrestrial archives and produced in meteorites or planetary surfaces provide a powerful tool or even can be considered as the only diagnostics. The half-lives of the different radionuclides then correspond to the different time intervals the dating system is sensitive to. For example, the $^{14}$C activity concentration is sensitive to the last $\sim\!20$\,kyr, $^{53}$Mn to the last $\sim\!15$\,Myr, and $^{40}$K to the entire age of the \acrshort{ss}. For all these studies, a precise knowledge of the cosmogenic production rates is mandatory, and such production rates can only be determined based on an accurate, reliable, and consistent cross-section database for the relevant nuclear reactions~\cite{Smithetal2019}.

\subsubsection{Space exploration}
\label{sec:motiv_spaceexploration}
Uncertainties in predicting radiation-related health effects due to exposure to the space radiation environment are one of the major challenges for human spaceflight beyond Earth orbit \cite{Schwadron2014, Rizzo2023, Sihver2021}. In shielded environments, light ions (i.e., isotopes of H and He) and neutrons make the largest contributions to the dose equivalent received by astronauts. The largest uncertainties in predicting radiation doses and the associated health risks stem from a limited understanding of radiation biology and from disagreements in transport codes \cite{Fogtman2023, Chancellor2018, Durante2011}. The latter is primarily due to inadequate knowledge about the light-ion production cross-sections \cite{Heinbockel2006, Slaba2017}, which also constitute the largest gap in currently available nuclear data \cite{Norbury2012}. It is therefore imperative that these cross-sections be measured to place space radiation protection on a solid foundation.

\subsubsection{Hadrontherapy}
\label{sec:motiv_hadrontherapy}
Hadrontherapy treats deep-seated tumours using charged particle beams, such as protons and $^{12}$C. Indeed, these particles exhibit a favourable depth-dose distribution in tissue, characterised by a peak in energy deposition (the Bragg peak) near their end range, which coincides with the tumour’s location. Additionally, C and O ions demonstrate enhanced biological effectiveness, making them suitable for treating radio-resistant tumours.
However, nuclear interactions between the ion beam and patient tissues can result in the fragmentation of projectiles and/or target nuclei. These interactions must be carefully considered when designing \acrfull{tps}.
Currently, there is a significant lack of experimental data on nuclear fragmentation involving light fragments ($Z<10$) within the energy range commonly used in hadrontherapy of 80\,MeV/n to 400\,MeV/n. Such data would be invaluable for further optimisation of \acrshort{tps} in hadron therapy~\cite{Battistoni2021}.

\subsection{Further astroparticle physics cases affected by cross-section uncertainties}
\label{sec:physics_case_other}

Two additional examples where cross-sections data bridge the astrophysics and high-energy physics communities are presented in this section.
Indeed, not only better cross-section data enhance our ability to interpret \acrshort{cr} data, but also to deepen our understanding of dense and compact astrophysical objects and environments.

\subsubsection{Origin of ultra-high energy cosmic-rays}
\label{sec:physicscase_UHECRs}

Unveiling the composition and spectrum of \acrshort{uhecr}s would reveal the most energetic accelerators in the Universe and the nucleosynthesis processes in the most hostile environments (e.g., the \acrfull{agn}, compact binary mergers, collapsars). Mostly nuclei up to $A=56$ are considered for the interpretation of current data, but heavier nuclei could play a role for the observed trend of the mass composition at the highest energies \cite{Zhang:2024sjp}. The results from the Pierre Auger Observatory and \acrshort{ta} (\acrlong{ta}) have stimulated several studies focusing on the nuleosynthesis \cite{Farrar:2024zsm}, acceleration and multi-messenger secondary emissions \cite{2020JCAP...04..045D, PierreAuger:2022atd}.
Recent multi-messenger observations \cite{2017ApJ...848L..12A, 2018Sci...361..147I} strongly indicate that the discovery of the sources is imminent, but the cross-sections are still one of the limiting factors \cite{2019JCAP...11..007M}.
\acrshort{uhecr}s generate extensive air showers, whose precise modelling hinges on accurate hadronic cross-sections, well beyond energies accessible at current accelerators. Uncertainties in these cross-sections propagate directly into the interpretation of shower development, energy spectra and composition. Improved data are therefore crucial to refining theoretical frameworks and reducing systematic uncertainties in \acrshort{uhecr} observations at facilities like the Pierre Auger Observatory and \acrshort{ta}.

\subsubsection{The equation-of-state of neutron stars and femtoscopy}
\label{sec:physicscase_femto}

Neutron stars are the most compact material objects
in the Universe, with extreme conditions of pressure and density, possibly undergoing phase transitions during their evolution. These extreme properties can be probed by astronomical observations \cite{Chatziioannou:2024tjq}. The outer core of neutron stars consists of matter composed by nucleons, electrons and muons, in a strongly interacting regime, where sophisticated models describing the correlations among two and three nucleons are necessary. The inner core is the least understood, and there could be charged mesons, such as pions or kaons in a Bose condensate, or other heavier baryons
with strangeness (e.g., hyperons). At the highest densities, quarks might be deconfined. 
The recent \acrshort{nicer} (\acrlong{nicer}) observations of pulsed X-ray emission from millisecond pulsars \cite{2024ApJ...974..295D, 2024ApJ...974..294S, 2024ApJ...971L..20C, 2024ApJ...961...62V} provide the first data to constrain the equation-of-state from the reconstructed radius--mass diagrams. While the exact matter content and interactions in the neutron star are not the only ingredients of the modelling, they are one of the limiting factors.
Femtoscopy probes hadron–hadron interactions, including multi-body forces, at distance scales unresolvable by direct scattering experiments. Precise cross-section measurements for particles containing strange quarks are particularly relevant for modelling the composition and equation-of-state of neutron stars, where strangeness and multi-body forces can drastically alter matter at high densities.

\section{Direct detection \texorpdfstring{\acrshort{cr}}{CR} experiments in a high-precision era: highlights and frontiers}
\label{sec:CRdata}

Direct detection \acrshort{cr} experiments -- as opposed to indirect detection via air-showers arising from interactions of \acrshort{cr}s in the Earth atmosphere, or other techniques (not covered here) -- have probed the tens of MeV to hundreds of TeV energy range.
Almost all \acrshort{cr} data collected so far are from the Earth neighbourhood and are thus affected by the solar modulation cycle~\cite{2013LRSP...10....3P, 2015LRSP...12....4H, 2020LRSP...17....2P, 2023LRSP...20....2U}. The modulation of fluxes is a few percents above hundreds of GeV/n, but critical at low energy, where the tens of MeV/n \acrfull{toa} energies measured by balloons or satellites correspond to $\lesssim$\,GeV/n \acrfull{is} energies. The only \acrshort{is} data, and very significant ones, are those of the Voyager~1 and 2 detectors, launched in 1977, which crossed the heliopause in August~2012 and November~2018 respectively, providing for the first time (and probably for a long time) the only direct measurements outside the solar cavity. These tiny detectors (hundreds of cm$^3$) provided fluxes of elements in the range of a few tens to a few hundreds of MeV/n (\acrshort{is} energies) with a precision ranging from 5\% for H up to 10\% for Fe. The data for $e^-+e^+$ are at even lower energies and with a precision of $10\%$ \cite{2013Sci...341..150S, 2016ApJ...831...18C, 2019NatAs...3.1013S}.

In the last two decades, large-acceptance sophisticated particle detectors have been launched onboard satellites and space stations for long-duration missions, allowing precision measurements of species-resolved \acrshort{cr} spectra in the energy range from GeV to hundreds of TeV (\acrshort{toa} energies). Some of these detectors have run over 10~years or more -- e.g., \acrshort{pamela} (\acrlong{pamela}) and \acrshort{ams} (\acrlong{ams}) -- and provided measurements of specie-resolved monthly and daily fluxes of particles and nuclei~\cite{PAMELA:time-ele-pos-PhysRevLett.116.241105, PAMELA:monthly-protons-Martucci_2018, AMS:daily-p-PhysRevLett.127.271102, AMS:daily-He-PhysRevLett.128.231102, AMS:daily-ele-PhysRevLett.130.161001, AMS:daily-positron-PhysRevLett.131.151002, AMS:monthly-antiproton-PhysRevLett.accepted.Dec.2024, AMS:monthly-nuclei-PhysRevLett.accepted.Dec.2024} crucial for space weather, space radiation and solar modulation studies.

To measure \acrshort{cr}s above GeV energies, two experimental approaches can be distinguished: magnetic spectrometers, as \acrshort{bess} (\acrlong{bess}) on 9 balloon flights from 1993 to 2002~\cite{BESS:MITCHELL200431}, \acrshort{bess}-Polar-I (flown on a balloon in 2004) and \acrshort{bess}-Polar-II (flown on a balloon in 2007)~\cite{BESS-Polar:ABE2017806}, \acrshort{pamela} (satellite mission 2006--2016)~\cite{2017NCimR..40..473P, 2017PPN....48..710G}, \acrshort{ams} (operating on the \acrfull{iss} since 2011)~\cite{AMS:PhysRep2021}, \acrshort{helix}~\cite{HELIX:Coutu_2024} (\acrlong{helix}, first balloon flight in May 2024 for $\sim\!6$\,days);
and calorimeters, as \emph{Fermi}-\acrshort{lat} (\acrlong{lat}, satellite operating since 2008)~\cite{Fermi-LAT:Atwood_2009}, \acrshort{dampe} (\acrlong{dampe}, satellite operating since 2015)~\cite{DampeMission}, \acrshort{calet} (\acrlong{calet}, operating on the \acrshort{iss} since 2015)~\cite{CALET}, NUCLEON (satellite operated between 2014 and 2017)~\cite{2021BRASP..85..353T}
and \acrshort{iss}-\acrshort{cream} (\acrlong{cream}, operated on the \acrshort{iss} between 2017 and 2019)~\cite{ISS-CREAM:Choi2022}. Magnetic spectrometers can distinguish particles from antiparticles, and hence they can measure the spectra of positrons, \antiproton, and search for antinuclei in \acrshort{cr}s. They measure the rigidity, i.e., the ratio between momentum and charge, of the incoming \acrshort{cr} particle. Their \acrfull{mdr}, defined as the rigidity at which the relative rigidity resolution is equal to 1, is set by the intensity of the magnetic field provided by the magnet and the lever arm of the instrument. The largest magnetic spectrometer ever deployed, \acrshort{ams}, reaches an \acrshort{mdr} of 2\,TV for protons, 3.2\,TV for He nuclei, and 3.5\,TV for heavier nuclei~\cite{AMS:PhysRep2021}.

Magnetic spectrometers combined with detectors able to measure the velocity, such as \acrfull{tof} systems or \acrfull{rich} counters, have the ability to measure isotopic compositions of nuclei by reconstructing the mass from the rigidity and the velocity measurements. Current experiments employing this technique, \acrshort{ams} and \acrshort{helix}, can measure isotopic fluxes with accuracies of $\sim\!10\%$, up to a kinetic energy per nucleon of about 10\,GeV/n~\cite{AMS:RICH-GIOVACCHINI2023168434, HELIX:2024icrc.confE.121J, HELIX:Coutu_2024}.
Both \acrshort{pamela} and \acrshort{ams} have been equipped with an electromagnetic calorimeter to accurately determine the energy of electrons and positrons, with a resolution of a few percent over the entire energy range of their measurements. In the \acrshort{ams} experiment, the energy of electrons and positrons is determined with accuracies $\lesssim 2$\% from 30\,GeV to 500\,GeV and $<$4\% above, up to 3\,TeV~\cite{KOUNINE2017110}. 
The maximum energy that can be measured by electromagnetic calorimeters is mainly determined by their depth. The depth of the \acrshort{ams} calorimeter is 17~radiation lengths, allowing measurements of energies of electrons and positrons up to 3\,TeV~\cite{KOUNINE2017110}.
Because of the rapid decrease with increasing energy of \acrshort{cr} fluxes (see Fig.~\ref{fig:CRspectrum}), flux measurements at the highest energies require, in addition to an extended upper energy, larger acceptance detectors operated over long-duration missions.

The past decade marked the era of active deployment of large-area calorimetric \acrshort{cr} experiments in space. In~2015, the \acrshort{dampe}~\cite{DampeMission} satellite was launched into orbit and the \acrshort{calet}~\cite{Torii:2011zza} detector was delivered to the \acrshort{iss}. Both instruments feature deep total-absorption calorimeters, with an integrated detector thickness of $\gtrsim30$ radiation lengths. The acceptance for electron detection with \acrshort{dampe} and \acrshort{calet} is $\sim\!0.3\,\mathrm{m^2\cdot sr}$ and $\sim\!0.12\,\mathrm{m^2\cdot sr}$, respectively. Due to their thick fine-segmented calorimeters, the two detectors have excellent energy resolutions of about 1.2\% and 2\%, respectively, at $>\!100$\,GeV energies.
With their relatively large acceptance, this allows the combined electron and positron spectrum to be probed up to $\sim\!10$\,TeV energies. Next, the \acrshort{iss}-\acrshort{cream}~\cite{ISS-CREAM:Choi2022} detector -- legacy of the balloon-flight \acrshort{cream}~\cite{2011ApJ...728..122Y} -- is another calorimetric experiment that was deployed on the \acrshort{iss} in 2017. Unlike \acrshort{dampe} and \acrshort{calet}, it utilises a sampling calorimeter alike the one of \acrshort{ams}, with scintillating fibres interleaved with passive tungsten layers, having a total thickness of 21 radiation lengths, and a geometric factor of $\sim\!0.27\,\mathrm{m^2\cdot sr}$, close to the one of \acrshort{dampe}.
Thanks to their relatively large acceptance, calorimetric experiments like \acrshort{dampe}, \acrshort{calet} and \acrshort{iss}-\acrshort{cream} are capable of measuring individual \acrshort{cr} nuclei spectra up to hundreds of TeV.

\begin{figure}[t]
    \centering
\includegraphics[width=\linewidth]{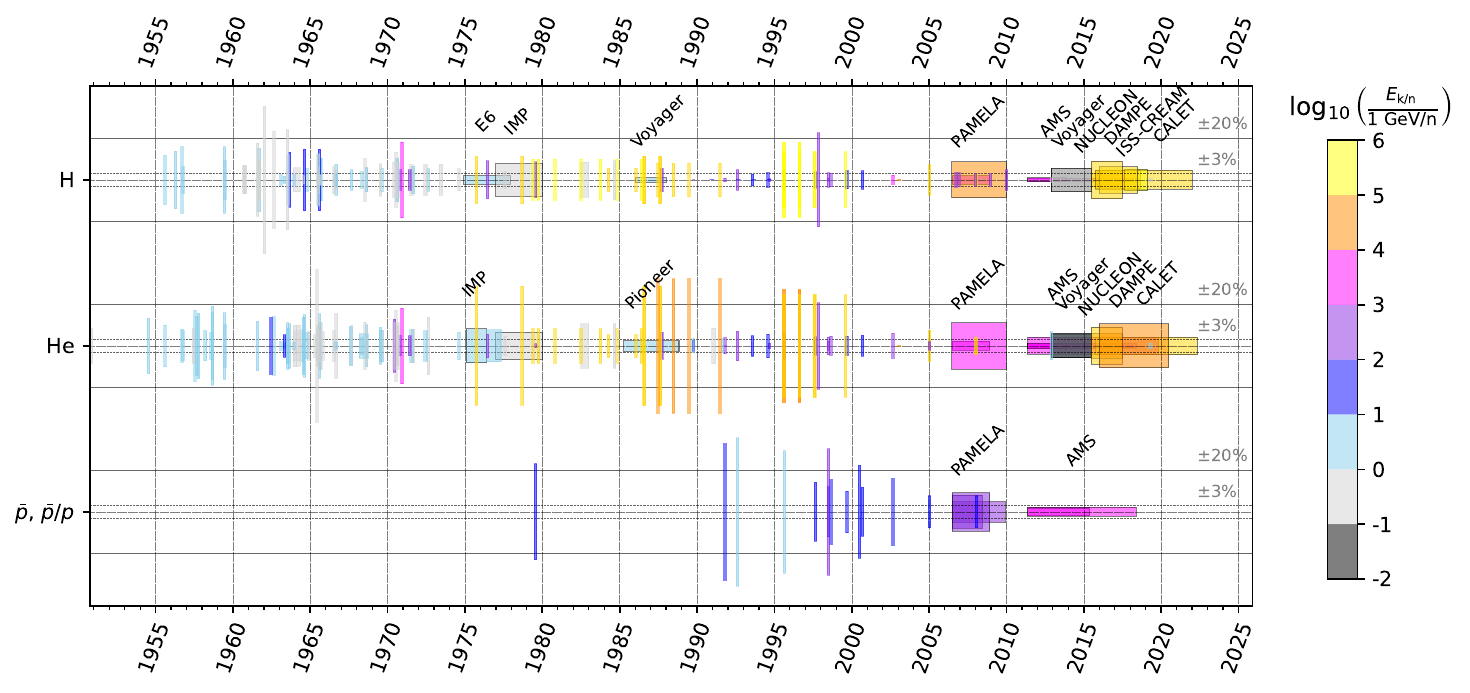}
    \caption{For H, He and \antiproton (top to bottom), timeline of the highest energy decade (colour-coded) and best precision reached (height of the bars) in \acrshort{cr} experiments, based on the data compiled in the \acrshort{crdb}~\cite{2014A&A...569A..32M, 2020Univ....6..102M, 2023EPJC...83..971M}. The width of the bars indicates the integration time of the \acrshort{cr} experiments: very thin widths correspond to balloon flights (few days flights at most), and larger widths correspond to satellite or space (or more rarely ground-based) experiments. The best precision is always achieved at low energies, not at the highest (colour-coded) energy reached. The name of experiments for which datasets were collected over several months or years is indicated on top of the relevant period; some experiments like \acrshort{ams} have several datasets published with overlapping periods (same start date but longer integration time), but their name is indicated only once to avoid overlapping text. Note that the Voyager data after 2015 are the only datasets outside the solar cavity.}
    \label{fig:CRdata_light}
\end{figure}

In this section, the status in terms of energy range and precision, and the recent progresses made by \acrshort{cr} experiments, are discussed for all \acrshort{gcr} species (Sec.~\ref{sec:CR_current}). Then, the near and far future projects are listed (Sec.~\ref{sec:CR_future}). The discussion is accompanied by a timeline of all datasets available for these various species. Figures~\ref{fig:CRdata_light} to \ref{fig:CRdata_secprim} illustrate the increase of precision and upper energy of the \acrshort{cr} experiments and data over time. Figure \ref{fig:CRexperiments} also provides a summary view of ongoing and future experiments.

\subsection{Energy range and precision of current data}
\label{sec:CR_current}

\subsubsection{Proton and He fluxes}
\label{sec:CR_pHe}
\acrshort{ams} has provided precision measurements of the proton spectrum as a function of rigidity from 0.5\,GV to 1.8\,TV, with accuracies of $1$\% at 100\,GV and $<5\%$ beyond 1\,TV, and for the He spectrum from 1.92 to 3\,TV with uncertainties of $1$\% at 100\,GV and $<4$\% at 1\,TV~\cite{AMS:PhysRep2021}.
A progressive spectral hardening has been observed around 200\,GV, with a different rigidity dependency between proton and He, confirming earlier observations by \acrshort{pamela}~\cite{2011Sci...332...69A} and \acrshort{cream}~\cite{2011ApJ...728..122Y, 2017ApJ...839....5Y}.
The proton-to-helium ratio decreases with increasing rigidity~\cite{AMS:PhysRep2021}.
Results of \acrshort{calet} and \acrshort{dampe} confirm the spectral hardening in both proton and He \acrshort{cr}s~\cite{CALET:2023nif, CALET:2022vro, Alemanno:2021gpb, DAMPE:2019gys} and the decrease of the proton-to-helium ratio up to about 10\,TeV/n. All experiments consistently show that the proton spectral index is about 0.1 softer than that of He, with no significant structures in the proton-to-helium flux ratio.
While both calorimetric experiments demonstrate a higher value of the spectral break position, compared to \acrshort{ams}, at about 500\,GeV for protons, the results for all the experiments are compatible within the uncertainties.

At higher energies, \acrshort{calet}, \acrshort{dampe} and \acrshort{iss}-\acrshort{cream} demonstrate a softening structure in proton and He spectra at about 10\,TeV/n~\cite{CALET:2023nif, CALET:2022vro, Alemanno:2021gpb, DAMPE:2019gys, ISS-CREAM:Choi2022}, as previously indicated by \acrshort{cream}~\cite{2017ApJ...839....5Y, 2011ApJ...728..122Y}. Recent updates from \acrshort{dampe} and \acrshort{calet} indicate that the positions of both hardening and softening structures favour the charge (rigidity) dependence of the breaks, although mass (energy per nucleon) dependence is not ruled out~\cite{CALET:2023nif, arshia_icrc2023}. It has to be noted that while calorimetric experiments measure particle kinetic energy, the conversion to energy per nucleon requires knowledge of the isotopic composition, which is normally taken from available measurements at low-energy, below few GeV/n, and extrapolated to higher energies.
The limited knowledge of isotopic compositions is considered as an additional source of systematic uncertainty in the interpretation of calorimetric data on nuclei. At even higher energies, a hint of a new structure -- hardening at about 150\,TeV -- is seen in the recent data of \acrshort{dampe}~\cite{arshia_icrc2023} and \acrshort{iss}-\acrshort{cream}~\cite{ISS-CREAM:Choi2022}. The \acrshort{dampe} measurement of combined p+He spectrum, profiting from higher statistics and a cleaner event selection, reaches 0.5\,PeV~\cite{DAMPE:2023pjt}. \acrshort{iss}-\acrshort{cream} results reach even higher energy, 0.65\,PeV for proton and $\sim\!1$\,PeV for He, although with much larger uncertainties~\cite{ISS-CREAM:Choi2023zdx}.

\subsubsection{Antiprotons and searches for antinuclei}
\label{sec:CR_antinuclei}

Antimatter is a tiny component in \acrshort{cr}s. In the GeV--TeV energy range, there is only 1~antiproton per 10000~protons.
Balloon-borne (e.g., \acrshort{bess}-Polar-I and II flights \cite{2008PhLB..670..103A, 2012PhRvL.108e1102A}) and satellites experiments (e.g., \acrshort{pamela} \cite{2009PhRvL.102e1101A, PAMELA:2010kea, 2013JETPL..96..621A}) have measured the \antiproton spectrum. The 6.5\,yr of \acrshort{ams} data provide the most precise information, with accuracies $<\!4\%$ in the rigidity range 1\,GV to 100\,GV and $\sim\!40\%$ at 500\,GV. Statistical uncertainties still dominate above 125\,GV~\cite{AMS:PhysRep2021}.
The low-energy \antiproton spectrum will soon be explored by the \acrshort{gaps} (\acrlong{gaps}) experiment~\cite{GAPS:ICRC2023}, see Sec.~\ref{sec:CRexp_GAPS}.
\acrshort{gaps} will measure the \antiproton flux in the kinetic energy per nucleon range from 0.07 to 0.21\,GeV/n, where signals from hidden sector \acrshort{dm} models are expected~\cite{2023APh...14502791R}.

So far, no observation of antinuclei heavier than \antiproton has been confirmed in \acrshort{cr}s.
The \acrshort{bess}, \acrshort{bess}-Polar-I and \acrshort{bess}-Polar-II magnetic spectrometer series of balloon flights, have extensively searched for \antideuteron in the kinetic energy per nucleon range from 0.163 to 1.\,GeV/n, setting the best upper limit on the \antideuteron flux in this energy range with the \acrshort{bess}-Polar-II flight at $6.7\cdot 10^{-5}\, (\mathrm{m}^2\,\mathrm{s}\;\mathrm{sr}\;\mathrm{GeV/n})^{-1}$ at 95\% CL~\cite{BESS-Polar-II:antiD-PRL2024}.
The \acrshort{bess} collaboration has also set the current best upper limit on the \antihelium to He flux ratio, in the rigidity range from 1 to 14\,GV, by combining the results from the \acrshort{bess}, \acrshort{bess}-Polar I and \acrshort{bess}-Polar-II flights. This limit is $6.9\cdot 10^{-8}$ at 95\% CL~\cite{BESS-Polar:ABE2017806}.
So far, \acrshort{ams} has reported few \antideuteron and \antihelium candidates, both \antiheliumthree and \antiheliumfour~\cite{kounin:PAW2024}, still needing further studies before an observation can be confirmed.

\subsubsection{Electrons and positrons fluxes}
\label{sec:CR_leptons}
\begin{figure}
    \centering
\includegraphics[width=\linewidth]{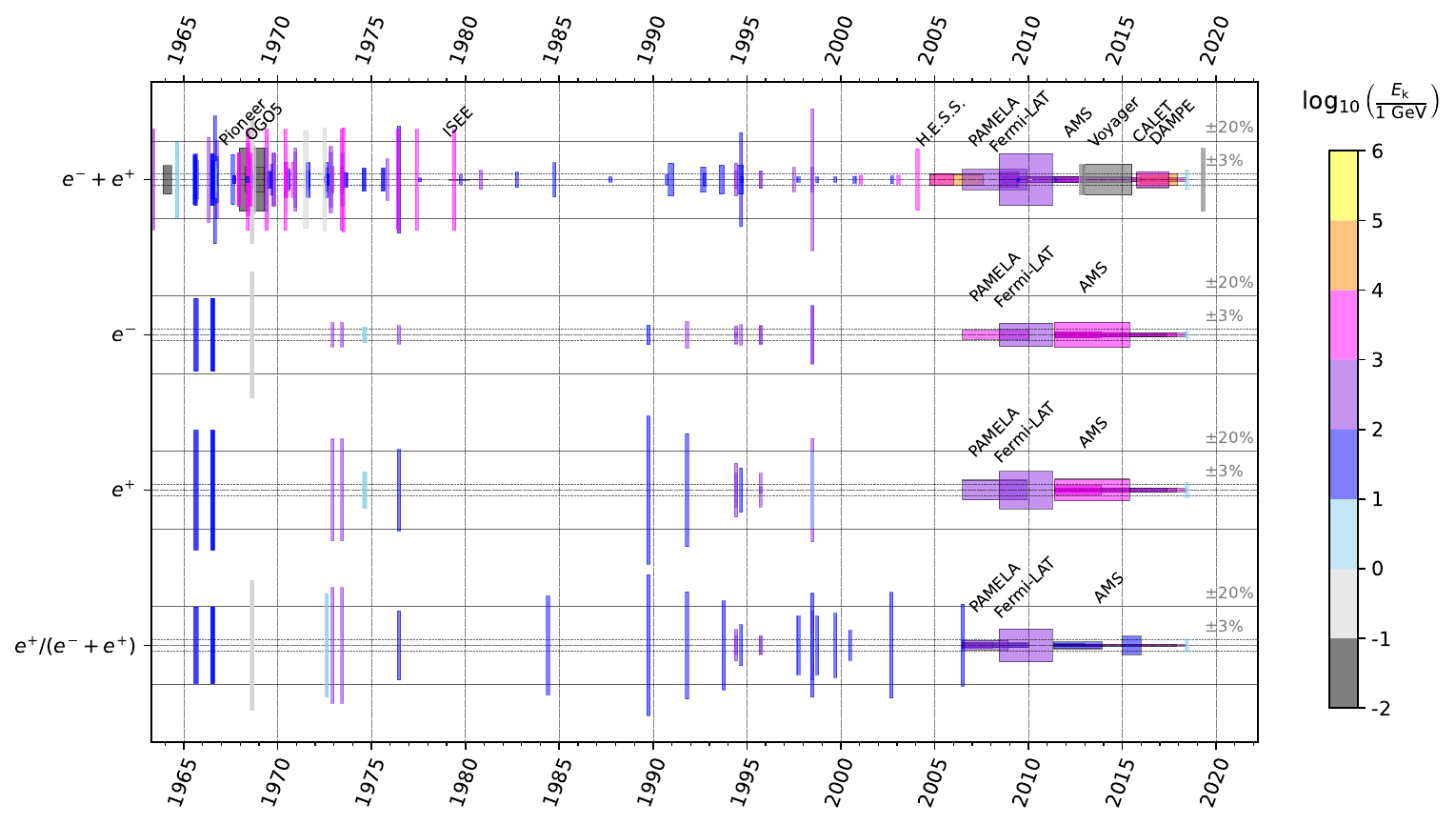}
    \caption{Same as Fig.~\ref{fig:CRdata_light} but for leptons (including the positron fraction, bottom row).}
  \label{fig:CRdata_leptons}
\end{figure}

To efficiently separate positrons from protons and \antiproton from electrons, magnetic spectrometers are combined with electromagnetic calorimeters, as in \acrshort{pamela} and \acrshort{ams}, and with transition radiation detectors, as in \acrshort{ams}. Electromagnetic calorimeters, beside distinguishing electron-like from proton-like particles, allow the determination of the energy of electrons and positrons with a resolution of a few percents.
The measurements of the separate spectra of positron and electrons by magnetic spectrometers, \acrshort{pamela} up to 300\,GeV, and \acrshort{ams} up to 1\,TeV, have ascertained that the rise of the positron fraction above $\sim\!10$\,GeV observed by earlier experiments is due to an excess of high-energy positrons. The high-precision of the \acrshort{ams} measurements (4\% at 100\,GeV and $<\!10\%$ up to 500\,GeV for the published results based on 6.5~years of data, still dominated by statistical uncertainties above 30\,GeV) has also revealed a rapid decrease of the positron flux above $\sim\!300$\,GeV compatible with an exponential energy cut-off in the TeV energy range~\cite{AMS:positrons2019, AMS:PhysRep2021}.
\acrshort{ams} has also released a high-precision measurement of the electron spectrum in the energy range from 0.5\,GeV to 1.4\,TeV ($\leq$2\% up to 130\,GeV and $<\!5\%$ up to 500\,GeV), observing a hardening above $\sim\!40$\,GeV but no high-energy cut-off~\cite{AMS:electrons2019, AMS:PhysRep2021}.
Analysis of the arrival directions of positrons and electrons with the \acrshort{ams} data has shown that both are compatible with the hypothesis of an isotropic flux, with upper limits on the amplitude of dipole anisotropy of 0.019 for positrons and 0.005 for electrons at 95\% CL above 16\,GeV~\cite{AMS:PhysRep2021, AMS:ele_pos_antip_ICRC2023}.

Results of the \acrshort{cr} electron plus positron spectrum from \acrshort{calet} and \acrshort{dampe}, reaching 4.6 and 7.5\,TeV, respectively, reveal a remarkable softening at $\sim\!1$\,TeV, consistent between the two experiments~\cite{DAMPE:2017fbg, CALET:2023emo}. While it is difficult to corroborate whether the observed spectral structure is due to the transition from a multiple source population to an individual \acrshort{gcr} accelerator, further measurements towards 10\,TeV and higher energies will be crucial to clarify the \acrshort{cr} electron picture~\cite{Thoudam:2024grg}. One of the key challenges in the realisation of such measurement with the existing experiments, such as \acrshort{dampe}, is the rejection of the overwhelming proton background contamination, which increases with energy~\cite{Droz:2021wnh}.

\subsubsection{Heavy elemental fluxes (Z=3--30)}
\label{sec:CR_HN}

\begin{figure}
    \centering
    \includegraphics[width=\linewidth]{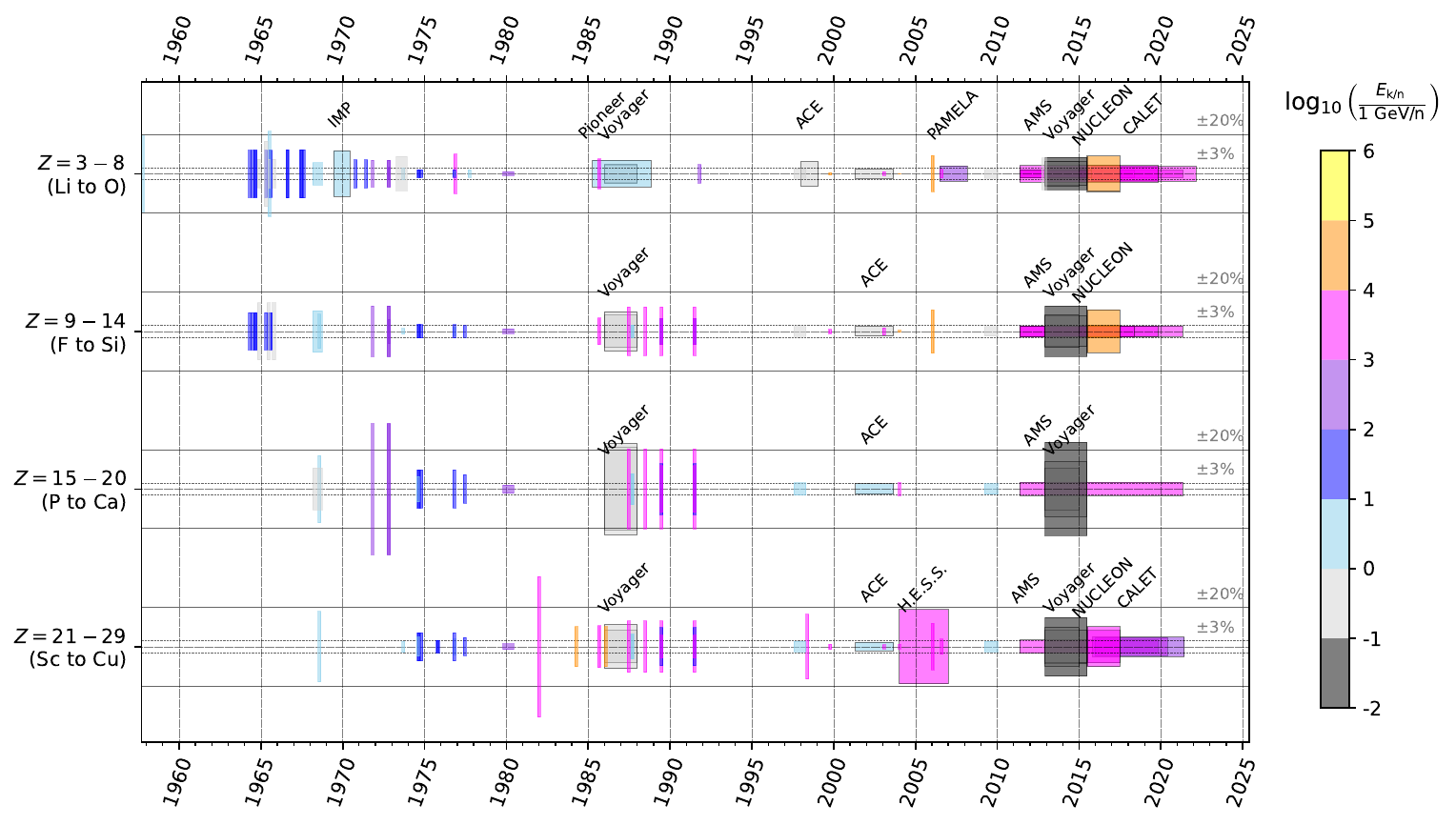}
    \caption{Same as Fig.~\ref{fig:CRdata_light}, but for individual elements of broad groups (for the sake of compactness) of heavy nuclei with $Z<30$. For each of these groups, the best precision and highest energy among the elements measured is reported: not all elements in these groups have been measured, and not all experiments have the capability to measure all elements of a given group, usually because of too low abundances.}
  \label{fig:CRdata_HN}
\end{figure}

For $Z\leq 30$, \acrshort{acecris} (\acrlong{acecris}) \cite{2009ApJ...698.1666G,2013ApJ...770..117L} and Voyager have provided fluxes below hundreds of MeV/n at 3--5\% precision, while \acrshort{ams}, \acrshort{calet} and \acrshort{dampe} are measuring the individual spectra of nuclei in the hundreds of GeV to multi-TeV region.
So far, \acrshort{ams} has published the rigidity spectra of all nuclei from He to Si, of S, and of Fe nuclei in the rigidity range from $\sim\!2$\,GV to 3\,TV~\cite{AMS:PhysRep2021, AMS:F-PRL2021, AMS:PRL-NaAlN2021, AMS:Fe-PRL2021, AMS:S-PRL2023}, with typical accuracies at 100\,GV of 3\% to 4\% for nuclei from Li to O, and 4\% to 6\% for heavier nuclei. Before \acrshort{ams}, the previous experiment which provided a comprehensive measurement of such a large range of nuclei was the \acrshort{heao} satellite, flown between 1979 and 1980, with Be to Ni fluxes from 0.6 to 35\,GeV/n at precision of $\sim\!10\%$~\cite{1990A&A...233...96E}).
At 1\,TV the \acrshort{ams} measurements of the primary nuclei, C, O, Ne, Mg and Si, have accuracies of 6\% to 7\%. For the less abundant Ne, Mg and Si nuclei, the statistical errors still dominate above 1.2\,TV~\cite{AMS:S-PRL2023}. The \acrshort{ams} measurement of the Fe spectrum at 1\,TV has an accuracy of 10\%, but it is still dominated by statistical errors above 300\,GV~\cite{AMS:Fe-PRL2021}.

\paragraph{Slopes for primary and secondary species}
\acrshort{ams} has found that above 60\,GV the rigidity spectra of C and O are identical to He~\cite{AMS:He-C-O-PRL2017}, and that the rigidity spectra of the heavier primaries Ne, Mg, and Si are distinctly different from the He spectrum~\cite{AMS:NeMgSi-PRL2020}. Above 86.5\,GV, Ne, Mg, and Si spectra have identical rigidity dependencies, they progressively harden above 200\,GV, but less than the He, C and O spectra~\cite{AMS:NeMgSi-PRL2020}. Instead, the Fe spectrum above 80.5\,GV follows the same rigidity dependence as the light primary He, C and O~\cite{AMS:Fe-PRL2021}.
\acrshort{ams} has found that light secondary nuclei, Li, Be and B, have identical rigidity dependencies above 30\,GV, and that they harden above 200\,GV with twice the hardening observed for C and O~\cite{AMS:LiBEB-PRL2018} hinting at a propagation origin (break in the diffusion coefficient) of the spectral hardenings.
The accuracy of the latest \acrshort{ams} measurement of the F spectrum, recently updated with 10-year dataset, is 6\% at 100\,GV and 18\% at $\sim\!1$\,TV, dominated by statistical errors above $\sim\!90$\,GV~\cite{AMS:S-PRL2023}.
\acrshort{ams} has found that the rigidity spectrum of the heavier secondary F nuclei is different from the spectra of the light secondary Li, Be and B, and that the secondary-to-primary ratio F/Si is significantly different from the light secondary-to-primary ratios, B/O or B/C~\cite{AMS:F-PRL2021}. Alternative interpretations have been proposed for this observation, as the presence of a primary F component~\cite{Boschini_2022} or to spatially dependent diffusion~\cite{2023PhRvD.107f3020Z}. However, model calculations of the expected secondary F spectrum with similar accuracy as the \acrshort{ams} measurement are currently out of reach because of uncertainties or lack of measurements of relevant nuclear fragmentation cross-sections~\cite{2024PhRvC.109f4914G}.

\paragraph{Multi-TeV domain}
The extension of B, C, O and Fe nuclei measurements to the multi-TeV domain has been recently advanced by \acrshort{calet} and \acrshort{dampe}. The \acrshort{calet} results on B~\cite{calet_boron_2022}, C and O~\cite{calet_CO_2020} reach about 3\,TeV/n, indicating a spectral hardening in both \acrshort{cr} primaries (C and O) and \acrshort{cr} secondaries (B) at around 200\,GeV/n, consistent with \acrshort{ams}. The \acrshort{calet} B/C and B/O ratios, similar to \acrshort{ams}, confirm that the break in secondaries is about twice as large as in primaries. Notably, while \acrshort{calet} B, C and O fluxes show an overall shape consistency with \acrshort{ams}, there is an apparent discrepancy in normalisation, with \acrshort{calet} fluxes being $\sim\!20\%$ lower than \acrshort{ams} fluxes. This difference is not accounted for by systematic uncertainties and, as was reported at this conference \cite{XSCRC2024}, is not attributed to the choice of the hadronic model (\acrshort{mc} generator) used for the interpretation of \acrshort{calet} data (see also Sec.~\ref{sec:XSforCRexp}). At the same time, the normalisation errors cancel out in the B/C and B/O ratio calculations, resulting in very good normalisation match of \acrshort{calet} with \acrshort{ams}.
Also, recent \acrshort{dampe} results on the B/C and B/O flux ratios confirm the B/C and B/O breaks and
find a hardening position at $100\pm10$\,GeV/n~\cite{dampe_bc_bo_2022}. This result is consistent with a hypothesis of a $\sim\!200$\,GV universal \acrshort{cr} hardening, and agrees well with the accurate measurements of spectral breaks in \acrshort{ams} data~\cite{ams_proton_2015, ams_helium_2015, AMS:LiBEB-PRL2018, AMS:PhysRep2021}.
For higher mass elements, \acrshort{calet}'s Fe spectrum reaches 2\,TeV/n and is consistent with a power law behaviour with coefficient $\gamma\!\sim\!2.6$ and no indication of hardening~\cite{CALET:2021fks}. \acrshort{calet}'s Fe spectral shape is very similar to the \acrshort{ams} result, but is 20\% lower in normalisation.
The \acrshort{calet} Ni flux~\cite{CALET:Ni-PRL2022} measurement is still limited to below 240\,GeV/n. It shows no structure in the spectrum along with a flat Ni/Fe ratio of about 0.06 in the entire energy range of the measurement, consistent with the expectation of similarity in acceleration mechanisms of primary \acrshort{gcr}s.

\subsubsection{Ultra-heavy elemental fluxes (\texorpdfstring{$Z>30$}{Z>30})}
\label{sec:CR_UHN}

Fluxes above $Z=30$ are about $10^4$ to $10^5$ lower than Fe \cite{2014ApJ...788...18B}, with the rarest actinides ($Z\geq90$) about $ 10^7$ times less abundant than Fe~\cite{1981Natur.291...45F} -- see the bottom panel of Fig.~\ref{fig:CRabundances}.
Standard techniques used for measurements of $Z<30$ \acrshort{cr}s can still be pushed to cover the $30\leq Z\leq60$ region \cite{2014ApJ...788...18B}, but they are not able yet to provide fluxes, only ratios: the \acrshort{acecris} satellite (Si scintillators) collected data for more than 20 years to unveil the isotopic content of \acrshort{gcr} elements $Z=30\text{--}38$ at a few hundreds of MeV/n \cite{2022ApJ...936...13B}; \acrshort{calet} has recently published elemental fractions up to $Z\leq44$ \cite{2025ApJ...988..148A}, and the Super\acrshort{tiger} balloon-borne experiment is measuring elemental fractions up to $Z\leq56$ \cite{2022AdSpR..70.2666W}.
For heavier species, passive detectors are exposed for long durations (several years), and chemical modifications made in a solid state nuclear track detector (by passing \acrshort{cr}s) are etched in a chemical agent to reconstruct the charge and velocity of the \acrshort{cr}. Very few datasets exist from $\lesssim$ 1990s experiments, with integrated measurements around GeV/n \acrshort{cr} energies. These datasets are from Ariel~6~\cite{1987ApJ...314..739F}, \acrshort{heaohne}~\cite{1989ApJ...346..997B}, \acrshort{uhcreldef}~\cite{2012ApJ...747...40D}, skylab~\cite{1978ApJ...220..719S} and Trek~\cite{1998Natur.396...50W}.
Some recent original data come from
\acrshort{olimpiya} \cite{2022AdSpR..70.2674A, 2022PAN....85..446A}, where olivine crystals contained in stony-iron meteorites (pallasites) are used as \acrshort{cr} detectors. They give, however, only integral measurements over energies and irradiation time over up to hundreds of Myr.

\begin{figure}
    \centering
\includegraphics[width=\linewidth]{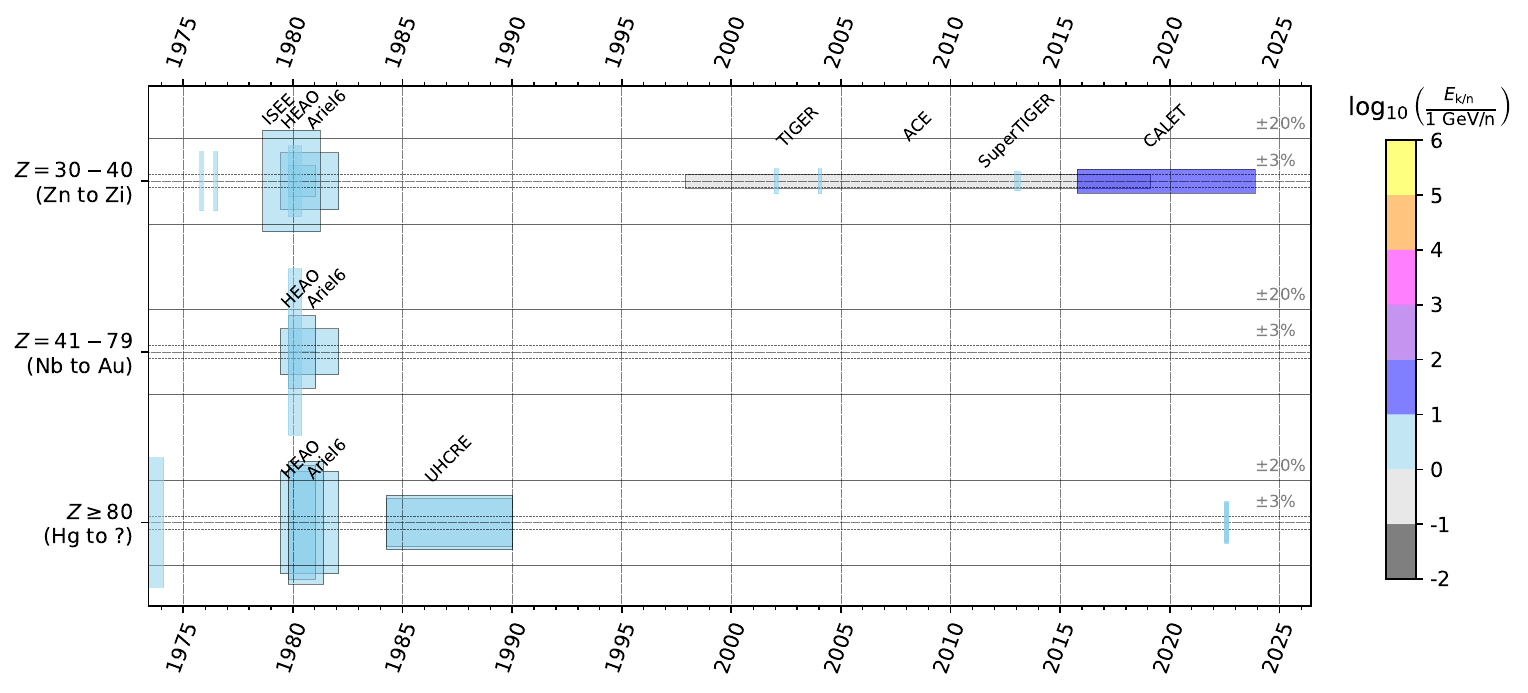}
    \caption{Same as Fig.~\ref{fig:CRdata_light} but for ultra-heavy nuclei. For the latter, only ratios are measured for individual elements below $Z=40$, and pairs or range of elements for $Z>40$. The question mark in the last group ($Z\geq80$, Hg to ?) highlights the fact that the current situation is not completely clear regarding the heaviest \acrshort{cr} species detected; the 2023 data points come from the \acrshort{olimpiya} stony-iron meteorites (see text for details).}
  \label{fig:CRdata_UHN}
\end{figure}

\begin{figure}
    \centering
    \includegraphics[width=\linewidth]{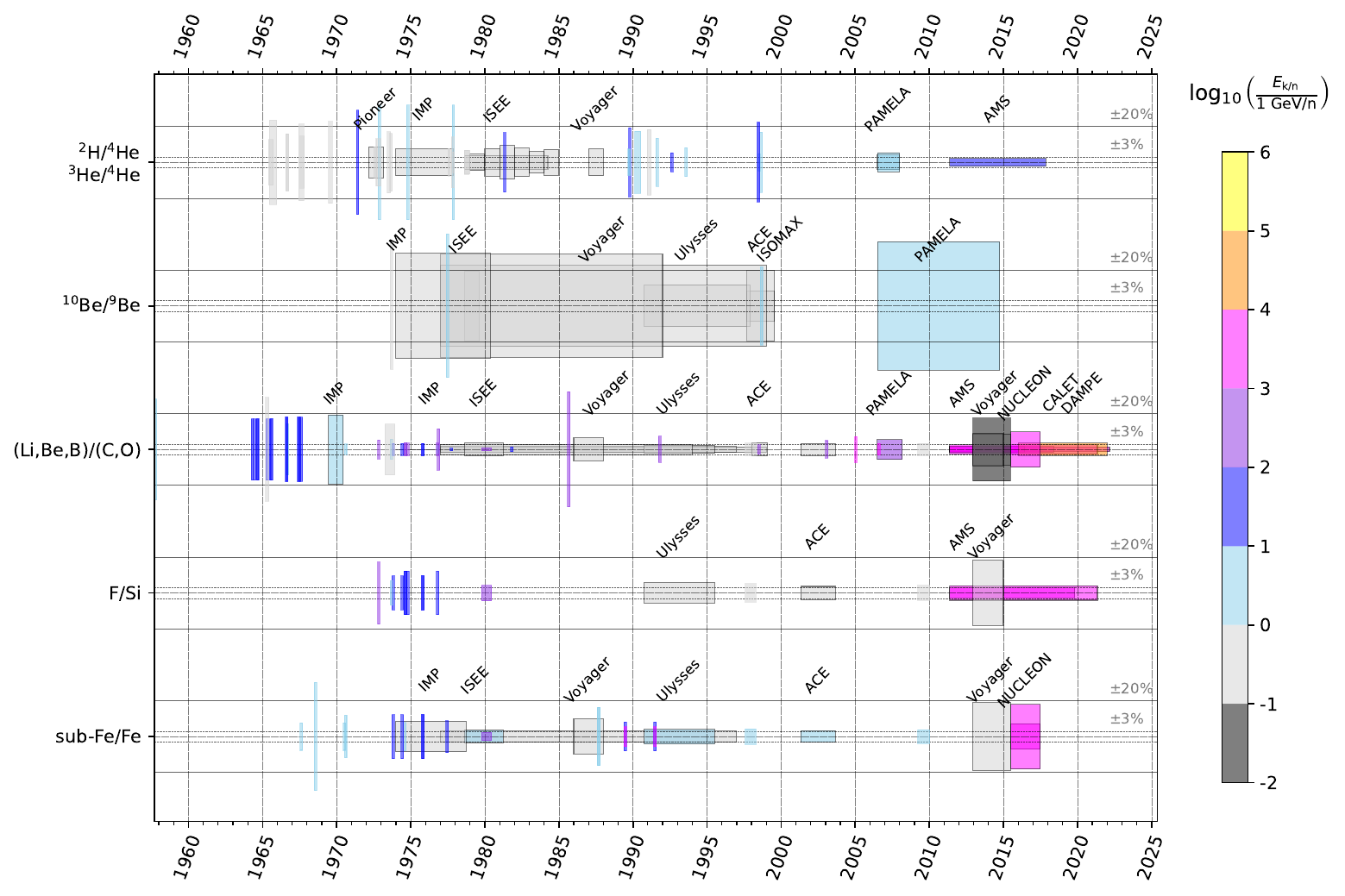}
    \caption{Same as Fig.~\ref{fig:CRdata_light} but for secondary-to-primary ratios currently used for \acrshort{gcr} analyses. For compactness, we grouped together \deuteron/$^4$He and $^3$He/$^4$He (top panel) and ratios of Li, Be, and B to C or O (third panel), but the precision and datasets are not exactly the same for these individual ratios (e.g., \deuteron is particularly difficult to separate from the dominant \proton, while a good charge separation is needed to isolate Li from the much more abundant He); sub-Fe (in the sub-Fe/Fe ratio, bottom panel) corresponds to $Z=21\text{--}23$ (or sometimes $Z=21\text{--}25$) and was used in almost all past experiments because of their limited charge resolution/statistics, but \acrshort{ams} will provide individual fluxes and ratios for this charge range. The \acrshort{gcr} clock ratio $^{10}$Be/$^{9}$Be (second panel) is the best measured ratio to date compared to other \acrshort{gcr} clocks and other relevant radioactive \acrshort{gcr} species (see text for details).}
  \label{fig:CRdata_secprim}
\end{figure}

\subsubsection{Isotopic fluxes and ratios}
\label{sec:CR_isotopes}
Elements in \acrshort{gcr}s are almost all mixtures of two or more isotopes. Isotopic composition measurements are still scarce and limited to low energies, below about 2\,GeV/n in most cases.

\paragraph{Light isotopes}
Recent measurements of H to Be isotopes have superseded all previous measurements in terms of precision and energy coverage (see Fig.~\ref{fig:CRdata_secprim}).
The so-called quartet isotopes include the dominant $^1\mathrm{H}$ and $^4\mathrm{He}$ species of primary origin, and $^2\mathrm{H}$ and $^3\mathrm{He}$ isotopes expected to be of secondary origin. \acrshort{pamela} has measured them from a few hundreds of MeV/n to GeV/n at a precision of 10\%, pushing their analysis to get ratios of some Li, Be, and B isotopes \cite{2016ApJ...818...68A, 2018ApJ...862..141M, 2021Univ....7..183N, 2023BRASP..87..863B, 2024PAN....87...71B}.
\acrshort{ams} has measured $^2\mathrm{H}$, $^3\mathrm{He}$ and $^4\mathrm{He}$ fluxes in the rigidity range 1.9\,GV to 21\,GV with accuracies of 3\%, $\lesssim3\%$, and $\lesssim1\%$, respectively~\cite{AMS:Deuteron-PRL2024}.
While the $^3\mathrm{He}$/$^4\mathrm{He}$ flux ratio exhibits a typical secondary-to-primary rigidity dependence, $\propto R^{-0.289\pm 0.003}$, the $^2\mathrm{H}$/$^4\mathrm{He}$ ratio follows a distinct power law $\propto R^{-0.108\pm 0.005}$.
\acrshort{ams} has also published the measurements of $^6\mathrm{Li}$ and $^7\mathrm{Li}$ fluxes, roughly in equal amount in \acrshort{cr}s, from 1.9\,GV to 25\,GV with accuracies $\sim\!3\%$ at 10\,GV, which do not support the hypothesis of a primary component in $^7\mathrm{Li}$~\cite{AMS:Li_isotopes}.

\acrshort{cr} Be nuclei are secondaries, composed of three isotopes ($^7\mathrm{Be}$, $^9\mathrm{Be}$ and $^{10}\mathrm{Be}$), with $^{10}\mathrm{Be}$ decaying to $^{10}\mathrm{B}$ with a half-life of 1.39\,Myr, probing the residence time in the Galaxy.
Current measurements of the $^{10}$Be/$^9$Be ratio include low-energy data at 100\,MeV/n with a 20--30\% uncertainty (\acrshort{acecris} \cite{2001ApJ...563..768Y}, Ulysses \cite{1998ApJ...501L..59C} and Voyager \cite{1999ICRC....3...41L}), and a couple of GeV/n data points from both \acrshort{pamela} \cite{2021Univ....7..183N, 2023BRASP..87..863B} and the \acrshort{isomax} balloon-borne superconducting spectrometer \cite{2004ApJ...611..892H}, with a much poorer precision.
\acrshort{ams} has presented at recent conferences preliminary measurements of the $^7\mathrm{Be}$, $^9\mathrm{Be}$ and $^{10}\mathrm{Be}$ fluxes as functions of kinetic energy per nucleon, ranging from 0.4\,GeV/n to 12\,GeV/n~\cite{ICRC2023:CRD_Rapporteur}. The accuracy of the \acrshort{ams} preliminary measurement of $^{10}\mathrm{Be}$/$^9\mathrm{Be}$ is $\sim\!10\%$. Complementary results are also expected from \acrshort{helix}~\cite{HELIX:Coutu_2024} (see Sec.~\ref{sec:CRexp_HELIX}).

All these great and recent experimental achievements, however, cannot be exploited at full potential yet, as the interpretation of the light isotopes are particularly plagued by scarce nuclear data and large uncertainties~\cite{2022A&A...668A...7M, Maurin:10Beto9Be_L2022}.

\paragraph{Heavier isotopes}
For isotopes in $Z=6\text{--}30$, only ratios are measured, and only at a very low energy below a few hundreds of MeV/n, with a precision of 10--20\%. These measurements are all from pre-1980's balloon flights and pre-2000's space experiments (\acrshort{acecris}, \acrshort{crres}, \acrshort{isee}, Ulysses, Trek, Voyager) integrating the signal over many years. A systematic survey of these isotopes is hence missing, with the main efforts concentrated on key isotopic ratios: \acrshort{gcr} clocks (i.e., $\beta$-unstable species similar to $^{10}$Be) via the ratios $^{10}$Be/$^{9}$Be (see above), $^{26}$Al/$^{27}$Al \cite{1983ICRC....9..147W, 1994ApJ...430L..69L, 1998ApJ...497L..85S, 2001ApJ...563..768Y}, $^{36}$Cl/Cl \cite{1985ICRC....2...84W, 1998ApJ...509L..97C, 2001ApJ...563..768Y} and $^{54}$Mn/Mn \cite{1995AdSpR..15a..25W, 1996A&A...314..785H, 1996ApJ...468..679W, 1997ApJ...481..241D, 1997ApJ...488..454L, 2001ApJ...563..768Y}; nucleosynthesis clocks (i.e., unstable species constraining the time elapsed between their nucleosynthesis and acceleration, expected to be $\sim\!$\,Myr) with electron-capture unstable species ($^{55}$Fe, $^{57}$Co, $^{59}$Ni, and their daughter) \cite{1993ApJ...405..567L, 1995AdSpR..15a..25W, 1997ApJ...475L..61C, 1997ApJ...488..454L, 1999ApJ...523L..61W} or the $\beta$-unstable $^{60}$Fe recently detected by \acrshort{acecris} (cumulating $\sim\!17$\,yrs of data) \cite{2016Sci...352..677B}; electron-capture decay species sensitive to \acrshort{gcr} re-acceleration (ratios of $^{49}$V and $^{51}$Cr and their daughters) \cite{1993ApJ...405..567L, 1997ApJ...488..454L, 1999ICRC....3...33C}; source abundance anomalies (i.e., isotopes whose \acrshort{gcr} abundance departs from \acrshort{ss} ratios) with the striking $^{22}$Ne/$^{20}$Ne anomaly \cite{1980ApJ...235L..95M, 1981PhRvL..46..682W, 1988A&A...193...69F, 1996ApJ...466..457D, 1997ICRC....3..381C, 1997ApJ...476..766W, 2005ApJ...634..351B} and possibly $^{58}$Fe/$^{56}$Fe \cite{1997ApJ...475L..61C, 2001AdSpR..27..773W}.

For elements beyond Ni, the only datasets are $Z\leq38$ ratios of isotopes to their elements provided by the \acrshort{acecris} experiment at a few hundreds of MeV/n with a precision $\sim\!50\%$, from data cumulated over more than 20~years~\cite{2022ApJ...936...13B}. While the number of unstable isotopes grows steadily with mass, and could help to shed further light to the processes discussed above, the difficulties of pursuing such measurements due to the very low abundances and the limitation of cross-section models in this range leaves this region as uncharted territories for now.

\subsection{Ongoing and future projects: energy, mass, isotopes, antinuclei and precision frontiers}
\label{sec:CR_future}

\begin{figure}
\centering
\includegraphics[width=0.95\linewidth]{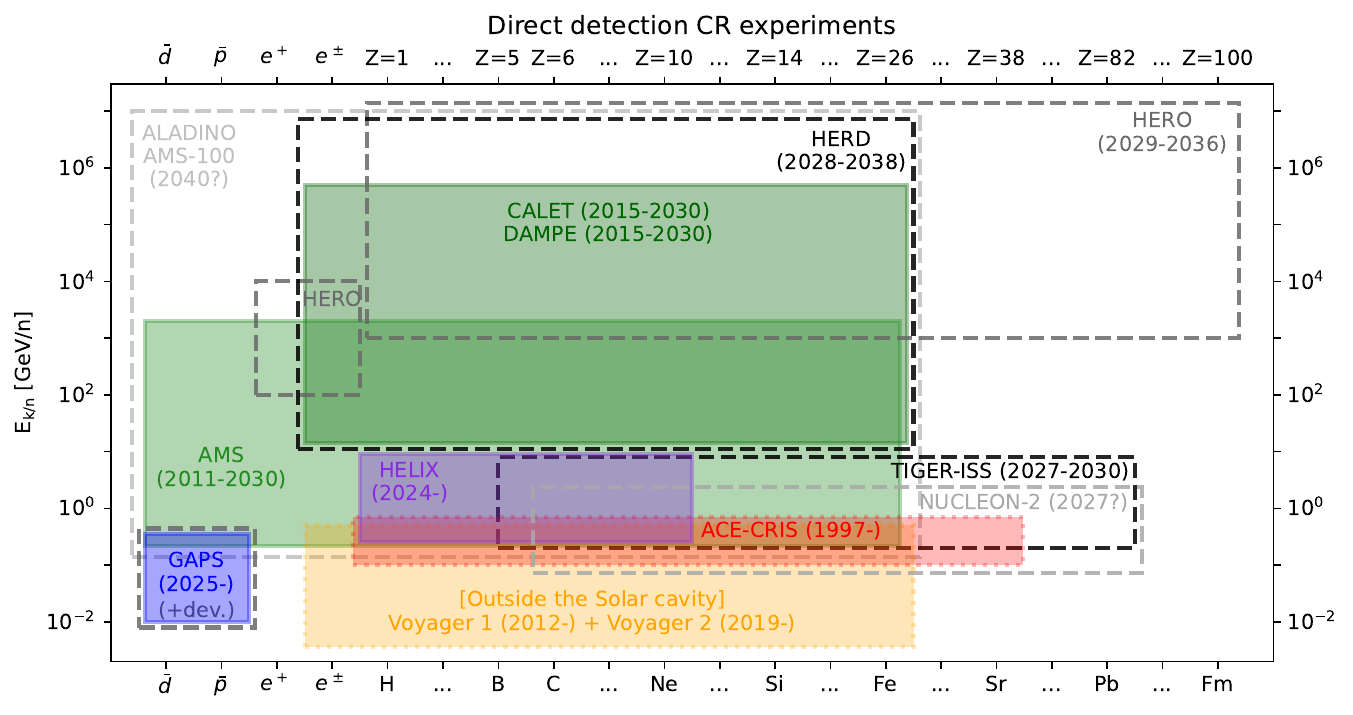}
\caption{Species ($x$-axis) and energy ($y$-axis) reach of direct detection \acrshort{cr} experiments. Shown are (i) experiments that have outlasted by far their initial physics programme but are still functioning (orange and red dotted line boxes), (ii) ongoing experiments (solid line filled boxes), and (iii) forthcoming or future projects (dashed line empty boxes). See Sec.~\ref{sec:CR_future} for details.}
\label{fig:CRexperiments}
\end{figure}

In this section, a quick overview of operational experiments and future projects is presented. \acrshort{acecris} and Voyager satellites, that have recently provided very useful data sets (see previous section), are not covered: these detectors have outlasted their initial programmes by far, and although they are still taking data, it is not clear if their recent relevant results (\acrshort{is} spectra for Voyager and $Z=30\text{--}40$ data for \acrshort{acecris}) could be surpassed, extended or reveal new surprises. In the coming years, space experiments will mostly target the energy frontier, thanks to upgrades (\acrshort{ams}), longer data taking periods (\acrshort{dampe} and \acrshort{calet}) or larger acceptance detectors like \acrshort{herd} (\acrlong{herd}) and \acrshort{hero} (\acrlong{hero}). The mass frontier will be explored by \acrshort{tiger}-\acrshort{iss} and NUCLEON-2. Balloon-borne experiments are targeting the isotope (\acrshort{helix}) and antinuclei (\acrshort{gaps} and others) frontiers. For the next decades, the sub-percent precision, energy, isotope and anti-matter frontiers will all be targeted at once, with the ambitious but very uncertain projects \acrshort{aladino} (\acrlong{aladino}) and \acrshort{ams}-100. The reach in terms of species, energy, and precision of these current and future experiments is discussed below and summarised in Fig.~\ref{fig:CRexperiments}. We do not report or discuss interstellar probe projects to measure very-low energy \acrshort{is} spectra by the end of the century~\cite{2023BAAS...55c.169H}.

\subsubsection{\texorpdfstring{\acrshort{ams}}{AMS} (2011--2030): prospects and upgrade}
\acrshort{ams} will operate for the entire \acrshort{iss} lifetime, through at least 2030.
By the end of the mission, \acrshort{ams} will provide the rigidity spectra for nuclei up to Ni and up to TV energies at least (to 3.7\,TV rigidities), and measure the isotope fluxes in the 0.4--12\,GeV/n range for light nuclei.
An upgrade of the detector is foreseen in early 2026 by adding a double-layer of silicon micro-strip detectors at the top of the instrument. This will increase the geometrical acceptance by a factor 3 and add two charge measurement points with almost no material above.
This allows measuring the fluxes of nuclei between S and Fe with similar accuracies as lighter nuclei, by collecting more statistics (3~times faster with the upgrade), and by improving the rejection of background originating from interactions with material above the first charge measurement point. The upgrade will also allow extending the measurements of the positron flux up to 2\,TeV, that of the electron flux up to 3\,TeV, and to improve the accuracy of the \antiproton flux~\cite{AMS:ele_pos_antip_ICRC2023}.

\subsubsection{\texorpdfstring{\acrshort{dampe}}{DAMPE} and \texorpdfstring{\acrshort{calet}}{CALET} (2015--2030): sub-PeV energy frontier}

All sub-systems of \acrshort{dampe} remain in excellent condition
and the satellite is expected to continue data-taking for at least a few more years. With more accumulated data, it will be able to reach a few hundred TeV for individual hadronic \acrshort{cr} spectral measurements and at least 10\,TeV for electrons. Similarly, the \acrshort{calet} mission demonstrated stable performance, leading to the extension of its lifetime on \acrshort{iss} until 2030, with no special operations or interventions scheduled.

\subsubsection{\texorpdfstring{\acrshort{helix}}{HELIX} (first balloon flight in 2024): \texorpdfstring{$^{10}$Be/$^9$Be}{10Be/9Be}}
\label{sec:CRexp_HELIX}

\acrshort{helix} is aimed at measuring spectra and composition of light isotopes from He to Ne nuclei, thanks to a combination of a 1\,T superconducting magnet, a high-resolution \acrshort{tof} system, and a \acrshort{rich} detector~\cite{HELIX:Coutu_2024}. The first balloon flight was successfully conducted from Kiruna, Sweden to northern Canada, for 6~days from May~28 to June~3,~2024\footnote{\url{https://stratocat.com.ar/fichas-e/2024/KRN-20240528.htm}}. An anticipated longer flight in Antarctica will yield measurements up to 10\,GeV/n. Compared to previous \acrshort{isomax} (\acrlong{isomax}) measurements~\cite{2004ApJ...611..892H}, \acrshort{helix} will enable, in particular, sampling the secondary production of the \acrshort{gcr} clock $^{10}$Be from a larger volume of the Galaxy, to provide more stringent constraints on the halo size of the Galaxy (crucial for \acrshort{dm} searches).

\subsubsection{\texorpdfstring{\acrshort{gaps}}{GAPS} (balloon-flight ready) and other future designs (\texorpdfstring{\acrshort{grams}}{GRAMS}, \texorpdfstring{\acrshort{phescami}}{PHeSCAMI}): low-energy antinuclei}
\label{sec:CRexp_GAPS}

\paragraph{\texorpdfstring{\acrfull{gaps}}{GAPS}}

The \acrshort{gaps} experiment is optimised for \acrshort{cr} antinuclei~\cite{Aramaki_2016} at low-energy ($<0.25$\,GeV/n). The experiment consists of ten planes of semiconducting Si(Li) strip detectors surrounded by a plastic scintillator \acrshort{tof} system. \acrshort{gaps} will undertake a series of Antarctic long-duration balloon flights, and is ready for its first flight during the 2025/26 balloon season.
\acrshort{gaps} relies on a novel particle identification technique based on exotic atom formation and decay~\cite{Aramaki_2016}, in which antinuclei slow down and eventually annihilate within the detector.
The identification of antinuclei uses the simultaneous occurrence in a narrow time window of X-rays of characteristic energy and nuclear annihilation products, providing high rejection power to suppress non-antiparticle background and identify the antinucleus species.
This exotic atom detector design yields a large grasp compared to typical magnetic spectrometers, and allows for identifying \antiproton, \antideuteron and \antihelium \acrshort{cr}s. \acrshort{gaps} will provide a precision \antiproton spectrum for the first time in the low-energy range below $0.25$\,GeV/n~\cite{2023APh...14502791R}, and has a sensitivity to \antideuteron that is about two orders of magnitude better than the current \acrshort{bess} limits. Though the instrument is optimised for \antideuteron, the exotic atom detection technique is also sensitive to \antihelium signatures~\cite{2021APh...13002580S}. Due to the higher charge, the \antihelium analysis is even less affected by \antiproton backgrounds than the \antideuteron analysis, which allows for a competitive \antihelium sensitivity in the low-velocity range.

\paragraph{\texorpdfstring{\acrfull{grams}}{GRAMS}}

The \acrshort{grams} experiment is a novel instrument designed to simultaneously target both astrophysical $\gamma$ rays with MeV energies and antimatter signatures of \acrshort{dm}~\cite{Aramaki:2019bpi}. The \acrshort{grams} instrument consists of a \acrfull{lartpc} surrounded by plastic scintillators. The \acrshort{lartpc} is segmented into cells to localise the signal, an advanced approach to minimise coincident background events in the large-scale \acrshort{lartpc} detector.
The \acrshort{grams} concept potentially allows for a larger instrument since argon is naturally abundant and low-cost, compared to current experiments that rely on semiconductors or scintillation detectors. \acrshort{grams} is proposed to begin as a balloon-based experiment, as a step forward to a satellite mission.
\acrshort{grams} has been developed to become a next-generation search for antimatter signatures of \acrshort{dm}. The detection concept resembles \acrshort{gaps}'s, relying on exotic atom capture and decay. However, as the \acrshort{lartpc} detector can provide an excellent 3-dimensional particle tracking capability, with nearly no dead volume inside the detector, the detection efficiency can be significantly improved while reducing the ambiguity of antimatter measurements, which is crucial for discovering rare events. A prototype flight called Mini\acrshort{grams} is planned for 2025/26.

\paragraph{\texorpdfstring{\acrfull{phescami}}{PHeSCAMI}}

The goal of the \acrshort{phescami} project is to study the signatures offered by a high-pressure He target for the identification of \antideuteron in space. Exotic atoms are produced by stopping \antiproton/\antideuteron in He gas. The identification uses the delayed annihilation of antinuclei in He to identify cosmic antimatter species. The typical lifetime for stopped \antideuteron in matter is of the order of picoseconds, similar to that of stopped \antiproton. However, the existence of long-lived (of the order of microseconds) metastable states for stopped \antiproton in He targets has been measured~\cite{PhysRevLett.67.1246}. These metastable states in He have also been measured for other heavy negative particles, such as pions and kaons~\cite{1992PhRvA..45.6202N, 1989PhRvL..63.1590Y}. The theoretical description of this effect predicts that the lifetimes of these metastable states increase quadratically with the reduced mass of the system, i.e., a larger delay of the annihilation signature is expected for \antideuteron than for \antiproton capture in He~\cite{osti_4029624, PhysRevLett.23.63, PhysRev.188.187, PhysRevA.6.2488, PhysRevA.51.2870}.
The project is still at the development stage~\cite{instruments8010003}, but a prototype could be flown as a payload for an Antarctic stratospheric balloon in the coming years.

\subsubsection{\texorpdfstring{\acrshort{tiger}}{TIGER}-\texorpdfstring{\acrshort{iss}}{ISS} (2027--2030) and NUCLEON-2 (2027?): ultra-heavy nuclei}
\acrshort{tiger}-\acrshort{iss}, scheduled for launch in 2027 \cite{2024icrc.confE.171R}, is the next step of the \acrshort{tiger} and Super\acrshort{tiger} project, whose dataset from the second flight, 32~days in 2019--2020, is still being analysed \cite{2021cosp...43E1335S, 2023HEAD...2030305R}. It is designed to measure the abundances of rare ultra-heavy nuclei of energies above 350\,MeV/n from B ($Z=5$) up to Pb ($Z=82$). The key instrumental difference of \acrshort{tiger}-\acrshort{iss}, compared to Super\acrshort{tiger}, is the replacement of scintillator-based detectors by Si strip detectors, to avoid scintillator saturation effects, improving the charge resolution capability~\cite{2024icrc.confE.171R}. The instrument will have a geometrical factor of 1.3\,m$^2\cdot$\,sr. In less than one year of operation, \acrshort{tiger}-\acrshort{iss} will collect as many events as the 55-day Super\acrshort{tiger}-1 balloon flight~\cite{2024icrc.confE.171R}. One of the main advantages of an \acrshort{iss}-based configuration, compared to balloon flights, is that the results will be free of systematic effects related to nuclear interactions in the atmosphere.

The NUCLEON-2 satellite mission, to measure nuclei from C ($Z=6$) to Pb and isotopes from $Z=6$ to $Z=66$ in the energy range from 100\,MeV/n to 3\,GeV/n, is currently under development~\cite{Bulatov2019,Vasiliev:2021gvs}.
The detector design achieves a geometric factor of 0.8\,m$^2\cdot$\,sr with 48~hexagonal modules made of stacks of 40~silicon detectors, composed of 4~double layers of micro-strip tracking devices interleaved by 3~stacks of 10~calorimetric sensors each.
NUCLEON-2 will determine nuclei charge and mass by multiple measurements of $E-dE$ along the nucleus trajectory inside the detector, until it stops. 
The isotope identification performance has been studied on a prototype detector, using test beams of \texorpdfstring{$^{40}_{18}$Ar}{40Ar} at the \acrshort{jinr} (\acrlong{jinr}) Nuclotron and \texorpdfstring{$^{159}_{18}$Xe}{159Xe} at the \acrshort{cern} \acrshort{sps}. To accurately calibrate the isotope mass measurement method, further tests are foreseen in the framework of the DPS project \cite{Vasilev:2021glj, Kurganov:2023jfy} at the \acrshort{nica} (\acrlong{nica} accelerator complex \cite{NICA} in Dubna.
NUCLEON-2 is expected to operate on a Russian commercial satellite at 400~km altitude for at least five years.

\subsubsection{\texorpdfstring{\acrshort{herd}}{HERD} (2028--2038) and \acrshort{hero} (2029--2036): going to the Knee}
\label{sec:CRexp_HERD_GERO}
Extension of direct \acrshort{cr} measurements towards the~\emph{Knee} requires a significant enhancement of the instruments' geometric acceptance.

The \acrshort{herd} mission, expected to be deployed on the Chinese Space Station by 2028, is a leap forward for calorimetric experiments~\cite{Kyratzis:2022rxa}. It will feature the first 3-D imaging cubic calorimeter, which allows accepting particles from 5~sides. This enables achieving the geometric acceptance of about 2\,m$^2\cdot$\,sr -- nearly one order of magnitude more than the current largest calorimeters (\acrshort{dampe} and \acrshort{iss}-\acrshort{cream}). With a thickness of about 55~radiation lengths and 3~nuclear interaction lengths, \acrshort{herd} will probe \acrshort{cr} protons and ions up to 10\,PeV, and \acrshort{cr} electrons up to 100\,TeV. Among the scientific goals of \acrshort{herd} are the measurement of the \acrshort{cr} elemental composition and spectra up to the~\emph{Knee}, and the search for indirect \acrshort{dm} signatures in \acrshort{cr} electrons and $\gamma$-ray spectra.
The response to hadrons of a large-scale CaloCube, similar to \acrshort{herd}, has been recently investigated using data collected at the \acrshort{cern} \acrshort{sps} accelerator, at energies of a few hundreds of GeV~\cite{2024NIMPA106169079A}.

Similarly, the \acrshort{hero} project will employ a heavy ionisation calorimeter in a 4$\pi$ acceptance design to reach a geometric acceptance of at least 12\,m$^2\cdot$\,sr for protons, and at least 16\,m$^2\cdot$\,sr for nuclei and electrons~\cite{Karmanov2019, Podorozhny2024}.
The calorimeter will be surrounded by multi-layer silicon detectors, able to measure absolute charges up to $Z\!\sim\!100$. \acrshort{hero} will measure the spectra of proton and nuclei in the energy range from 1\,TeV to 10\,PeV, and electron plus positron and $\gamma$-ray energy spectra from 100\,GeV to 10\,TeV~\cite{Karmanov2019}.
The \acrshort{hero} mission will operate for at least 5 years on a Russian satellite. Designs with heavier calorimeters, reaching geometric acceptances up to 60\,m$^2\cdot$\,sr, are being considered to match the payload capability of the Russian heavy and super-heavy launch vehicles, currently under development and expected to be ready not earlier than 2029~\cite{Kurganov2023}.

\subsubsection{\texorpdfstring{\acrshort{aladino}}{ALADInO} and \texorpdfstring{\acrshort{ams}}{AMS}-100 (beyond 2040): sub-percent precision and energies up to the Knee}
The qualitative leap forward on direct \acrshort{cr} measurements in space is expected with the deployment of large high-temperature superconducting magnets. Currently, two conceptual designs based on large-acceptance magnetic spectrometers equipped with deep 3D imaging cubic calorimeters are being developed, \acrshort{aladino}~\cite{2021ExA....51.1299B, ALADInO:2022ntw} and \acrshort{ams}-100~\cite{Schael:2019lvx}. The combination of a magnetic spectrometer with a calorimeter allows their cross-calibration and precise determination of rigidity and energy scales as in \acrshort{ams}. Both \acrshort{aladino} and \acrshort{ams}-100 are designed to be placed at the Earth Lagrange Point 2, to maintain a stable cold environment for the magnet operation. Either of the two instruments is anticipated to start science operation not earlier than 2040~\cite{Schael:2019lvx, ALADInO:2022ntw}.

The \acrshort{ams}-100 instrument has a solenoidal magnetic field configuration with a magnetic spectrometer acceptance of 100\,m$^{2}\cdot$\,sr, reaching an \acrshort{mdr} of 100\,TV, and a calorimeter of 70 radiation lengths and 4 interaction lengths with an acceptance of at least 30\,m$^{2}\cdot$\,sr. \acrshort{ams}-100 is expected to measure the energy spectrum of electrons and positrons up to 20\,TeV and 10\,TeV, respectively, the rigidity spectrum of \antiproton up to 10\,TV, and the energy spectra of nuclei (up to at least Ni) up to 10\,PeV. \acrshort{ams}-100 features also a \acrshort{tof} system with a 20\,ps time resolution, allowing to search for \antihelium and to measure \antideuteron in the energy range from 0.1\,GeV/n to 8\,GeV/n, with a sensitivity of $3 \times 10^{-11}(\mathrm{\,m^2\,s~sr~GeV/n})^{-1}$ in 10 years of data taking. \acrshort{ams}-100 will also be able to perform detailed studies of diffuse $\gamma$-ray emission and $\gamma$-ray sources up to 10\,TeV, with the ability of resolving structures with angular resolution comparable to modern X-ray telescopes.

With a much smaller payload mass (6.5~tons compared to the 40~tons of \acrshort{ams}-100), \acrshort{aladino} has a magnetic spectrometer acceptance larger than 10\,m$^{2}\cdot$\,sr, with a toroidal magnetic field setup and a calorimeter of similar acceptance, with 61 radiation lengths and 3.5~interaction lengths. \acrshort{aladino} reaches a \acrshort{mdr} better than 20|,TV, and will measure the energy of electrons and positrons up to 10\,TeV with 2\% resolution. It is expected to measure the rigidity spectrum of \antiproton up to 10\,TV, the energy spectra of proton and He nuclei up to 10\,PeV, and those of heavier nuclei (up to at least Ni) up to 1\,PeV \cite{2021ExA....51.1299B}. The \acrshort{aladino} setup will also include a \acrshort{tof} with time resolution better than 100~ps, allowing to measure the \antideuteron flux up to 4\,GeV/n and to search for \antihelium with a sensitivity better than $10^{-10}(\mathrm{\,m^2\,s~sr~GeV/n})^{-1}$ in the first 5~years of operation.

\section{Cross-section needs for GCRs: current {\em vs.} sought precision and energies}
\label{sec:XSneeds}

In this section, the reactions in terms of projectiles, targets and products, energy coverage, and the cross-section precision needed to be able to fully take advantage of current high-precision \acrshort{cr} data are detailed. These desired precisions are not the same for all reactions and all energies, as \acrshort{cr} fluxes in which these reactions are involved are not measured with the same precisions, as discussed in Sec.~\ref{sec:CRdata}. 

Technically, the propagation of uncertainties goes through the \acrshort{gcr} transport equation, and propagating them back to an observed \acrshort{gcr} flux precision -- to finally derive a desired nuclear data precision -- is not completely straightforward. 
These questions have been investigated recently in depth for the production cross-sections of \acrshort{gcr} nuclei and antinuclei. Indeed, both are pivotal to take full advantage of current \acrshort{cr} data, e.g., for the \acrshort{dm} searches discussed in Sec.~\ref{sec:physics_cases}. After introducing the transport equation and several definitions (Sec.~\ref{sec:CRtransport}), the reactions needed for the two above cases, i.e., nuclei and antinuclei production, are detailed (Sects~\ref{sec:XS_prodnuc} and~\ref{sec:xs_prodanti}). Many other reactions (inelastic, annihilation, etc.), relevant for \acrshort{gcr} studies or by \acrshort{cr} experiments themselves to deliver their promised precision, are also reviewed (Sec.~\ref{sec:xs_other}), in particular with respect to their status and their contribution to the error budget.

\subsection{The key transport equation for \texorpdfstring{\acrshort{gcr}s}{GCRs}: definitions and relevant cross-sections}
\label{sec:CRtransport}

For a \acrshort{gcr} species $j$, the central object of interest is the time, space and momentum-dependent function $\psi^j(t,{\bf x}, p)$, which is the ensemble and angle average (over realisations of magnetic inhomogeneities and momentum direction $\Omega_{\bf p}$, respectively) of the single-particle distribution function $f^j(t,{\bf x}, {\bf p})$ entering the Vlasov–Boltzmann equation. The function $\psi^j$ is connected to the flux by $\Phi^j = (v/(4\pi))\psi^j$, where $v$ is the particle's speed. It obeys the transport equation (see, e.g.,~\cite{Strong:2007nh}):
\begin{eqnarray}
&&\label{eq:comppropeq}
\frac{\partial \psi^j}{\partial t}-\nabla \cdot \left(\overline{\overline D}\nabla \psi^j \right)+ \vec{u}\cdot \nabla \psi^j-\frac{1}{3}\nabla \cdot\vec{u}\left(p\frac{\partial \psi^j}{\partial p}\right)-\frac{1}{p^2}\frac{\partial}{\partial p}\left(p^2{D_{pp}}\frac{\partial \psi^j}{\partial p}\right)=\\
&& q^j{+\frac{1}{p^2}\frac{\partial}{\partial p}\left[p^2 \left(\frac{\dd p}{\dd t}\right)\psi^j\right]}{-\Gamma_{\rm tot}^j\,\psi^j}{+\sum_i \psi^i\otimes\Gamma^{i\to j}}\,.\nonumber
\end{eqnarray}
The left-hand side of Eq.~\eqref{eq:comppropeq} contains transport terms of collisionless origin, i.e., from the scattering onto the electromagnetic field irregularities.
The second to the fifth term correspond to the spatial diffusion with diffusion tensor $\overline{\overline D}(\vec{\bf x},E)$, a convective term associated to the \acrshort{ism} plasma velocity field $\vec{u}({\bf x})$, adiabatic energy changes, and reacceleration (i.e., magnetic inhomogeneity diffusion in momentum space) controlled by $D_{pp}$. The latter two terms are notably involved in the so-called {\it first} and {\it second order Fermi acceleration}, respectively. The right-hand side, besides a possible primary source term $q^j$ (first term), accounts for {\it collisional effects}, described by terms representing continuous losses (second term), catastrophic sinks (third term) and possible secondary sources (last term). Catastrophic sinks include both decays, if the species is unstable with lifetime at rest $\tau_{\rm dec}^j$, and inelastic interactions over all possible targets $t$ of the \acrshort{ism} with density $n^t_{\rm ISM}$, quantified by the cross-section $\sigma^{j+t}_{\rm inel}$, so that
  \begin{equation}
\Gamma_{\rm tot}^j=\frac{1}{\gamma\,\tau_{\rm dec}^j}+\sum_{t} n^t_{\rm ISM}\,v\,\sigma^{j+t}_{\rm inel}\;.
\label{eq:transport_sink}
\end{equation}
In practice, only H and He have significant densities in the \acrshort{ism} to be targets relevant above the percent level.
The last term in Eq.~\eqref{eq:comppropeq} represents a generic integral operator acting on $\psi^i$, which is more easily expressed in terms of the kinetic energy $E_{\rm k}=E-m$, rather than momentum variables\footnote{Note that $\psi(E_{\rm k})=\psi(p(E_{\rm k})) \cdot \dd p/\dd E_{\rm k}=\beta^{-1}\cdot\psi(p(E_{\rm k}))$, since $p^2=(m+E_{\rm k})^2-m^2=2mE_{\rm k}+E_{\rm k}^2$.}, so that
  \begin{equation}
\psi^i\otimes\Gamma^{i\to j}=\sum_{t} n_{\rm ISM}^t \int \dd E_{\rm k}^i \,v\, \frac{\dd \sigma_{\rm prod}^{i+t\to j}}{\dd E_{\rm k}^j}(E_{\rm k}^i,E_{\rm k}^j) \psi^i(E_{\rm k}^i)\,,\label{eq:spallsource}
\end{equation}
where we introduced the differential cross-sections to produce the secondary particle $j$, in the collision of the primary $i$ with the target $t$. They can be written as a function of the kinetic energy of the primary parent, $E_{\rm k}^i$, and of the secondary daughter, $E_{\rm k}^j$, as
  \begin{equation}
\frac{\dd \sigma_{\rm prod}^{i+t\to j+{\rm X}}}{\dd E_{\rm k}^j}(E_{\rm k}^i,E_{\rm k}^j) = \sigma^{i+t}_{\rm inel}(E_{\rm k}^i)\;\frac{\dd {\cal N}^{i+t\to j+{\rm X}}}{\dd E_{\rm k}^j}(E_{\rm k}^i,E_{\rm k}^j)\,,\label{eq:generalparam}
\end{equation}
with $\dd {\cal N}/\dd E_{\rm k}^j$ the multiplicity spectrum of the species $j$ in the collision of $i$ with $t$. Since data are typically scarce, regularities motivating semi-empirical formulae turn out to be useful in interpolating between and extrapolating beyond measurements, or to estimate cross-sections involving nuclei for which no measurement exists.

As an example of some of these regularities, away from the thresholds,
$\dd {\cal N}/\dd E_{\rm k}^j$ is only weakly dependent on $E_{\rm k}^i$ and depends on $i$ and $t$ mostly via a normalisation. To a good approximation, in spallation reactions, the kinetic energy per nucleon $E_{\rm k/n}=E_{\rm k}/A$, where $A$ is the mass number, is conserved (we come back to this in Sec.~\ref{sec:xs_other}), so that
  \begin{equation}
  \label{eq:SHA}
\frac{\dd {\cal N}^{i+t\to j+{\rm X}}}{\dd E_{\rm k}^j}(E_{\rm k}^i,E_{\rm k}^j)\simeq \kappa^{i+t\to j+{\rm X}}\delta\left(\frac{E_{\rm k}}{A_j}-\frac{E_{\rm k}^i}{A_i}\right)\,.
\end{equation}

Fragmentation and spallation cross-sections are hence strategic ingredients in numerous astroparticle physics processes related to the acceleration, propagation and detection of cosmic particles, both charged and neutrals (photons, neutrinos). In astrophysical settings, they limit for instance the maximum acceleration energy attainable in a source, and for the propagation from the source to the detector, they enter both as energy-loss channels and, above all, as source channels of the so-called {\it secondary} species~\cite{Serpico:2023yag}. In order to isolate these effects from other interesting and poorly known astrophysical aspects, however, one should reduce the current cross-section uncertainties below the differences spanned by several viable astrophysical scenarios.

\subsection{Isotopic production cross-sections}
\label{sec:XS_prodnuc}

The history of cross-sections and \acrshort{gcr}s goes a very long way. Indeed, nuclear/particle and \acrshort{cr} topics were one and the same until the 1950s, before becoming two communities going their separate ways and addressing different questions. With the flight of many balloons and space experiments from the 1950s to the 1970s, it was realised that the poor accuracy of the nuclear cross-sections was a limitation for the interpretation of their data (e.g., \cite{1979ICRC...14..146R}). The situation back then had strong similarities with the current one, with dedicated studies to identify the needed reactions and then the set-up of long term programs for these new measurements (that started in the 1980s). Most of these data are still of use and remain the most accurate for many reactions.

Below, the procedure devised in Refs.~\cite{2018PhRvC..98c4611G,2024PhRvC.109f4914G} is recalled to provide a priority list of nuclear production cross-sections to be measured and improved, in order to profit from the current \acrshort{cr} data precision.
Throughout this section, the straight-ahead approximation Eq.~\eqref{eq:SHA} is used, in which the kinetic energy per nucleon $E_{\rm k/n}$ is conserved in nuclear reactions. As a result, the quantities of interest are the total (and not the differential) production cross-section $\sigma_{\rm prod}^{\ijk}$ and their uncertainties, with $i$ the \acrshort{gcr} projectile, $j$ the \acrshort{ism} target and $k$ the fragment. Actually, for \acrshort{gcr} propagation studies, cumulative cross-sections are used, i.e.,
\begin{equation}
  \sigma_{\rm cumul}^{\ijk} =
    \sigma_{\rm prod}^{\ijk} + \sum_{\rm g\in ghosts}\sigma_{\rm prod}^{i+j\to g} \cdot {\cal B}r(g\to k)\;.
    \label{eq:ghosts}
\end{equation}
In this cumulative, the so-called {\em ghosts} are short-lived nuclei with half-live $\lesssim 100$\,kyr, i.e., decay times much shorter than the propagation time, ending their decay chain into fragment $k$ with a branching ratio ${\cal B}r(g\to k)$. 

In practice, a network of more than a thousand reactions is involved for $Z<30$, and many more for ultra-heavy nuclei. Nuclear data and codes need to provide both the direct and ghost production of any \acrshort{gcr} fragment.
To illustrate the severity of the situation and of the needs, a brief summary of the existing nuclear data and models and their limitation is provided.

\subsubsection{Status of nuclear data and models}
The \acrshort{ism} is made of $\sim\!90\%$ of H and $\sim\!10\%$ of He in number, with only traces of heavier elements. With the few percent precision of \acrshort{cr} data, the need to include the interactions with C and O in the \acrshort{ism} is getting closer and should be re-evaluated in the future. But for current data, the requirements on the production cross-section precision is typically the percent level on H and tens of percents on He (see next section). Actually, for the latter, very few data exist, and all \acrshort{gcr} studies rely on the scaling proposed more than 35 years ago in Ref.~\cite{1988PhRvC..37.1490F}, which has limitations, e.g., for the production of light elements from heavy projectiles \cite{2022A&A...668A...7M}.

\paragraph{Reference nuclear data in \texorpdfstring{\acrshort{gcr}}{GCR} studies} Nuclear data are obtained from two main techniques: (i) heavy ion beams on H effective targets, i.e., liquid hydrogen or CH$_2$ and C subtraction, and outgoing fragments identified by a spectrometer; (ii) targets of heavy elements irradiated by a proton beam with the cross-sections determined either by $\gamma$-spectrometry, whenever fragments are radioactive isotopes, or after chemical processing by mass spectroscopy for the long-lived and stable isotopes produced.

The current body of data and models (based on these data) used for contemporary \acrshort{gcr} studies is a combination of a patchy collection of reactions measured by the nuclear and particle physics community, with some data dating back to the 1950s, and systematic measurements led by different groups over the last 40 years; for a comprehensive compilation, see Ref.~\cite{MichelAndOtuka2014}.
One figurehead of the \acrshort{cr} community, Bill Webber, led and coordinated many efforts over two decades, 1980s to 2000s, using mostly $Z<26$ beams, to measure isotopic production cross-sections on liquid hydrogen, carbon and methylene CH$_2$ targets, in the energy range $\sim\!400$--800\,MeV/n~\cite{1982ApJ...260..894W, 1988PhRvC..37.1490F, 1990PhRvC..41..520W, 1990PhRvC..41..533W, 1990ICRC....3..428T, 1990PhRvC..41..547W, 1990PhRvC..41..566W, 1996PhRvC..53..347K, 1997ApJ...479..504C, 1997PhRvC..56..398K, 1997PhRvC..56.1536C, 1998PhRvC..58.3539W, 1998ApJ...508..940W, 1998ApJ...508..949W}.
Extensive efforts were also driven by the study of space-flight radiation shielding applications (see Sec.~\ref{sec:Transverse_SpaceRadiation}) and the study of cosmogenic isotopes \cite{2002AIPC..610..285Z, 2004AdSpR..34.1383C} (see Sec.~\ref{sec:Transverse_Cosmogenic}): for the former,
a large body of data, using beams of light to heavy species, were obtained in the 2000s by Zeitlin's group \cite{1997PhRvC..56..388Z, 2001PhRvC..64b4902Z, 2007PhRvC..76a4911Z, 2008PhRvC..77c4605Z, 2011PhRvC..83c4909Z}, but for charge-changing cross-sections only (not directly of interest for \acrshort{gcr} studies); for the latter, extensive measurements using proton~\cite{1998LPI....29.1189S} and neutron~\cite{2013NIMPB.294..470R} beams were carried out from the 1990s, and over two decades, by R.~Michel's group~\cite{1989Ana...114..287M, 1990NIMPB..52..588D, 1993NIMPB..82....9B, 1995NIMPB.103..183M, 1996NIMPB.114...91S, 1997NIMPB.129..153M, 1998NIMPB.145..449L, 2005NIMPB.229....1L, 2006NIMPA.562..760L, 2008NIMPB.266....2A}, and also by J.~Sisterson's group~\cite{1992LPI....23.1305S, 1994NIMPB..92..510S, 1997NIMPB.123..324S, 1998LPI....29.1234S, 2000LPI....31.1432S, 2002NIMPB.196..239K, 2006NIMPB.251....1S}.
To this list, a relatively recent and very useful body of high-precision Fe fragmentation cross-sections can also be added, down to Li fragments \cite{2007PhRvC..75d4603V, 2008PhRvC..78c4615T}. However, it is fair to say the reactions of interest for the nuclear physics community nowadays involve ultra-heavy species, highly deformed nuclei, and/or short-lived radioactive beams, which is not providing further data for \acrshort{gcr} science.

\paragraph{Nuclear codes status and perspectives}
To account for the lack of data for many reactions and energies, dedicated formulae were developed as early as the 1960s~\cite{1966ZNatA..21.1027R}. Parametric codes soon followed to describe both the inelastic (see Sec.~\ref{sec:XS_inelastic}) and production cross-sections, with the semi-empirical parametrisation in the \texttt{YIELDX} code of Silberberg and Tsao's group~\cite{1973ApJS...25..315S, 1973ApJS...25..335S, 1983ApJS...51..271L, 1985ApJS...58..873S, 1990PhR...191..351S, 1993PhRvC..47.1225S, 1993PhRvC..47.1257T, 1998ApJ...501..911S, 1998ApJ...501..920T} and in the \texttt{WNEW} code of Webber and coworkers~\cite{1990PhRvC..41..566W, 1998ApJ...508..940W, 1998ApJ...508..949W, 1998PhRvC..58.3539W, 2003ApJS..144..153W} and other efforts (e.g., \cite{Cucinotta:2022aar} -- we refer the reader to Sec.~5 of Ref.~\cite{DavidLeya2019} for a recent detailed review of nuclear models and the various interaction mechanisms). Both were developed in the 1980s and updated till the 2000s, and by comparing the prediction of these codes (fit on older data) to new nuclear data, the former was found to be better (resp. worse) than the latter for reactions without (resp. with) data~\cite{1990ICRC....3..420C, 1998ApJ...501..911S}.
These codes remain the underlying models (original \texttt{FORTRAN} code) of the widely used numerical \texttt{GALPROP} package~\cite{2022ApJS..262...30P} for the propagation of relativistic $Z\leq 30$ \acrshort{gcr}s. However, to improve on these models, \texttt{GALPROP} combines several parametric formulae re-normalised to data and direct fits~\cite{2001ICRC....5.1836M, 2003ApJ...586.1050M, 2003ICRC....4.1969M, 2004AdSpR..34.1288M, 2005AIPC..769.1612M} and also uses the parametrisation of~Ref.~\cite{2012A&A...539A..88C} for light isotopes.
The overall accuracy of these codes is difficult to assess, but is estimated to be in the 10\%-20\% range. Recent works have also shown the importance of continuously importing more recent nuclear data and sometimes less important \acrshort{gcr} production channels to keep improving these nuclear predictions~\cite{2018JCAP...01..055R, 2018JCAP...07..006E, 2022A&A...668A...7M, Maurin:10Beto9Be_L2022}. It is also worth noticing that these parametrisations need further improvements, as they assume energy-independent cross-sections above a few GeV/n, whereas inelastic cross-sections are known to rise~\cite{2011PhRvL.107u2002B, 2015PhRvD..92k4021B}.

Outside the \acrshort{gcr} community, other parametric codes exist (\texttt{EPACS} \cite{1990PhRvC..42.2546S, 2000PhRvC..61c4607S, 2012PhRvC..86a4601S}, \texttt{SPACS} \cite{2014PhRvC..90f4605S, 2016PhRvC..94c9901S}, \texttt{FRACS} \cite{2017PhRvC..95c4608M, 2018NuScT..29...96S}, \texttt{NUCFRAG} \cite{1986NIMPB..18..225W, 1987CoPhC..47..281B, 1994NIMPB..94...95W, 2012NIMPA.678...21A}, \texttt{TALYS} \cite{2023EPJA...59..131K}) as well as \acrshort{mc} simulation codes and event generators (\texttt{FLUKA} \cite{2011NIMPB.269.2850B}, \texttt{MCNP6} with \texttt{CEM} and \texttt{LAQGSM} \cite{2011EPJP..126...49M, 2014NIMPA.764...59M, 2015NIMPB.356..135K}, \texttt{PHITS} \cite{1995PhRvC..52.2620N, 2010AdSpR..45..892S, 2014cosp...40E3081S}, \texttt{SHIELD-HIT} \cite{2012PMB....57.4369H}, etc.). These models are benchmarked and compared with overall fair agreement (see also Sec.~\ref{sec:xscode_spaceradiation}). The \texttt{Geant4} framework~\cite{2010NIMPB.268..604P} also provides many options, including specific cascade models in the above list, that can be used, combined or compared.
In terms of accuracy, these transport codes cannot replace the above-discussed ones tailored for \acrshort{gcr} studies, but they could probably bridge some gaps in the data for specific regimes, despite this requiring dedicated studies and careful evaluations.

Overall, there is no free lunch with nuclear cross-section data and codes. Without new data, the margin of progress is probably thin, and will be based on painful compilations of missed data in the literature and possible updates and systematic benchmarking of existing codes. Machine learning techniques could possibly bring some improvements~\cite{2020ChPhC..44a4104M, 2022ChPhC..46g4104M, 2022JPhG...49h5102P, 2024PhRvC.109e4610I}, but this remains to be proven. Gathering new high-precision nuclear data seems to be the only path to go forward.

\subsubsection{From \texorpdfstring{\acrshort{gcr}}{GCRs} data precision to desired cross-section precision}

The impact of the production cross-section $\sigma_{abc}\equiv\sigma_{\rm prod}^{\abc}$, on the flux $\psi^j$ of a given \acrshort{gcr} isotope or element $j$, is quantified in terms of the relative difference between the standard or reference flux calculation,
$\psi^j_{\rm ref}$, and the calculation where this cross-section is set to zero, $\psi^j_{\sigma_{abc}=0}$. We thus define the coefficients
   \begin{equation}
   \label{eq:coeffs_fabc}
      f^j_{abc}(E_{\rm k/n}) \equiv 1-\frac{\psi^j_{\sigma_{abc}=\,0}(E_{\rm k/n})}{\psi^j_{\rm ref}(E_{\rm k/n})}\;,
   \end{equation}
whose ranking is equivalent to rank the most important production cross-sections~\cite{2018PhRvC..98c4611G}. These coefficients vary with energy, because the various \acrshort{gcr} progenitors contributing to $j$ have different energy dependence, owing to the energy-dependent solution of the transport equation and, obviously, to the energy dependence of $\sigma_{abc}$ itself. Moreover, if we decompose $\psi^j_{\rm tot}$ into a primary and secondary origin, i.e., $\psi^j_{\rm tot}\equiv\psi^j_{\rm prim} + \psi^j_{\rm sec}$, then by definition $\psi^j_{\rm prim}$ does not depend on the production cross-sections, and $f_{abc}$ is roughly the contributing fraction of the reaction $a+b\to c$ to $\psi^j_{\rm sec}$~\cite{2018PhRvC..98c4611G}.

As shown in Ref.~\cite{2018PhRvC..98c4611G,2024PhRvC.109f4914G}, the $f_{abc}^j$ coefficients enable to link the cross-section uncertainties to the predicted flux uncertainties. Different plausible assumptions on the presence or absence of correlations between these cross-section datasets (i.e., in practice, on the modelling of the cross-sections based on these data) lead to different error propagation formulae. We report two noteworthy cases below, dropping the energy dependence for simplicity:
\begin{eqnarray}
\left(\frac{\Delta \psi^j_{\rm tot}}{\psi_{\rm tot}}\right)^{\rm mix} &\approx\;& f^j_{\rm sec}\, \sum_a
\sqrt{\sum_{b,c} \left(f^j_{abc} \frac{\Delta\sigma_{abc}}{\sigma_{abc}}\right)^2}\,,
\label{eq:errpropag_mix}\\
\left(\frac{\Delta \psi^j_{\rm tot}}{\psi^j_{\rm tot}}\right)^{\rm multi} &\approx\;& f^j_{\rm sec}\, \sqrt{\sum_{a,b} \frac{(\C_{ab}^j)^2}{N_{ab}}} \quad{\rm with}\quad (C^j_{ab})^2 \equiv \sum_{k} (f^j_{abc})^2 \frac{\sigma_{ab}}{\sigma_{abc}}\,,
\label{eq:errpropag_multi}
\end{eqnarray}
where $f^j_{\rm sec}=\psi^j_{\rm sec}/\psi^j_{\rm tot}$ is the secondary fraction of the flux $j$ considered, $\Delta\sigma_{abc}/\sigma_{abc}$ the relative uncertainty of the nuclear cross-section, $N_{ab}$ is the number of $a+b$ reactions considered in a new measurement campaign, and $\sigma_{ab}=\sigma_{\rm inel}^{a+b}$ is the inelastic cross-section of reaction $a+b$ (discussed in Sec.~\ref{sec:XS_inelastic}).
These formulae can be used to decide which reactions need to be measured in order to reach a relative precision on the modelled flux, $\psi^j$, better than $\epsilon$. The two cases are useful in the following situations:
\begin{itemize}
    \item {\em Eq.~\eqref{eq:errpropag_mix} for rough estimates of improvements brought by new measurements}: this case assumes that the data gathered so far -- and nuclear models based on these data -- have uncorrelated uncertainties for fragments of the same projectile, but correlated uncertainties for different projectiles. This formula can be used to illustrate how the flux uncertainties can be brought back below the sought precision~$\epsilon$, when a growing number of the most important reactions are perfectly measured (e.g., see Fig.~3 in~\cite{2024PhRvC.109f4914G});
    \item {\em Eq.~\eqref{eq:errpropag_multi} to determine the number of reactions $N_{ab}$ and beam time to reach $(\Delta\psi/\psi)<\epsilon$}: this case assumes a multinomial distribution of the measured fragments $c$ in the $a+b$ reaction, an approximation that fails for light fragments because of the multiplicity. As shown in Ref.~\cite{2024PhRvC.109f4914G}, demanding for all measured reaction that
    \begin{equation}
     N^j_{ab} \geq (f^j_{\rm sec}/\epsilon)^2 \cdot C^j_{ab} \cdot \left(\sum_{a,b} C^j_{ab}\right)
    \end{equation}
    is the optimal scheme to minimise the beam time, as it minimises the number of $N_{\rm tot}=\sum_{a,b}N_{ab}$ reactions that must be measured. The use of this equation is illustrated below to determine a wish list of measurements.
\end{itemize}

\subsubsection{A game-changing wish list for \texorpdfstring{$Z\leq 30$}{Z <= 30}}
\label{sec:xs_nucforecasts}
The above formulae can be used to set up a wish list for any \acrshort{gcr} species, energy, and \acrshort{cr} data precision. However, this is a tedious and incremental process, requiring a careful review and update of the best nuclear data available. So far, only \acrshort{gcr} fluxes from Li to Si have been analysed in detail~\cite{2018PhRvC..98c4611G,2024PhRvC.109f4914G}, relying on the \texttt{USINE} propagation code~\cite{2020CoPhC.24706942M}, and production cross-section parametrisations from \texttt{GALPROP}~\cite {2022ApJS..262...30P} updated for Li, Be, B and F isotopic production as described in Ref.~\cite{2022A&A...668A...7M,2024A&A...688A..17F}. The study of light nuclei ($Z<3$) and heavier ones ($14<Z<30$) are not published yet, but preliminary results are shown in Table~\ref{table:xs_ranked}.

\begin{table}
\footnotesize
\caption{Wish list of individual reactions sorted according to their flux impact $f_{abc}^j$, Eq.~\eqref{eq:coeffs_fabc}, in percent, at 10.6~GeV/n, for \acrshort{cr} secondary fluxes (only $f_{abc}^j>1\%$ are shown). We highlight reactions with short-lived fragments~({\bf bold}), reactions without nuclear data~($^\dagger$), and the ranking after which the cumulative is $>50\%$~($^\star$). Adapted from Tables V, VI, VII, and XI of Ref.~\cite{2024PhRvC.109f4914G} for $j$ equals Li, Be, B, and F respectively, and preliminary analysis for the rest.}
\label{table:xs_ranked}
\begin{tabular}{p{0.115\textwidth}p{0.02\textwidth}}
\hline\hline
\multicolumn{2}{c}{\vspace{-3mm}}\\
\multicolumn{2}{c}{Deuterium}\\
\hline
\multicolumn{2}{c}{\vspace{-2.7mm}}\\
$\!\!\rm ^{4}He\!+\!H\!\to\!^{2}$$\rm H$     &  38.           \\[-1mm]
$\!\!\rm ^{16}O\!+\!H\!\to\!^{2}$$\rm H$     &  9.0           \\[-1mm]
$\!\!\rm ^{3}He\!+\!H\!\to\!^{2}$$\rm H$     &  7.9$^\star$           \\[-1mm]
$\!\!\rm ^{4}He\!+\!He\!\to\!^{2}$$\rm H$    &  6.5           \\[-1mm]
$\!\!\rm ^{12}C\!+\!H\!\to\!^{2}$$\rm H$     &  5.8           \\[-1mm]
$\!\!\rm ^{4}He\!+\!H\!\to\!^{3}$$\rm He$    &  5.2           \\[-1mm]
$\!\!\rm ^{1}H\!+\!He\!\to\!^{2}$$\rm H$     &  4.7           \\[-1mm]
$\!\!\rm ^{56}Fe\!+\!H\!\to\!^{2}$$\rm H$    &  4.0$^\dagger$ \\[-1mm]
$\!\!\rm ^{28}Si\!+\!H\!\to\!^{2}$$\rm H$    &  2.6$^\dagger$ \\[-1mm]
$\!\!\rm ^{24}Mg\!+\!H\!\to\!^{2}$$\rm H$    &  2.1$^\dagger$ \\[-1mm]
$\!\!\rm ^{16}O\!+\!He\!\to\!^{2}$$\rm H$    &  1.5           \\[-1mm]
$\!\!\rm ^{20}Ne\!+\!H\!\to\!^{2}$$\rm H$    &  1.2$^\dagger$ \\[-1mm]
$\!\!\rm ^{16}O\!+\!H\!\to\!^{15}$$\rm N$    &  1.1           \\
\hline
\multicolumn{2}{c}{}\\[0.25cm]
\hline\hline
\multicolumn{2}{c}{\vspace{-2.4mm}}\\
\multicolumn{2}{c}{$^{3}$He}\\
\hline
\multicolumn{2}{c}{\vspace{-2.7mm}}\\
$\!\!\rm ^{4}He\!+\!H\!\to\!^{3}$$\rm He$    &  32. \\[-1mm]
$\!\!\rm \mathbf{^{4}He\!+\!H\!\to\!^{3}}$$\rm \mathbf{H}$     &  32.$^\star$ \\[-1mm]
$\!\!\rm ^{4}He\!+\!He\!\to\!^{3}$$\rm He$   &  5.5 \\[-1mm]
$\!\!\rm \mathbf{^{4}He\!+\!He\!\to\!^{3}}$$\rm \mathbf{H}$    &  5.5 \\[-1mm]
$\!\!\rm ^{16}O\!+\!H\!\to\!^{3}$$\rm He$    &  2.6 \\[-1mm]
$\!\!\rm \mathbf{^{16}O\!+\!H\!\to\!^{3}}$$\rm \mathbf{H}$     &  2.6 \\[-1mm]
$\!\!\rm ^{12}C\!+\!H\!\to\!^{3}$$\rm He$    &  2.1 \\[-1mm]
$\!\!\rm \mathbf{^{12}C\!+\!H\!\to\!^{3}}$$\rm \mathbf{H}$     &  2.1 \\[-1mm]
$\!\!\rm ^{56}Fe\!+\!H\!\to\!^{3}$$\rm He$   &  1.0 \\[-1mm]
$\!\!\rm \mathbf{^{56}Fe\!+\!H\!\to\!^{3}}$$\rm \mathbf{H}$    &  1.0 \\
\hline
\end{tabular}
\begin{tabular}{p{0.115\textwidth}p{0.025\textwidth}}
\hline\hline
\multicolumn{2}{c}{\vspace{-3mm}}\\
\multicolumn{2}{c}{Lithium}\\
\hline
\multicolumn{2}{c}{\vspace{-2.7mm}}\\
$\!\!\rm ^{16}O\!+\!H\!\to\!^{6}$$\rm Li$   &  15.2           \\[-1mm]
$\!\!\rm ^{12}C\!+\!H\!\to\!^{6}$$\rm Li$   &  12.5           \\[-1mm]
$\!\!\rm ^{12}C\!+\!H\!\to\!^{7}$$\rm Li$   &  9.93           \\[-1mm]
$\!\!\rm ^{16}O\!+\!H\!\to\!^{7}$$\rm Li$   &  9.74           \\[-1mm]
$\!\!\rm ^{11}B\!+\!H\!\to\!^{7}$$\rm Li$   &  2.92           \\[-1mm]
$\!\!\rm ^{16}O\!+\!He\!\to\!^{6}$$\rm Li$  &  2.86$^\dagger$ \\[-1mm]
$\!\!\rm ^{12}C\!+\!He\!\to\!^{6}$$\rm Li$  &  2.14$^\dagger$ \\[-1mm]
$\!\!\rm ^{7}Li\!+\!H\!\to\!^{6}$$\rm Li$   &  2.11           \\[-1mm]
$\!\!\rm ^{13}C\!+\!H\!\to\!^{7}$$\rm Li$   &  2.05$^\dagger$ \\[-1mm]
$\!\!\rm ^{56}Fe\!+\!H\!\to\!^{7}$$\rm Li$  &  2.03           \\[-1mm]
$\!\!\rm ^{15}N\!+\!H\!\to\!^{7}$$\rm Li$   &  1.95           \\[-1mm]
$\!\!\rm ^{16}O\!+\!H\!\to\!^{15}$$\rm N$   &  1.88           \\[-1mm]
$\!\!\rm ^{16}O\!+\!He\!\to\!^{7}$$\rm Li$  &  1.82$^\dagger$ \\[-1mm]
$\!\!\rm ^{56}Fe\!+\!H\!\to\!^{6}$$\rm Li$  &  1.74           \\[-1mm]
$\!\!\rm ^{12}C\!+\!He\!\to\!^{7}$$\rm Li$  &  1.71$^\dagger$$^\star$ \\[-1mm]
$\!\!\rm \mathbf{^{16}O\!+\!H\!\to\!^{13}}$${\rm\mathbf O}$ & 1.70 \\[-1mm]
$\!\!\rm ^{24}Mg\!+\!H\!\to\!^{6}$$\rm Li$ &   1.69$^\dagger$      \\[-1mm]
$\!\!\rm ^{28}Si\!+\!H\!\to\!^{6}$$\rm Li$ &   1.69$^\dagger$      \\[-1mm]
$\!\!\rm ^{13}C\!+\!H\!\to\!^{6}$$\rm Li$  &   1.68$^\dagger$      \\[-1mm]
$\!\!\rm \mathbf{^{16}O\!+\!H\!\to\!^{15}}$${\rm\mathbf O}$ & 1.68 \\[-1mm]
$\!\!\rm ^{16}O\!+\!H\!\to\!^{12}$$\rm C$  &   1.51                \\[-1mm]
$\!\!\rm ^{24}Mg\!+\!H\!\to\!^{7}$$\rm Li$ &   1.50$^\dagger$      \\[-1mm]
$\!\!\rm ^{28}Si\!+\!H\!\to\!^{7}$$\rm Li$ &   1.50$^\dagger$      \\[-1mm]
$\!\!\rm ^{10}B\!+\!H\!\to\!^{6}$$\rm Li$  &   1.41$^\dagger$      \\[-1mm]
$\!\!\rm ^{14}N\!+\!H\!\to\!^{6}$$\rm Li$  &   1.39                \\[-1mm]
$\!\!\rm ^{15}N\!+\!H\!\to\!^{6}$$\rm Li$  &   1.37                \\[-1mm]
$\!\!\rm ^{16}O\!+\!H\!\to\!^{14}$$\rm N$  &   1.27                \\[-1mm]
$\!\!\rm ^{20}Ne\!+\!H\!\to\!^{6}$$\rm Li$ &   1.16$^\dagger$      \\[-1mm]
$\!\!\rm ^{12}C\!+\!H\!\to\!^{11}$$\rm B$  &   1.15                \\[-1mm]
$\!\!\rm ^{7}Be\!+\!H\!\to\!^{6}$$\rm Li$  &   1.14$^\dagger$      \\[-1mm]
$\!\!\rm \mathbf{^{12}C\!+\!H\!\to\!^{11}}$${\rm\mathbf C}$ & 1.11 \\[-1mm]
$\!\!\rm ^{16}O\!+\!H\!\to\!^{13}$$\rm C$  &   1.10                \\[-1mm]
$\!\!\rm ^{20}Ne\!+\!H\!\to\!^{7}$$\rm Li$ &   1.02$^\dagger$        \\
\hline
\multicolumn{2}{c}{}\\[-0.5cm]
\end{tabular}
\begin{tabular}{p{0.115\textwidth}p{0.025\textwidth}}
\hline\hline
\multicolumn{2}{c}{\vspace{-3mm}}\\
\multicolumn{2}{c}{Beryllium}\\
\hline
\multicolumn{2}{c}{\vspace{-2.7mm}}\\
$\!\!\rm ^{16}O\!+\!H\!\to\!^{7}$$\rm Be$   &  16.8           \\[-1mm]
$\!\!\rm ^{12}C\!+\!H\!\to\!^{7}$$\rm Be$   &  14.5           \\[-1mm]
$\!\!\rm ^{12}C\!+\!H\!\to\!^{9}$$\rm Be$   &  8.25           \\[-1mm]
$\!\!\rm ^{16}O\!+\!H\!\to\!^{9}$$\rm Be$   &  6.09           \\[-1mm]
$\!\!\rm ^{16}O\!+\!He\!\to\!^{7}$$\rm Be$  &  2.87$^\dagger$ \\[-1mm]
$\!\!\rm ^{28}Si\!+\!H\!\to\!^{7}$$\rm Be$  &  2.72           \\[-1mm]
$\!\!\rm ^{12}C\!+\!H\!\to\!^{10}$$\rm Be$  &  2.63           \\[-1mm]
$\!\!\rm ^{24}Mg\!+\!H\!\to\!^{7}$$\rm Be$  &  2.58           \\[-1mm]
$\!\!\rm ^{14}N\!+\!H\!\to\!^{7}$$\rm Be$   &  2.26           \\[-1mm]
$\!\!\rm ^{12}C\!+\!He\!\to\!^{7}$$\rm Be$  &  2.26$^\dagger$ \\[-1mm]
$\!\!\rm ^{11}B\!+\!H\!\to\!^{9}$$\rm Be$   &  2.22           \\[-1mm]
$\!\!\rm ^{16}O\!+\!H\!\to\!^{10}$$\rm Be$  &  1.69           \\[-1mm]
$\!\!\rm ^{20}Ne\!+\!H\!\to\!^{7}$$\rm Be$  &  1.67$^\dagger$ \\[-1mm]
$\!\!\rm ^{56}Fe\!+\!H\!\to\!^{7}$$\rm Be$  &  1.66$^\star$           \\[-1mm]
$\!\!\rm ^{16}O\!+\!H\!\to\!^{12}$$\rm C$  &   1.61                \\[-1mm]
$\!\!\rm ^{10}B\!+\!H\!\to\!^{9}$$\rm Be$  &   1.53$^\dagger$      \\[-1mm]
$\!\!\rm ^{24}Mg\!+\!H\!\to\!^{9}$$\rm Be$ &   1.52                \\[-1mm]
$\!\!\rm ^{16}O\!+\!H\!\to\!^{15}$$\rm N$  &   1.41                \\[-1mm]
$\!\!\rm ^{56}Fe\!+\!H\!\to\!^{9}$$\rm Be$ &   1.33                \\[-1mm]
$\!\!\rm ^{12}C\!+\!He\!\to\!^{9}$$\rm Be$ &   1.29$^\dagger$      \\[-1mm]
$\!\!\rm ^{12}C\!+\!H\!\to\!^{11}$$\rm B$  &   1.28                \\[-1mm]
$\!\!\rm \mathbf{^{16}O\!+\!H\!\to\!^{15}}$${\rm\mathbf O}$ & 1.26 \\[-1mm]
$\!\!\rm \mathbf{^{12}C\!+\!H\!\to\!^{11}}$${\rm\mathbf C}$ & 1.24 \\[-1mm]
$\!\!\rm ^{15}N\!+\!H\!\to\!^{9}$$\rm Be$  &   1.22                \\[-1mm]
$\!\!\rm ^{16}O\!+\!H\!\to\!^{14}$$\rm N$  &   1.20                \\[-1mm]
$\!\!\rm ^{11}B\!+\!H\!\to\!^{10}$$\rm Be$ &   1.10                \\[-1mm]
$\!\!\rm ^{15}N\!+\!H\!\to\!^{7}$$\rm Be$  &   1.08                \\[-1mm]
$\!\!\rm \mathbf{^{16}O\!+\!H\!\to\!^{13}}$${\rm\mathbf O}$ & 1.08 \\[-1mm]
$\!\!\rm ^{16}O\!+\!He\!\to\!^{9}$$\rm Be$ &   1.05$^\dagger$      \\[-1mm]
$\!\!\rm ^{11}B\!+\!H\!\to\!^{7}$$\rm Be$  &   1.04                \\
\hline
\end{tabular}
\begin{tabular}{p{0.115\textwidth}p{0.025\textwidth}}
\hline\hline
\multicolumn{2}{c}{\vspace{-3mm}}\\
\multicolumn{2}{c}{Boron}\\
\hline
\multicolumn{2}{c}{\vspace{-2.7mm}}\\
$\!\!\rm ^{12}C\!+\!H\!\to\!^{11}$$\rm B$                   & 17.2 \\[-1mm]
$\!\!\rm \mathbf{^{12}C\!+\!H\!\to\!^{11}}$${\rm\mathbf C}$ & 16.6 \\[-1mm]
$\!\!\rm ^{16}O\!+\!H\!\to\!^{11}$$\rm B$                   & 10.8 \\[-1mm]
$\!\!\rm ^{12}C\!+\!H\!\to\!^{10}$$\rm B$                   & 7.85 \\[-1mm]
$\!\!\rm ^{16}O\!+\!H\!\to\!^{10}$$\rm B$                   & 7.05 \\[-1mm]
$\!\!\rm \mathbf{^{16}O\!+\!H\!\to\!^{11}}$${\rm\mathbf C}$ & 4.42 \\[-1mm]
$\!\!\rm ^{11}B\!+\!H\!\to\!^{10}$$\rm B$                   & 4.12$^\star$ \\[-1mm]
$\!\!\rm ^{12}C\!+\!He\!\to\!^{11}$$\rm B$ &   2.46$^\dagger$       \\[-1mm]
$\!\!\rm ^{16}O\!+\!H\!\to\!^{12}$$\rm C$  &   2.45                 \\[-1mm]
$\!\!\rm ^{15}N\!+\!H\!\to\!^{11}$$\rm B$  &   2.25                 \\[-1mm]
$\!\!\rm \mathbf{^{12}C\!+\!He\!\to\!^{11}}$${\rm\mathbf C}$ & 2.20$^\dagger$ \\[-1mm]
$\!\!\rm ^{16}O\!+\!H\!\to\!^{15}$$\rm N$  &   1.84                 \\[-1mm]
$\!\!\rm ^{16}O\!+\!He\!\to\!^{11}$$\rm B$ &   1.66$^\dagger$       \\[-1mm]
$\!\!\rm \mathbf{^{16}O\!+\!H\!\to\!^{15}}$${\rm\mathbf O}$ &  1.64 \\[-1mm]
$\!\!\rm ^{14}N\!+\!H\!\to\!^{11}$$\rm B$  &   1.64                 \\[-1mm]
$\!\!\rm ^{13}C\!+\!H\!\to\!^{11}$$\rm B$  &   1.52$^\dagger$       \\[-1mm]
$\!\!\rm \mathbf{^{12}C\!+\!H\!\to\!^{10}}$${\rm\mathbf C}$ &  1.45 \\[-1mm]
$\!\!\rm \mathbf{^{16}O\!+\!H\!\to\!^{13}}$${\rm\mathbf O}$ &  1.32 \\[-1mm]
$\!\!\rm ^{16}O\!+\!H\!\to\!^{14}$$\rm N$  &   1.31                 \\[-1mm]
$\!\!\rm ^{12}C\!+\!H\!\to\!^{10}$$\rm Be$ &   1.16                 \\[-1mm]
$\!\!\rm ^{12}C\!+\!He\!\to\!^{10}$$\rm B$ &   1.12$^\dagger$       \\[-1mm]
$\!\!\rm ^{16}O\!+\!He\!\to\!^{10}$$\rm B$ &   1.10$^\dagger$       \\[-1mm]
$\!\!\rm ^{24}Mg\!+\!H\!\to\!^{11}$$\rm B$ &   1.01$^\dagger$       \\
\hline
\end{tabular}
\begin{tabular}{p{0.12\textwidth}p{0.025\textwidth}}
\hline\hline
\multicolumn{2}{c}{\vspace{-3mm}}\\
\multicolumn{2}{c}{Fluorine}\\
\hline
\multicolumn{2}{c}{\vspace{-2.7mm}}\\
$\!\!\rm \mathbf{^{20}Ne\!+\!H\!\to\!^{19}}$${\rm\mathbf Ne}$ & 16.8  \\[-1mm]
$\!\!\rm ^{20}Ne\!+\!H\!\to\!^{19}$$\rm F$                    & 13.9  \\[-1mm]
$\!\!\rm ^{24}Mg\!+\!H\!\to\!^{19}$$\rm F$                    & 8.80  \\[-1mm]
$\!\!\rm ^{28}Si\!+\!H\!\to\!^{19}$$\rm F$                    & 8.12  \\[-1mm]
$\!\!\rm ^{22}Ne\!+\!H\!\to\!^{19}$$\rm F$                    & 6.13  \\[-1mm]
$\!\!\rm \mathbf{^{24}Mg\!+\!H\!\to\!^{19}}$${\rm\mathbf Ne}$ & 3.48  \\[-1mm]
$\!\!\rm ^{56}Fe\!+\!H\!\to\!^{19}$$\rm F$                    & 3.25  \\[-1mm]
$\!\!\rm ^{21}Ne\!+\!H\!\to\!^{19}$$\rm F$                    & 3.17  \\[-1mm]
$\!\!\rm \mathbf{^{28}Si\!+\!H\!\to\!^{19}}$${\rm\mathbf Ne}$ & 2.99  \\[-1mm]
$\!\!\rm ^{23}Na\!+\!H\!\to\!^{19}$$\rm F$                    & 2.30  \\[-1mm]
$\!\!\rm ^{25}Mg\!+\!H\!\to\!^{19}$$\rm F$                    & 2.24$^\star$  \\[-1mm]
$\!\!\rm ^{26}Mg\!+\!H\!\to\!^{19}$$\rm F$                    & 2.19  \\[-1mm]
$\!\!\rm \mathbf{^{20}Ne\!+\!He\!\to\!^{19}}$${\rm\mathbf Ne}$& 1.97$^\dagger$ \\[-1mm]
$\!\!\rm ^{20}Ne\!+\!He\!\to\!^{19}$$\rm F$                   & 1.73$^\dagger$ \\[-1mm]
$\!\!\rm ^{27}Al\!+\!H\!\to\!^{19}$$\rm F$                    & 1.65  \\[-1mm]
$\!\!\rm ^{24}Mg\!+\!H\!\to\!^{20}$$\rm Ne$                   & 1.36  \\[-1mm]
$\!\!\rm \mathbf{^{22}Ne\!+\!H\!\to\!^{19}}$${\rm\mathbf O}$  & 1.29  \\[-1mm]
$\!\!\rm ^{28}Si\!+\!He\!\to\!^{19}$$\rm F$                   & 1.21$^\dagger$ \\[-1mm]
$\!\!\rm ^{24}Mg\!+\!He\!\to\!^{19}$$\rm F$                   & 1.19$^\dagger$ \\[-1mm]
$\!\!\rm ^{24}Mg\!+\!H\!\to\!^{21}$$\rm Ne$                   & 1.08  \\[-1mm]
$\!\!\rm ^{24}Mg\!+\!H\!\to\!^{23}$$\rm Na$                   & 1.08  \\[-1mm]
$\!\!\rm ^{28}Si\!+\!H\!\to\!^{27}$$\rm Al$                   & 1.05  \\[-1mm]
$\!\!\rm ^{56}Fe\!+\!He\!\to\!^{19}$$\rm F$                   & 1.03$^\dagger$ \\[-1mm]
$\!\!\rm \mathbf{^{21}Ne\!+\!H\!\to\!^{19}}$${\rm\mathbf Ne}$ & 1.03$^\dagger$   \\
\hline
\end{tabular}
\begin{tabular}{p{0.118\textwidth}p{0.025\textwidth}}
\hline\hline
\multicolumn{2}{c}{\vspace{-3mm}}\\
\multicolumn{2}{c}{Scandium}\\
\hline
\multicolumn{2}{c}{\vspace{-2.7mm}}\\
$\!\!\rm ^{56}Fe\!+\!H\!\to\!^{45}$$\rm Sc$ &   43.9                  \\[-1mm]
$\!\!\rm \mathbf{^{56}Fe\!+\!H\!\to\!^{45}}$${\rm\mathbf Ti}$ & 6.35$^\star$  \\[-1mm]
$\!\!\rm ^{56}Fe\!+\!He\!\to\!^{45}$$\rm Sc$&   4.95$^\dagger$        \\[-1mm]
$\!\!\rm \mathbf{^{56}Fe\!+\!H\!\to\!^{45}}$${\rm\mathbf Ca}$ & 3.27  \\[-1mm]
$\!\!\rm ^{47}Ti\!+\!H\!\to\!^{45}$$\rm Sc$ &   3.24$^\dagger$        \\[-1mm]
$\!\!\rm ^{54}Fe\!+\!H\!\to\!^{45}$$\rm Sc$ &   3.07$^\dagger$        \\[-1mm]
$\!\!\rm ^{56}Fe\!+\!H\!\to\!^{55}$$\rm Fe$ &   2.76                  \\[-1mm]
$\!\!\rm ^{56}Fe\!+\!H\!\to\!^{47}$$\rm Ti$ &   2.63                  \\[-1mm]
$\!\!\rm ^{46}Ti\!+\!H\!\to\!^{45}$$\rm Sc$ &   2.51$^\dagger$        \\[-1mm]
$\!\!\rm ^{48}Ti\!+\!H\!\to\!^{45}$$\rm Sc$ &   2.45$^\dagger$        \\[-1mm]
$\!\!\rm \mathbf{^{46}Ti\!+\!H\!\to\!^{45}}$${\rm\mathbf Ti}$ & 2.24$^\dagger$ \\[-1mm]
$\!\!\rm ^{52}Cr\!+\!H\!\to\!^{45}$$\rm Sc$ &   2.02                  \\[-1mm]
$\!\!\rm ^{56}Fe\!+\!H\!\to\!^{49}$$\rm V$  &   2.02                  \\[-1mm]
$\!\!\rm ^{56}Fe\!+\!H\!\to\!^{46}$$\rm Ti$ &   1.98                  \\[-1mm]
$\!\!\rm ^{56}Fe\!+\!H\!\to\!^{52}$$\rm Cr$ &   1.93                  \\[-1mm]
$\!\!\rm ^{56}Fe\!+\!H\!\to\!^{51}$$\rm Cr$ &   1.93                  \\[-1mm]
$\!\!\rm ^{56}Fe\!+\!H\!\to\!^{54}$$\rm Mn$ &   1.91                  \\[-1mm]
$\!\!\rm ^{49}V\!+\!H\!\to\!^{45}$$\rm Sc$  &   1.90$^\dagger$        \\[-1mm]
$\!\!\rm \mathbf{^{54}Fe\!+\!H\!\to\!^{45}}$${\rm\mathbf Ti}$ & 1.84$^\dagger$ \\[-1mm]
$\!\!\rm ^{56}Fe\!+\!H\!\to\!^{53}$$\rm Mn$ &   1.79                  \\[-1mm]
$\!\!\rm ^{56}Fe\!+\!H\!\to\!^{55}$$\rm Mn$ &   1.53                  \\[-1mm]
$\!\!\rm ^{51}Cr\!+\!H\!\to\!^{45}$$\rm Sc$ &   1.49$^\dagger$        \\[-1mm]
$\!\!\rm ^{55}Mn\!+\!H\!\to\!^{45}$$\rm Sc$ &   1.47                  \\[-1mm]
$\!\!\rm ^{55}Fe\!+\!H\!\to\!^{45}$$\rm Sc$ &   1.40$^\dagger$        \\[-1mm]
$\!\!\rm ^{56}Fe\!+\!H\!\to\!^{50}$$\rm V$  &   1.32                  \\[-1mm]
$\!\!\rm ^{56}Fe\!+\!H\!\to\!^{50}$$\rm Cr$ &   1.30                  \\[-1mm]
$\!\!\rm ^{53}Mn\!+\!H\!\to\!^{45}$$\rm Sc$ &   1.25$^\dagger$        \\[-1mm]
$\!\!\rm ^{56}Fe\!+\!H\!\to\!^{48}$$\rm Ti$ &   1.22                  \\[-1mm]
$\!\!\rm ^{50}V\!+\!H\!\to\!^{45}$$\rm Sc$  &   1.11$^\dagger$        \\[-1mm]
$\!\!\rm ^{50}Cr\!+\!H\!\to\!^{45}$$\rm Sc$ &   1.00                  \\
\hline
\multicolumn{2}{c}{}\\[-0.75cm]
\end{tabular}
\begin{tabular}{p{0.118\textwidth}p{0.022\textwidth}}
\hline\hline
\multicolumn{2}{c}{\vspace{-3mm}}\\
\multicolumn{2}{c}{Titanium}\\
\hline
\multicolumn{2}{c}{\vspace{-2.7mm}}\\
$\!\!\rm ^{56}Fe\!+\!H\!\to\!^{47}$$\rm Ti$ &   17.3                   \\[-1mm]
$\!\!\rm ^{56}Fe\!+\!H\!\to\!^{48}$$\rm Ti$ &   11.1                   \\[-1mm]
$\!\!\rm ^{56}Fe\!+\!H\!\to\!^{46}$$\rm Ti$ &   10.5                   \\[-1mm]
$\!\!\rm \mathbf{^{56}Fe\!+\!H\!\to\!^{48}}$${\rm\mathbf V}$  & 8.71   \\[-1mm]
$\!\!\rm \mathbf{^{56}Fe\!+\!H\!\to\!^{46}}$${\rm\mathbf Sc}$ & 4.34$^\star$   \\[-1mm]
$\!\!\rm ^{56}Fe\!+\!H\!\to\!^{49}$$\rm Ti$ &   3.36                   \\[-1mm]
$\!\!\rm ^{56}Fe\!+\!H\!\to\!^{55}$$\rm Fe$ &   2.80                   \\[-1mm]
$\!\!\rm ^{56}Fe\!+\!H\!\to\!^{49}$$\rm V$  &   2.63                   \\[-1mm]
$\!\!\rm ^{56}Fe\!+\!H\!\to\!^{51}$$\rm Cr$ &   2.08                   \\[-1mm]
$\!\!\rm ^{56}Fe\!+\!H\!\to\!^{54}$$\rm Mn$ &   2.02                   \\[-1mm]
$\!\!\rm ^{56}Fe\!+\!He\!\to\!^{47}$$\rm Ti$&   1.94$^\dagger$         \\[-1mm]
$\!\!\rm ^{56}Fe\!+\!H\!\to\!^{53}$$\rm Mn$ &   1.93                   \\[-1mm]
$\!\!\rm \mathbf{^{56}Fe\!+\!H\!\to\!^{47}}$${\rm\mathbf V}$  & 1.93   \\[-1mm]
$\!\!\rm ^{56}Fe\!+\!H\!\to\!^{52}$$\rm Cr$ &   1.86                   \\[-1mm]
$\!\!\rm \mathbf{^{56}Fe\!+\!H\!\to\!^{47}}$${\rm\mathbf Sc}$ & 1.75   \\[-1mm]
$\!\!\rm ^{56}Fe\!+\!H\!\to\!^{50}$$\rm V$  &   1.70                   \\[-1mm]
$\!\!\rm ^{56}Fe\!+\!H\!\to\!^{55}$$\rm Mn$ &   1.53                   \\[-1mm]
$\!\!\rm ^{54}Fe\!+\!H\!\to\!^{46}$$\rm Ti$ &   1.36$^\dagger$         \\[-1mm]
$\!\!\rm \mathbf{^{54}Fe\!+\!H\!\to\!^{48}}$${\rm\mathbf V}$  & 1.33$^\dagger$ \\[-1mm]
$\!\!\rm ^{54}Fe\!+\!H\!\to\!^{47}$$\rm Ti$ &   1.29$^\dagger$         \\[-1mm]
$\!\!\rm ^{56}Fe\!+\!H\!\to\!^{50}$$\rm Cr$ &   1.25                   \\[-1mm]
$\!\!\rm ^{56}Fe\!+\!He\!\to\!^{48}$$\rm Ti$&   1.24$^\dagger$         \\[-1mm]
$\!\!\rm ^{48}Ti\!+\!H\!\to\!^{47}$$\rm Ti$ &   1.23$^\dagger$         \\[-1mm]
$\!\!\rm ^{56}Fe\!+\!He\!\to\!^{46}$$\rm Ti$&   1.17$^\dagger$         \\[-1mm]
$\!\!\rm ^{47}Ti\!+\!H\!\to\!^{46}$$\rm Ti$ &   1.09$^\dagger$        \\[-1mm]
$\!\!\rm \mathbf{^{56}Fe\!+\!He\!\to\!^{48}}$${\rm\mathbf V}$ & 1.02$^\dagger$ \\
\hline
\end{tabular}
\begin{tabular}{p{0.118\textwidth}p{0.023\textwidth}}
\hline\hline
\multicolumn{2}{c}{\vspace{-3mm}}\\
\multicolumn{2}{c}{Vanadium}\\
\hline
\multicolumn{2}{c}{\vspace{-2.7mm}}\\
$\!\!\rm ^{56}Fe\!+\!H\!\to\!^{49}$$\rm V$ &   30.7                    \\[-1mm]
$\!\!\rm ^{56}Fe\!+\!H\!\to\!^{50}$$\rm V$ &   22.3$^\star$                    \\[-1mm]
$\!\!\rm ^{56}Fe\!+\!H\!\to\!^{51}$$\rm V$ &   5.66                    \\[-1mm]
$\!\!\rm \mathbf{^{56}Fe\!+\!H\!\to\!^{51}}$${\rm\mathbf Ti}$ & 3.93   \\[-1mm]
$\!\!\rm ^{56}Fe\!+\!He\!\to\!^{49}$$\rm V$&   3.59$^\dagger$          \\[-1mm]
$\!\!\rm \mathbf{^{56}Fe\!+\!H\!\to\!^{49}}$${\rm\mathbf Cr}$ & 3.26   \\[-1mm]
$\!\!\rm ^{56}Fe\!+\!H\!\to\!^{52}$$\rm Cr$&   2.77                    \\[-1mm]
$\!\!\rm ^{56}Fe\!+\!He\!\to\!^{50}$$\rm V$&   2.61$^\dagger$          \\[-1mm]
$\!\!\rm ^{54}Fe\!+\!H\!\to\!^{49}$$\rm V$ &   2.58$^\dagger$          \\[-1mm]
$\!\!\rm ^{56}Fe\!+\!H\!\to\!^{55}$$\rm Fe$&   2.50                    \\[-1mm]
$\!\!\rm ^{56}Fe\!+\!H\!\to\!^{51}$$\rm Cr$&   2.17                    \\[-1mm]
$\!\!\rm ^{52}Cr\!+\!H\!\to\!^{49}$$\rm V$ &   2.04                    \\[-1mm]
$\!\!\rm ^{56}Fe\!+\!H\!\to\!^{54}$$\rm Mn$&   1.79                    \\[-1mm]
$\!\!\rm ^{56}Fe\!+\!H\!\to\!^{53}$$\rm Mn$&   1.73                    \\[-1mm]
$\!\!\rm ^{52}Cr\!+\!H\!\to\!^{50}$$\rm V$ &   1.72                    \\[-1mm]
$\!\!\rm ^{52}Cr\!+\!H\!\to\!^{51}$$\rm V$ &   1.69                    \\[-1mm]
$\!\!\rm ^{56}Fe\!+\!H\!\to\!^{55}$$\rm Mn$&   1.59                    \\[-1mm]
$\!\!\rm ^{55}Mn\!+\!H\!\to\!^{49}$$\rm V$ &   1.34                    \\[-1mm]
$\!\!\rm ^{51}Cr\!+\!H\!\to\!^{49}$$\rm V$ &   1.30$^\dagger$          \\[-1mm]
$\!\!\rm ^{51}Cr\!+\!H\!\to\!^{50}$$\rm V$ &   1.23$^\dagger$          \\[-1mm]
$\!\!\rm ^{50}V\!+\!H\!\to\!^{49}$$\rm V$  &   1.18$^\dagger$          \\[-1mm]
$\!\!\rm ^{54}Fe\!+\!H\!\to\!^{50}$$\rm V$ &   1.14$^\dagger$          \\[-1mm]
$\!\!\rm ^{55}Fe\!+\!H\!\to\!^{49}$$\rm V$ &   1.07$^\dagger$          \\[-1mm]
$\!\!\rm ^{56}Fe\!+\!H\!\to\!^{50}$$\rm Cr$&   1.03                    \\[-1mm]
$\!\!\rm ^{53}Mn\!+\!H\!\to\!^{49}$$\rm V$ &   1.01$^\dagger$          \\
\hline
\end{tabular}
\begin{tabular}{p{0.118\textwidth}p{0.023\textwidth}}
\hline\hline
\multicolumn{2}{c}{\vspace{-3mm}}\\
\multicolumn{2}{c}{Chromium}\\
\hline
\multicolumn{2}{c}{\vspace{-2.7mm}}\\
$\!\!\rm ^{56}Fe\!+\!H\!\to\!^{52}$$\rm Cr$ &   22.7                   \\[-1mm]
$\!\!\rm ^{56}Fe\!+\!H\!\to\!^{51}$$\rm Cr$ &   21.2                   \\[-1mm]
$\!\!\rm ^{56}Fe\!+\!H\!\to\!^{50}$$\rm Cr$ &   12.9$^\star$                   \\[-1mm]
$\!\!\rm ^{56}Fe\!+\!H\!\to\!^{53}$$\rm Cr$ &   5.96                   \\[-1mm]
$\!\!\rm \mathbf{^{56}Fe\!+\!H\!\to\!^{52}}$${\rm\mathbf Mn}$ & 4.08   \\[-1mm]
$\!\!\rm ^{56}Fe\!+\!H\!\to\!^{55}$$\rm Fe$ &   3.11                   \\[-1mm]
$\!\!\rm ^{56}Fe\!+\!He\!\to\!^{52}$$\rm Cr$&   2.81$^\dagger$         \\[-1mm]
$\!\!\rm ^{56}Fe\!+\!He\!\to\!^{51}$$\rm Cr$&   2.62$^\dagger$         \\[-1mm]
$\!\!\rm ^{56}Fe\!+\!H\!\to\!^{53}$$\rm Mn$ &   2.59                   \\[-1mm]
$\!\!\rm ^{56}Fe\!+\!H\!\to\!^{54}$$\rm Mn$ &   2.31                   \\[-1mm]
$\!\!\rm \mathbf{^{54}Fe\!+\!H\!\to\!^{52}}$${\rm\mathbf Mn}$ & 2.23   \\[-1mm]
$\!\!\rm ^{56}Fe\!+\!H\!\to\!^{55}$$\rm Mn$ &   1.96                   \\[-1mm]
$\!\!\rm ^{54}Fe\!+\!H\!\to\!^{51}$$\rm Cr$ &   1.88                   \\[-1mm]
$\!\!\rm ^{56}Fe\!+\!H\!\to\!^{54}$$\rm Cr$ &   1.86                   \\[-1mm]
$\!\!\rm ^{52}Cr\!+\!H\!\to\!^{51}$$\rm Cr$ &   1.62                   \\[-1mm]
$\!\!\rm ^{56}Fe\!+\!He\!\to\!^{50}$$\rm Cr$&   1.59$^\dagger$         \\[-1mm]
$\!\!\rm ^{54}Fe\!+\!H\!\to\!^{50}$$\rm Cr$ &   1.21                   \\[-1mm]
$\!\!\rm ^{55}Mn\!+\!H\!\to\!\!^{52}$$\rm Cr$ &   1.11                   \\
\hline
\end{tabular}
\begin{tabular}{p{0.132\textwidth}p{0.023\textwidth}}
\hline\hline
\multicolumn{2}{c}{\vspace{-3mm}}\\
\multicolumn{2}{c}{Manganese}\\
\hline
\multicolumn{2}{c}{\vspace{-2.7mm}}\\
$\!\!\rm ^{56}Fe\!+\!H\!\to\!^{53}$$\rm Mn$ &   30.2                   \\[-1mm]
$\!\!\rm ^{56}Fe\!+\!H\!\to\!^{55}$$\rm Mn$ &   27.2$^\star$                   \\[-1mm]
$\!\!\rm ^{56}Fe\!+\!H\!\to\!^{54}$$\rm Mn$ &   18.6                   \\[-1mm]
$\!\!\rm ^{54}Fe\!+\!H\!\to\!^{53}$$\rm Mn$ &   4.78                   \\[-1mm]
$\!\!\rm ^{56}Fe\!+\!He\!\to\!^{53}$$\rm Mn$&   4.00$^\dagger$         \\[-1mm]
$\!\!\rm ^{56}Fe\!+\!He\!\to\!^{55}$$\rm Mn$&   3.62$^\dagger$         \\[-1mm]
$\!\!\rm \mathbf{^{54}Fe\!+\!H\!\to\!^{53}}$${\rm\mathbf Fe}$ &   3.06 \\[-1mm]
$\!\!\rm \mathbf{^{56}Fe\!+\!H\!\to\!^{53}}$${\rm\mathbf Fe}$ &   2.64 \\[-1mm]
$\!\!\rm ^{56}Fe\!+\!He\!\to\!^{54}$$\rm Mn$&   2.47$^\dagger$         \\[-1mm]
$\!\!\rm ^{56}Fe\!+\!H\!\to\!^{55}$$\rm Fe$ &   2.33                   \\[-1mm]
$\!\!\rm ^{55}Fe\!+\!H\!\to\!^{53}$$\rm Mn$ &   1.30$^\dagger$         \\[-1mm]
$\!\!\rm ^{55}Mn\!+\!H\!\to\!^{54}$$\rm Mn$ &   1.03                   \\
\hline
\end{tabular}
\end{table}

\begin{table}
\small
\centering
\caption{Required number of interactions to be recorded, ordered by increasing charge and mass of the projectiles, in order to reach a modelling precision $\lesssim 1\%$ on \acrshort{gcr} fluxes Li, Be, B and F. Adapted from Table~IV of Ref.~\cite{2024PhRvC.109f4914G}.\vspace{1.7mm}}
\label{tab:ninter}
\begin{tabular}{r@{+}lcc}
\hline\hline
\multicolumn{2}{c}{\vspace{-3.5mm}}\\
\multicolumn{2}{c}{Reaction} & $N_{\text{int}}$ \\
\hline
\multicolumn{2}{c}{\vspace{-3.mm}}\\
  $^{7}\text{Li}$  & $\text{H}$  & 5k \\ 
  $^{10}\text{B}$  & $\text{H}$  & 5k \\ 
  $^{11}\text{B}$  & $\text{H}$  & 10k\\ 
  $^{12}\text{C}$  & $\text{H}$  & 50k\\ 
  $^{12}\text{C}$  & $\text{He}$ & 10k\\ 
  $^{13}\text{C}$  & $\text{H}$  & 5k \\ 
  $^{14}\text{N}$  & $\text{H}$  & 10k\\ 
  $^{15}\text{N}$  & $\text{H}$  & 10k\\ 
  $^{16}\text{O}$  & $\text{H}$  & 60k\\ 
  $^{16}\text{O}$  & $\text{He}$ & 20k\\ 
\hline
\end{tabular}
\quad
\begin{tabular}{r@{+}lcc}
\hline\hline
\multicolumn{2}{c}{\vspace{-3.5mm}}\\
\multicolumn{2}{c}{Reaction} & $N_{\text{int}}$ \\
\hline
\multicolumn{2}{c}{\vspace{-3.mm}}\\
  $^{20}\text{Ne}$ & $\text{H}$  & 50k\\ 
  $^{20}\text{Ne}$ & $\text{He}$ & 10k\\ 
  $^{21}\text{Ne}$ & $\text{H}$  & 10k\\ 
  $^{22}\text{Ne}$ & $\text{H}$  & 20k\\ 
  $^{22}\text{Ne}$ & $\text{He}$ & 5k \\ 
  $^{23}\text{Na}$ & $\text{H}$  & 10k\\ 
  $^{24}\text{Mg}$ & $\text{H}$  & 50k\\ 
  $^{24}\text{Mg}$ & $\text{He}$ & 10k\\ 
  $^{25}\text{Mg}$ & $\text{H}$  & 10k\\ 
  $^{26}\text{Mg}$ & $\text{H}$  & 10k\\ 
  $^{27}\text{Al}$ & $\text{H}$  & 10k\\ 
\hline
\end{tabular}
\quad
\begin{tabular}{r@{+}lcc}
\hline\hline
\multicolumn{2}{c}{\vspace{-3.5mm}}\\
\multicolumn{2}{c}{Reaction} & $N_{\text{int}}$ \\
\hline
\multicolumn{2}{c}{\vspace{-3.mm}}\\
  $^{28}\text{Si}$ & $\text{H}$  & 50k\\ 
  $^{28}\text{Si}$ & $\text{He}$ & 10k\\ 
  $^{29}\text{Si}$ & $\text{H}$  & 5k \\ 
  $^{32}\text{S}$  & $\text{H}$  & 5k \\ 
  $^{56}\text{Fe}$ & $\text{H}$  & 30k\\ 
  $^{56}\text{Fe}$ & $\text{He}$ & 10k\\ 
\hline
\end{tabular}
\end{table}

As discussed in Sec.~\ref{sec:CR_isotopes} and shown in Fig.~\ref{fig:CRabundances}, the most informative \acrshort{gcr} nuclei are the secondary species, namely \deuteron, $^3$He, LiBeB (and their isotopes), F, sub-Fe $Z=21\text{--}25$ and also some \acrshort{gcr} clocks. The precision of their measurements in \acrshort{cr}s is illustrated in Fig.~\ref{fig:CRdata_secprim}. Actually, the consistency of a pure secondary origin of the \deuteron~\cite{2022ChPhC..46i5102W, 2023PhRvD.107l3008G, 2024ApJ...974L..14Y, 2024PhRvD.110l3030L}, Li~\cite{2020ApJ...889..167B, 2021JCAP...03..099D, 2021JCAP...07..010D, 2021PhRvD.103j3016K, 2022A&A...668A...7M} and F~\cite{Boschini_2022, 2023PhRvD.107f3020Z, 2024A&A...688A..17F} fluxes is particularly debated in the literature in the light of recent \acrshort{ams} data, respectively Refs.~\cite{AMS:Deuteron-PRL2024}, \cite{AMS:LiBEB-PRL2018} and~\cite{AMS:F-PRL2021}. The $^{10}$Be case is also an issue~\cite{Maurin:10Beto9Be_L2022, 2023MNRAS.526..160J, 2024PhRvD.109h3036Z, 2025JCAP...02..062D}, and the production cross-sections for $Z=21\text{--}25$ elements will become a new focal point as soon as \acrshort{ams} will release its data.
As highlighted in Sec.~\ref{sec:physics_cases}, these species have a key role in determining the transport parameters, and also to calculate accurate backgrounds for \acrshort{dm} searches: they are also maximally sensitive to the production cross-sections, as the relevant fluxes are directly proportional to them. Mixed species, like N, Na or Al, come second in terms of priority, and require fewer reactions to reach the same modelling precision; indeed, their secondary production, overall, is only a fraction of the total flux. Finally, purely primary species like H, He, O, Si and Fe are just not impacted at all by these production cross-section uncertainties and are irrelevant in this context.
There are two extreme and complementary situations regarding how to carry out new nuclear data measurement campaigns, which are motivated by past measurements and existing experimental setups, impacting the choice of the wish list. In all cases, the energy range of interest is from a few hundreds of MeV/n up to a few GeV/n (same projectile and fragment energy) and ideally up to a few tens of GeV/n for a few reactions, in order to test the expected mild energy dependence of the cross-sections.

\begin{itemize}
    \item {\em high-precision measurement ($\lesssim 1\%$) of a few specific production cross-sections} over the energy range $\sim 0.1$--10\,GeV/n: in that case, the goal is to determine the energy dependence of the most important reactions, as available in the ranked list shown in Table~\ref{table:xs_ranked}. In there, the reactions for Li, Be, B and F are taken from Ref.~\cite{2024PhRvC.109f4914G}, and those for \deuteron and $^{3}$He isotopes and $Z=21\text{--}25$ elements from a preliminary analysis following the same steps. The cumulative weight of tens to hundreds of reactions with individual $f_{abc}\lesssim 1\%$ can reach $\lesssim 20\%$, with most of these reactions having no data. Moreover, even more important reactions sometimes have inconsistent data, or only one or two energy points below a few hundreds of MeV/n, still in the rising or resonance part of the cross-section (i.e., before having reached their asymptotic high-energy value).

    \item {\em high-precision measurement of all fragments of many reactions at once} at a unique energy: in that case, the recommendation is to evaluate the number of each reaction to be measured in order to reach a desired flux precision of $\epsilon$. This is done by using Eq.~\eqref{eq:errpropag_multi} and the $f_{abc}^j$ coefficients (reported for Li to Si fragments in Ref.~\cite{2024PhRvC.109f4914G}). As an illustration,  Table~\ref{tab:ninter} reports the number of reactions required to reach a $\approx$1\% uncertainty (i.e., below the best \acrshort{ams} $\sim\!3\%$ accuracy) on Li, Be, B and F \acrshort{gcr} data. These numbers were prepared for the test case of measurements at NA61/\acrshort{shine} (\acrlong{shine}), where systematic uncertainties are $\lesssim 0.5\%$~\cite{NA61SHINE:2015bad}; see Sec.~\ref{sec:HEP_NA61} and Refs.~\cite{2019arXiv190907136U, 2022icrc.confE.102A} for more details on the pilot run.
\end{itemize}

\begin{figure}
\centering
\includegraphics[width=\linewidth]{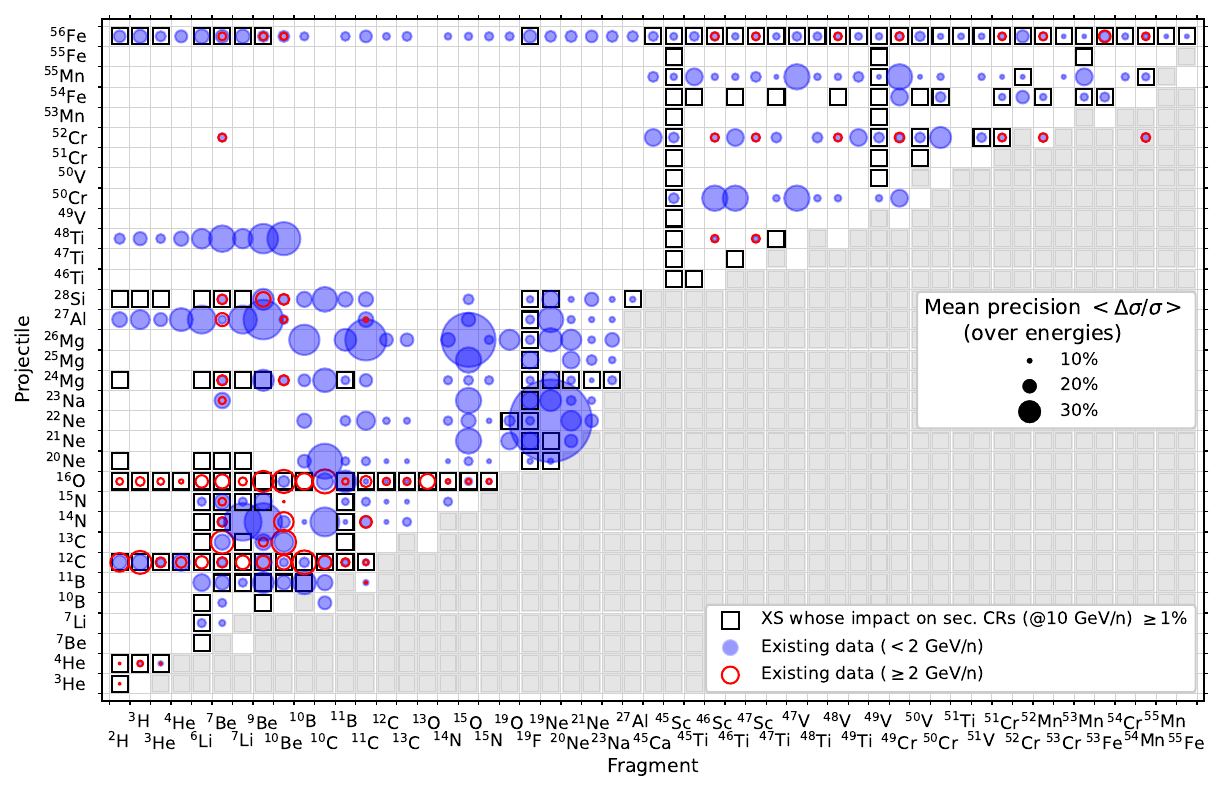}
\caption{Illustration of the existing nuclear data below (blue discs) and above (red circles) 2\,GeV/n, and their relative precision (size of the circles). We restrict ourselves to the matrix of projectiles ($y$-axis) and fragments ($x$-axis) formed from the reactions (black empty squares) contributing to at least 1\% of the flux of \acrshort{gcr} secondary species $Z<30$, as listed in Table~\ref{table:xs_ranked}). The grey zone shows forbidden production regions ($A_f>A_p$): the fact that some nuclear data are reported for $^{52}$Cr into $^{54}$Mn, illustrates that some measured cross-sections come from projectiles in natural abundances (i.e., a mix of several isotopes, some heavier than the one reported) instead of single isotopes reported in this figure (for simplicity).}
\label{fig:nucdata_precision}
\end{figure}

In Tables~\ref{table:xs_ranked} and \ref{tab:ninter},
the most abundant \acrshort{gcr} isotopes ($^4$He, $^{12}$C, $^{16}$O, $^{20}$Ne, $^{24}$Mg, $^{28}$Si, $^{56}$Fe) are recognised as the most important progenitors of the ranked reactions. The key target is H, but reactions on He contribute to $\sim\!10\text{--}15\%$ of the \acrshort{gcr} fluxes overall. Concerning the fragments, as seen in Table~\ref{table:xs_ranked}, the main channels always involve direct production of the isotopes, as well as unstable short-lived parents of the \acrshort{gcr} element under investigation. However, intermediate steps reactions, like the production of C and O isotopes from heavier nuclei always show up: this explains why the strategy of measuring all fragments for all reactions of interest (Table \ref{tab:ninter}) is always the best option, if experimentally possible, to decrease the overall uncertainties on the modelled \acrshort{gcr} fluxes.
Figure~\ref{fig:nucdata_precision} illustrates, for the most relevant reactions identified with black squares (matching those listed in Table~\ref{table:xs_ranked}), that (i) many reactions either have no data (empty squares) or data below 2\,GeV/n only (blue discs and no red circles), and (ii) for the cases where data exist in the asymptotic regime above 2\,GeV/n (red circles), their uncertainty is typically at the 20\% level (as captured by the size of the circles).

\begin{figure}
   \includegraphics[width=0.302\textwidth]{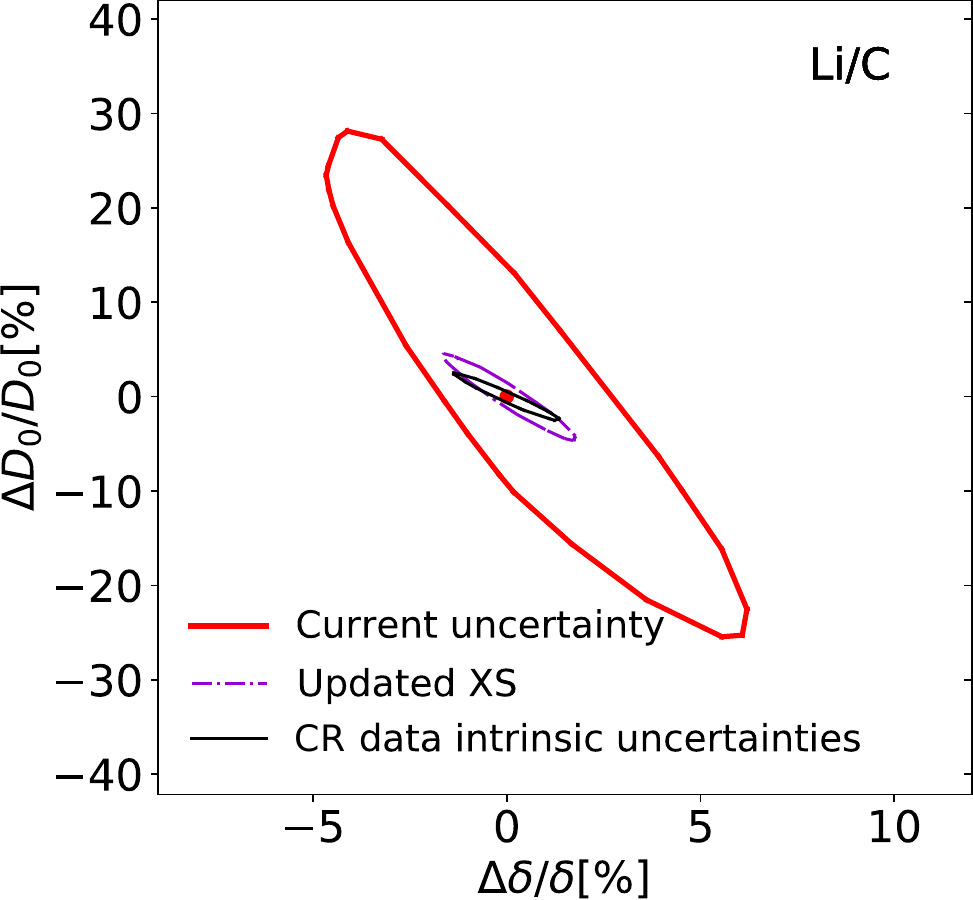}
    \includegraphics[trim={0 3.2mm 0 1.5mm},clip,width=0.322\textwidth]{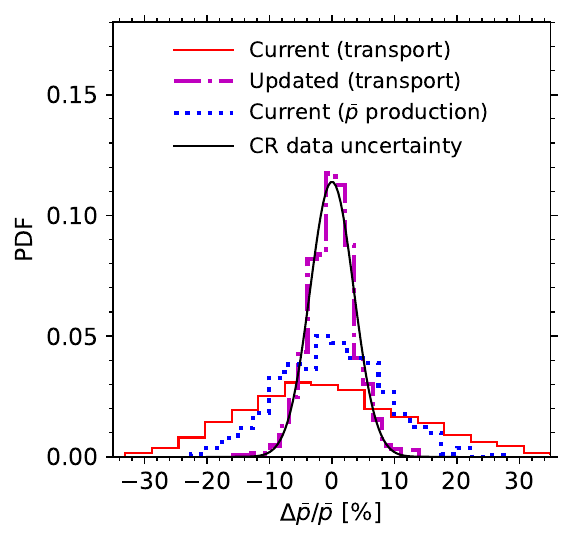}
    \includegraphics[trim={0 2mm 0 2.5mm},clip,width=0.322\textwidth]{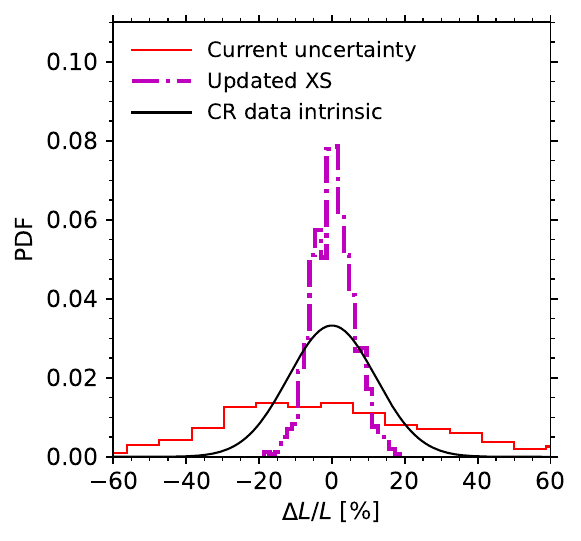}
   \caption{Forecast of the impact of new cross-section measurement campaigns on the normalisation and slope of the spatial diffusion coefficient entering the \acrshort{gcr} transport Eq.~\eqref{eq:comppropeq}, namely $D_0$--$\delta$ (left panel), \antiproton background calculation at 10\,GV (middle panel) and diffusive halo size $L$ determination (right panel). Each figure shows $1\sigma$ contours or distributions for current nuclear data uncertainties (red solid lines) and newly measured cross-sections according to Table~\ref{tab:ninter} (magenta dashed lines), along with the irreducible/intrinsic uncertainty from current \acrshort{cr} data (solid black line). For \antiproton, we also show the uncertainties related to their direct production cross-section (blue dotted line), see Sec.~\ref{sec:xs_prodanti}. Adapted from \cite{2024PhRvC.109f4914G}.}
   \label{fig:forecasts_nuc}
\end{figure}

To further prove that these measurements are worth doing, we can use the wish-list reactions given in Table~\ref{tab:ninter} to sample the cross-sections before and after the new measurements. By repeating \acrshort{gcr} analyses of secondary-to-primary ratios \cite{Weinrich:2020ftb}, the radioactive clock ratio $^{10}$Be/$^9$Be \cite{Maurin:10Beto9Be_L2022} and \antiproton background calculations \cite{Boudaud:2019efq}, we can forecast the impact of these new measurements \cite{2024PhRvC.109f4914G}. Figure~\ref{fig:forecasts_nuc}, adapted from \cite{2024PhRvC.109f4914G}, shows that the improvements on several key \acrshort{gcr} parameters is drastic, and a sure game changer for the field.

\subsubsection{Uncharted needs for \texorpdfstring{$Z>30$}{Z>30} reactions}

The situation is far less clear for ultra-heavy nuclei. As illustrated in Fig.~\ref{fig:CRdata_UHN} and anticipated in Sec.~\ref{sec:CR_UHN}, the \acrshort{gcr} data in this mass range are very scarce and were mostly taken several decades ago. However, in the range $Z=30\text{--}40$, the interpretation of the recent high-precision \acrshort{acecris} data~\cite{2022ApJ...936...13B}, the forthcoming Super\acrshort{tiger}~\cite{2021cosp...43E1335S, 2023HEAD...2030305R}, and the future \acrshort{tiger}-\acrshort{iss}~\cite{2024icrc.confE.171R} and \acrshort{hero}~\cite{Kurganov2023} data (see Sec.~\ref{sec:CRexp_HERD_GERO}), will also hit the cross-section uncertainties bottleneck~\cite{1993ApJ...403..644C, 1996ApJ...470.1218W, 1998NewAR..42..277W}. Besides, even the interpretation of the past $Z>40$ data might also be limited by these uncertainties. The abundance pattern in Fig.~\ref{fig:CRabundances} shows no dominant primary species (contrarily to C, O, Si and Fe for $Z\leq30$), meaning that instead of a few dominant progenitors (as for $Z\leq30$ species), many reactions from a broad range of elements will contribute equally in the modelled fluxes.

The efforts to perform new measurements have already started at Brookhaven, see Sec.~\ref{sec:Facility_Brookhaven}. However, as for the lighter nuclei, we must carry out a systematic analysis of the reactions to be measured with high priority. The \texttt{YIELDX} code~\cite{1998ApJ...501..911S, 1998ApJ...501..920T}, which gives the best results for unmeasured reactions, is appropriate for such study. But a systematic compilation of existing nuclear data is mandatory, to first renormalise the code before applying the predictions. Also, the number of short-live nuclei grows with $A$, so the decay branches of \acrshort{cr}s, compiled more than four decades ago in Ref.~\cite{1984ApJS...56..369L}, must first be updated from recent nuclear properties~\cite{2021ChPhC..45c0001K, 2021ChPhC..45c0002H, 2021ChPhC..45c0003W}, in order to have the full list of ghost nuclei entering Eq.~\eqref{eq:ghosts}.


\subsection{Production cross-sections relevant for indirect DM searches}
\label{sec:xs_prodanti}

As discussed in Sec.~\ref{sec:CR_antinuclei} and~\ref{sec:CR_leptons}, the indirect search for \acrshort{dm} in antimatter \acrshort{gcr}s is pursued since decades. \acrshort{cr} \antiproton, as well as antinuclei, positrons and $\gamma$ rays, are very sensitive probes for \acrshort{dm} annihilation or decay in our Galaxy, as discussed in Sec.~\ref{sec:physics_cases}. 
The first data on antimatter were collected by balloon-borne detectors~\cite{BESS:MITCHELL200431, BESS-Polar-II:antiD-PRL2024}, then followed by satellite~\cite{2017NCimR..40..473P, PAMELA:2010kea, PAMELA:2013vxg}, and space-based experiments, especially by \acrshort{ams} on the \acrshort{iss}~\cite{AMS:PhysRep2021, AMS:positrons2019, 2016PhRvL.117i1103A}.
In particular, the discovery of the rise of the positron fraction data above 10\,GeV, found by \acrshort{pamela}~\cite{PAMELA:2013vxg} and Fermi-\acrshort{lat}~\cite{2012PhRvL.108a1103A}, and confirmed with unprecedented precision by \acrshort{ams}~\cite{PhysRevLett.113.121101}, have been the subject of a broad theoretical debate.
In fact, these very high-energy positrons cannot be explained with the secondary production alone, but may originate from primary sources, such as pulsar wind nebulae~\cite{Hooper:2008kg, DiMauro:2014iia, Orusa:2021tts}, supernova remnants~\cite{Mertsch:2020ldv} and \acrshort{dm} annihilation or decay~\cite{DiMauro:2015jxa}.

The modelling of all the above fluxes suffer from several uncertainties, including propagation uncertainties driven by nuclear cross-section uncertainties (see previous section).
The calculation of exotic (primary) fluxes suffer in particular from the diffusive halo size uncertainty (see Section \ref{sec:CRtransport}), while the calculation of background (secondary) fluxes are dominated by the production cross-sections uncertainties.
The cross-sections entering the computations of the secondary flux, which acts as a background when searching for an exotic component, are the singly differential production cross-sections, $\dd\sigma^{\ijk + X}(E^i,E^k)/\dd E^k$, of a \acrshort{gcr} projectile $i$ (with energy $E^i$) interacting on the \acrshort{ism} target $j$ to produce a \acrshort{gcr} species 
$k$ with energy~$E^k$. In the Galaxy, the set-up is that of a fixed-target (\acrshort{ism}-like) experiment, and the secondary source spectrum
is computed from an integration of the inclusive cross-section, $\dd\sigma^{\ijk+X}(E^i,E^k)/\dd E^k$, over all the \acrshort{gcr} spectrum energies $E^i$, as seen in Eq.~\eqref{eq:spallsource}. However, the above cross-section derives from the measured, more fundamental and convenient, double differential Lorentz invariant cross-section:
\begin{equation}
\sigma_{\rm inv} = E \frac{\dd^3\sigma}{\dd p^3}= \frac{E}{\pi} \frac{\dd^2\sigma}{\dd p_{\rm L} \dd p_{\rm T}^2}\,,
\end{equation}
with $E$, $p_{\rm L}$ and $p_{\rm T}$ the energy, longitudinal and transverse momentum of the outgoing species $f$. The radial and Feynman scaling variables $x_R = E^*/E_{\text{max}}^*$ and $x_{\rm F} = 2p_{\rm L}^*/\sqrt{s}$, where $E^*$ and $p_{\rm L}^*$ are the energy and longitudinal momentum in the centre-of-mass frame, are also used. The centre-of-mass energy and \acrshort{gcr} projectile energies are linked by $\sqrt{s}=(m_i^2+m_j^2+2E_im_j)^{1/2}$, where $m_i$ and $E_i$ are the mass and total energy of the \acrshort{gcr} projectile, and $m_j$ the mass of the \acrshort{ism} target (at rest).

The production cross-sections involved in the calculation of secondary \antiproton are discussed in Sec.~\ref{sec:xs_prodpbar}, those for antinuclei in Sec.~\ref{sec:xs_proddbar}, and those for positrons and $\gamma$ rays in Sec.~\ref{sec:xs_prodpositrons}. For each \acrshort{gcr} species, the current status of nuclear data is presented, and then the reactions -- in terms of projectiles and targets, the energy range, and the cross-section precision --, needed to fully exploit current and near future \acrshort{cr} data for \acrshort{dm} searches, are listed. The role of nuclear codes is commented in Sec.~\ref{sec:xs_prodsummary}, along with a synthetic view of a wish list in Table~\ref{tab:measurements}.

\subsubsection{Antiprotons: status and game-changing measurements}
\label{sec:xs_prodpbar}

Concerning \antiproton, the \pp channel dominates the secondary production, alongside the contributions from \Hep, \pHe and \HeHe, either in the \acrshort{gcr} projectile or as \acrshort{ism} target, the rest coming mostly from interactions of heavier abundant \acrshort{gcr} species (CNO, NeMgSi, and Fe) on H~\cite{diMauro:2014zea, Donato:2017ywo, Boudaud:2019efq, Korsmeier_2018, Boudaud:2019efq}.
The fixed-target NA49 experiment collected data on the \pp channel at $\sqrt{s}=17.3$\,GeV \cite{Anticic_2009}, and the NA61 experiment at $\sqrt{s}$=7.7, 8.8, 12.3 and 17.3\,GeV, corresponding to beam proton energies $E_{\rm k}=31$, 40, 80 and 158\,GeV, respectively \cite{Aduszkiewicz:2017sei}. Lower-energy data are available at $\sqrt{s}=6.1$ and 6.7\,GeV~\cite{Dekkers1965}, and at $\sqrt{s}=6.15$\,GeV~\cite{Allaby:1970jt}.
Data from the \acrshort{brahms} (\acrlong{brahms}) experiment have been taken in \pp collisions at $\sqrt{s}=200$\,GeV \cite{Arsene:2007jd}.
Data on p$^{4}$He have been recorded by the \acrshort{lhcb} (\acrlong{lhcb}) collaboration at \acrshort{cern} in fixed-target mode, using the \acrshort{smog} (\acrlong{smog}) device~\cite{LHCb-PAPER-2018-031} with a proton beam momentum of 6.5\,TeV/c (corresponding to $\sqrt{s} = 110$\,GeV).
More recently the \acrshort{amber} (\acrlong{amber}) collaboration, at the \acrshort{cern} \acrshort{sps} M2 beam line, has collected data with a proton beam impinging on a liquid $^{4}$He target at six different momenta, from 60\,GeV/c to 250\,GeV/c (corresponding to $\sqrt{s} = 10.7$\,GeV and 21.7\,GeV).
Data on \pC have also been collected by NA49 \cite{NA49:2012jna} at $\sqrt{s}=17.3$\,GeV. A full discussion on all the \antiproton cross-section data and their role in the context of cosmic source spectra can be found in Ref.~\cite{Korsmeier_2018}.

The current modelling of reactions on He of heavier \acrshort{gcr} projectiles is based on a rescaling of the models derived on the \pp reaction channels, and denoted $\sigma_{\rm inv}^{\antiproton}$ hereafter. The latter is separated in a prompt and a delayed emission originating from the decay of strange hadrons, labelled $\Lambda$ in the following.
 Assuming $\sigma_{\rm inv}^{\antiproton\;\rm delayed} =\sigma_{\rm inv}^{\antineutron\;\rm delayed}$ \cite{Winkler:2017xor}, one can write:
\begin{equation}
 \sigma_{\rm inv}^{\antiproton} = \sigma_{\rm inv}^{\antiproton,\antineutron\;\rm prompt} + \sigma_{\rm inv}^{\antiproton,\antineutron\;\rm delayed} =\sigma_{\rm inv}^{\antiproton\;\rm prompt} (2+\Delta_{\rm IS}+ 2 \Delta_{\Lambda})\,,
 \label{eq:def_pbarinv}
\end{equation}
with the isospin enhancement and the hyperon factors defined as:
\begin{equation}
    \Delta_{\rm IS} = \frac{\sigma_{\rm inv}^{\antineutron\;\rm prompt}}{\sigma_{\rm inv}^{\antiproton\;\rm prompt}} -1 \quad {\rm and}\quad
    \Delta_\Lambda = \frac{\sigma_{\rm inv}^{\antiproton\;\rm delayed}}{\sigma_{\rm inv}^{\antiproton\;\rm prompt}}\,.
    \label{eq:Delta}
\end{equation}
The currently available data are insufficient to establish distinct parametrisations for the above individual cross-sections. The parametrisations are therefore rescaled to the prompt emission and to the enhancement terms
$\Delta_{\rm IS}$ and $\Delta_{\Lambda}$.
The dominant production comes from the prompt emission, with significant additional contributions from hyperon-induced channels like $\overline{\Lambda}$ and $\overline{\Sigma}$. The \antineutron contribution requires the knowledge of a possible isospin asymmetry, that could induce a possible enhancement $\Delta_{\rm IS}$ of \antineutron over \antiproton production.

The total secondary \acrshort{gcr} $\antiproton$ production uncertainty ranges about 15--20\%~\cite{Winkler:2017xor, Korsmeier_2018, Boudaud:2019efq}. It is primarily driven by the $\proton+\proton\rightarrow \antiproton+{\rm X}$ cross-section and receives contributions from all nuclei channels, dominated by the ones involving He.
To enhance the accuracy of current models -- to be on par with \acrshort{cr} data precision (see Sec.~\ref{sec:CR_antinuclei} and Fig.~\ref{fig:forecasts_nuc}) --, a set of key \antiproton production measurements of the Lorentz-invariant fully differential cross-section ($\sigma_{\rm inv}$) is essential for:
\begin{itemize}
    \item {\em Prompt emission from \pp at better than 3\% precision}: first and foremost, new measurements should focus on reducing uncertainties of $\proton+\proton\rightarrow\antiproton+{\rm X}$ across the $\sqrt{s} = 5\text{--}100$\,GeV range, and cover regions with $p_{\rm T} \lesssim1$\,GeV/c and $|x_{\rm F}| \leq 0.3$;

    \item {\em Production on He target with uncertainties below 5\%:} additionally, measurements in \pHe reactions are needed. The first-ever data on the inclusive cross-section $\proton+\helium\rightarrow \antiproton+{\rm X}$ were collected by the \acrshort{lhcb} collaboration at \acrshort{cern}, using proton beams with $E_{\rm k} = 6.5$\,TeV and a fixed He target (see Sec.~\ref{sec:HEP_LHCb}). These data were analysed in \cite{Korsmeier_2018}, although the centre-of-mass energy of the provided data is higher than the energy of the \antiproton measured by \acrshort{ams} \cite{2016PhRvL.117i1103A}. An extensive coverage of $\sigma_{\rm inv}$ in the $\sqrt{s} = 5\text{--}100$\,GeV range for \pHe would allow an independent parametrisation for this channel.

    \item {\em Isospin enhancement $\Delta_{\rm IS}$ with uncertainties below 5\%}: an improved determination of the isospin asymmetry, affecting the contributions of \antineutron, is needed.
    The potential isospin asymmetry is particularly significant, as approximately half of the \antiproton in \acrshort{gcr}s originate from the decay of long-lived \antineutron. However, due to the limited experimental data on \antineutron production, this contribution can only be inferred using symmetry arguments. Preliminary results from the NA49 experiment \cite{Fischer:2003xh} suggest that \antineutron production in \pp collisions exceeds \antiproton production, indicating the presence of an asymmetry; see also Sec.~\ref{sec:HEP_AMBER} and Fig.~\ref{fig:antiproton_experiments} for future measurements from \acrshort{amber} and \acrshort{lhcb}. To evaluate this asymmetry accurately, we critically need data on $\sigma_{\rm inv}$ for \antineutron production in \pp collisions, or for \antiproton production in \pp and pn collisions in the $\sqrt{s} = 5\text{--}100$\,GeV range.

    \item {\em Strange hadrons factor $\Delta_\Lambda$ with uncertainties below 10\%}:  the contributions from strange hadron decays, in particular the total $\overline{\Lambda}$ production in both \pp and \pHe reactions in the $\sqrt{s} = 5\text{--}100$\,GeV range, should be measured, with uncertainties at the 10\% level \cite{Winkler:2017xor}.
\end{itemize}
For all these reactions, pushing the lower limit of $\sqrt{s}$ to values close to the \antiproton production threshold \mbox{($\sim\!3.8$\,GeV)} will be also extremely helpful for interpreting upcoming low-energy \acrshort{cr} data from \acrshort{gaps} \cite{Aramaki_2016}. In order to get the \acrshort{gcr} \antiproton modelling uncertainties below a few percent, two conditions are required: first, the above quoted \antiproton production precisions must be achieved for all production channels; second, significant improvement must be made on the nuclear production cross-sections, also responsible for dominant uncertainties on the \acrshort{gcr} \antiproton flux modelling (see Sec.~\ref{sec:xs_nucforecasts}).

\subsubsection{Antideuterons and \antihelium: coalescence-driven uncertainties}
\label{sec:xs_proddbar}

While not detected in \acrshort{cr}s yet, \antideuteron are expected to be even more sensitive probes than \antiproton for \acrshort{dm} searches. Compared to possible \acrshort{dm} production with a thermal cross-section, the secondary contribution of \antideuteron is suppressed by a factor 10 or more at kinetic energy per nucleon below 1\,GeV/n~\cite{Donato_2000, Donato:2008yx, Cirelli_2014}. This is due to the fact that the secondary production has to satisfy the baryonic number conservation, and thus \acrshort{gcr} \proton must have a total energy in the lab frame of at least $17 m_{\rm p}$ to produce an \antideuteron (the corresponding value for \antiproton is $7 m_{\rm p}$).

\paragraph{Formation of antinuclei}
The \acrshort{gcr} spectra of \antideuteron and \antihelium are typically calculated using the so-called coalescence models~\cite{PhysRevC.21.1301, PhysRev.129.836, PhysRevC.59.1585} both for secondary production and \acrshort{dm} contributions. These models assume that individual \antiproton and \antineutron form antinuclei when their relative momentum falls below a certain threshold, referred to as the coalescence momentum. A first consequence is that the model describing \antideuteron and \antihelium astrophysical production is directly impacted by the \antiproton production cross-section uncertainties described in the previous section. This coalescence parameter cannot be derived from first principles and varies depending on the production mechanism of the nucleus, such as whether it originates from different final states in \acrshort{dm} annihilation or hadronic interactions. Alternative coalescence models, employing a quantum-mechanical Wigner function formalism~\cite{Blum:2017qnn, Bellini:2018epz, Kachelriess:2019taq, Kachelriess:2020uoh, Bellini:2020cbj, Mahlein2023, Mahlein2024}, form the basis of recent advancements in calculating (anti)nuclei production from hadronic interactions, using \acrshort{mc} simulations.
We also stress that, the coalescence models are not the only approach to produce light nuclei in hadronic interactions. They can be produced by the statistical hadronisation of hot quark matter \cite{PhysRevC.21.1301} which can be produced even in light systems~\cite{ALICE:2016fzo}. This should be taken into account in particular in the description of \pp data. Further studies and data are needed to understand if this could replace the coalescence model, or if both mechanisms play a role at the same time.

\paragraph{Existing nuclear data}
The dataset for \antideuteron production from \pp collisions, which is relevant for the secondary flux, is quite rich thanks to the \acrshort{alice} experiment, which provided data between 900\,GeV and 13\,TeV~\cite{ALICE:2021ovi, ALICE:2015wav, ALICE:2017xrp, ALICE:2019dgz, ALICE:2020foi, ALICE:2021mfm}, as described in Sec.~\ref{sec:HEP_ALICE}. However, near the production threshold, only the Serpukov data at $p_{\rm lab} = 70$\,GeV/c \cite{Abramov:1986ti} are available.
For the production relevant for \antideuteron primary flux, there are currently two data points from \acrshort{argus} at the Upsilon mass resonances 1S, 2S, 4S and continuum $\sim\!10$\,GeV \cite{ARGUS:1985cfz}, and one data point from \acrshort{aleph} at the $Z$ boson resonance \cite{ALEPH:2006qoi}. The latter is typically used to tune the \acrshort{dm} production of \antideuteron, because the $Z$ boson production from $e^+e^-$ annihilation is assumed to be similar to the \acrshort{dm} particle annihilation process.

\paragraph{Antideuteron and \antihelium flux modelling uncertainty}
For primary antinuclei fluxes, Ref.~\cite{DiMauro:2024kml} showed that once the coalescence models are tuned on the \acrshort{aleph} data, the predictions for the \antideuteron yield agree within 10\% for high mass \acrshort{dm}, despite the very different assumptions used in the simple coalescence and the Wigner function approach. For low mass \acrshort{dm}, only the assumption of a decay into W$^+$W$^-$ pairs gives a comparable prediction. In contrast, a decay of \acrshort{dm} into $b\bar{b}$ pairs shows a significant enhancement of factor two for the Wigner function formalism using the Argonne~v18 wave function.
This indicates that the theoretical uncertainties in \antideuteron production are no longer a major limitation, at least for high-mass \acrshort{dm} and several channels for low-mass \acrshort{dm}: the main limiting factor remains the error on the \acrshort{aleph} data, of the order of $30\%$.
While the Wigner function formalism is free of this parameter dependence, its dependence on the size of the emission source induces a similar constraint to its predictive power.
The typical value of the coalescence momentum $p_{\rm coal}$ (see \cite{2018PhRvD..97j3011K} for definitions)
found when fitting \acrshort{aleph} data is $0.15\text{--}0.21$\,GeV/c (see Refs.~\cite{Donato:2008yx, Kachelriess:2020uoh, 2018PhRvD..97j3011K, Ibarra:2012cc, Fornengo:2013osa}). This leads to
a conservative $\lesssim 60\%$ uncertainty on the \antideuteron primary flux prediction for most \acrshort{dm} masses and channels, folding into a factor of a few for the flux of primary antinuclei.

For secondary antinuclei fluxes, values $p_{\rm coal}>0.2$\,GeV/c are found when using \acrshort{alice} data for \pp collisions.
Changing the coalescence momentum from the above 0.15 to 0.21\,GeV/c would introduce uncertainties in the \antideuteron spectra of approximately a factor of 3.
However, the scarcity of current data on \antideuteron from $e^+e^-$ and the lack of reliable data at low energy for \pp collisions (we recall that only the Serpukov data are available), makes difficult to study a possible dependence of the coalescence momentum value according to the physical process and the centre-of-mass energy: considering these Serpukov data leads to a strong decrease in the \antideuteron production, which can be interpreted as an energy dependence of $p_{\rm coal}$ \cite{Coral2018}. As a result, coalescence-related uncertainties for the secondary flux are a factor of a few for \antideuteron, and up to an order of magnitude for antinuclei~\cite{Shukla:2020bql}.

\paragraph{Desired nuclear data}
Collecting data on the production of \antideuteron and \antihelium in \pp collisions at $\sqrt{s}<100$\,GeV would greatly improve the theoretical predictions for their secondary production. This could be achieved by measuring $\sigma_{\rm inv}$ over a large kinematic range in $p_{\rm T}$ and $x_{\rm F}$ or, at least, the integrated multiplicity in the energy range $\sqrt{s}\in[10,100]$\,GeV. Owing to a lower threshold, production from p\antiproton collisions is also of interest: its contribution amounts to a few percents only at a few GV (see the pink dashed line in Fig.~\ref{fig:XS_pbardbar_impact}), but any data would be useful to check that the correct magnitude is used for this cross-section, i.e., to check that this production channel is not underestimated.
In addition to this, more precise measurements for the antinuclei production from $e^+e^-$ collisions are mandatory to verify whether the coalescence models and their parameters change according to the underlying physical process.
In the light of the recent \acrshort{ams} claims on possible observation of several \antihelium candidates, new scenarios for the enhanced production of \antihelium have been proposed, including the decay of $\overline{\Lambda}_b$ baryons produced by the DM annihilation into $b\bar{b}$ pairs \cite{Winkler:2020ltd}, though it is disputed~\cite{Kachelriess:2021vrh, Winkler:2021cmt}. Preliminary results from \acrshort{lhcb} seem to disfavour these models \cite{LHCb-CONF-2024-005}, but further measurements of the \antihelium production from antibaryons decay could help better constrain the expected \antihelium flux.

\subsubsection{Positrons and \texorpdfstring{$\gamma$}{gamma}-rays: improvement needed}
\label{sec:xs_prodpositrons}

The production cross-sections of positrons above 1\,GeV in the \pp channel, the primary channel for secondary positron production, were recently derived in an accurate model presented in Ref.~\cite{Orusa_2022}. This parametrisation was directly tuned using available measurements at various $\sqrt{s}$ values, from 3\,GeV to 10\,TeV, specifically from NA49~\cite{2005_NA49,NA49_2010}, NA61~\cite{Aduszkiewicz:2017sei, NA61SHINE:2015haq}, \acrshort{alice}~\cite{2011_ALICE}, \acrshort{cms}~\cite{2012_CMS,2017_CMS} and a collection of older data~\cite{Antinucci1973}.
While the empirical framework for the \pp production cross-section is provided with an uncertainty of about 5-7\%, there is room for improvement in the treatment of other nuclear channels.
Indeed, no data on reactions involving He have ever been taken.
For reactions beyond \pp, the cross-sections employed in Ref.~\cite{Orusa_2022} rely on rescaling of the \pp reaction channels. This rescaling is tuned on data from \pC collisions collected by NA49~\cite{NA49_2007} at $\sqrt{s}=17.3$\,GeV and by NA61/\acrshort{shine}~\cite{NA61SHINE:2015bad} at $\sqrt{s}=7.7$\,GeV.
Precise measurements of the Lorentz invariant fully differential cross-sections of $\pi^{\pm}$ and $K^{\pm}$ from \pHe collisions in the $\sqrt{s}=5\text{--}100$\,GeV range, with a primary focus on $\sqrt{s}=10\text{--}20$\,GeV, would allow us to achieve improvements. These measurements should cover a broad kinematic range, with $p_{\rm T} \lesssim 1$\,GeV/c and extensive coverage in $x_{\rm F}$, aiming for uncertainties at the 5\% level. Such data would enable proper modelling of individual reaction channels involving He, eliminating the need for simple rescaling approaches.
For positron energies $E_{\rm k}\lesssim 1$\,GeV, the cross-section data are missing, and the computation of the $e^+$ source spectrum relies on extrapolations.

Most of the $\gamma$ rays produced by hadronic interactions and detected by Fermi-\acrshort{lat}~\cite{Fermi-LAT:2022byn} originate from the $\pi^0 \rightarrow \gamma \gamma$ decay, which results from hadronic interactions, with \pp being the main production channel.
A new model for the Lorentz-invariant cross-section of $\pi^0$ production was proposed in \cite{orusa2023new}, with uncertainties ranging between 10\% and 20\%. This model was developed using the limited available data on total cross-sections of $\pi^0$\cite{dermer1986binary}, \acrshort{lhcf} (\acrlong{lhcf}) data in the high-energy regime \cite{LHCf:2015rcj}, and is strongly based on the previous analyses of the $e^\pm$ cross-section from Ref.~\cite{Orusa_2022}.
New data on the Lorentz-invariant cross-section of $\pi^0$ production are necessary to reduce the uncertainty in $\sigma(i+j \rightarrow \pi^0 + X)$ to 5\%, aligning it with the statistical uncertainties of Fermi-\acrshort{lat}. Specifically, measurements of $\pi^0$ production in the $\sqrt{s} = 5\text{--}1000$\,GeV range, covering a broad kinematic range with $p_{\rm T} \lesssim 1$\,GeV/c and extensive coverage in $x_{\rm F}$, for both \pp and \pHe collisions, would significantly reduce these uncertainties. Even for \pp collisions, the model in \cite{orusa2023new} depends on results obtained for $\pi^\pm$ to describe the $p_{\rm T}$ and $x_{\rm F}$ dependencies of the cross-section.
The larger $\sqrt{s}$ range with respect to $e^+$ and \antiproton is required by the energy range at which $\gamma$ rays are measured by experiments like \acrshort{magic} (\acrlong{magic})~\cite{MAGIC:2014zas}, \acrshort{hess} (\acrlong{hess})~\cite{DeNaurois:2020bac}, \acrshort{hawc} (\acrlong{hawc}) ~\cite{HAWC:2020hrt}, \acrshort{lhaaso} (\acrlong{lhaaso})~\cite{LHAASO:2019qtb} and the upcoming \acrshort{ctao} (\acrlong{ctao})~\cite{CTAConsortium:2013ofs}.

\subsubsection{Summary and wish list}
\label{sec:xs_prodsummary}
\begin{figure}
	\centering
	\includegraphics[width=7.5cm]{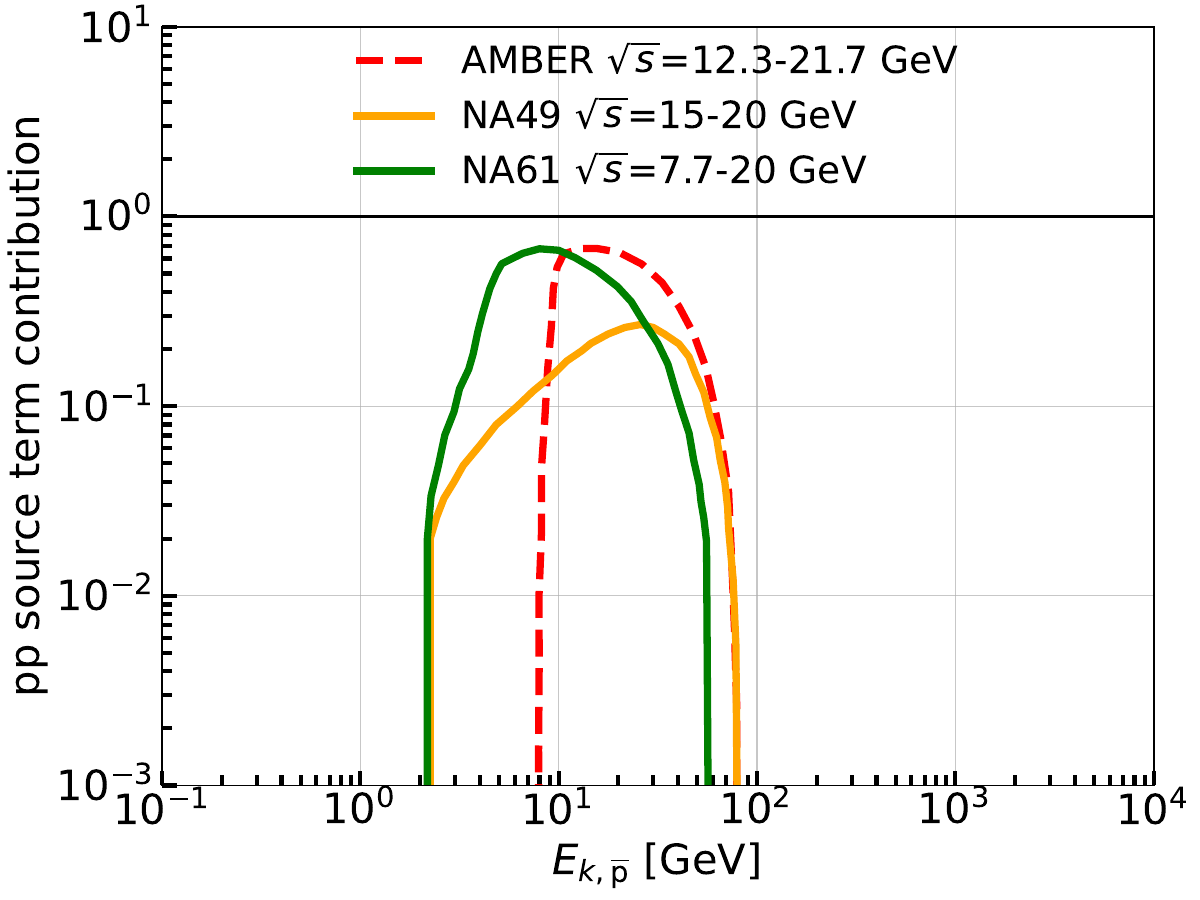}\\
	\includegraphics[width=7.5cm]{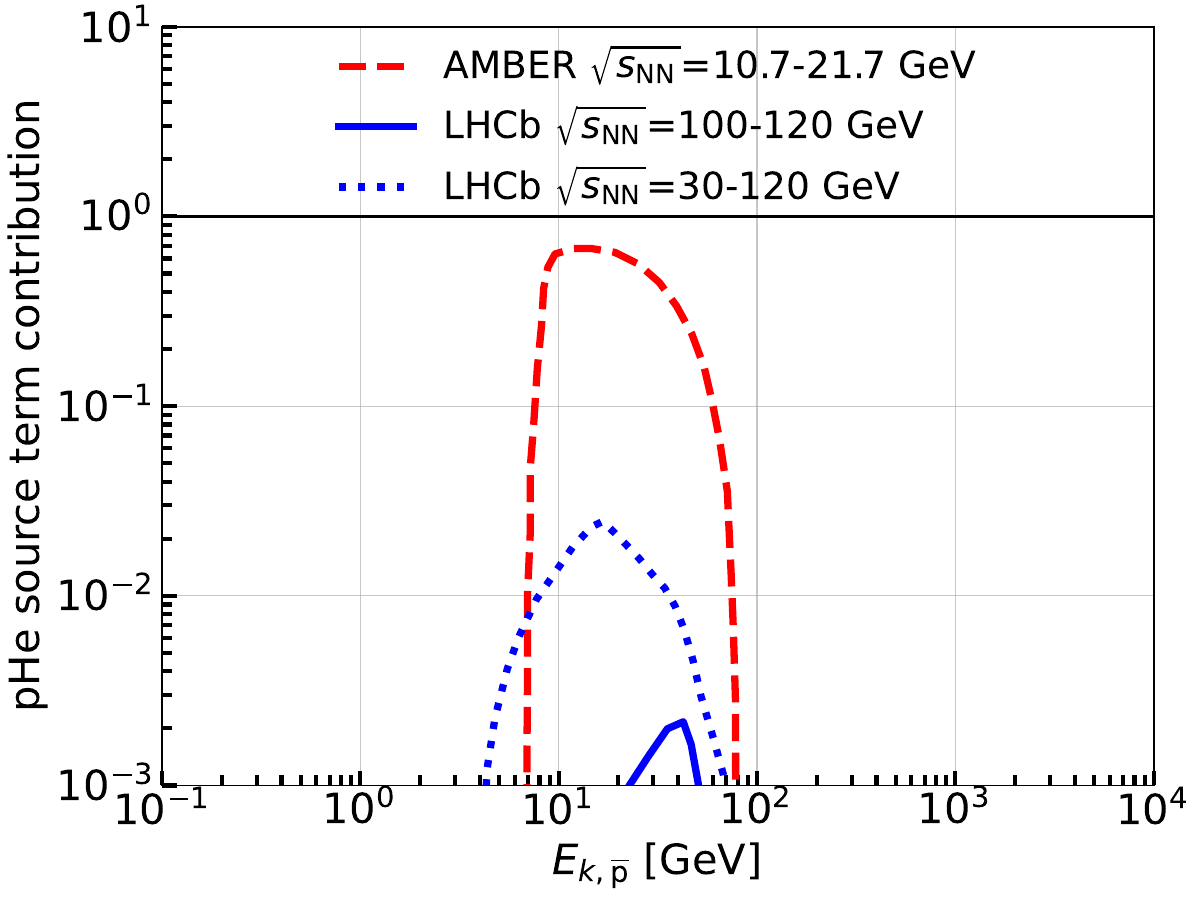}
    \includegraphics[width=7.5cm]{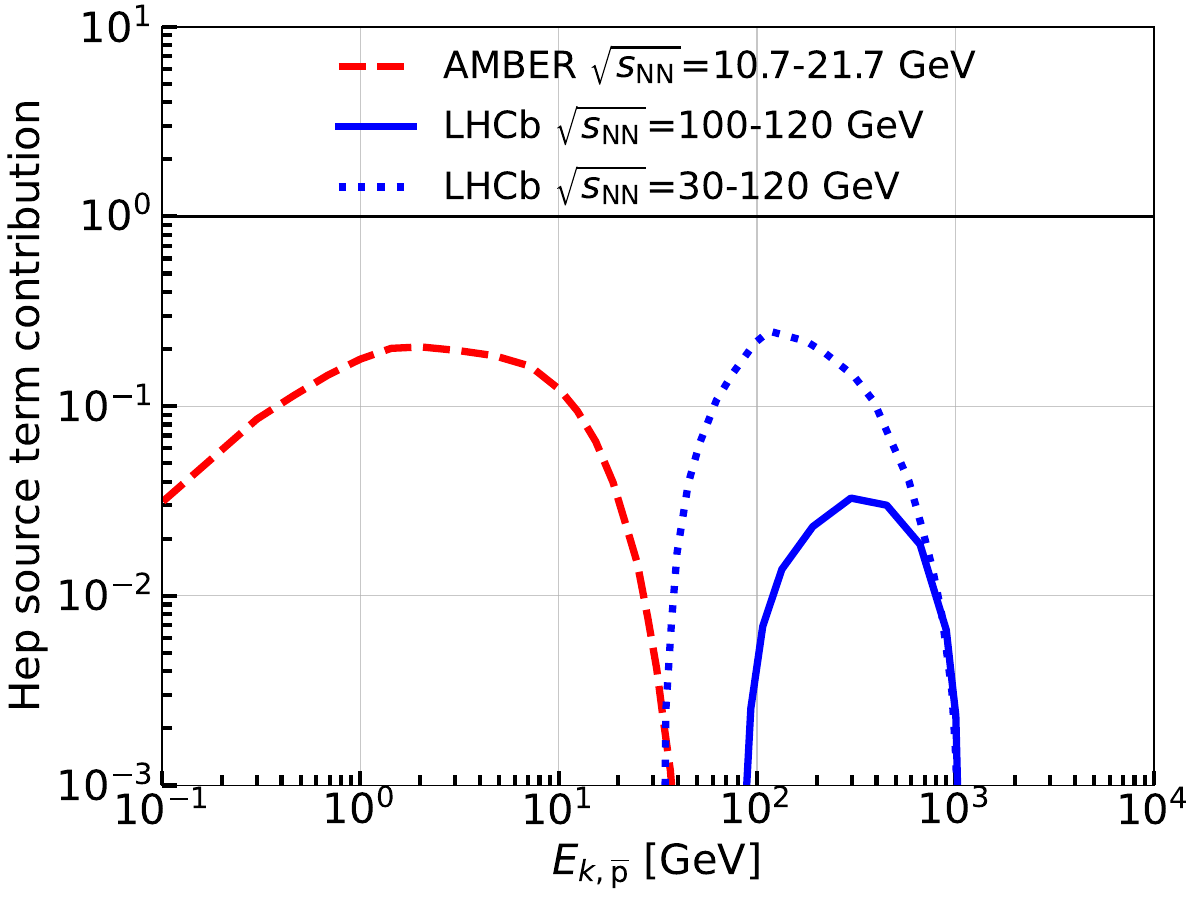}
	\caption{Fraction of the \pp (top), \pHe (bottom left) and \Hep (bottom right) source terms originating from the kinematic parameter space of the cross-sections covered by different experiments. The contributions are normalised to the total source term of each channel. Solid lines represent experiments with data already collected and publicly available (NA49 \cite{Anticic_2009}, NA61/\acrshort{shine} \cite{Aduszkiewicz:2017sei} and \acrshort{lhcb}~\cite{CERN-THESIS-2021-313, SMOG2_paper}). Dashed lines indicate predictions for data that have been collected but are not yet publicly accessible (\acrshort{amber}). Dotted lines correspond to future predictions for potential data-taking campaigns (\acrshort{lhcb}).}
  \label{fig:pbar_production}
\end{figure}

We report the wish list of the measurements discussed throughout this section In Table~\ref{tab:measurements}. The needs and precision are not the same for various \acrshort{gcr} species. The most pressing physics case is for \antiproton, where new nuclear data are needed now. Figure \ref{fig:pbar_production} illustrates the fraction of the source term, as defined in Eq.~\eqref{eq:spallsource}, covered by current and forthcoming data. As the source spectrum implies an integration of the cross-section over the kinematic phase-space of the produced \antiproton and a convolution with the projectile, i.e., incident \acrshort{gcr}, energy, a plethora of data with different $\sqrt{s}$ is needed for a very precise determination of the source spectrum. From left to right, we report the contribution 
to the \pp, \pHe and \Hep source terms covered by available data (NA49 \cite{Anticic_2009}, NA61/\acrshort{shine} \cite{Aduszkiewicz:2017sei} and \acrshort{lhcb}~\cite{CERN-THESIS-2021-313, SMOG2_paper}), assuming the cross-sections are constant in a $\sqrt{s}$ interval around the provided results or the foreseen campaigns.
The contributions are normalised to the total source term of each channel.
Possible extensions brought by data that have been collected but are not yet publicly accessible (\acrshort{amber}, dashed lines), or by potential data-taking campaigns (\acrshort{lhcb}, dotted lines), are also indicated (see Sec.~\ref{sec:facilities_LHC} for more details). It is worth underlining that, while data at lower $\sqrt{s}$ cover a larger fraction of the total produced \antiproton, following the power-spectrum decrease of the incident \acrshort{gcr} fluxes (see Fig.~\ref{fig:CRspectrum}), data at higher $\sqrt{s}$ allow the violation of the Feynman scaling to be constrained. For example, in Ref.~\cite{Korsmeier_2018}, it was shown how the pioneering measurement by \acrshort{lhcb} for antiprotons produced in pHe collisions~\cite{LHCb-PAPER-2018-031} was able to discriminate between two different parametrisations for the invariant antiproton production cross-section.
The \pp channel is satisfactorily covered only for \antiproton kinetic energies in the 5--30\,GeV, corresponding to the low-energy range of \acrshort{ams} \acrshort{cr} data; there are no high-precision \acrshort{cr} data for \antiproton below a few GeV, in a range where the future \acrshort{gaps} data will take data. For the He channels, the situation is far from optimal, and campaigns undertaken by \acrshort{amber} and by \acrshort{lhcb} would be very desirable. 
Second, improving the coalescence factor for \antideuteron is a necessity for the coming years, where one could expect their detection by ongoing and future \acrshort{cr} experiments. Third, the impact of nuclear data uncertainties is less critical for the physics cases associated to $e^+$ and $\gamma$ rays, but bringing nuclear data precision at the level of the current and forthcoming \acrshort{cr} data precision remains desired.

\begin{table}
    \setlength{\tabcolsep}{6pt}
    \centering
    \caption{Summary of the wish list of production cross-sections for \acrshort{gcr}s that can be indirect probes of particle \acrshort{dm}.
    Here, $n_{\rm tot}$ is the integrated multiplicity. The most pressing need is for \antiproton, whose interpretation is already limited by cross-section uncertainties, but forthcoming \acrshort{cr} data for \antideuteron, and possible \antihelium events from \acrshort{ams}, call for new cross-section measurements for these species. See text for the detailed motivations.\vspace{1.5mm}}
    \label{tab:measurements}
    \centering
    \begin{tabular}{clcccc}
        \toprule
        \textbf{Particle} & \textbf{Reaction} & \textbf{Measurement} & \textbf{$\sqrt{s}$} & \textbf{Sought precision}\\
        \midrule
        \multirow{6}{*}{\antiproton} & $\proton+\proton\rightarrow\antiproton+{\rm X}$ & \multirow{6}{*}{$\sigma_{\rm inv}$} & \multirow{6}{*}{5 to 100\,GeV} & $<3\%$\\
         & $\proton+\helium\rightarrow\antiproton+{\rm X}$ & & & $<5\%$\\
         & $\proton+\proton\rightarrow\overline{\Lambda}+{\rm X}$ & & & $<10\%$\\
         & $\proton+\helium\rightarrow\overline{\Lambda}+{\rm X}$ & & & $<10\%$\\
         & $\proton+\proton\rightarrow \antineutron+{\rm X}$ & & & $<5\%$ \\
         & $\proton+\neutron\rightarrow\antiproton+{\rm X}$ & & & $<5\%$\\
        \midrule
        \multirow{2}{*}{\antideuteron} & $\proton+\proton\rightarrow\antideuteron+{\rm X}$ & $\sigma_{\rm inv}/n_{\rm tot}$ & 5 to 100\,GeV & (any data)\\
        & $\proton+\helium\rightarrow\antideuteron+{\rm X}$ & $\sigma_{\rm inv}/n_{\rm tot}$ & 5 to 100\,GeV & (any data)\\
         & $\antiproton+\proton\rightarrow\antideuteron+{\rm X}$ & $\sigma_{\rm inv}$ & 2 to 10\,GeV & (any data) \\
        \midrule
        \multirow{1}{*}{\antihelium} & $\proton+\proton\rightarrow\antihelium+{\rm X}$ & $\sigma_{\rm inv}/n_{\rm tot}$ & 5 to 100\,GeV & (any data)\\
        \midrule
        \multirow{2}{*}{$e^\pm$} & $\proton+\helium\rightarrow\pi^\pm+{\rm X}$ & \multirow{2}{*}{$\sigma_{\rm inv}$} & \multirow{2}{*}{5 to 100\,GeV} & $<5\%$\\
        & $\proton+\helium\rightarrow K^\pm+{\rm X}$ & & & $<5\%$ \\
        \midrule
        \multirow{2}{*}{$\gamma$} & $\proton+\proton\rightarrow\pi^0+{\rm X}$ & \multirow{2}{*}{$\sigma_{\rm inv}$} & \multirow{2}{*}{5 to 1000\,GeV} & $<5\%$\\
        & $\proton+\helium\rightarrow\pi^0+{\rm X}$ & & & $<5\%$\\
        \bottomrule
    \end{tabular}
\end{table}

These potential new measurements, encompassing various reactions and elements, are instrumental in refining and validating \acrshort{mc} \cite{Kelner:2008ke,delaTorreLuque:2022vhm,Kamae:2006bf,Kachelriess:2019ifk,Ostapchenko:2010vb,Kafexhiu:2014cua,Bhatt:2020dxk}, the latter being crucial in simulating particle interactions and secondary production processes. By tuning \acrshort{mc}s against experimental data, it is possible to improve their predictive accuracy, particularly for key observables such as cross-sections, particle multiplicities and energy spectra.

\begin{figure}
	\centering
	\includegraphics[width=8.6cm]{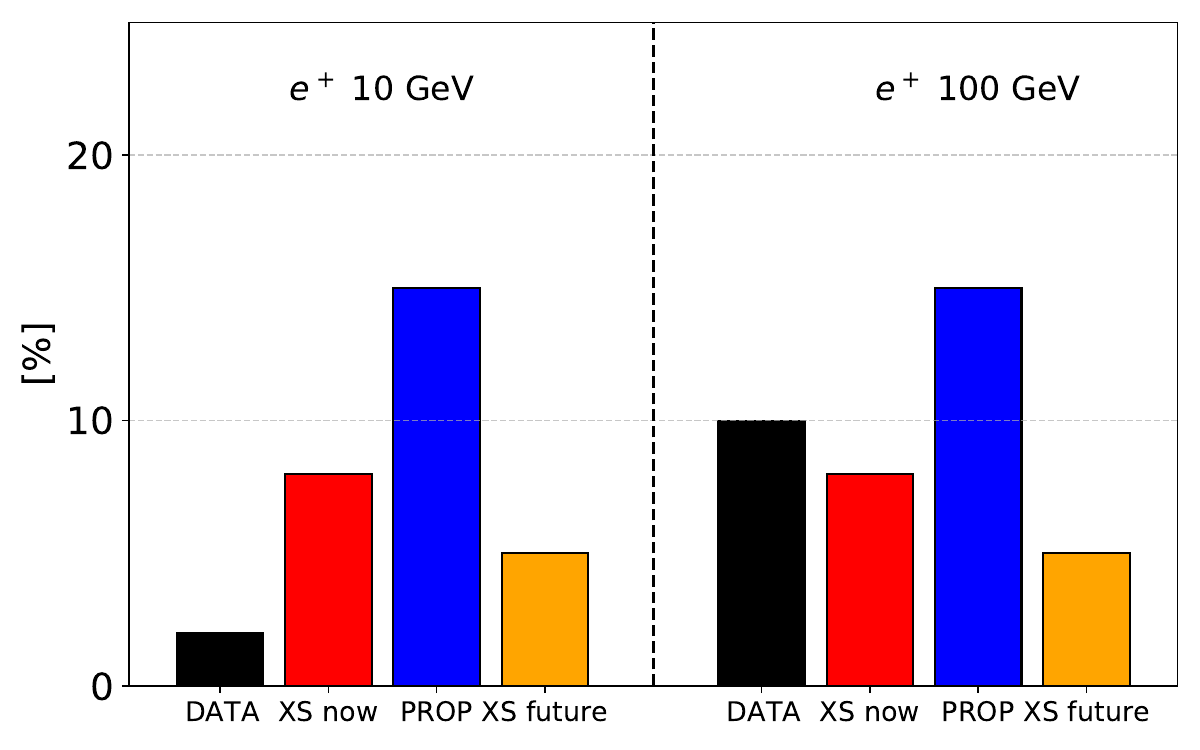}
	\includegraphics[width=7.4cm]{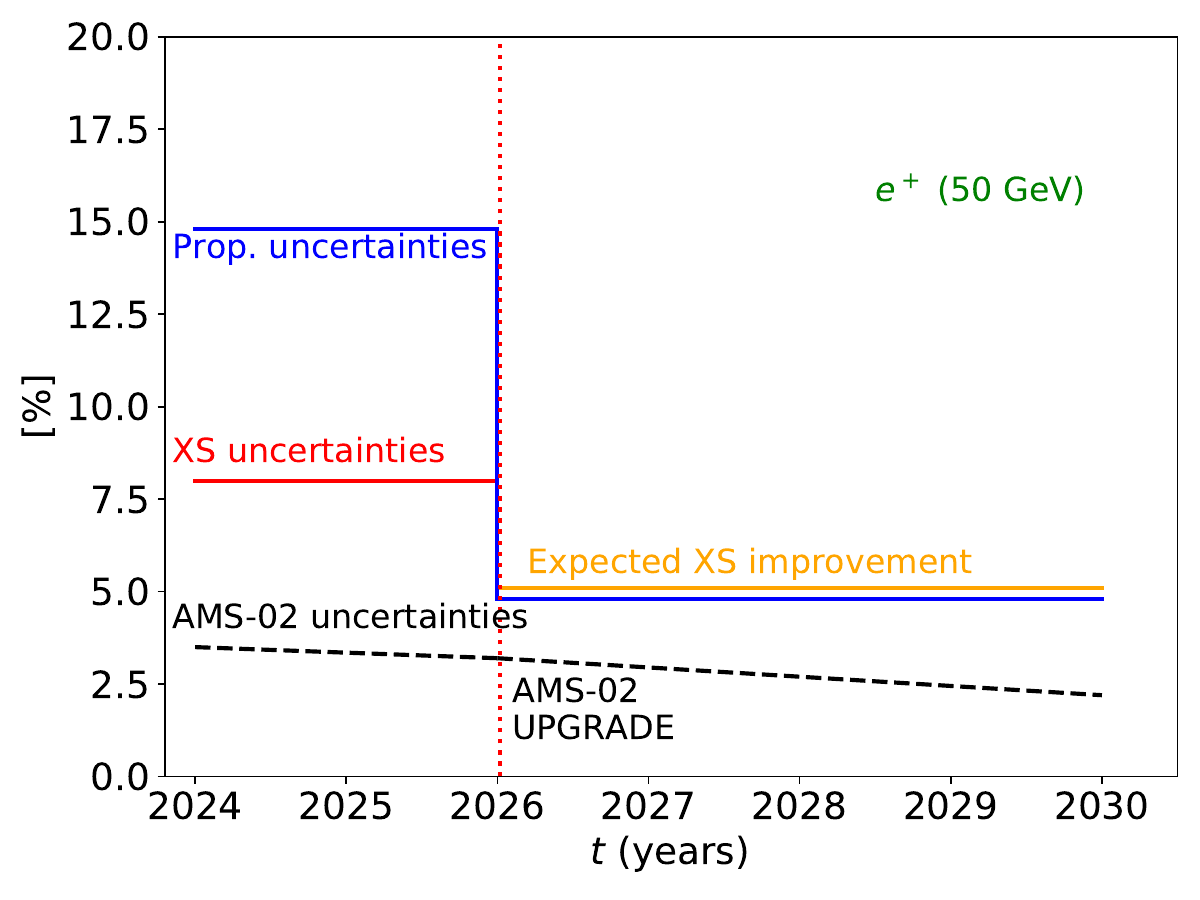}
    \includegraphics[width=8.6cm]{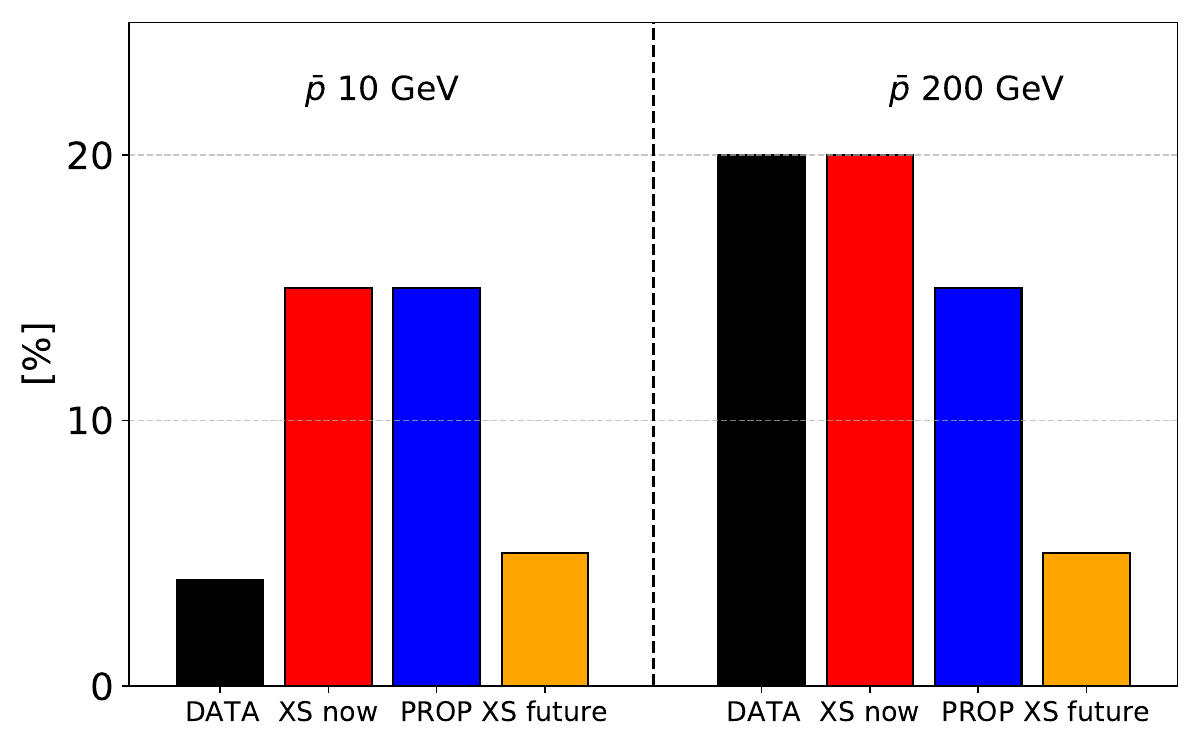}
	\includegraphics[width=7.4cm]{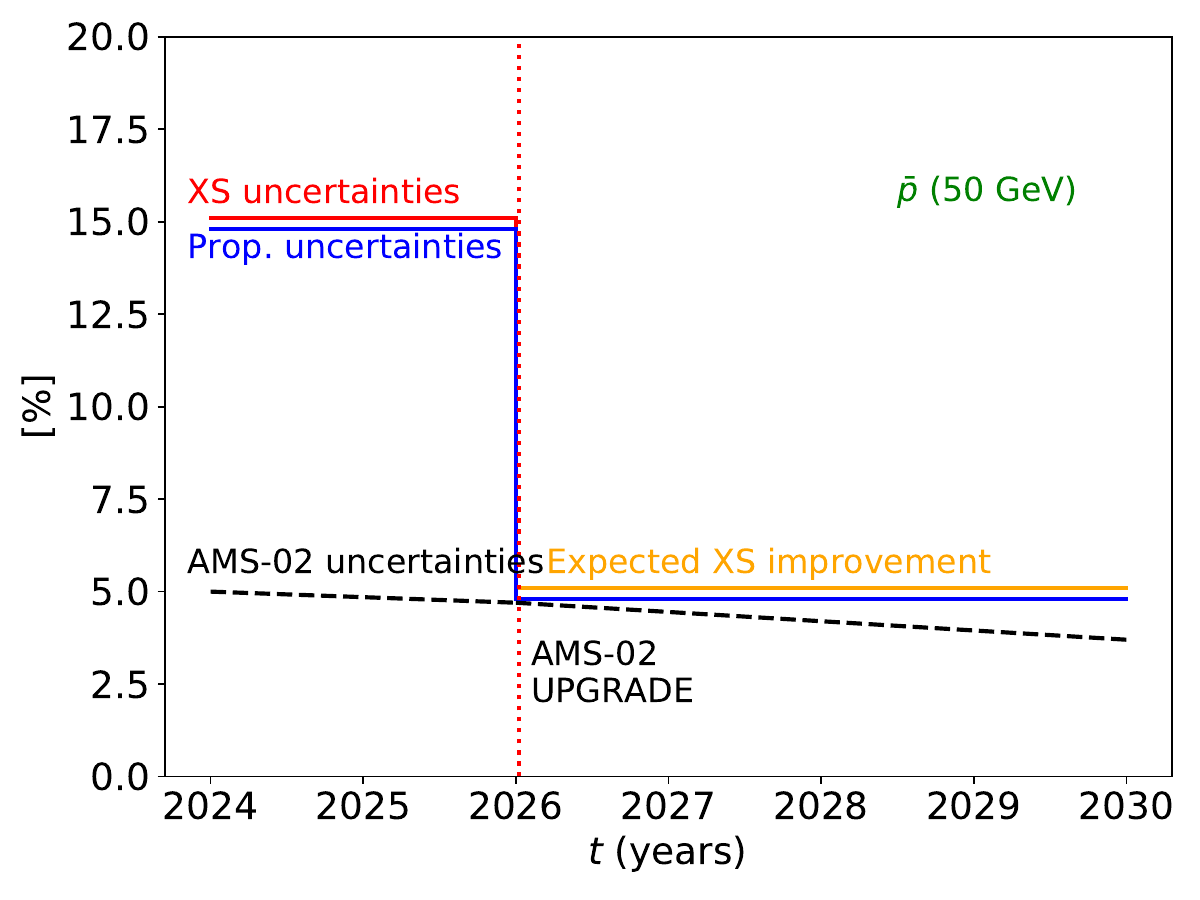}
	\caption{\label{fig:XS_cases}Left: the current \acrshort{ams} experimental \antiproton errors at 10 and 200\,GeV, alongside cross-section and propagation uncertainties. Right: prospects for the future, showing when and how cross-section uncertainties might reach levels comparable to \acrshort{ams} data.}
\end{figure}

To conclude this section, as illustrative cases, Fig.~\ref{fig:XS_cases} presents the current and future experimental errors for the \acrshort{ams} flux data and cross-sections for \antiproton and electrons, together with the theoretical errors related to propagation. Current cross-section data uncertainties are $15\%$ for \antiproton and $8\%$ for positrons, while the current \acrshort{ams} data errors are taken from \cite{AMS:PhysRep2021}. The propagation uncertainties are taken to be about $15\%$ across all the energies \cite{DiMauro:2024kml, Genolini:2021doh}.
Currently, the propagation and cross-section uncertainties are much larger than the \acrshort{ams} flux errors. This is true for both positrons and \antiproton and for the most relevant energies for propagation and new physics studies.
In the future, with the \acrshort{ams} upgrade, the \acrshort{cr} flux errors could reach about $4\%$ for \antiproton and $2\%$ for positrons at 50\,GeV. The envisioned improvements in the $e^+$ and \antiproton production cross-sections could reduce significantly the theoretical errors, at a level close to the \acrshort{ams} ones. Moreover, as explained in Sec.~\ref{sec:xs_nucforecasts}, the envisioned improvements in the nuclear cross-sections will bring to a significant reduction of the propagation uncertainties.

\subsection{Other relevant cross-sections for data interpretation and experiments}
\label{sec:xs_other}

Accurately modelling the propagation of \acrshort{gcr} (anti-)nuclei from their source to their detection site requires precise knowledge of all cross-sections governing their interaction with the matter they encounter. Furthermore, and for antinuclei in particular, annihilation cross-sections are important not only to propagation but also to experiments. Some instruments purposely built for the detection of antinuclei, such as \acrshort{gaps}~\cite{GAPS:ICRC2023}, rely on the characteristic patterns of secondary particles, mostly pions and $\gamma$ rays, created upon annihilation, and therefore require knowledge of the multiplicities and energy spectra of these secondaries.

\subsubsection{Contributions to $\sigma_{\rm tot}$ (inelastic, quasi-elastic, etc.) for nuclei and antinuclei}

The total cross-section for the interaction of (anti-)nuclei with matter is given by the sum of the cross-sections for the different physical processes that can occur. In the following, we present decompositions of the cross-sections for nuclei and antinuclei that are consistent with each other, and briefly discuss the inconsistent use of nomenclature in the literature.

\paragraph{Contributions for nuclei}
Following the notation used in Ref.~\cite{NA61SHINE:2019aip}, we decompose the total cross-section into
\begin{equation}
    \sigma_{\rm tot} = \sigma_{\rm el} + (\sigma_{\quasiel} + \sigma_{\rm prod}) = \sigma_{\rm el} + \sigma_{\rm inel},
    \label{eq:def_tot-inel-nuclei}\,.
\end{equation}
Considering a generic reaction for a projectile $i$ on a target $j$, the various subscripts in the above equation correspond to total (i.e., $i+j\to X$), elastic ($i+j\to i+j$), the sum of the quasi-elastic ($i+j\to i+j+X$) and production ($i+j\to X~{\rm not}~i$) reactions, combined into the total inelastic cross-section, $\sigma_{\rm inel}$. Other notations are also used in the literature, namely $\sigma_{\rm R}$ (reaction) for $\sigma_{\rm inel}$ and $\sigma_{\rm abs}$ (absorption) for $\sigma_{\rm prod}$. However, as stressed in Ref.~\cite{NA61SHINE:2019aip}, not all measurements and experiments use the same terminology for these processes (e.g., $\sigma_{\rm abs}$ has been used for $\sigma_{\rm prod}$ and $\sigma_{\rm inel}$), leading to some confusion. In the \acrshort{gcr} community, the modelling for propagation studies relies on $\sigma_{\rm inel}$, but the different roles of $\sigma_{\quasiel}$ and $\sigma_{\rm prod}$ is probably overlooked -- as illustrated in Fig.~6 of Ref.~\cite{2018PhRvC..98c4611G} for C projectiles, where $\sigma_{\rm inel}$ and $\sigma_{\rm prod}$ data are treated on the same footing.

\paragraph{Contributions for antinuclei}
As for nuclei, the total antinuclei interaction cross-section can be decomposed into
\begin{equation}
    \sigma_{\rm tot} = \sigma_{\rm el} + \left( \sigma_{\quasiel} + [\sigma_{\rm ann} +\sigma_{\rm prod}]\right)=  \sigma_{\rm el} + (\sigma_{\quasiel} + \sigma_{\rm abs}) = \sigma_{\rm el} + \sigma_{\rm inel},
    \label{eq:def_tot-inel-antinuclei}
\end{equation}
where the subscripts stand for total, quasi-elastic, annihilating, production, absorption and inelastic cross-sections, respectively. Compared to the total cross-section for nuclei, Eq.~\eqref{eq:def_tot-inel-nuclei}, annihilation is a third inelastic contribution to take into account: at high energy (above a few tens of GeV), $\sigma_{\rm ann}\approx0$, so that $\sigma_{\rm inel}\approx\sigma_{\quasiel}+\sigma_{\rm prod}$. In \acrshort{gcr} publications, $\sigma_{\rm abs}$ is often denoted $\sigma_{\rm ann}$, while $\sigma_{\quasiel}$ is denoted $\sigma_{\rm non-ann}$ or $\sigma_{\rm NAR}$ (for non-annihilating rescattering).

\subsubsection{Inelastic and other relevant cross-sections for interpreting \texorpdfstring{\acrshort{gcr}}{GCR} nuclear data}
\label{sec:XS_inelastic}

The low-energy part of the \acrshort{gcr} spectrum is shaped by energy losses, dominant below a few hundred~MeV/n, and by inelastic interactions --- the second term in Eq.~\eqref{eq:transport_sink}. The latter is the second most important ingredient, after production cross-sections, required to model \acrshort{gcr} fluxes with percent-level precision.

\paragraph{Relevance of elastic, quasi-elastic and production contributions in Eq.~\eqref{eq:def_tot-inel-nuclei}}

First, in all \acrshort{gcr} propagation codes, elastic interactions are neglected. However, a recent analysis pointed out that their impact is $\sim\!1\%$ for protons below GeV energies \cite{2024PhRvD.109l3024E}. For consistency, this effect should thus be taken into account in the modelling, given the percent-level precision of measured proton fluxes. 

Second, the correct way to model \acrshort{gcr} transport should be to use $\sigma_{\rm inel}$ (i.e., including $\sigma_{\quasiel}$) {\em and} consider the energy redistribution of the surviving projectile towards lower energies. The latter effect is modelled via the differential cross-section $\dd\sigma_{\quasiel}(E_{\rm in},E_{\rm out})/\dd E_{\rm out}$, which is non-zero for $E_{\rm out}\lesssim E_{\rm in}$ only, and which satisfies $\int_0^\infty (\dd\sigma_{\quasiel}/\dd E_{\rm out})\;\dd E_{\rm out}=\sigma_{\quasiel}(E_{\rm in})$. This redistribution is important for antinuclei (see next section), but is expected to be sub-dominant for nuclei: indeed, \acrshort{gcr} nuclei shifted to lower energies add up to a larger flux (at these lower energies), owing to a power-law behaviour observed down to $\sim\!100$\,MeV/n \cite{2013Sci...341..150S,2016ApJ...831...18C}. This energy redistribution is not considered in current propagation codes and should be quantified, as it may impact the flux, in the GeV/n regime, at the few-percent level.

\paragraph{Parametrisations and codes for $\sigma_{\rm inel}$ (often denoted $\sigma_{\rm R}$ in the literature)}

The simplest approximation for this cross-section considers the total surface of the projectile $i$ and the target $j$, i.e., $\sigma_{\rm inel}\propto (A_i^{1/3}+A_j^{1/3})^2$. A first correction was proposed in the 1950s in Ref.~\cite{Bradt1950},
\begin{equation}
    \sigma_{\rm inel} = \pi r_0 \left(A_i^{1/3}+A_j^{1/3} -b_0\right)^2\,,
\end{equation}
with $r_0$ the effective radius of the nucleus and $b_0$ the overlapping or transparency parameter. To account for the energy dependence of $\sigma_{\rm inel}$ at low energy, several improvements and refinements have been proposed since~\cite{1983ApJS...51..271L, 1993PhRvC..47.1225S, 1989NuPhA.491..130S, Barshay1974, Barshay1975, Barashenkov1994, Wellish1996, 1996NIMPB.117..347T, 1999NIMPB.155..349T, 2003PhRvC..67f4605I, 2007JPSJ...76d4201I, Sihver2009, 2014PhRvC..89f7602S, 2014NIMPB.334...34S}, based on the inclusion of more nuclear effects (Coulomb barrier, Pauli blocking, etc.), and thanks to growing nuclear data sets. Several of the above parametrisations are tailored for interactions on protons, scaled to He targets following Ref.~\cite{1988PhRvC..37.1490F}, while others provide full parametrisations for nucleon--nucleon interactions. For light projectile--target systems, whose nuclear structure is quite different, specific modifications of the above formulae were proposed \cite{Barashenkov1994, Wellish1996, 1999NIMPB.155..349T} (see, in particular, Ref.~\cite{2022ApJS..262...30P} for comparisons of these predictions).
Some of the above cases are available in general-purpose tools (e.g., \texttt{Geant4} and \texttt{PHITS}). Alternative approaches are also being developed, for instance based on the Glauber theory (see, e.g., Refs.~\cite{1989CoPhC..54..125S, 2003PhRvC..67e4607S, 2009PhRvC..79f1601T, 2019NuPhA.989...21F}), and in nuclear and \acrshort{mc} codes (\texttt{FLUKA} \cite{2004AdSpR..34.1302A}, \texttt{CEM} event generator in \texttt{MCNP6}~\cite{2015NIMPB.356..135K}, the \texttt{Liège intranuclear-cascade model}~\cite{2017PhRvC..96e4602R}) -- the modelling of the elastic and differential cross-sections is included in some of these codes.

\begin{figure}
	\centering
	\includegraphics[width=0.48\textwidth]{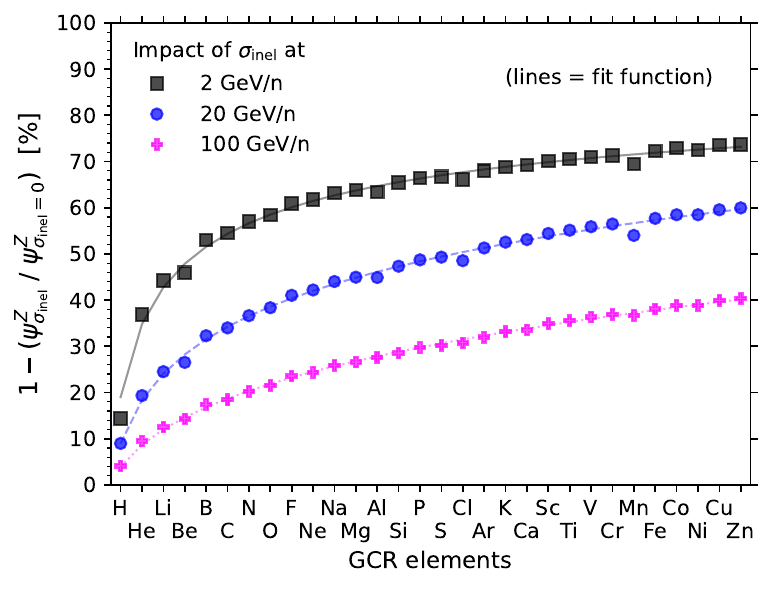}
   	\includegraphics[width=0.49\textwidth]{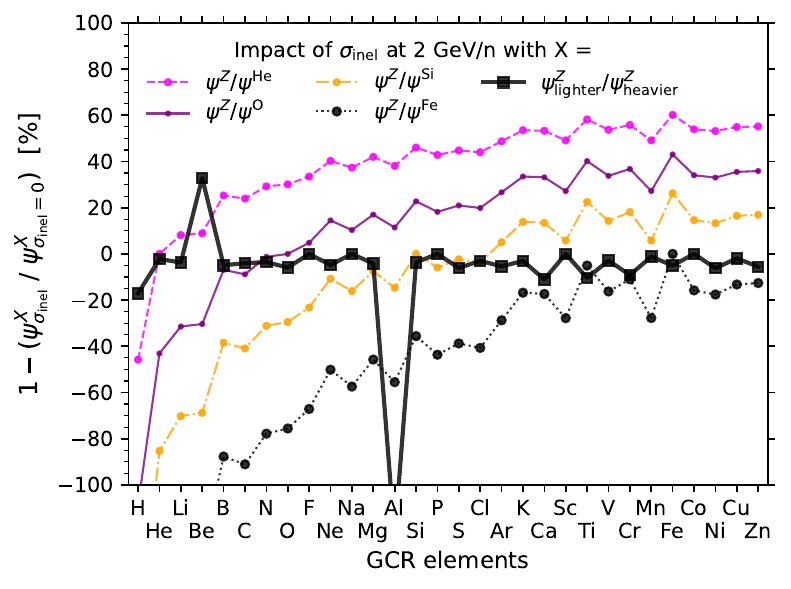}
	\caption{Impact of destructive interactions, Eq.~\eqref{eq:impact_xsinel}, on the flux of \acrshort{gcr} with $Z \le 30$. Left: impact on elemental fluxes at 2\,GeV/n (black squares), 20\,GeV/n (blue discs) and 100\,GeV/n (magenta crosses). The lines correspond to the fit function given in Eq.~\eqref{eq:xsinel_fitfunction}. Right: impact on the ratio of two elemental fluxes (coloured lines with discs) or the lighter-to-heavier isotopic flux ratio of the corresponding element (solid black line and squares). For the latter, the spikes for Be and Al are related to the presence of a $\beta$-unstable isotope in the ratios $X=(^{7}$Be/$^{10}$Be) and $X=(^{26}$Al/$^{27}$Al) because the fluxes of radioactive isotopes are insensitive to destruction at low energy. In this plot, the impact of inelastic interactions on flux ratios is shown at 2\,GeV/n only. See text for discussion.}
    \label{fig:XS_inel_impact}
\end{figure}

Over the years, systematic comparisons between the proposed cross-sections have been carried out \cite{1998ApJ...501..911S,2012AdSpR..49..812S,2014NIMPB.334...34S,2015NIMPB.356..135K,2021NJPh...23j1201L}; see, in particular, Ref.~\cite{2021NJPh...23j1201L} for the most recent comparison of a variety of parametrisations.
Overall, the nucleon--nucleon \acrshort{nasa} parametrisations~\cite{1996NIMPB.117..347T}, with special cases for light systems \cite{1999NIMPB.155..349T}, are always among the most successful to match all data sets: they are thus the ones used below for estimating the error budget in \acrshort{gcr} fluxes. It is worth stressing that many nuclear data reach a $\lesssim 5\%$ precision, but as their coverage is not complete in terms of reaction (see Fig.~\ref{fig:XS_inel_impact}), the precision of the models is unclear, because their spread in the asymptotic high-energy region can be as large as $20\%$ (e.g., \cite{2022JInst..1703012Z}).

\paragraph{From \acrshort{gcr} data precision to desired $\Delta\sigma_{\rm inel}/\sigma_{\rm inel}$ precision}
To estimate the precision needed on $\sigma_{\rm inel}$ for propagation modelling, we first quantify their impact on fluxes and flux ratios. We define the impact on the isotopic flux $\psi^j$ to be
\begin{equation}
    I_{\rm inel}^{j} = 1-\frac{\psi^
    j}{\psi^j_{\sigma_{\rm inel}^{j+\rm (H,He)}=0}},
    \label{eq:impact_xsinel}
\end{equation}
i.e., the relative difference between the modelled flux with and without $\sigma_{\rm inel}$. For simplicity, the \acrshort{ism} targets are not considered separately. This relative difference is calculated with \texttt{USINE}, and is shown in the left panel of Fig.~\ref{fig:XS_inel_impact} for several energies and elements with $Z\leq30$ (fluxes are not directly measured for heavier elements, see Sec.~\ref{sec:CR_UHN}, so they are not shown). The flux of individual isotopes (also not shown) have similar $\sigma_{\rm inel}$, hence exhibit similar impacts (only isotopic fluxes for $Z<6$ have been reported so far, see Sect.~\ref{sec:CR_isotopes}). Overall, the impact of $\sigma_{\rm inel}$ on fluxes decreases with energy (from black squares to pink crosses) -- as the timescale of escape from the Galaxy becomes much shorter than the inelastic cross-section timescale ($t_{\rm diff}/t_{\rm inel}\propto 1/R^\delta$, with $\delta$ the diffusion slope) --, and increases with the \acrshort{gcr} mass (as $\sigma_{\rm inel}\propto A^{2/3}$).

It is also useful to show the impact of destruction on elemental and isotopic flux ratios, the latter being often published in experiments (as they minimise the systematics) and also used in \acrshort{gcr} phenomenology analyses. The thin lines (disc symbols) in the right panel of Fig.~\ref{fig:XS_inel_impact} show the impact of $\sigma_{\rm inel}$ on elemental ratios: the best way to mitigate this impact (and that of $\Delta\sigma_{\rm inel}$) is to consider adjacent charges, i.e., $\psi^{Z}/\psi^{Z_{\rm ref}}$ with $|Z-Z_{\rm ref}|\lesssim3$. For isotopes, most of the \acrshort{cr} data consists of ratios at low energy, and the solid black line (square symbols) shows that these ratios strongly mitigate the impact of $\sigma_{\rm inel}$, and thus of $\Delta\sigma_{\rm inel}$. The only, but very important, exception is for ratios involving a $\beta$-unstable isotope (e.g., $^{10}$Be in $^{10}$Be/$^9$Be). In that case, $\sigma_{\rm inel}$ does not impact the flux of the radioactive species (whose transport is dominated by its decay time), while the stable isotope is fully impacted (no mitigation). As a result, the impact of $\sigma_{\rm inel}$ on this ratio is directly that of the stable isotope.

The precision needed for $\sigma_{\rm inel}$ to reach a desired flux modelling precision is given by
\begin{equation}
 \left(\frac{\Delta \psi^Z}{\psi^Z} \right)^{\rm modelled}(E_{\rm k/n}) =
 \left(\frac{\Delta \sigma_{\rm inel}^Z}{\sigma_{\rm inel}^Z}\right) \times \left|I_{\rm inel}(Z,E_{\rm k/n})\right|\,.
 \label{eq:xsinel_forecast}
\end{equation}
With the best \acrshort{cr} data precision at the few-percent level for many elements and some very light isotopes, a modelling precision of $1\%$ should be targeted, i.e., nuclear data with
$\Delta \sigma_{\rm inel}/\sigma_{\rm inel} \lesssim 1/|I_{\rm inel}|$ are needed.
To calculate this number for any species and energy, $I_{\rm inel}$ can be parametrised as
\begin{equation}
 I_{\rm inel}^{Z+\rm (H,He)}(E_{\rm k/n}) =
\left[\left(a_0+a_1E_{\rm k/n}+a_2\log_{10}\left(E_{\rm k/n}\right)\right)\times \log_{10}(Z+0.45)\times Z^{b_0+b_1E_{\rm k/n}+b_2\log_{10}\left(E_{\rm k/n}\right)}\right]\,,
\label{eq:xsinel_fitfunction}
\end{equation}
with $a_{(0,1,2)} = (0.49565, 0.00069, -0.104312)$, $ b_{(0,1,2)} = (-0.254173, 0.000252, 0.070613)$ and $E_{\rm k/n}$ in GeV/n, as shown by the thin lines in the left panel of Fig.~\ref{fig:XS_inel_impact}. 
Finally, to get the elemental flux at the desired precision, the uncertainty on the isotopic inelastic cross-sections is further weighted by the contribution of each isotope $j$ to the flux of the element $Z$:
\begin{equation}
\left(\frac{\Delta\sigma_{\rm inel}^j}{\sigma_{\rm inel}^j} \right) =
 \left(\frac{\Delta \sigma_{\rm inel}^Z}{\sigma_{\rm inel}^Z}\right) \times \left(\frac{\psi^Z}{\psi^j}\right)^{\rm GCR data}\,.
 \label{eq:xsinel_forecast_isotopes}
\end{equation}

\begin{sidewaysfigure}
	\centering
	\includegraphics[width=\textwidth]{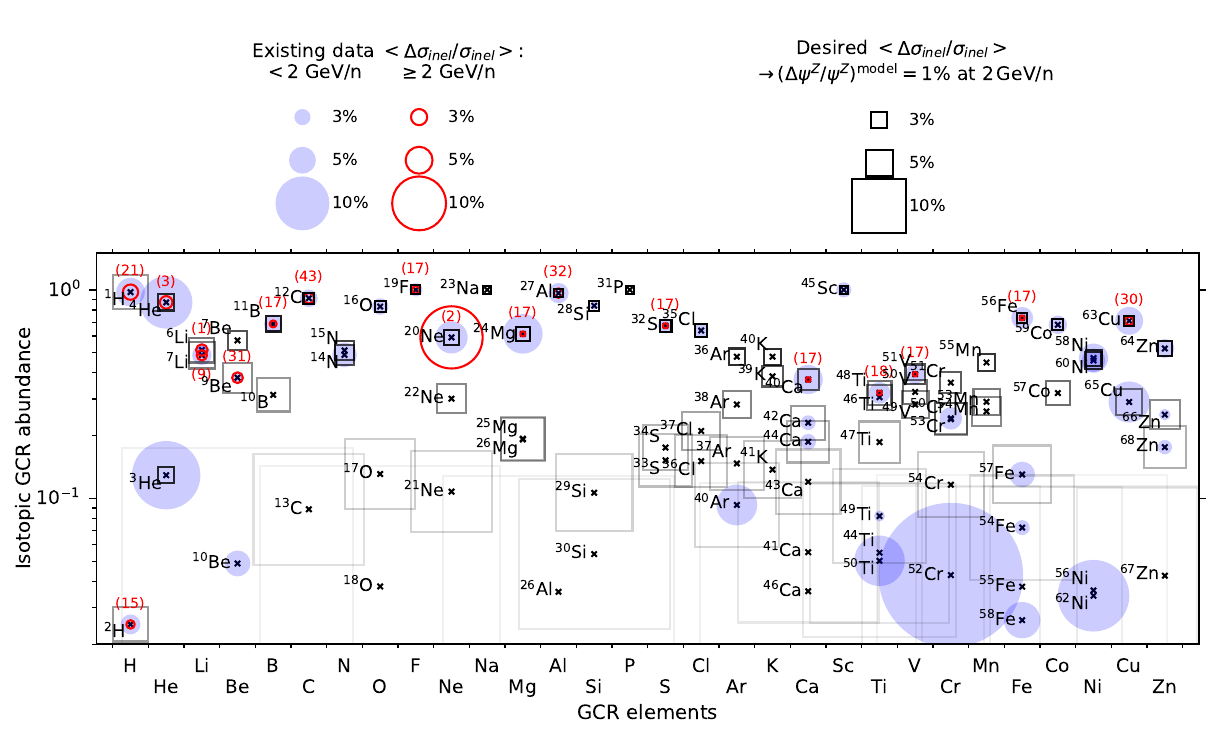}
	\caption{For all meta-stable isotopes in \acrshort{gcr}s (sorted by increasing $Z$ on the $x$-axis), the plot shows: (i) the relative abundance of each isotope in \acrshort{gcr}s ($\times$ symbol), calculated from the weighted average over all data extracted from \acrshort{crdb}; (ii) the currently existing nuclear data and their average precision below (blue discs) and above 2\,GeV/n (red circles); (iii) the cross-section precision needed to get a $1\%$ modelling precision (black/grey squares) on $Z\leq2$ isotopic fluxes and $Z\leq30$ elemental fluxes, as obtained from Eqs.~\eqref{eq:xsinel_forecast}, \eqref{eq:xsinel_fitfunction} and \eqref{eq:xsinel_forecast_isotopes}. To illustrate the scarcity of data in the high-energy asymptotic regime, the number of available data sets is indicated in red in parentheses. Higher-precision (or any) data are needed for each isotope for which the red or blue circles are larger than the black/grey squares, or if there are no red circles at all. See text for discussion.\label{fig:XS_inel_data}}
\end{sidewaysfigure}

These values are shown in Fig.~\ref{fig:XS_inel_data}, where the needs are compared to the current precision of (or merely whether there exists) nuclear data for all relevant isotopes.
The $y$-axis shows the \acrshort{gcr} relative abundance of relevant isotopes, averaged over all data found in \acrshort{crdb}. The desired nuclear data precision calculation is weighted by the above isotopic fraction, Eq~\eqref{eq:xsinel_forecast_isotopes}.
Some inelastic cross-section data used in Fig.~\ref{fig:XS_inel_data} were extracted from the \texttt{EXFOR} database~\cite{Otukaetal2014, 2018NIMPA.888...31Z, 2022JInst..1703012Z}: pN scattering measurements below 1\,GeV/n, as compiled in Refs.~\cite{Bauhoff:1986gcb, Carlson:1996ofz} and, above 2\,GeV/n, very few data points for pd \cite{Chen:1956zz, Bugg:1966zz}, p on He, Be, Al and Cu \cite{Jaros:1977it, Dietrich:2002swm}. The rest of the data above 2\,GeV/n were retrieved from the largest set of Ref.~\cite{Bobchenko:1979hp}, complemented by Refs.~\cite{Ashmore:1960zz, Bellettini:1966zz, Igo:1967osa, Denisov:1971jb, Ganguli:1973rja, Denisov:1973zv, Carroll:1978hc, Fumuro:1979yw, Azimov:1981by}, and more recent measurements from 50 to 900\,GeV \cite{Dudkin:1994ui, SELEX:1999cvg, Carvalho:2003pza}, including the very recent NA61/\acrshort{shine} data points \cite{NA61SHINE:2019aip}.

Overall, the existing nuclear data points exhibit precisions below a few percent, but it is somehow difficult to believe that data taken from the 1960s to the 1980s are as precise as the data reported very recently from NA61/\acrshort{shine}~\cite{NA61SHINE:2019aip}. Moreover, most data come from interaction with elements in natural abundances, not always dominated by a single isotope. Also, the most abundant natural and \acrshort{gcr} isotopes are not always the same.
The main conclusion from this plot is that the data for $\sigma_{\rm inel}$ above a few GeV/n (i.e., in the asymptotic regime) remain scarce for many important isotopes/elements (e.g., Be, N, O and heavier species) while, for the modelling, a precision of $\lesssim 3\%$ is needed for all leading-order \acrshort{gcr} isotopes.

\paragraph{Beyond the straight-ahead approximation for nuclear production}
For secondary production, all propagation calculations rely on the straight-ahead approximation, Eq.~\eqref{eq:SHA}, where the kinetic energy per nucleon of the fragment is conserved. However, non-zero recoil velocities of the fragments \cite{2007PhRvC..75d4603V} lead to a broadening of their energy distributions. Besides the intrinsic interest of connecting this velocity to reaction mechanisms and to the internal nucleonic motion (e.g., \cite{2004PhRvC..69f4601G, 2023PhyS...98h5301M}), this broadening, if not taken into account, affects the precision of the modelled fluxes. As studied in Ref.~\cite{1995ApJ...451..275T}, it can be parametrised by a Gaussian distribution of the momentum transfer, with a parabolic dependence on the fragment mass. For the B/C ratio, the impact of relaxing the straight-ahead assumption was found to be $\lesssim 6\%$, peaking at $\sim\!1$\,GeV/n; similar conclusions were reached using an improved modelling of the recoil-velocity distributions~\cite{2003ApJ...589..217K}. This implies that the effect must be incorporated when comparing modelled secondary fluxes to \acrshort{ams} data. In particular, it should also be re-investigated and assessed for different secondary elements. A priori, the precision of current data should be sufficient for modelling this effect, given its small impact on the calculated fluxes. This hypothesis, however, must be confirmed with a dedicated study.

\paragraph{Electron attachment and stripping cross-sections for electron capture decay}

Beside spontaneous $\beta$-decay, \acrfull{ec} decay -- whereby an orbital electron is captured by a proton in the nucleus and the latter is transformed into a neutron -- is another catastrophic loss to consider in the \acrshort{gcr} transport equation. It involves a competition between three timescales \cite{1984ApJS...56..369L, 1985Ap&SS.114..365L}: $t_{\rm attachment}$, $t_{\rm stripping}$ and $t_{\rm EC}$, which refer to the timing of the attachment and stripping of electrons in the \acrshort{ism}, and to the \acrshort{ec} decay, respectively. Precise \acrshort{ec}-decay time can be challenging to disentangle from $\beta^+$ decay experimentally, as the daughter nucleus is the same in both decays, and fully ionised species (in the laboratory) are needed to disentangle the two channels. According to available data \cite{1978PhDT........12W, 1979PhDT........67C}, the attachment and stripping cross-sections vary roughly as ($Z^5$, $E^{1.5}$) and ($Z^{-2}$, $E^0$), respectively. This explains why, above a few hundred MeV/n, \acrshort{gcr} species are fully ionised and why heavy species at low energy are the most likely to pick up electrons. For this reason, the latter have, among others, been used as (re-)acceleration clocks (see the discussion and references in Sec.~\ref{sec:CR_isotopes}, and see also, e.g., Ref.~\cite{2017ApJ...851..109B}).
It is worth stressing that the current attachment and stripping cross-sections are based on parametrisations from the 1980s \cite{1984ApJS...56..369L, 1985Ap&SS.114..365L} fit on data from the 1970s \cite{1978PhDT........12W, 1979PhDT........67C}. So, in principle, they should be updated and their uncertainties re-estimated. However, in Ref.~\cite{
2024icrc.confE..66B}, the impact of \acrshort{ec} in the associated fluxes and flux ratios -- which is maximal at $\lesssim 1$\,GeV/n -- was estimated to be at most at the precision of the current data (\acrshort{ams} and Super\acrshort{tiger}). Therefore, there seems to be no urgency for new measurements of these cross-sections, unless one can assess that they are not known at better than a $50\%$ precision.

\subsubsection{Inelastic and non-annihilating cross-sections for interpreting \texorpdfstring{\acrshort{gcr}}{GCR} antinuclei data}
\label{sec:XS_inelastic_antinuc}

The precision of available data and required improvements for the three inelastic processes of Eq.~\eqref{eq:def_tot-inel-antinuclei} are discussed below. As underlined in Sec.~\ref{sec:XS_inelastic}, we do not discuss elastic scattering, as it peaks in the forward direction (and hence results in negligible energy losses) and has a $\lesssim 1\%$ impact on antinuclei fluxes.

\begin{figure}
	\centering
	\includegraphics[width=0.55\textwidth]{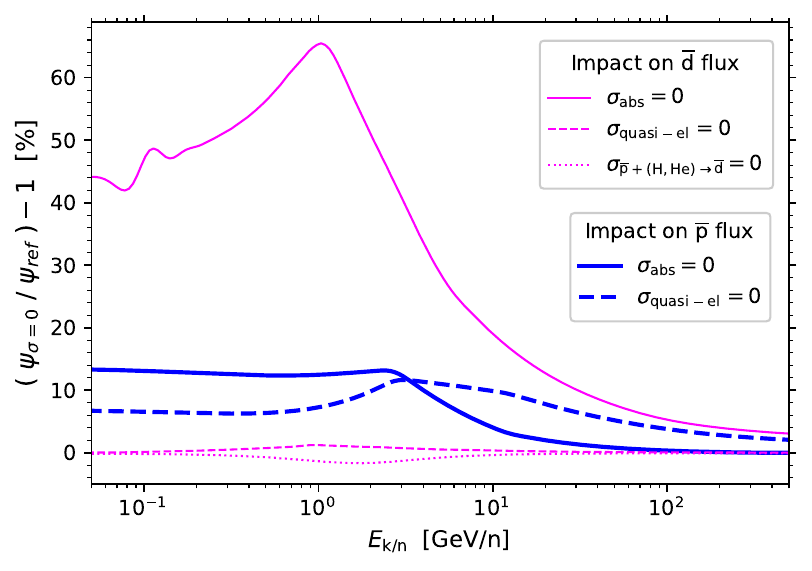}
	\caption{Impact of $\sigma_{\rm abs}$ (solid lines)and $\sigma_{\quasiel}$ (dashed lines) on \antideuteron (thin magenta) and \antiproton (thick blue) fluxes as a function of $E_{\rm k/n}$. For \antideuteron, the specific contribution to the flux of the production via \antiproton is also shown, as discussed in Sec.~\ref{sec:xs_prodpbar}.
    \label{fig:XS_pbardbar_impact}}
\end{figure}

\paragraph{Desired precision on inelastic and quasi-elastic cross-sections for \texorpdfstring{\acrshort{gcr}}{GCR} antinuclei}
Figure~\ref{fig:XS_pbardbar_impact} shows the impact of the above cross-sections on the flux of \antiproton and \antideuteron calculated with the \texttt{USINE} code. It does so in the form of the relative difference,
\begin{equation}
  I^{\overline{\rm p}, \overline{\rm d}}_\sigma = \psi^{\overline{\rm p}, \overline{\rm d}}_{\sigma=0}/\psi^{\overline{\rm p}, \overline{\rm d}}_{\rm ref} -1\,,
\end{equation}
for the reaction cross-section, $\sigma_{\rm inel}$ (solid lines), and for quasi-elastic scattering, $\sigma_{\quasiel}$ (dashed lines), using the parametrisations of Refs.~\cite{2005PhRvD..71h3013D, 2003EPJA...16...27D, 2003PhRvD..68i4017D, 2003EPJA...18..597D, Donato:2008yx} and assuming that $\dd\sigma_{\quasiel}/\dd E=\sigma_{\quasiel}/E_{\rm k/n}$. The impact of $\sigma_{\rm abs}$ on \antiproton and \antideuteron is similar to its effect on p and \deuteron shown in Fig.~\ref{fig:XS_inel_impact}: it is highest at low rigidities and increases with the atomic number, $A$. Actually, the impact of $\sigma_{\rm abs}$ (resp. $\sigma_{\quasiel}$) increases (resp. decreases) with $A$, so that $\sigma_{\rm abs}$ is the dominant source of uncertainty for $A\geq2$.
To model the \antiproton flux at the precision of \acrshort{ams} data, we thus need the uncertainty of both $\Delta\sigma_{\rm abs}/\sigma_{\rm abs}$ and $\Delta\sigma_{\quasiel}/\sigma_{\quasiel}$ to be much better than $(\Delta\psi/\psi)^{\rm data}/I_\sigma$, i.e., much smaller than $20\%$. For $(\Delta\psi/\psi)^{\rm modelled}\lesssim1\%$, this translates into $\Delta\sigma/\sigma\lesssim 5\%$. 
For some forthcoming \acrshort{cr} experiments, and as part of a long-term effort towards the detection of antinuclei (e.g., for \acrshort{dm} searches), we must already begin validating and improving the existing~\cite{1997APh.....6..379M, 2011PhLB..705..235U} inelastic cross-section parametrisations (e.g., used in \texttt{Geant4}) for antinuclei--\proton and antinuclei--antinuclei. While not critical for current \acrshort{gcr} modelling, new data for the differential cross-section $\dd\sigma_{\rm non-ann}/\dd E$ and any data for $\sigma_{\rm non-ann}$ for \antideuteron (none exist at the moment) should prove very useful as well in the coming years.

\paragraph{Status of total and elastic scattering data}

Total-reaction ($\sigma_{\rm tot}$) and elastic ($\sigma_{\rm el}$) cross-sections for \antiproton--\proton scattering were measured at \acrshort{cern}'s \acrshort{lear} (\acrlong{lear})~\cite{Lemaire1985, Bugg1987}, \acrshort{isr} (\acrlong{isr})~\cite{Carboni1982, Owen1983}, \acrshort{ps} (\acrlong{ps})~\cite{Amaldi1964} and \acrshort{sps} (\acrlong{sps})~\cite{Arnison1983, Owen1983, Bozzo1984, Alner1986}, at Fermilab~\cite{Carroll1976, Carroll1979, Abe1994, Avila1999}, at \acrshort{lbnl} (\acrlong{lbnl})~\cite{Coombes1958, Elioff1959, Cork1962, Elioff1962}, at \acrshort{bnl} (\acrlong{bnl})~\cite{Lindenbaum1961, Galbraith1965, Abrams1970} and at \acrshort{ihep} (\acrlong{ihep})~\cite{Allaby1969, Denisov1971}, over a wide range of energies. Most early fixed-target experiments were performed at \antiproton beam momenta between 200\,MeV/c and 900\,MeV/c, and between 4\,GeV/c and 370\,GeV/c. Only one experiment covered the range from 575\,MeV/c to 5.35\,GeV/c. For a compilation of all these data, see Ref.~\cite{Workman:2022ynf}. Many of the older measurements suffer from poorly understood (or reported) systematic uncertainties, and few have statistical uncertainties better than 5\%. Later experiments at colliders were more precise but focused on much higher energies ($31\,{\rm GeV} \le \sqrt{s} \le 1.8$\,TeV, corresponding to fixed-target projectile momenta between 512\,GeV/c and 1727\,TeV/c), where the difference between \antiproton--\proton and \proton--\proton scattering is negligible.

\paragraph{Status of inelastic scattering data}

Less data is available on inelastic \antiproton--nucleus scattering ($\sigma_{\rm non-ann}$). Measurements were performed on deuterons (deuterium) for projectile momenta between 1\,GeV/c and 370\,GeV/c at \acrshort{lbnl}~\cite{Elioff1962}, at \acrshort{bnl}~\cite{Galbraith1965, Abrams1970} and at Fermilab \cite{Carroll1976, Carroll1979}. Total-reaction cross-sections have also been published at a few selected energies (i.e., usually not more than two or three per experiment) for He \cite{Balestra1985, Balestra1988}, Li \cite{Denisov:1973zv}, Be \cite{Cork1957, Denisov:1973zv}, C \cite{Cork1957, Denisov:1973zv, Aihara1981}, O~\cite{Agnew1957}, Al \cite{Denisov:1973zv, Aihara1981}, Cu \cite{Denisov:1973zv, Aihara1981, Agnew1957}, Ag \cite{Agnew1957}, Sn \cite{Denisov:1973zv}, Pb \cite{Cork1957, Denisov:1973zv, Agnew1957} and U \cite{Denisov:1973zv}. These, however, often suffer from poorly understood uncertainties and limited statistical significance. Comprehensive measurements are therefore needed to improve the ability to model the interaction of \antiproton with, for example, detectors and the shielding surrounding them.

Besides \antiproton, the first measurement of the inelastic cross-section for \antideuteron--nucleus interactions at low energies was performed by the \acrshort{alice} experiment at momenta between 300\,MeV/c and 4\,GeV/c~\cite{ALICE:dbar2020}. The only previously published measurements were those of the \antideuteron absorption cross-sections in Li, C, Al, Cu, and Pb at fixed momenta of 25\,GeV/c~\cite{Binon1970} and 13.3\,GeV/c~\cite{Denisov1971b}. In a unique experimental approach, the \acrshort{alice} collaboration used their detector as interaction target for \antideuteron created in pPb collisions at $\sqrtsnn = 5.02$\,TeV.\footnote{In collisions involving nuclei $a$ and $b$ with four-momenta $p_a$ and $p_b$, respectively, it is conventional to introduce another variable in addition to the centre-of-mass energy $\sqrt{s}=\sqrt{(p_a+p_b)^2}$, denoted the \textit{centre-of-mass energy per nucleon pair} $\sqrtsnn\equiv\sqrt{(p_a/A_a+p_b/A_b)^2}$, $A_a$ and $A_b$ being the respective nuclear mass numbers. For collisions involving equal nuclei, one has $\sqrtsnn=\sqrt{s}/A$, so that this new variable coincides with $\sqrt{s}$ for nucleon--nucleon collisions. 
In the relativistic limit $E\approx p \gg m$,
\begin{equation}
\sqrtsnn\approx 2\sqrt{\frac{E_aE_b}{A_aA_b}}=\sqrt{s_{\rm pp}}\sqrt{\frac{Z_aZ_b}{A_aA_b}}\,,
\end{equation}
where the latter equality assumes that the energy of a nucleus of atomic number $Z_a$ is $E_a=Z_a\,E_p$, $E_p$ being the energy that a proton reaches in the same accelerator; the expression thus relates $\sqrtsnn$ to $\sqrt{s_{\rm pp}}$, the centre-of-mass energy for two colliding protons in the same machine.} This approach, however, did not allow them to determine the cross-section for the interaction with a specific material. Instead, they measured $\sigma_{\rm inel}(\antideuteron + \langle A \rangle)$, where $\langle A \rangle$ is the average mass number, with corresponding average nuclear charge, $\langle Z \rangle$, of the material seen by the particles; $\langle A \rangle$ and $\langle Z \rangle$ come from simulations with an accurate model of the detector. They benchmarked their method with \antiproton, achieving good agreement with the \texttt{Geant4} parametrisation of antinucleus--nucleus interactions. For \antideuteron, they observed reasonable agreement above 1\,GeV/c but discrepancies of up to a factor 2.1 at lower momenta.

This near complete lack of experimental data must therefore be addressed. The \acrshort{alice} measurements have only hinted at a discrepancy between experimental data and theoretical models at lower momenta. They are unfortunately of limited use for improving the latter, because they do not allow to extract element-specific values.
Using the same experimental technique, \acrshort{alice} also performed the first ever measurements of the inelastic interaction cross-sections for $\antitriton$~\cite{ALICE:3Hbar2024} and $^3\antihelium$~\cite{ALICE2022II}, though with larger uncertainties than for \antideuteron. So far, there are no other measurements for these isotopes, and none at all for \antiheliumfour.
While conclusively detecting \antideuteron and \antihelium in \acrshort{gcr} will probably only be achieved with the next generation of experiments, the \antideuteron and \antihelium event candidates observed by \acrshort{ams} \cite{kounin:PAW2024} show that cross-section measurements related to these species will be needed.

\subsubsection{Nuclear cross-section needs for \texorpdfstring{\acrshort{cr}}{CR} detectors}
\label{sec:XSforCRexp}
Apparent discrepancies exist among \acrshort{cr} nuclei flux measurements from different experiments. The recent precision measurement of B, C, O and Fe from \acrshort{ams}~\cite{AMS:He-C-O-PRL2017, AMS:LiBEB-PRL2018, AMS:Fe-PRL2021} and~\acrshort{calet}~\cite{calet_CO_2020, calet_boron_2022, CALET:2021fks} exhibit an overall normalisation discrepancy of about $20\%$ (as discussed in Sec.~\ref{sec:CR_HN}).
In contrast, preliminary B flux measurements by \acrshort{dampe}~\cite{DAMPE:2024qwc} show no such discrepancy relative to \acrshort{ams}. Earlier measurements from \acrshort{heao}~\cite{1990A&A...233...96E}, \acrshort{pamela}~\cite{2014ApJ...791...93A}, \acrshort{crn} (\acrlong{crn})~\cite{1991ApJ...374..356M} and \acrshort{cream}~\cite{2009ApJ...707..593A} seemingly show a similar $\sim\!20\%$ offset with respect to the B, C, O and Fe data of \acrshort{ams}. Disagreements are also observed between \acrshort{ams} and \acrshort{heao}, \acrshort{cream} and \acrshort{crn}  regarding Ne, Mg and Si fluxes~\cite{AMS:NeMgSi-PRL2020}. 
Interestingly, no evident normalisation difference is observed among the H and He fluxes measured by \acrshort{ams}~\cite{AMS:He-C-O-PRL2017, ams_proton_2015}, \acrshort{calet}~\cite{CALET:2022vro, CALET:2023nif} and \acrshort{dampe}~\cite{DAMPE:2019gys, Alemanno:2021gpb}. 
The existing discrepancies, particularly in the recent high-precision datasets from \acrshort{ams}, \acrshort{dampe} and \acrshort{calet}, pose a challenge to the understanding of \acrshort{cr} propagation.

While it is difficult to conclude what could be the origin of such discrepancies -- given the large differences in the employed experimental and data-analysis techniques in different experiments --, an important and often predominant source of systematic error is due to the limited understanding of nuclear interactions in the detector materials.
In general, the \acrshort{cr} flux estimation relies on the knowledge of the experiment's geometric factor and fragmentation probabilities of the traversing \acrshort{cr} nuclei, which are calculated using simulation codes such as \texttt{Geant4}~\cite{Agostinelli2003} or \texttt{FLUKA}~\cite{Battistoni2015, Ahdida2022}. These codes model the transport of nuclei through the detection volumes, simulating their interactions with detector materials and the subsequent production of secondary particles.
However, in such codes, the interaction of heavy \acrshort{cr} nuclei ($Z>2$) with detector materials (such as C, Al, Si, and others) is modelled with a sparse and scattered dataset of nucleus--nucleus cross-section measurements. Often, nuclear models are extrapolated in regions where no experimental data exists. The lack of such nuclear data translates into a systematic uncertainty in the \acrshort{cr} measurements.
As described in Sec.~\ref{sec:facilities_CRexps}, important efforts have been made by the \acrshort{ams}~\cite{AMSXS} and \acrshort{dampe}~\cite{DAMPEXS} experiments to determine the interaction of \acrshort{cr} nuclear species in their detectors, though additional measurements are still required. 

Current (\acrshort{ams}, \acrshort{calet} and \acrshort{dampe}), and future experiments (\acrshort{herd}, \acrshort{hero}, \acrshort{tiger}-\acrshort{iss}, etc.), can significantly benefit from precision measurements of the nucleus--nucleus inelastic cross-sections. The knowledge of the cross-sections of materials used in calorimeters is of particular importance, because they make up most of the mass (and thus radiation length) of experiments. For instance, \acrshort{bgo} (\acrlong{bgo}) and \acrshort{lyso} (\acrlong{lyso}) scintillating crystals are the major constituents of the \acrshort{dampe} and \acrshort{herd} calorimeters, respectively, with other materials (mostly scintillators and absorbers) used in the calorimeters of some other experiments.

\subsubsection{Annihilation cross-section needs for antinuclei \texorpdfstring{\acrshort{cr}}{CR} detectors}
The annihilation cross-section ($\sigma_{\rm ann}$) is particularly relevant to experiments that rely on the detection of annihilation products for identifying antinuclei, like \acrshort{gaps}~\cite{GAPS:ICRC2023}. Such experiments do not only require precise knowledge of $\sigma_{\rm ann}$ to calculate fluxes but also of the multiplicities and energy distributions of the created secondary particles. Some data exists for \antiproton annihilation in H at rest \cite{Doser1988, Amsler2014} and in flight \cite{Agnew1957, Agnew1960, Loken1963, Boeckmann1966, Cline1971, Brueckner1990} but much of the energy range of interest to \acrshort{cr} experiments remains uncovered and many older experiments have large (between about 5\% and $>$20\%) and sometimes unknown uncertainties. Only some newer measurements for in-flight annihilation reach precisions $<5\%$ (e.g., \cite{Brueckner1990}). Limited experimental data for other targets are available \cite{Chamberlain1959, Agnew1960, McCaughey1986, Markiel1988, Cugnon1989, Riedlberger1989, Sudov1993, Ahmad1993, Zharenov2023}, suffering from a similarly limited precision, and agreement with theoretical models varies. For \antiproton annihilation in C, for example, significant deviations from model expectations were observed below 500\,MeV/c, which led to renewed interest in theoretical calculations \cite{Wronska2016,Aghai2018}. In addition to the problem of poorly known cross-sections, the multiplicities and energy distributions of secondaries emerging from the annihilation were only measured for a selection of targets and energies. The available data do not agree well with model predictions: see, for example, \texttt{Geant4} and \texttt{FLUKA}, which deviate from each other by as much as 25\% \cite{Amsler2014}.

Due to the complexity of such experiments, even fewer measurements were performed for \antideuteron \cite{Andreyev1990} and none at all for \antihelium. While the recent measurements by \acrshort{alice} (see Section~\ref{sec:XS_inelastic}) have shown that inelastic and absorption cross-sections for \antideuteron and \antihelium can in principle be measured above a certain threshold energy using unconventional techniques, the feasibility of performing experiments at the low energies required to probe the annihilation process currently remains questionable. With the availability of facilities like \acrshort{cern}'s Antiproton Decelerator and \acrshort{lear}, the situation is much better for \antiproton, for which comprehensive measurements of the most relevant $\sigma_{\rm ann}$ and of the secondary particles created during annihilation, including multiplicities and energy distributions, should be conducted. Better data on the \antineutron--\proton \cite{Gunderson1981, Armstrong1987, Iazzi200} and \antineutron--nucleus \cite{Astrua2002} annihilation cross-sections and secondaries would allow improving theoretical models of the annihilation process. Finally, if the annihilation of \antideuteron and \antihelium could be probed in ways similar to the \acrshort{alice} approach, even data with relatively large uncertainties would be useful for model validation.

\subsubsection{Summary and wish list}
In this section, all cross-sections (not production) relevant for \acrshort{gcr} data interpretation, and also for \acrshort{cr} experiment analyses, were carefully reviewed. The most pressing needs are gathered in Table~\ref{tab:antinuclei-measurements}.

Related to the first item, i.e., to better constrain \acrshort{gcr} propagation models, inelastic cross-sections for nuclei on H are needed at the few percent level -- for energies from a few hundreds of MeV/n to several tens of GeV/n -- for all leading isotopes in \acrshort{gcr}s; measurements on He targets are also needed, but at a lesser precision ($\lesssim10\%$). Figure~\ref{fig:XS_inel_data} provides the current status of the nuclear data on H along with the desired precision for all the \acrshort{cr} isotopes. For the interpretation of \antiproton data, absorption and quasi-elastic cross-sections for \antiproton--\proton scattering need to be precise at the $\sim\!5\%$ level for energies between 1 and 10\,GeV; slightly less accurate measurements for \antiproton--He are also desired. While data at this level of precision exists for the total and elastic cross-sections (with some caveats), direct measurements of the absorption cross-sections would be highly useful because of their large impact on the \antiproton flux (see Figure~\ref{fig:XS_pbardbar_impact}). 
If the quest for \antideuteron detection in forthcoming \acrshort{cr} experiments succeeds, having data on the absorption cross-sections for $\antideuteron + {\rm p}$ and $\antideuteron + {\rm He}$ at energies between \unit[100]{MeV/n} and \unit[50]{GeV/n} will no longer be optional; the required precision will depend on how precisely these future experiments will be able to measure fluxes.
In light of the potential detection of \antihelium nuclei by \acrshort{ams}, experiments should also be devised to measure cross-sections for 
\antihelium--\proton and \antihelium--He scattering at similar projectile kinetic energies.

A second item on the wish list is the extension of cross-sections to target materials relevant to instrumentation. First, new nuclear data are needed for inelastic interactions of \acrshort{gcr} nuclei above GeV/n energies on C, Al, Si and Cu targets, and also on elements constituting \acrshort{bgo} and \acrshort{lyso} crystals (because their composition is variable) and other commonly used calorimeter materials. Dedicated studies are required to assess the possible mass-dependence of the cross-section uncertainty impact on the \acrshort{cr} data precision. This impact is detector-dependent, but roughly, a better than 10\% precision is needed.
Second, cross-sections of \antiproton and antinuclei annihilating on C, Al, Si, Fe, and Cu for \acrshort{gaps}-like experiments, or inelastically scattering on N and O (prevalent in Earth's atmosphere), to aid the interpretation of data gathered by balloon-borne experiments, are required. The energy ranges and desired precisions are provided in Table~\ref{tab:antinuclei-measurements}.
For experiments relying on annihilation signatures, the precision to which $\sigma_{\rm ann}$ is known directly drives the uncertainty of the measured flux; a precision of $<$5\% for H and $<$10\% for other targets is therefore desirable. Since $\sigma_{\rm ann}$ is largest for annihilation at rest, experiments should probe energies below 10\,MeV/n and, if possible, extend to about 500\,MeV/n, above which the cross-section becomes small enough for its uncertainty to not significantly impact flux calculations. Realistically, these measurements can only be performed comprehensively for \antiproton. Data for \antineutron, \antideuteron and \antihelium are not a priority for the next ten years, but any measurements would certainly be crucial to \acrshort{gaps}-like next-generation experiments. Any data for \antideuteron and \antihelium on \proton and He targets would also help to validate and improve interaction models and, therefore, reduce uncertainties compared to the scenario where only \antiproton data is available and extrapolated to heavier projectiles.

\begin{table}
    \centering
    \caption{Summary of highest-priority measurements for inelastic nucleus--nucleus and antinucleus--nucleus interactions (see definitions in Eq.~\eqref{eq:def_tot-inel-nuclei} and Eq.~\eqref{eq:def_tot-inel-antinuclei}, respectively): (i) interactions of \acrshort{gcr} isotopes (for $Z\leq30$ species only, summarised from Fig.~\ref{fig:XS_inel_impact}) and antinuclei with H and He targets are required for propagation modelling, and thus for the interpretation of data gathered by \acrshort{cr} experiments --- the needed precision decreases with energy; (ii) interactions of \acrshort{gcr} elements and \antiproton with heavier targets are needed for modelling the interaction with the detector material (C, Al, Si, Fe, and Cu) and with Earth's atmosphere (N and O) for the specific case of balloon-borne experiments; (iii)~$\sigma_{\rm ann}$ and the total multiplicity, $n_{\rm tot}$, are required for experiments relying on annihilation signatures for identifying antinuclei (e.g., \acrshort{gaps}), with $\sigma_{\rm ann}$ measured at rest up to 500\,MeV (with a stronger emphasis on lower energies), and $n_{\rm tot}$ and the energy spectra of the emerging secondaries measured particularly for $^{1,2,3}$H, $^{3,4}$He, $\gamma$ and $\pi^\pm$. In this table, the projectile kinetic energies per nucleon, $\mathrm{E_{\rm k/n}}$, for a (hypothetical) fixed-target experiment are quoted.\vspace{1mm}}
    \label{tab:antinuclei-measurements}
    \begin{tabular}{lccc}
        \toprule
        \textbf{Reaction} & \textbf{Measurements} & \textbf{Projectile $\mathrm{E_{\rm k/n}}$} & \textbf{Precision}\\
        \midrule
        \!\!$(\proton,\deuteron) + {\rm H}$ & \multirow{6}{*}{$\sigma_{\rm inel}$, $\sigma_{\rm prod}$} & \multirow{6}{*}{1 to 10\,GeV/n} & $<10\%$ \\        
        \!\!$(\proton,\deuteron) + {\rm He}$ & & & $\lesssim50\%$ \\        
        \!\!$(^3{\rm He}, ^4{\rm He}) + {\rm H}$ & & & $<5\%$ \\
        \!\!$(^3{\rm He}, ^4{\rm He}) + {\rm He}$ & & & $\lesssim50\%$ \\
        \!\!$(^{6}{\rm Li}, ^{7}{\rm Be}, ^{11}{\rm B}\dots ^{56}{\rm Fe}\dots ^{64}{\rm Zn}) + {\rm H}$ & & & $<1\%$ \\
        \!\!$(^{6}{\rm Li}, ^{7}{\rm Be}, ^{11}{\rm B}\dots ^{56}{\rm Fe}\dots ^{64}{\rm Zn}) + {\rm He}$ & & & $\lesssim10\%$ \\
         \addlinespace[6pt]
        $\antiproton + \proton$ & $\sigma_{\rm abs}$, $\sigma_{\quasiel}$ & \multirow{4}{*}{0.1 to 50\,GeV/n} & $<5\%$\\
         $\antiproton + \helium$ & $\sigma_{\rm abs}$, $\sigma_{\quasiel}$ &  & $<50\%$\\
        $\antideuteron + (\proton,\helium)$ & $\sigma_{\rm abs}$, $\sigma_{\quasiel}$ & & (any data)\\
       $\antihelium + (\proton,\helium)$ & $\sigma_{\rm abs}$ &  & (any data)\\
       \midrule
        \!\!${\rm (p, He, C\dots Fe)} + {\rm (C,N,O,Al,Si,Fe,Cu)}$ & $\sigma_{\rm inel}$ & 0.1 to 1000\,GeV/n & $\lesssim10\%$ \\
        $\antiproton + {\rm (C,N,O,Al,Si,Fe,Cu)}$ & $\sigma_{\rm abs}$ & 0.1 to 50\,GeV/n & 10\% \\
        \midrule
        $\antiproton + \proton$ & \multirow{3}{*}{$\sigma_{\rm ann}$, $n_{\rm tot}$} & \multirow{3}{*}{$<500$\,MeV/n} & $<$5\% \\
        $\antiproton + {\rm (C,Al,Si,Fe,Cu)}$  &  &  & $<$10\%\\
        $\antiproton + {\rm n}$ & & & (any data) \\
        \addlinespace[6pt]
        $\antineutron + {\rm any}$ & \multirow{3}{*}{$\sigma_{\rm ann}$, $n_{\rm tot}$} & \multirow{3}{*}{$<500$\,MeV/n} & \multirow{3}{*}{(any data)}\\
        $\antideuteron + {\rm any}$ & & & \\
        $\antihelium + {\rm any}$ & & & \\
        \bottomrule
    \end{tabular}
\end{table}

\section{Main facilities and experiments for ongoing and future cross-section measurements}
\label{sec:facilities}

In this section, the ongoing and future efforts carried out by the \acrshort{hep} and nuclear physics communities to provide some cross-sections listed in Sec.~\ref{sec:XSneeds} (nuclear fragmentation, anti-matter production and interactions) are presented.
The variety of required reactions and energy ranges calls for a variety of facilities and experiments, whose properties and specifics define in turn the measurements they are/will be able to do.
These facilities include high-energy beams available at \acrshort{cern} from the \acrshort{lhc} (Sec.~\ref{sec:facilities_LHC}), the \acrshort{sps} and the \acrshort{ps} accelerators (Sec.~\ref{sec:facilities_SPS}), current and future nuclear physics multi-GeV accelerators (Sec.~\ref{sec:facilities_MultiGeV}), and space \acrshort{cr} experiments themselves (Sec.~\ref{sec:facilities_CRexps}). Other facilities, where in principle relevant measurements can be carried out, such as the \acrshort{kek} or Fermilab~\cite{Gori:2024zbs} accelerators, are not covered here (see also Ref.~\cite{Proceedings:2025uly} for a road map of the next generation accelerator facilities in the next decade). In the following, the most important characteristics of these facilities and experiments are highlighted, together with some recent results and planned measurements. The results obtained in the last years illustrate the successful emergence and growth of synergies between our communities, which need to be further strengthened in order to tackle the physics cases presented in Sec.~\ref{sec:physics_cases}.

\subsection{\texorpdfstring{\acrshort{cern}}{CERN} \texorpdfstring{\acrshort{lhc}}{LHC} experiments}
\label{sec:facilities_LHC}
Measurements of the \antiproton production cross-sections have been performed at various facilities and different collision energies. While historical experiments~\cite{Dekkers1965, Antinucci1973, Capiluppi1974, Alper1975} laid the groundwork for these studies, the precision of their results falls short of current requirements, about 5\%, as summarised in Table~\ref{tab:measurements}. In addition, the interpretation of their systematic effects, such as the subtraction of the \antiproton feed-down contribution, is not always clear. High-statistics collision samples are also needed for \antideuteron and \antihelium production measurements, as their production is rare and suppressed by at least a factor of 1000 for each additional antinucleon with respect to antiprotons~\cite{ALICE:2017xrp}. For light collision systems and collision energies below 100\,GeV, which are relevant for models of \acrshort{gcr} antinuclei production, only sparse data are available. Precise measurements of their production mechanisms have only been achieved at colliders with very large collision energy~\cite{Coral2018, Antipov1971, Bozzoli1978}. Finally, as discussed in Sec.~\ref{sec:xs_prodpositrons}, measurements of neutral particle production cross-sections are crucial to better constrain the $\gamma$-ray case.

\begin{figure}
    \centering
    \includegraphics[width=\linewidth]{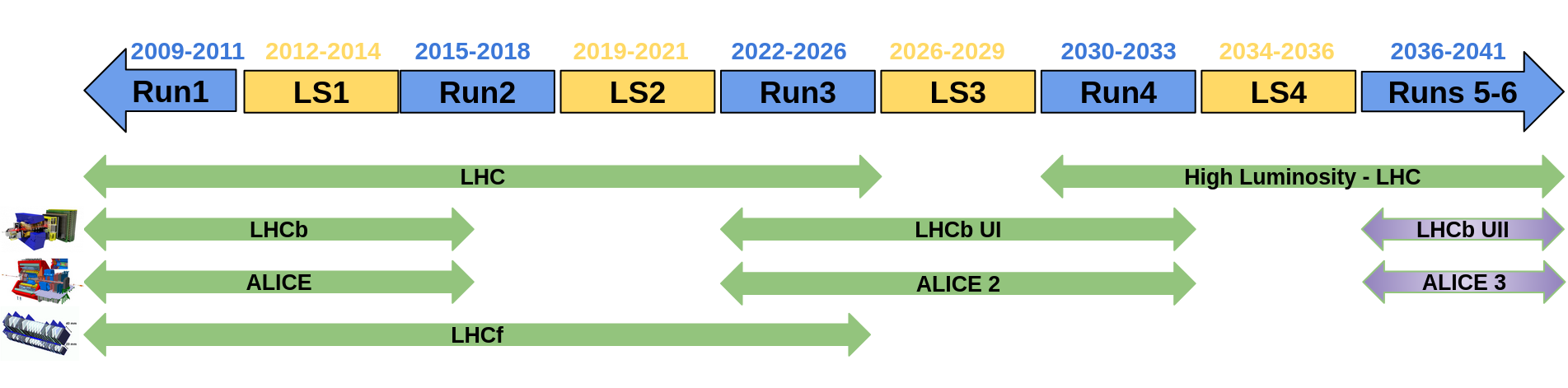}
    \caption{Timeline of the data acquisition (Run) and shutdown (\acrshort{ls}) periods for the \acrshort{cern} \acrshort{lhc} accelerator operations. The main phases of the \acrshort{alice}, \acrshort{lhcb} and \acrshort{lhcf} experiments (discussed in the main text) are also overlaid. Upgrades currently under review and hence not yet officially approved, \acrshort{lhcb}UII and \acrshort{alice}3, are indicated in light purple.}
    \label{fig:LHC_schedule}
\end{figure}

At \acrshort{cern}, the \acrshort{lhc} particle accelerator provides collisions of protons at multi-TeV energies, up to the value of 13.6\,TeV (recently achieved in 2022), and also of lead and lighter ions. As summarised in Fig.~\ref{fig:LHC_schedule}, the accelerator has operated since 2009, with several \acrfull{ls} periods, i.e., interruptions necessary to increase the achievable energy or luminosity. During those periods, experiments also have been significantly improved, as separately discussed in more details in the sections below.

\subsubsection[LHCb]{The \texorpdfstring{\acrshort{lhcb}}{LHCb} experiment}
\label{sec:HEP_LHCb}
The \acrshort{lhcb} experiment~\cite{LHCb} at \acrshort{cern} \acrshort{lhc} is a fully instrumented single-arm spectrometer, covering a pseudorapidity\footnote{Being $\theta$ a particle polar angle with respect to the beam axis, its pseudorapidity is defined as $\eta = - \ln[\tan(\theta/2)]$.} region, \mbox{$2 < \eta < 5$}, not accessible by other \acrshort{lhc} experiments. In addition, \acrshort{lhcb} has the unique possibility to operate in fixed-target mode, leveraging on the injection of gases in the \acrshort{lhc} beam pipe through the \acrshort{smog}~\cite{SMOG}. The energies covered by \acrshort{lhcb}-\acrshort{smog}, between 27 and 113\,GeV in the nucleon--nucleon centre-of-mass frame for \acrshort{lhc} beam energies ranging between 450\,GeV and 7\,TeV, bridge the gap between previous fixed-target experiments and the higher values by \acrshort{rhic} (\acrlong{rhic}) or \acrshort{lhc}. Starting from the \acrshort{lhc} Run~3, simultaneous operation of the upgraded experiment~\cite{LHCb-DP-2022-002} in collider and fixed-target mode has been proven~\cite{CERN-THESIS-2021-313, SMOG2_paper}, allowing the negative to central Feynman-\textit{x} values to be probed with very high-precision at a poorly explored energy scale. The \acrshort{smog} upgrade, \acrshort{smog}2, is equipped with a direct measurement of the gas flux, which reflects into a measurement of the collected luminosity with a 1--2\% expected uncertainty, and also non-noble gases, such as H and D, can be used. In all collision systems, \antiproton identification is ensured by \acrshort{rich} detectors, for momenta above 10\,GeV/c. The \acrshort{lhcb} collaboration has published first measurements in \pHe collisions for the prompt~\cite{LHCb-PAPER-2018-031} and feed-down from strange~\cite{LHCb-PAPER-2022-006} \antiproton production using data collected in 2016 at $\sqrt{s_{\text{NN}}} = 110.5$\,GeV, constraining for the first time -- at the relevant energy scales -- the extrapolation from H to He as a target. Another \pHe dataset at $\sqrt{s_{\text{NN}}} = 87$\,GeV was collected in 2016 with \acrshort{smog}, while with \acrshort{smog}2 high-statistics samples with injected H, D and He have been acquired at $\sqrt{s_{\text{NN}}} = 70.9$\,GeV and $\sqrt{s_{\text{NN}}} = 113$\,GeV in 2024. With these datasets, as summarised in Fig.~\ref{fig:antiproton_experiments}, the isospin-violating difference between \antiproton and \antineutron production will be constrained with expected uncertainties below 5\%, filling the gap between the expected \acrshort{amber} results and measurements at collider energies. Absolute cross-section measurements with similar precisions will also be repeated with all these gases. Finally, discussions are ongoing to explore the opportunity to develop machine optics such as to squeeze a 1\,TeV beam sufficiently, in order to close the \acrshort{lhcb} innermost detector and acquire data close to the lowest possible energy scale at \acrshort{lhc}~\cite{LHCb-PUB-2018-015}. This would provide a larger coverage of the \antiproton production phase space, and would allow comparing with results from lower-energy fixed-target experiments.

Although heavier antinuclei identification was initially not planned at \acrshort{lhcb}, methods have been recently developed to identify \antideuteron and \antihelium in the recorded Run~2 data. For \antideuteron, the \acrshort{tof} capabilities of the tracking detectors downstream of the \acrshort{lhcb} magnet are exploited, leading to identification of low-momentum (anti)deuterons~\cite{LHCb-FIGURE-2023-017, CERN-THESIS-2023-338}. Based on this new technique, measurements of the deuteron production cross-section, both absolute and relative to \antiproton, are ongoing. For \antihelium, the detector responses are used to build discriminators quantifying the energy loss. Despite initially only applied to \pp collision data~\cite{HeliumID_LHCb, LHCb-CONF-2023-002, LHCb-CONF-2024-005}, measurements are ongoing to constrain \antihelium production in pNe fixed-target data as well. The absolute cross-sections at 13.6\,TeV pp collision data, albeit at higher energy scales with respect to those needed for \acrshort{cr} experiments interpretation, will also be finalised soon. In the long-term future, a further upgrade of the \acrshort{lhcb} experiment is planned~\cite{LHCb-TDR-023, LHCb-TDR-026}, starting from the \acrshort{lhc} Run5. This will include a dedicated \acrshort{tof} detector, \acrshort{torch} (\acrlong{torch})~\cite{LHCb-TDR-023}, allowing direct identification of nuclei in a large momentum range.

\subsubsection[ALICE]{The \texorpdfstring{\acrshort{alice}}{ALICE} experiment}
\label{sec:HEP_ALICE}
The \acrshort{alice} experiment at \acrshort{cern} \acrshort{lhc} is a mid-rapidity experiment dedicated to nuclear physics in \pp and heavy-ion collisions. The detector consists of several subdetectors that allow particle tracking in the pseudorapidity range of $|\eta| \leq 0.9$. Particle identification is mainly based on energy-loss measurements in a large time-projection chamber, and is completed with a \acrshort{tof} measurement system for higher momenta~\cite{ALICE2008}.

The \acrshort{alice} experiment has measured \antiproton production in \pp, pPb, PbPb and XeXe collisions at several collision energies and particle multiplicities~\cite{2011_ALICE, ALICE2010, ALICE:2016dei, ALICE2020II, ALICE:2021lsv,ALICE:2024say}. Different antinuclei species and their ratios have also been measured, including \antideuteron, \antitriton, \antiheliumthree and \antiheliumfour~\cite{ALICE:2024say, ALICE:2019bnp, ALICE2022IV, Rasa2024}. Additionally, (anti)hypernuclei -- nuclei including a nucleon with strangeness -- productions have been measured~\cite{Bartsch2023}.
The variety of measurements allows detailed studies of the formation process of light nuclei, often described by the so-called coalescence model~\cite{PhysRevC.21.1301}. Parameters related to the coalescence probability of nuclei, with two and three nucleons, have been measured, and the experimental results have triggered studies beyond the classical coalescence model~\cite{ALICE:2021mfm, Mahlein2023}.

\acrshort{alice} has also performed momentum correlation studies \cite{ALICE:2020ibs,ALICE:2023sjd,ALICE:2025aur}, known as femtoscopy technique, of several hadron pairs. This allowed to constrain the size of the particle-emitting source in \pp collisions, which is a necessary input for the coalescence model based on the Wigner function formalism. These results, together with the above-mentioned precise measurements of \antiproton and \antideuteron production spectra, showed that \antideuteron yields can be successfully predicted by this model~\cite{Mahlein2024}. Moreover, \acrshort{alice} also used femtoscopy to study the residual strong interaction between hadron pairs and triplets, including hyperons. This provides essential inputs for astrophysics, for instance to constrain the equation of state of dense matter, and to understand better the composition of such dense systems, in particular the inner core of neutron stars~\cite{Fabbietti2020, ALICE:2022boj, Mihaylov:2023ahn}. Further details on this technique are given in Sec.~\ref{sec:femto}.

Besides the formation process of antinuclei, \acrshort{alice} also measured the absorption of \antideuteron and \antihelium in matter~\cite{ALICE:dbar2020, ALICE2022II}. These processes have been experimentally mostly unexplored, while they impact the survival probability of antinuclei produced in our Galaxy during their propagation from their sources to the Earth (see Sec.~\ref{sec:XS_inelastic_antinuc}).

An upgrade of the experiment, \acrshort{alice}3, is expected to start by the \acrshort{lhc} Run~5. This will include a more extensive rapidity coverage, allowing to probe antinuclei production out of the central rapidity regime~\cite{ALICE2022III}.

\subsubsection[LHCf]{The \texorpdfstring{\acrshort{lhcf}}{LHCf} experiment}
\label{sec:HEP_LHCf}
The \acrshort{lhcf} experiment \cite{LHCf:2008lfy} at \acrshort{cern} \acrshort{lhc} is made of two imaging and sampling calorimeters, located at a distance of 141.05\,m from Interaction Point 1, and covering a pseudorapidity region $\eta>8.4$. The experiment is dedicated to the precise measurement of forward neutral particle production in \pp and p--ion collisions, in order to provide calibration data to tune the hadronic interaction models used to simulate extensive air showers. These data provide indirect information on the event inelasticity, and on the fraction of the primary energy that goes into the electromagnetic and the hadronic channels. By changing the collision energy and the colliding ion, it is also possible to test the reliability of different scaling laws, and study the impact of mass number on forward production. Finally, the \acrshort{lhcf}--\acrshort{atlas} (\acrlong{atlas}) joint analysis gives access to an even higher level of information, for example by separating different mechanisms responsible for forward production (e.g., diffractive and non-diffractive), and by studying central-forward correlation or exclusive production mechanisms (like one-pion exchange or $\Delta$ resonance).

So far, the experiment has acquired data in \pp collisions between $\sqrt{s} = 0.9$\ and $13.6$\,TeV and pPb collisions at $\sqrt{s_{\rm NN}} = 5.02$ and $8.16$\,TeV. The published results indicate a tension between models and data, which is particularly strong in the case of neutron production \cite{LHCf:2018gbv, LHCf:2020hjf}, and not negligible in the case of $\gamma$ \cite{LHCf:2017fnw}, $\pi^0$ \cite{LHCf:2015rcj} and $\eta$ \cite{LHCf:2023yam} production. Thanks to an ongoing improvement of the reconstruction algorithm, it will be possible to measure $K^0_s$ and possibly $\Lambda^0$ forward production, from the data acquired in \pp collisions at $\sqrt{s} = 13.6$\,TeV. In parallel, the \acrshort{lhcf}--\acrshort{atlas} joint analysis will give a better insight in the understanding of production mechanisms, leading to greater constraints for the calibration of hadronic interaction models. 
The LHCf measurements can be used to constrain the production cross-sections entering the calculation of the astrophysical $\gamma$-ray background (from \acrshort{gcr}s on the \acrshort{ism}), as shown in Ref.~\cite{orusa2023new}. Indeed, $\pi^0$ and $\eta$ are the main contributions to the $\gamma$-ray flux, and the inclusive $\gamma$ production measured by the experiment can be used as a benchmark.

\subsection{\texorpdfstring{\acrshort{cern}}{CERN} \texorpdfstring{\acrshort{sps}}{SPS} and \texorpdfstring{\acrshort{ps}}{PS} experiments}
\label{sec:facilities_SPS}
Before circulating in \acrshort{lhc}, particles are accelerated at \acrshort{cern} by lower-energy machines. The \acrshort{ps} accelerator is one of the first acceleration stages, and can reach a maximum energy of 26\,GeV. It delivers particles to the \acrshort{sps}, which has a maximum energy of 450\,GeV for nuclei, and also provides secondary beams of nuclei to experiments from 10 to 158\,GeV energy per nucleon. To this purpose, a primary beam of $^{208}$Pb is extracted onto a beryllium target, and nuclear fragments are guided to the experimental area. The rigidity acceptance of the beam line can be adjusted to select the specific mass-to-charge ratio of the desired nuclei~\cite{Buenerd:622246, Strobele:2012zz, NA61SHINE:2024rzv}.
The \acrshort{sps} also delivers secondary hadrons at momenta up to 400\,GeV/c, depending on the beam line\footnote{\url{https://sba.web.cern.ch/sba/BeamsAndAreas/H2/H2_presentation.html}}: on the H2 beam line (for NA61/\acrshort{shine}), secondary hadron beams up to 400\,GeV/c can be produced; on the M2 beam line (for \acrshort{amber}), 280\,GeV/c is the maximum hadron beam momentum.

\subsubsection[AMBER]{The \texorpdfstring{\acrshort{amber}}{AMBER} experiment}
\label{sec:HEP_AMBER}
The \acrshort{amber} experiment at the M2 secondary beam line of \acrshort{cern} \acrshort{sps} is a fixed-target experiment that started data-taking in 2023 as the successor of the long-standing \acrshort{compass} (\acrlong{compass}) experiment. Within the first approved phase of the experiment, from 2023 to around 2031, \acrshort{amber} reuses and upgrades the 2-stage magnetic spectrometer from \acrshort{compass}, to perform measurements of the \antiproton production cross-section, of the charge radius of the proton, and of parton-distribution functions of pions and kaons via the Drell-Yan process~\cite{Adams:2676885}.
Measurements of \antiproton production in collisions of protons on H, D and He targets took place in 2023 and 2024.
The experimental setup includes two differential Cherenkov counters with achromatic ring focus to identify protons in the mixed hadron beam, a cryogenic target filled with the target gas, and the \acrshort{amber} spectrometer to characterise the secondary particles created in the interaction.
In order to measure the momentum of the secondary particles, the \acrshort{amber} spectrometer consists of around 300 tracking detector planes that measure the tracks of charged particles traversing the two spectrometer magnets, with a bending strength of up to $1$\,Tm and 4\,Tm, respectively. Additionally, a \acrshort{rich} detector and muon detectors allow particle identification over an extensive momentum range. Antiprotons with a total momentum between $10$\,GeV/c and $60$\,GeV/c, and transverse momentum up to $2$\,GeV/c, are identified.
In 2023, the collaboration used a cryogenic target filled with He to performing the \antiproton production measurement in \pHe collisions at six different collision energies, between $\sqrt{s_{\mathrm{NN}}} = 10.7$\,GeV and $\sqrt{s_{\mathrm{NN}}} = 21.7$\,GeV. In 2024, a new cryogenic target was built to allow the usage of flammable gases, such as H and D. For both targets, the collaboration recorded collisions at $12.3$\,GeV, $17.3$\,GeV and $21.7$\,GeV, with an identical spectrometer setup.

\begin{figure}
    \centering
    \includegraphics[width=0.7\linewidth]{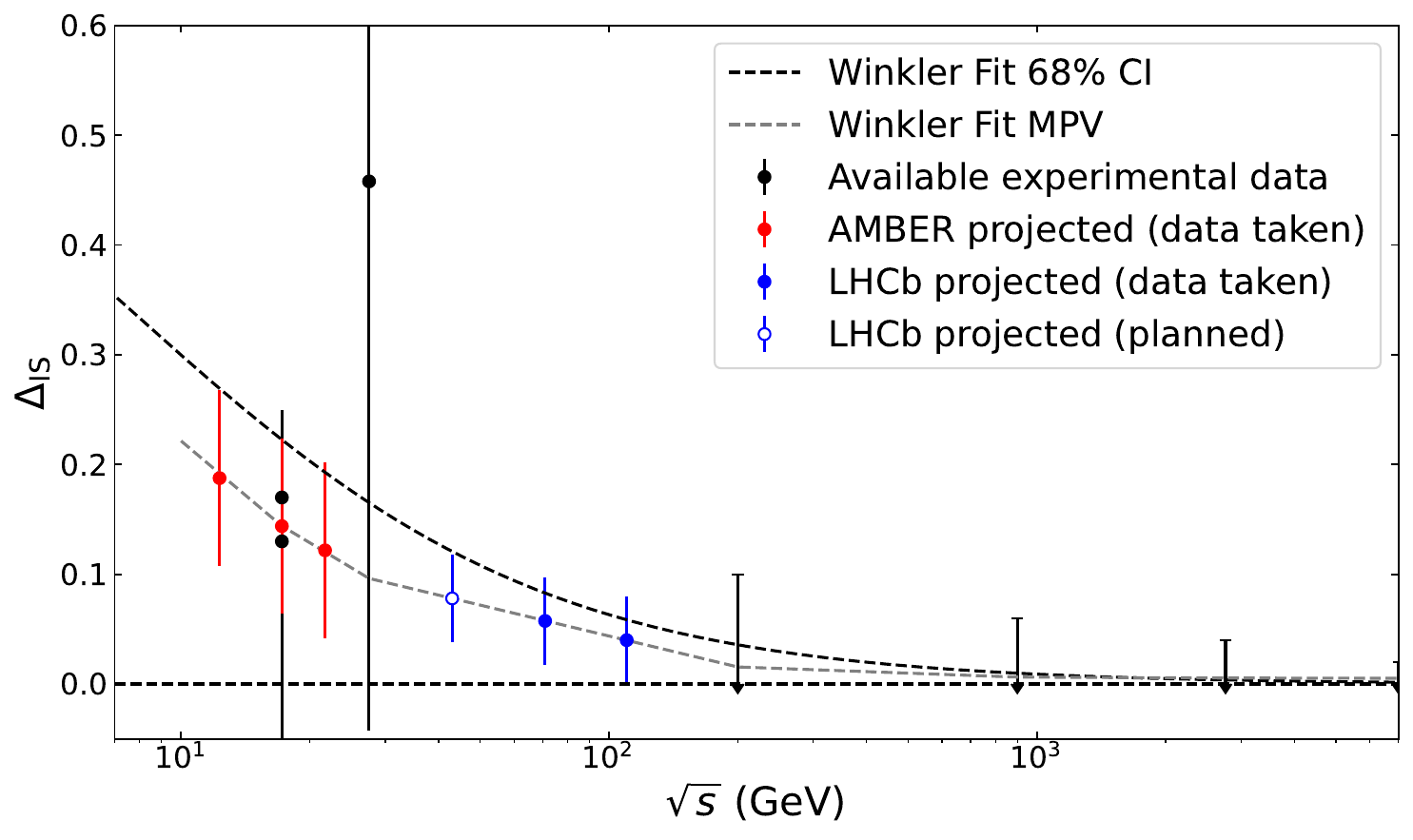}
    \caption{Impact of the \acrshort{amber} and \acrshort{lhcb} measurements on the uncertainties of a potential isospin asymmetric production of \antiproton and \antineutron; see Eqs.~\eqref{eq:def_pbarinv} and \eqref{eq:Delta}. Data points are filled in case of data already taken and empty in case of planned for the upcoming years. The black dashed line indicates the current uncertainty on $\Delta_{\rm IS}$ given the current data and model of the isospin asymmetry by M.~Winkler~\cite{Winkler:2017xor}. The projected uncertainties for the individual measurement points from \acrshort{lhcb} and \acrshort{amber} are estimated based on assumed measurement uncertainties of $1\%$ and $2\%$ for the ratio of \antiproton production cross-sections in p--D and p--H, respectively. Additionally, the limited phase-space coverage of both experiments reduces the sensitivity to the isospin asymmetry, which is primarily located in the target-fragmentation region (negative Feynman-x values, $x_{\rm F}$). As a conservative estimate, the given uncertainties account for sensitivities of the experiments only down to $x_{\rm F} = 0$. Other potential modelling uncertainties, such as an explicit $x_{\rm F}$ dependence of the isospin asymmetry, are not considered in this estimate.}
    \label{fig:antiproton_experiments}
\end{figure}

Besides providing \antiproton production cross-sections for the different targets on the level of $5\%$ relative uncertainty, one dedicated goal of the measurements is to investigate the possible isospin asymmetry of \antiproton in \pp and pn collisions (see Sec.~\ref{sec:xs_prodpbar}), as suggested by data from NA49~\cite{Fischer:2003xh}, by comparison of the \antiproton production in proton--hydrogen and proton--deuterium. The expected uncertainties on the individual cross-sections should allow a measurement of the isospin asymmetry, $\Delta_{\rm IS}=f^{\antineutron}_0 / f^{\antiproton}_0-1$, at the $10\%$ level for the three collision energies. In the case of a measurable asymmetry, the measurement of the different collision energies would additionally constrain the collision-energy dependence of the effect.
Figure~\ref{fig:antiproton_experiments} illustrates the impact of the \acrshort{amber} and \acrshort{lhcb} measurements on the isospin asymmetry in \antiproton production, given an arbitrary asymmetry (based on the parametrisation of M. Winkler~\cite{Winkler:2017xor}).

In the future, the \acrshort{amber} spectrometer will undergo several upgrades and improvements to operate the spectrometer at around 10--100 times higher read-out rates~\cite{Zemko2021}. This improvement would allow the measurement of rare particles, such as \antideuteron and \antihelium. However, dedicated nuclei identification is needed for these studies, and is currently under investigation.

\subsubsection[NA61]{The NA61/\texorpdfstring{\acrshort{shine}}{SHINE} experiment}
\label{sec:HEP_NA61}
The fixed-target experiment NA61/\acrshort{shine} at the \acrshort{cern} \acrshort{sps} is a hadron spectrometer capable of studying collisions of hadrons with different targets, over a wide range of incident beam momenta~\cite{na61}. It is the successor to the NA49 experiment, which pioneered \antiproton production cross-section measurements at $\sqrtsnn = 17.3$\,GeV in \pp and \pC collisions, covering nearly the full phase space of created \antiproton. Using a deuterium beam, NA49 also provided a first measurement of a potential isospin asymmetric production of \antiproton and \antineutron, by comparing the flipped reaction of \antiproton in \pp and pn collisions. The recorded data hints at an isospin asymmetry of up to $50\%$ in central production ($x_{\rm F} = 0$).

NA61/\acrshort{shine} consists of different subdetectors for particle identification. It has already recorded \pp interactions with beam momenta from 13 to 400\,GeV/c, and has also collected data for other hadron interactions, including pC, $\pi^\pm$C, ArSc, pPb, BeBe, XeLa and PbPb at different energies.
During the \acrshort{cern} \acrshort{ls}2, upgrades to the time projection chamber backend electronics resulted in improvements in the specific energy loss ($\text{d}E/\text{d}x$) resolution. Essential for future \antideuteron production measurements is the new data acquisition system, with about 20~times faster rate and new \acrshort{tof} detectors with improved time resolution~\cite{Gazdzicki:2692088, Fields:2739340}.
NA49 and NA61/\acrshort{shine} have published several relevant
data~\cite{NA49:2012jna, Aduszkiewicz:2017sei} for tuning models of the production of 
\acrshort{gcr} species, and also for models describing \acrshort{cr}-induced air showers~\cite{NA61SHINE:2017vqs, NA61SHINE:2022tiz}.

\paragraph{Antiproton cross-sections and coalescence for antinuclei}

The published measurements of light nuclei in \pp and various nucleus--nucleus data sets can be used to study the production of light ions at the threshold. These measurements will complement the NA49~\cite{dbarna49, Anticic:2016ckv} and \acrshort{alice} results, and allow testing coalescence and thermal models in a different regime.
Extended data-taking with an upgraded NA61/\acrshort{shine} experiment, relevant to understanding cosmic antinuclei, is already planned before 2026. A \pp dataset of approximately 600~M events, collected with a beam energy of 300\,GeV, will provide new measurements of \pp correlations, \antiproton and deuterons. This proposed dataset will feature significantly reduced systematic and statistical uncertainties, enhancing the ability to discriminate between different nuclear formation models. It is also anticipated that, for the first time, \antideuteron will be identified in this range, crucial for the cosmic antideuteron interpretation. Combining these new measurements will enable building, testing and validating data-driven \deuteron and \antideuteron production models in the energy range most relevant to \acrshort{gcr}s. This will reduce uncertainties in the modelling of the astrophysical background of \antideuteron.

\paragraph{Nuclear fragmentation}
NA61/\acrshort{shine} conducted a first pilot run of carbon fragmentation measurement at 13.5 $A$\,GeV in 2018, demonstrating that the measurements are possible~\cite{NA61SHINE:2024rzv}. For this type of measurement, the primary $^{208}$Pb is extracted from the \acrshort{sps} and fragmented in collisions with a 160\,mm-long beryllium plate in the H2 beam line. The resulting nuclear fragments of a chosen rigidity are guided to the NA61/\acrshort{shine} experiment, where the projectile isotopes are identified via a measurement of the particle charge and \acrshort{tof} over a length of approximately $240$\,m. Moreover, the collaboration collected data of the fragmentation of nuclei from Li to Si at 12.5\,GeV/n at the end of 2024, and is currently analysing them. The collected data were inspired by the interactions listed in Tab.~\ref{tab:ninter}, and they will provide a comprehensive set of cross-sections in the lower triangular region of Fig.~\ref{fig:nucdata_precision}.
High-precision measurements of $\sigma_{\rm inel}$ and $\sigma_{\rm prod}$, as defined in Eq.~\eqref{eq:def_tot-inel-nuclei}, necessary for \acrshort{gcr} data interpretation (see Sec.~\ref{sec:XS_inelastic}), is also possible, as illustrated in Ref.~\cite{NA61SHINE:2019aip}.
In the future, these measurements of nuclear fragmentation (and inelastic cross-sections) with NA61/\acrshort{shine} can potentially be extended up to Fe and performed at different energies.

\subsubsection{\texorpdfstring{\acrshort{cern}}{CERN} n\_\!TOF for neutron-related cross-sections}

The n\_\!\acrshort{tof} neutron time-of-flight facility~\cite{ntof} is located at \acrshort{cern}. A 20\,GeV/c proton beam from the \acrshort{ps} accelerator is shot on a thick lead target, generating neutron beams covering kinetic energies ranging from the thermal region to several GeV. The facility was optimised for high-precision measurements on radioactive materials, due to a very low duty cycle (repetition rate less than 0.5\,Hz) and very long flight-paths, from 20 to 180\,m.
In recent years, developments have been made to measure cross-sections of reactions leading to the emission of charged particles, often abbreviated in (n,cp) channels, using silicon detectors. Preliminary results were obtained for neutron energies up to a few MeV~\cite{Barbagallo2019}. A limitation to reach higher energies can be ascribed to the very strong $\gamma$ flash that comes with every neutron pulse and blinds most detectors for a short time. A development has started to use gaseous detectors \acrshort{ppac}s (\acrlong{ppac}) for such measurements, since these detectors are much less sensitive to the $\gamma$ flash.

This facility could be taken advantage of to carry out measurements on various targets (C, O, etc.), crucial for cosmogenic studies. Indeed, neutron-induced reactions are the dominant contributors to the formation, for instance, of $^{10}$Be (see Fig.~\ref{fig:cosmogenic_1} in Sec.~\ref{sec:Transverse_Cosmogenic}).

\subsection{Multi-GeV facilities for nuclear cross-sections}
\label{sec:facilities_MultiGeV}
Nuclear fragmentation cross-sections are critical down to energies of a few hundreds of MeV/n, hence high-precision measurements are also needed in this energy range. Below the energies of the \acrshort{cern} complex (covered in sections~\ref{sec:facilities_LHC} and~\ref{sec:facilities_SPS}), many facilities exist. However, the requirement of energy beams above $\sim\!100$\,MeV/n drastically limits the list of facilities where direct relevant measurements may be conducted: the \acrshort{cnao} (\acrlong{cnao}) reaches a few hundreds of MeV/n for proton and carbon beams (Sec.~\ref{sec:Facility_CNAO}), the \acrshort{nsrl} at \acrshort{bnl} offers a variety of beams in the GeV/n range (Sec.~\ref{sec:Facility_Brookhaven}), the \acrshort{gsi}/\acrshort{fair} (\acrlong{fair} at \acrlong{gsi}) offers various beams in the 1--2\,GeV/n range~\cite{FAIR} (Sec.~\ref{sec:Facility_FAIR}), and the upcoming \acrshort{hiaf} (\acrlong{hiaf}) in China should deliver heavy beams up to 2.9\,GeV/n (Sec.~\ref{sec:Facility_HIAF}). 

Below, the most salient features of these facilities are presented, including the detection systems adapted to high-precision fragmentation cross-section measurement. In Table~\ref{tab:beams}, the maximum energy and most relevant beams that could be suitable to carry out the nuclear fragmentation programme of Sec.~\ref{sec:XS_prodnuc} are listed. Note that the \acrshort{nica} complex~\cite{NICA} in Dubna, designed to produce heavy ions with an energy of 1--3.9\,GeV/n, is not discussed in this paper, since no appropriate detection system is known at the time of writing.

\begin{table}
    \centering
    \small
    \caption{Overview of maximum beam energies (in GeV/n), for a few selected nuclei of interest for \acrshort{gcr} fragmentation measurements (see Sec.~\ref{sec:XS_prodnuc}), at the facilities described in Sect. \ref{sec:facilities}. See the text for more details and discussions about possible measurements. The next-to-last column (secondary beams) indicates whether any of the above beams can be fragmented and filtered, thus offering a high-purity secondary beam with an energy close to the primary beam; for NA61, except for p, all species listed are from a primary beam of $^{208}$Pb (see Sec.~\ref{sec:facilities_SPS}), but other primary nuclear beams (e.g., O) are possible. The last column (experiment type) corresponds to the detection system used at the date of this paper --- it might evolve in the future: {\em spectrometry} refers to the identification in mass and charge of each heavy fragments; {\em activation} refers to the determination of concentrations of elements from $\gamma$-ray emitters. \vspace{1.5mm}}
    \label{tab:beams}
    \begin{tabular}{cccccccccccc}
    \toprule
        \textbf{Facility}  & \multicolumn{8}{c}{\textbf{Selected beams and maximum $\mathbf{E_{\rm k/n}}$ (GeV/n)}} & \textbf{Secondary} & \textbf{Experiment}\\
         & p & \isotope[4]{He} & \isotope[7]{Li} & \isotope[12]{C} & \isotope[16]{O} & \isotope[20]{Ne} & \isotope[28]{Si} & \isotope[56]{Fe} &\textbf{beams} &   \textbf{type}  \\
        \midrule
        \acrshort{cnao}   & 0.25  &      &      & 0.4  &      &      &      &      & no & Spectrometry           \\
        \acrshort{nsrl}  & 2.5 & 1.5 &      & 1.0 & 1.0 & 1.0 & 1.0 & 1.0 & no & Activation             \\
        \acrshort{fair} & 4.5 & 2.0 & 1.55 & 2.0 & 2.0 & 2.0 & 2.0 & 1.75 & yes & Spectrometry \\
        \acrshort{hiaf}     & 6.5 & 2.9 & 2.4 & 2.9 & 2.9 & 2.9 & 2.9 & 2.7 & yes & Spectrometry \\
        \acrshort{sps} & 400 & 158 & 158 & 158 & 158 & 158 & 158 & 158 & yes & Spectrometry \\
        \bottomrule
    \end{tabular}
\end{table}

\subsubsection{\texorpdfstring{\acrshort{cnao}}{CNAO} (Italy)}
\label{sec:Facility_CNAO}

The \acrshort{cnao} facility, located in Pavia (Italy), is a hadrontherapy centre using a synchrotron accelerator. It is equipped with an experimental room that can provide clinical ion beams \cite{Rossi2015}. Currently, it can provide protons and \isotope[12]{C} ions with energies between 60--250\,MeV (protons) and 120--400\,MeV/n for \isotope[12]{C}, and will soon be able to provide other ion types: \isotope[4]{He}, \isotope[7]{Li}, \isotope[16]{O}. The accelerator can deliver up to $10^{10}$~p per spill (equivalent to 2~nA) and $4.10^8$~C ions per spill (equivalent to 0.4~nA). The spills are delivered within 1~s, with a time of 2~s between each spill. The size of the pencil beam is around 10 mm (full width half-maximum), and thanks to the scanning magnets, it can irradiate a field up to 200$\times$200 mm$^2$ at the isocenter. 

Researchers are using this facility to measure a variety of cross-sections relevant to hadrontherapy (see details in Sec.~\ref{sec:Transverse_Hadrontherapy}), and future measurements will take advantage of the additional ion beams available. No energy or intensity upgrades are planned for the next 5 years.

\subsubsection{Brookhaven (USA)}
\label{sec:Facility_Brookhaven}

The \acrshort{rhic} accelerator~\cite{RHIC_2016} at \acrshort{bnl}, New York, is a high-energy collider (up to 100\,GeV/n for gold and to 250\,GeV/n for protons). However, \acrshort{rhic} is dedicated to the physics of quark and gluon plasma and no experimental hall is currently equipped or foreseen to host fixed-target experiments focused on the identification of heavy fragments. 

Also part of \acrshort{bnl}, the \acrshort{nsrl} (\acrlong{nsrl}) is an irradiation facility that takes advantage of the \acrshort{bnl} \acrshort{ags} (\acrlong{ags}) to provide protons and heavy ions at space-relevant energies~\cite{2016LSSR....9....2S} (see more in Sec.~\ref{sec:Transverse_SpaceRadiation}): protons up to 2.5\,GeV, \isotope[4]{He} up to 1.5\,GeV/n, and several heavy nuclei (\isotope[12]{C}, \isotope[16]{O}, \isotope[20]{Ne}, \isotope[40]{Ar}, \isotope[56]{Fe}) up to 1\,GeV/n. The maximum intensities are on the order of a few $10^9$ per second, except for protons, where the intensity can reach $2.2\times10^{11}$/s. A series of quadrupole magnets is available to shape the beam to cover large areas or to irradiate multiple targets at once. \acrshort{nsrl} can provide measurements over the entire energy range with a single experimental setup, a feature which contributes to reduce systematic uncertainty. The \acrshort{nsrl} proton beam flux is measured in situ with a precision of 3.6\%.

After irradiation at \acrshort{nsrl}, scientists perform on-site activation measurements for short-lived daughter products using local \acrfull{hpge}, while for longer-lived isotopes the irradiated targets can be sent back to \acrshort{nasa} 
\acrshort{gntf} (\acrlong{gntf}) at \acrshort{gsfc} (\acrlong{gsfc}) for decay analysis using high-precision Ge-based $\gamma$-spectrometers. The detection efficiency is calibrated using \acrshort{nist}-traceable (\acrlong{nist}) sources that have activities known with a precision of 3\% or better. Cross-sections can be determined from the activated target measurements using basic nuclear physics coupled with the relevant nuclear decay parameters.
Recent examples of such experiments are the measurement of spallation cross-sections of \isotope[\textrm{nat}]{Cu}~\cite{PEPLOWSKI2021122067}, \isotope[\textrm{nat}]{Cr} and \isotope[\textrm{nat}]{Mn}, from 200\,MeV to 2.5\,GeV protons.

\subsubsection{\texorpdfstring{\acrshort{gsi}}{GSI} (Germany): past measurements and \texorpdfstring{\acrshort{fair}}{FAIR} in 2027}
\label{sec:Facility_FAIR}

\acrshort{gsi} offers a large variety of beams with a maximum energy defined by the magnetic rigidity of its synchrotron, 18\,Tm presently. This corresponds to 2\,GeV/n energies for light systems such as C and O. Currently, 28~different primary beams are available at \acrshort{gsi}, from p to \isotope[238]{U}, covering the most abundant nuclei in the \acrshort{ism}. In addition to these primary beams, hundreds of radioactive species can be proposed as secondary beams using the \acrshort{frs} (\acrlong{frs}), a low-transmission, high-resolution recoil magnetic spectrometer~\cite{Geissel92_FRS}.

The facility has a significant history with spallation cross-section measurements. Between 1996 and 2011, several such measurements were performed at \acrshort{gsi}, based on the inverse kinematics technique and using liquid hydrogen targets: \isotope[56]{Fe} at 300, 500, 750, 1000 and 1500\,MeV/n~\cite{2007PhRvC..75d4603V}; \isotope[136]Xe at 200 and 500~\cite{Paradela17}, and at 1000\,MeV/n~\cite{Napolitani07}, \isotope[197]{Au} at 800\,MeV/n~\cite{Rejmund01}, \isotope[208]Pb at 500\,MeV/n~\cite{Audouin06} and at 1000\,MeV/n~\cite{Enqvist01}, and \isotope[238]U at 1000\,MeV/n~\cite{Taieb01}. Most of these measurements relied on the complete identification (mass and charge) of the heavy fragment by the aforementioned \acrshort{frs}. Cross-sections were obtained with a typical uncertainty of 4\%, but it should be noted that the transmission of the spectrometer was estimated in a rather crude way (a sharp 15\,mrad cut-off), so error bars for large mass variations and/or low energies were possibly underestimated at the time. Another pair of experiments, on \isotope[56]{Fe}~\cite{LeGentil08} and \isotope[136]Xe~\cite{Gorbinet19}, both at 1000\,MeV/n, were conducted using the large-acceptance \acrshort{aladin} (\acrlong{aladin}) magnet, with the goal of simultaneously measuring the heavy residue and the light fragments and particles.

The \acrshort{fair} facility is currently in the final stage of construction and is expected to host its first experiments in 2027. Its powerhouse will be a new synchrotron, the SIS100~\cite{Spiller20_SIS100}, with a maximum magnetic rigidity of 100\,Tm. Depending on the chosen element, beams are expected to have a maximum intensity of the order of $10^{11}$ per spill, the slow extraction mode making this roughly equivalent to a constant intensity of $10^{11}$\,Hz. Taking over the role of the present \acrshort{frs}, the new recoil separator, the Super-\acrshort{frs}~\cite{Geissel03_SFRS}, will offer access to an even larger diversity of secondary beams, with increased intensities.

The first possibility to measure \acrshort{gcr} cross-sections is to use the Super-\acrshort{frs}, in a way similar to the spallation campaign mentioned above. It is important to note that the Super-\acrshort{frs} itself was designed with a maximal magnetic rigidity of 18\,Tm, the same as the \acrshort{frs}. However, the Super-\acrshort{frs} will offer improved transmission (cut-off of 40\,mrad in the horizontal direction and 20\,mrad in the vertical one) and, more importantly, better knowledge and flexibility of its optics. Therefore, Super-\acrshort{frs} could prove to be an excellent tool for measuring cross-sections with high accuracy, provided the efficiency can be assessed with sufficient precision. At the time of this writing, no liquid-hydrogen target is foreseen at the entrance of the Super-\acrshort{frs}, so early experiments should be conducted with CH$_2$ and C targets.

A second direction is to set up a measurement in the future \acrshort{hec} (\acrlong{hec}), where the new supraconducting, large-acceptance magnet \acrshort{glad} (\acrlong{glad}) will be installed to be the backbone of the R$^3$B (\acrlong{r3b}) setup. Exclusive experiments, similar to the SPALADIN (spallation at \acrshort{aladin}) measurements mentioned above, can be foreseen there. A liquid-hydrogen target is already available for this setup. A possible downside of such an experiment is that only reduced beam intensities can be used, as the full beam goes through the detection chain. Since the beams will be delivered to the \acrshort{hec} through the Super-\acrshort{frs}, the rigidity limit of 18\,Tm also applies.

A third, and much more hypothetic at the moment, possibility is the \acrfull{cbm} cave. There, the SIS100 beams will be delivered up to their maximal energy. However, \acrshort{cbm} is a fixed setup, designed to study high-multiplicity events, and focused on the identification of light hadrons and baryons. A spallation measurement would require a large modification of the \acrshort{cbm} setup.

\subsubsection{\texorpdfstring{\acrshort{hiaf}}{HIAF} (China) in 2026}
\label{sec:Facility_HIAF}

For over half a century, the \acrshort{hirfl} (\acrlong{hirfl})~\cite{Xia:2002xpu}, designed and operated by the \acrshort{imp} (\acrlong{imp}) of the Chinese Academy of Sciences, has played a pivotal role in advancing heavy-ion accelerator technology, heavy-ion physics, and their applications in China. To meet the increasing demands for higher beam intensity and beam energy in next-generation heavy-ion accelerators, \acrshort{imp} proposed the \acrshort{hiaf}(\acrlong{hiaf})~\cite{Zhou:2022pxl}, which is expected to become operational by the end of 2025.

\acrshort{hiaf} consists of a superconducting ion linear accelerator, a high-energy synchrotron booster, a high-energy radioactive isotope beam line, an experimental storage ring, and a few experimental setups. The \acrshort{bring} (\acrlong{bring}) is designed to achieve a maximum magnetic rigidity of 34\,Tm, enabling the delivery of a typical \isotope[16]{O} beam with the energy of 2.6\,GeV/n at an intensity of 6$\cdot 10^{11}$ particles per pulse (ppp) or a proton beam with the energy of 9.3\,GeV/n at 2$\cdot 10^{12}$ ppp~\cite{Zhou:2022pxl}. Beams extracted from the \acrshort{bring} are injected into the \acrshort{hfrs} (\acrlong{hfrs}), a powerful high-energy radioactive beam line. At the entrance of the \acrshort{hfrs}, the beams impinge on a target to produce \acrshort{ribs} (\acrlong{ribs}) via projectile fragmentation or in-flight fission reactions. The \acrshort{hfrs} is engineered to purify and separate \acrshort{ribs} with a maximum magnetic rigidity of 15\,Tm through a two-stage separation process (pre-separator and main-separator). This setup achieves an excellent removal rate of primary beams and an effective separation for nuclides from hydrogen to uranium. Notably, the current design of 15\,Tm magnetic rigidity is not the ultimate goal, as future upgrades aim to enhance the \acrshort{hfrs} to a maximum magnetic rigidity of 25\,Tm, which corresponds to the maximum energy of 2.9\,GeV/n for light nuclei with a mass-to-charge ratio of two.

Several methods are available for cross-section measurements at \acrshort{hiaf}. The first approach utilises the \acrshort{hfrs} itself. Isotopes are separated and purified in the pre-separator and the initial half of the main separator before impinging on a target. The second half of the main separator acts as a zero-degree spectrometer, enabling the identification of fragments after the reaction using the B$\rho$--\acrshort{tof}--$\Delta$E method, and fragmentation cross-section measurements. Recent design optimisations have increased the horizontal and longitudinal angular acceptances of the \acrshort{hfrs} to $\pm30$\,mrad and $\pm25$\,mrad, respectively~\cite{Sheng:2023ojn}. These enhancements allow the \acrshort{hfrs} to collect fragments after the target more effectively, thereby yielding more precise cross-section measurements. Similar to the Super-\acrshort{frs} in \acrshort{gsi}, initial experiments with the \acrshort{hfrs} do not support the hydrogen targets, and the C$-$CH$_2$ subtraction would be used instead.

The second method relies on the \acrshort{etf} (\acrlong{etf}) of the \acrshort{hfrs}. The \acrshort{hfrs} can function as a complete beam separator, delivering \acrshort{ribs} to the \acrshort{etf}, where the beams impinge on a target. A detector array at the \acrshort{etf} can identify reaction products and facilitate fragmentation cross-section measurements. Currently, \acrshort{imp} operates a radioactive isotope beam line, the second \acrshort{ribll} (\acrlong{ribll}) at \acrshort{hirfl}, which has successfully conducted fragmentation cross-section measurements at its external target facility~\cite{Mei:2023bho}. With the enhanced magnetic rigidity of the \acrshort{hfrs} compared to \acrshort{ribll}, future operations are expected to enable cross-section measurements in higher energy regions.

\subsection{GeV-to-TeV measurements from \texorpdfstring{\acrshort{cr}}{CR} experiments}
\label{sec:facilities_CRexps}

\acrshort{cr} experiments, as particle physics detectors in space, possess the capability to perform cross-section measurements. Although not primarily designed for this purpose, the need to understand and constrain challenging systematic uncertainties in their data provides a strong motivation for such measurements (see Sec.~\ref{sec:XSforCRexp}). A key advantage of \acrshort{cr} experiments is their natural access to a diverse range of beam species and energies (although the \acrshort{gcr} “beam luminosity” rapidly declines with energy). Recent results have demonstrated their potential to deliver important contributions to cross-section measurements, as described below.

\paragraph{Spectroscopic}
Accurate measurements of the individual spectra of nuclei require knowledge of nuclear fragmentation cross-sections with the detector material to reject background from fragmentation of heavier nuclei within the upper part of the detector. The modularity of modern experiments, such as \acrshort{ams}, allows to directly measure nuclei survival probabilities within the detector and rescale the \acrshort{mc} simulation accordingly to overcome the lack of measurements of nuclear fragmentation cross-sections. From the measured survival probabilities and the knowledge of the material within the \acrshort{ams} detector, the \acrshort{ams} collaboration has derived charge-changing inelastic cross-sections on a C target for projectile nuclei: He, Li, Be, B, C, N, O, Ne, Mg, Si, and Fe in the rigidity range from 2\,GV to 1\,TV~\cite{AMSXS, AMS:PhysRep2021, AMS:Fe-PRL2021}.

\begin{figure}
	\centering
	\includegraphics[width=8.1cm]{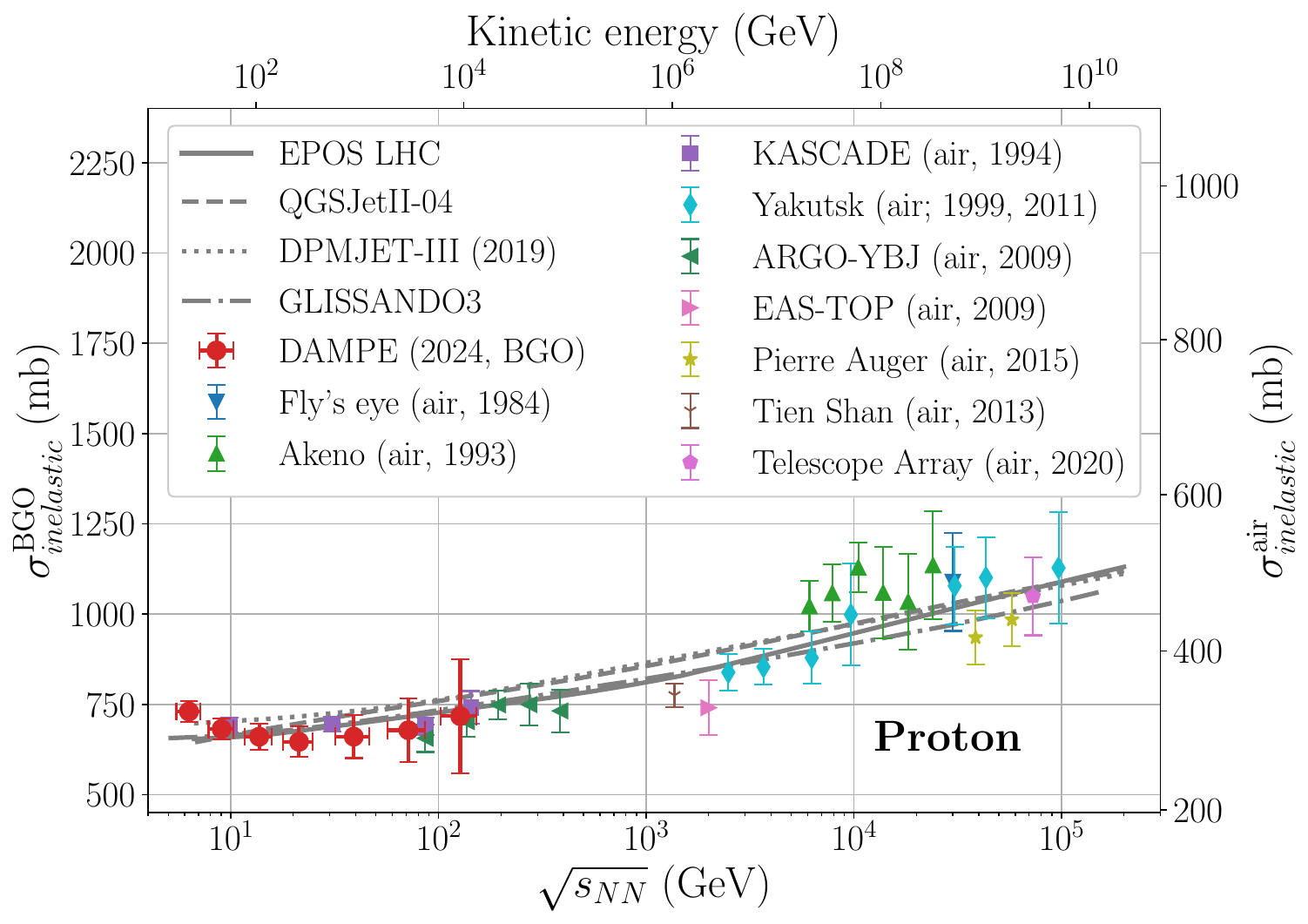}
	\includegraphics[width=8.1cm]{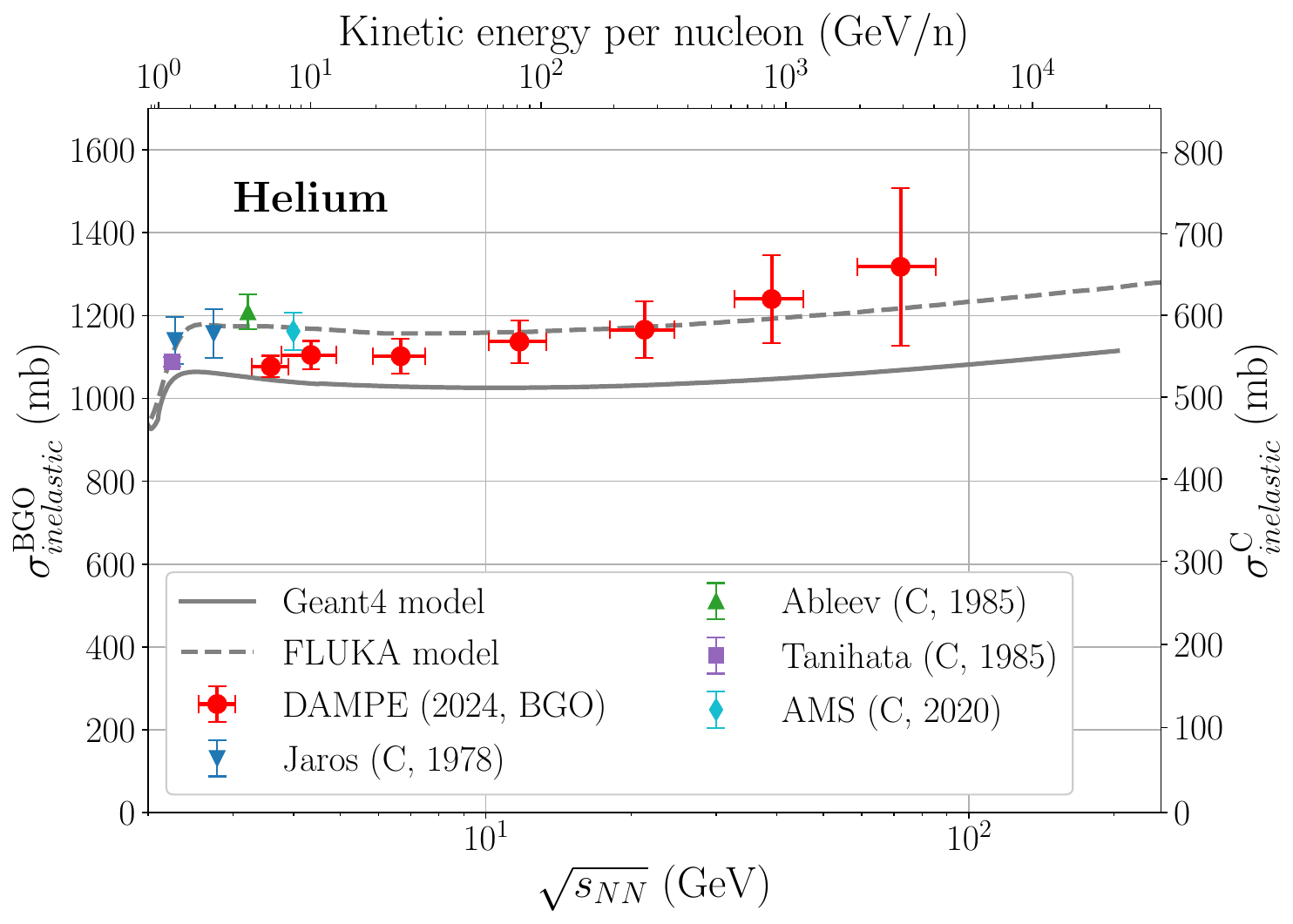}
	\caption{\label{fig:EASComparison}Left: measurements of the proton inelastic cross-section with calorimetric space-based experiments \cite{DAMPEXS}, respective to results by extensive air shower experiments \cite{FlysEye, EASTOP, 1999Yakutsk, 2011Yakutsk, Akeno, TienShan, Abreu2012PhRvL, Auger2015, TelescopeArray, ARGO, KASKADE} and model predictions \cite{EPOS, EPOS-LHC, dpmjet3_1, dpmjet3_2, dpmjet3_3, Glissando3}. Right: inelastic cross-section of He measured by \acrshort{dampe} \cite{DAMPEXS} and \acrshort{ams} \cite{AMSXS}, compared to results by accelerator experiments \cite{Jaros:1977it, Ableev85, Tanihata85} and model predictions \cite{EPOS, EPOS-LHC, dpmjet3_1, dpmjet3_2, dpmjet3_3, Glissando3}.}
\end{figure}

\paragraph{Calorimetric}
For calorimetric experiments, uncertainties from fragmentation cross-sections are secondary to those of the total inelastic cross-sections, which determines the acceptance and energy response of the detector. More accurate parametrisations of inelastic nucleus--nucleus cross-sections in the GeV--PeV regime are one of the main requirements needed to improve the accuracy of \acrshort{cr} nuclei fluxes at these energies. Due to a lack of measurements, models generally rely on the conversion of \pp cross-sections to different primaries and target materials using, e.g., the Glauber-Gribov approach~\cite{Glauber,Glauber0,Gribov1,Gribov2}. For heavy-target materials $\left(A\gtrsim 50\right)$, often present in calorimetric detectors, such conversions come with typical uncertainties of 10--20\%. Lowering these hadronic uncertainties would make \acrshort{cr} flux measurements more constraining, significantly improving \acrshort{cr} production and propagation models. The \acrshort{dampe} collaboration recently published a measurement of the inelastic cross-section of H (mostly protons) and He (mostly $^4$He) on a \acrshort{bgo} ($\text{Bi}_4\text{Ge}_3\text{O}_{12}$) target \cite{DAMPEXS}, see Figure~\ref{fig:EASComparison}. In the case of He, these are the first measurements in the kinetic energies range of 20\,GeV to 10\,TeV (lab frame) on any heavy-target material.
This pioneering measurement achieved three major objectives: it demonstrates the feasibility and potential of doing inelastic cross-section measurements with calorimetric space-based experiments; it enables improving the accuracy of \acrshort{cr} flux measurements; and it provides a base measurement from which the cross-section of other heavy-target materials can be extracted with model dependencies of only a few percent. The collaboration currently plans to extend the \acrshort{dampe} cross-section measurements to other nuclei, including C and O. On longer timescales, it is worth noting that the \acrshort{herd} mission~\cite{HERD}, planned to launch in 2027, will significantly enhance the statistics of high-energy \acrshort{cr} observations. With its calorimeters segmented in all three spatial dimensions, \acrshort{herd} data will enable to improve the accuracy of the current \acrshort{cr} cross-section measurements, and to extend them to higher energies.

\paragraph{Caveats of these measurements}
It is worth noting that for both the \acrshort{ams} and \acrshort{dampe} cross-section measurements, the projectile nuclei are \acrshort{gcr}s and therefore a mixture of two or more isotopes, with one dominating isotope in some cases. Moreover, only the charge of the projectile and of the final-state nuclei was measured, but not the isotopic composition, though with the \acrshort{ams} detector it is possible to measure the isotopic composition of the daughter nuclei (see Section \ref{sec:CR_isotopes}). In general, space detector have not been designed to perform cross-section measurements, and verifications with accelerator experiments where the same quantities can be measured in a more straightforward way are needed. This is illustrated with the discrepant charge-changing cross-section in C+C as inferred in \acrshort{ams} or recently measured in NA61/\acrshort{shine} \cite{NA61SHINE:2024rzv}.

\section{Overlapping cross-section needs from other communities, and further astro/\texorpdfstring{\acrshort{hep}}{HEP} synergies}
\label{sec:Transverse}

As briefly introduced in Sec.~\ref{sec:physics_case_transverse}, \acrshort{gcr} physics is an interdisciplinary domain, with connections to several adjacent fields of research. The long term evolution and stability of \acrshort{gcr} fluxes over Gyr timescales comes with its specific wish list of reactions and priorities (Sec.~\ref{sec:Transverse_Cosmogenic}).
Other science cases, more related to applied physics and societal topics, have yet other cross-section needs and priorities. Among them, space radiation protection is a topic of growing interest (Sec.~\ref{sec:Transverse_SpaceRadiation}). Hadrontherapy is a curious example, where \acrshort{gcr}s are completely absent, yet a significant overlap exists in terms of the cross-sections and energies of interest (Sec.~\ref{sec:Transverse_Hadrontherapy}). For these three topics, where some overlap exists with the reactions needed for \acrshort{gcr} studies, the reactions involved, the status of nuclear data or codes, and the needs in terms of reactions, energy, and precision are detailed below. 

To further illustrate the richness, similarities, and advantages of synergies between the \acrshort{hep} and astroparticle communities, the cases of ultra-high energy \acrshort{cr}s and femtoscopy (related to the equation of state of neutron stars) -- where even completely different cross-sections or regimes are explored --, are also briefly covered (Sec.~\ref{sec:Transverse_Other}).

\subsection{Cosmogenic production in meteorites}
\label{sec:Transverse_Cosmogenic}

As briefly introduced in Sec.~\ref{sec:motiv_cosmogenic}, cosmogenic studies focus on the nuclides stored in meteorite and terrestrial archives. Long-lived radioactive nuclides, in particular, provide a powerful tool to assess \acrshort{gcr} fluency variations over time.

Obtaining cosmogenic production rates is possible via physical model calculations, some type of empirical calibration using experimental meteorite data, or a combination of both. Here we focus on physical model calculations as they are widely used for cosmogenic nuclide studies. Though the following discussion is mainly focused on cosmogenic nuclide production in meteorites, most of the arguments are also applicable to terrestrial cosmogenic nuclides. The production rate $P^k$ (number of atoms per mass and time unit) of a cosmogenic nuclide of type $k$ in a meteoroid of radius $R$ at position $\vec{r}$ is given by:

\begin{equation}
\label{eq:CRE3}
P^{k}(\vec{r},\vec{c}_{\rm sample},\vec{C}_{\rm meteoroid},R) = \sum_{i} \sum_{j} \vec{c}^{j} \int_{E} \sigma_{\rm prod}^{i+j\to k}(E) \psi^{i}(\vec{r},\vec{C}_{\rm meteoroid},R,E) dE,
\end{equation}
with $R$ the pre-atmospheric radius of the meteoroid, $\vec{c}_{\rm sample}$ the chemical composition of the studied sample, and $\vec{C}_{\rm meteoroid}$ the chemical composition of the meteoroid. Note that $\vec{c}_{\rm sample}$ and $\vec{C}_{\rm meteoroid}$ can be different if, for example, a metallic inclusion in a stony meteorite is studied. The differential flux density of primary and secondary particles of type $i$ is $\psi^{i}(\vec{r},\vec{C}_{\rm meteoroid},R,E)$ -- particle per time, surface, and energy unit -- and depends on the pre-atmospheric geometry, the depth of the sample within the pre-atmospheric meteoroid, and the bulk chemical composition of the meteoroid $\vec{C}_{\rm meteoroid}$. The nuclear cross-section for the production of nuclide $k$ from chemical element $j$ by particles of type $i$ is $\sigma_{\rm prod}^{i+j\to k}(E)$, i.e., the same cross-section type as needed for \acrshort{gcr}s, and $\vec{c}^{j}$ represents the concentration of element $j$ in the meteoroid, assumed to be constant. For simplicity, it is usually assumed that the pre-atmospheric meteoroid was spherical and that the \acrshort{gcr} flux was temporally constant. It can be shown that the calculated production rates are not very sensitive to the assumption of a spherical shape (e.g., \cite{Prettymanetal2014}). In addition, there are earlier and still ongoing studies tackling the very important question of whether the \acrshort{gcr} fluence was temporal constant (for a discussion, see \cite{DavidLeya2019}).

While the chemical composition of the studied sample $\vec{c}_{\rm sample}$ and of the bulk meteoroid $\vec{C}_{\rm meteoroid}$ can easily be measured, the particle spectra and the relevant cross-sections are more problematic input parameters. Since the particle spectra $\psi^{i}(\vec{r},\vec{C}_{\rm meteoroid},R,E)$ -- that are typically calculated using \acrshort{mc} techniques -- are not the subject of this paper, we focus the discussion on the current status of knowledge for the relevant cross-sections.
Considering the possible target elements, the bulk composition of \textit{chondrites} (common meteorite type) closely matches the bulk composition of the Sun, except for a few elements that usually occur in the gas phase, e.g., H, C, Ni and He. Consequently, almost 95\% of a chondrite is made from only six elements. As an example, for a special chondrite type, called CI carbonaceous chondrite, the six major elements are (in percent by weight): C~(3.22\%), O~(46.5\%), Mg~(9.61\%), Si~(10.68\%), S~(5.41\%), Fe~(18.43\%) and~Ni (1.08\%). The other elements of the periodic table are all there but only in minor concentrations (e.g., \cite{PalmeJones2003}). Some of such minor elements, however, are major target elements for some cosmogenic nuclides. For example, cosmogenic Kr is produced from the minor amounts of Rb, Sr, Yr and Zr, cosmogenic Xe is produced from La and Ba, and cosmogenic $^{129}$I is produced from Te.

\subsubsection{Cross-sections for proton-induced reactions}
\label{pXS}
Decades of experimental effort have yielded most of the relevant cross-sections for proton-induced reactions, which are now relatively well known and compiled in various databases, e.g., \cite{Otukaetal2014}. However, we need to improve/adjust some relevant published cross-sections, due to the changed/adjusted standards used for the analysis by accelerator mass spectrometry ($^{10}$Be, $^{36}$Cl, $^{41}$Ca). For example, when the activity concentrations of the long-lived radionuclide $^{10}$Be are measured by accelerator mass spectrometry \cite{2023RvMP...95c5006K}, the $^{10}$Be/$^{9}$Be ratio of the sample is measured against the $^{10}$Be/$^{9}$Be ratio of a standard. If the $^{10}$Be/$^{9}$Be ratio of the standard is not as assumed, all samples measured against these standards end with wrong $^{10}$Be/$^{9}$Be ratios and therefore wrong cross-sections. Thanks to recent inter-laboratory comparisons, the differences among the different standards used in different laboratories are better resolved. Some standards, however, needed to be revised, and such data need either to be recalculated and/or to be remeasured. Such changes are sometimes in the range 10--15\% and are therefore relevant (e.g., \cite{Geppertetal2005, Mercheletal2011, Christletal2013, Leyaetal2021}). In addition, some relevant half-lives have recently been revised, e.g., for $^{10}$Be, $^{41}$Ca and $^{60}$Fe (e.g., \cite{Korschineketal2010, Jorgetal2012}) and therefore also such cross-sections need to be recalculated and/or redetermined.

For some relevant target product combinations, the cross-section database is still scarce. Examples are the production of $^{41}$Ca and $^{53}$Mn from Fe and Ni, the production of Kr isotopes from Rb, Sr, Yr and Zr, and the production of Xe isotopes from Ba and La. Importantly, the cross-section database for the production of $^{14}$C, which is a relevant target element for terrestrial and extraterrestrial applications, is still very scarce. This is probably due to the fact that extracting $^{14}$C and measuring $^{14}$C/$^{12}$C ratios via accelerator mass spectroscopy was for a long time very challenging. However, new instruments making $^{14}$C extractions and measurements more accurate and more precise have been developed (e.g., \cite{Slizetal2022, Tauseefetal2024}, which would make revisiting the $^{14}$C cross-sections worthwhile.

\subsubsection{Cross-sections for neutron-induced reactions}
For neutron-induced reactions the situation is different, as very few experimental data exist in the energy range of interest, from the reaction threshold up to a few GeV. To overcome this problem, some relevant neutron-induced cross-sections have been inferred from various thick target irradiation experiments \cite{Michel1985, Micheletal1989, Micheletal1993, Micheletal1995, Herpersetal1991, Leyaetal2000a}. For details of the procedure, see \cite{Leyaetal2000a, LeyaMichel2011}. In addition, there are very few directly measured neutron-induced cross-sections relevant for the study of meteorites and planetary surfaces (\cite{Sistersonetal2004, Sisterson2007}). For some relevant target-product combinations, there are not enough data to perform the adjustment procedure, and the neutron excitation functions must be calculated using various nuclear model codes (e.g., \texttt{TALYS} or \texttt{INCL++6}). Though some of these models have significantly been improved over the decades, the quality of the calculated cross-sections is often still not sufficient for high-quality studies of cosmogenic nuclide production in meteorites and planetary surfaces. As a consequence, currently the most limiting factor for the quality of the model calculations is the ill-known and sometimes missing knowledge of the neutron cross-sections. Some major improvements can be expected, if at least some relevant cross-sections could be measured directly.

\subsubsection{Cross-sections for \texorpdfstring{$^4$He}{4He}-induced reactions}
For $^4$He-induced reactions, the situation is even worse: there are essentially no experimental data for the relevant reactions, namely the production of cosmogenic species on O, Si, Fe\dots targets, and in the energy range of interest, from threshold up to a few GeV/n. So far, this posed no major problem, because the \acrshort{mc} codes used to calculate the differential particle spectra could not reliably consider $^{4}$He or other light charged particles. Due to this shortcoming, there was no real effort in measuring the cross-sections for $^4$He-induced reactions. Moreover, primary \acrshort{gcr} $^4$He ions and their secondary products were considered using a relatively crude approximation. However, this situation has just changed, and it is now possible, for the first time, to directly include the full interactions of these $^4$He ions \cite{Leyaetal2021}, and exciting new results are expected. However, the missing experimental cross-sections are a serious limitation.

\subsubsection{Cross-sections for muon-induced reactions}
Most studies of cosmogenic nuclides in extraterrestrial material assume that production is dominated by primary and secondary protons and secondary neutrons and that for some target--product combinations, secondary $^4$He-particles also contribute. However, a recent study argues that secondary charged pions might contribute more than 20$\%$ to the measured activity concentration of $^{10}$Be on the lunar surface \cite{Lietal2017}. This finding contrasts with the good results obtained by studies that describe depth profiles for a variety of cosmogenic nuclides on the lunar surface (including $^{10}$Be) without it (e.g., \cite{Leyaetal2021, Cechvalaetal2023}). Therefore, confirming or rejecting the secondary pions hypothesis is crucial. This is especially true, considering that the study of muons (originating from the decay of pions) on planetary surfaces is just becoming an important tool, in space missions, to study the water-ice composition, and the density and chemical composition of lunar or asteroidal surfaces (e.g., \cite{PintoandTiit2023}).

In contrast to cosmogenic nuclides in extraterrestrial matter, muon-induced reactions are very relevant for some terrestrial cosmogenic nuclide studies. Since muons are leptons -- and therefore do not interact via the strong force --, they penetrate much deeper into the ground than neutrons. Therefore, despite their short lifetime and their relatively small contribution ($\lt$2\%) to the total terrestrial nuclide production at the Earth surface, slow negative muons and fast muons are the dominant projectiles for nuclide production at depths larger than $\sim\!4$\,m. This makes muon-induced production very important, whenever the sample was buried more than 1 m deep and/or whenever terrestrial cosmogenic nuclides are used to study burial histories (for more information see, e.g., \cite{Leyaetal2021}). Currently, there are no cross-sections for muon-induced production of cosmogenic nuclides and all estimates are based on theoretical nuclear model codes with unverifiable quality. Consequently, studying and quantifying muon-induced production in terrestrial and extraterrestrial samples is one of the next important steps in the field. Some data are already available~\cite{Heisingeretal2002a, Heisingeretal2002b}, but the database is far from complete.

\subsubsection{Summary and wish list}
\begin{figure}
    \centering
    \includegraphics[width=.7\linewidth]{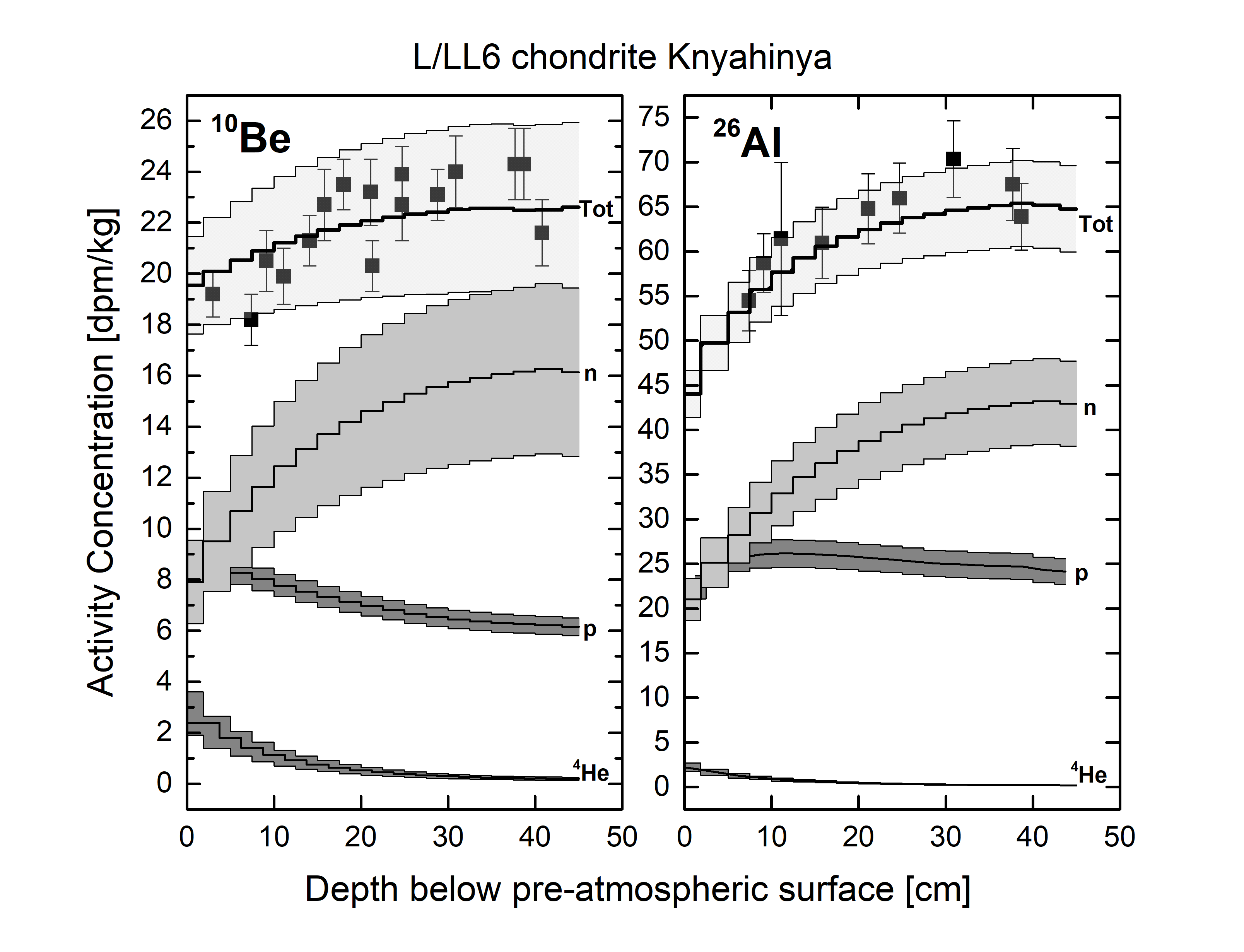}
    \caption{Production rates of $^{10}$Be (left panel) and $^{26}$Al (right panel) as a function of depth below the pre-atmospheric surface for the L/LL6 chondrite Knyahinya (solid symbols). The result of the physical model calculations are shown by the solid lines. The grey areas indicate the estimated uncertainties. For the model calculations, the contributions by protons, neutrons and $^{4}$He projectiles are distinguished. The total production rate is the sum of the three individual contributions. The experimental data are from Ref.~\cite{Grafetal1990}.}
    \label{fig:cosmogenic_1}
\end{figure}

The most important needs for modelling cosmogenic production are summarised in Table~\ref{tab:Cosmogenic-measurements}.
The main target elements are C, O, Mg, Al, Si, S, Fe and Ni, although for some product nuclides, other (and heavier) target elements are also important. Cross-sections from the respective reaction thresholds up to $\sim\!20$\,GeV are needed. Since most excitation functions show little energy dependencies above a few GeV, extrapolation towards higher energies is often possible and accurate enough.
Figure~\ref{fig:cosmogenic_1} shows measured and modelled cosmogenic production rates for $^{10}$Be (left panel) and $^{26}$Al (right panel) for the L/LL6 chondrite Knyahinya. The experimental data are from Ref.~\cite{Grafetal1990}, and for the model calculations, the contributions from protons, neutrons and $^{4}$He are distinguished. The estimated uncertainties for the individual contributions, but also for the total production rates, are given by the grey areas.
This plot helps to demonstrate the different priorities and precisions reported in Table~\ref{tab:Cosmogenic-measurements}.
\begin{itemize}
  \item The highest priority and precision is for neutron-induced reactions, because their contributions often dominate the total production in extraterrestrial samples; they are also often the sole contributions for terrestrial applications. In that respect, the scarce (or very often lacking) cross-section database is the limiting factor for most cosmogenic nuclide studies. This situation must urgently been improved. This is clearly seen in Fig.~\ref{fig:cosmogenic_1}, where the neutrons contribute to $\sim\!70\%$ and $\sim\!50\%$ of the total for $^{10}$Be and $^{26}$Al, respectively, and where the (so far large) uncertainties attributed to the neutron contributions clearly dominate the uncertainties for the total production rates.
  \item The second priority (and precision) is for protons-induced reactions, as illustrated in Fig.~\ref{fig:cosmogenic_1}, although this component can dominate the production of some nuclides (e.g., $^{36}$Cl and $^{41}$Ca from Fe and Ni). Most of the relevant data are relatively well known. However, some target-product combinations, highlighted in boldface in Table~\ref{tab:Cosmogenic-measurements}, need remeasurements, due to recent changes in standards and/or half-lives (as detailed in Sec.~\ref{pXS}). 
  \item The contribution of $^4$He-induced reactions is small, and cross-sections calculated from nuclear models should be reliable enough not to significantly enlarge the uncertainties for the production rates (darker shade in Fig.~\ref{fig:cosmogenic_1}). However, there are only very few experimental data, and new measurements (at mild precision) would more strongly support these conclusions.

  \item Finally, there is the open question about muon-induced reactions, whose contribution is not yet clear. Studying, for instance, $^{10}$Be cross-sections for some relevant target elements (at a mild precision) is a necessary first step to assess the impact and importance of these interactions. Muon production and muon-induced reactions will also be very important for future space missions (collecting extraterrestrial samples), and it is also not clear, whether existing models are accurate enough or if experimental cross-sections will be needed.
\end{itemize}

\begin{table}
    \small
    \setlength{\tabcolsep}{6pt}
    \centering
    \caption{Summary of the critical cross-section measurements for cosmogenic nuclide studies in terrestrial and extraterrestrial matter. The highest-priority measurements are for neutron-induced, then proton- and muon-induced, and then $^4$He-induced reactions. The most important reactions within these priorities are highlighted in bold.\vspace{1.5mm}}
    \label{tab:Cosmogenic-measurements}
    \begin{tabular}{cccccc}
        \toprule
        \textbf{Particle} & \textbf{Targets $\boldmath{j}$} & \textbf{Products $\boldmath{k}$} & \textbf{Measurements} & \textbf{Projectile $\boldmath{E_{\rm k}}$} & \textbf{Precision}\\
        \midrule
        \multirow{3}{*}{Neutrons} & C, \bf{O}, Mg, Al, Si & \triton, $^{3,4}$He, $^{10}$Be, $^{14}$C, $^{i}$Ne, & \multirow{3}{*}{$\sigma_{\rm prod}^{\neutron+j\to k}$}& Threshold & \\
        & Ca, Fe, Ni, Rb, Sr & $^{26}$Al, $^{36}$Cl, $^{i}$Ar, $^{41}$Ca, & & up to a few & $<5\%$ \\
        & Yr, Zr, Nb, Ba, La & $^{53}$Mn, $^{60}$Fe, $^{i}$Kr, $^{i}$Xe & & 100\,MeV & \\
        \midrule
        \multirow{3}{*}{Protons} &  \bf{O}, Mg, Al, Si & \bf{\triton}, \bf{$^{3}$He}, \bf{$^{10}$Be}, $^{14}$C & \multirow{3}{*}{$\sigma_{\rm prod}^{\proton+j\to k}$} & Threshold & \\
        & Ca, Fe, Ni, Rb, Sr & $^{53}$Mn, $^{60}$Fe &  & up to a & $<10\%$ \\
        & Yr, Zr, Nb, Ba, La & $^{i}$Kr, $^{i}$Xe & & few GeV & \\
        \midrule
        \multirow{3}{*}{Muons} & O, Mg, Al, Si & \triton, $^{3,4}$He, $^{10}$Be, $^{14}$C, & \multirow{3}{*}{$\sigma_{\rm prod}^{\mu+j\to k}$} & Threshold\\
        & Ca, Fe, Ni & $^{i}$Ne, $^{26}$Al, $^{36}$Cl, $^{i}$Ar, $^{41}$Ca,&  & up to a & $<20\%$ \\
        & & $^{53}$Mn, $^{60}$Fe, $^{i}$Kr, $^{i}$Xe &  & few GeV & \\
        \midrule
        \multirow{3}{*}{$^{4}$He} &  O, Mg, Al, Si & \triton, $^{3,4}$He, $^{10}$Be, $^{14}$C, & \multirow{3}{*}{$\sigma_{\rm prod}^{^4{\rm He}+j\to k}$} & Threshold \\
        & Ca, Fe, Ni & $^{i}$Ne, $^{26}$Al, $^{36}$Cl, $^{i}$Ar, $^{41}$Ca,&  & up to a & $<20\%$ \\
        & & $^{53}$Mn, $^{60}$Fe, $^{i}$Kr, $^{i}$Xe & & few GeV/n & \\
        \bottomrule
    \end{tabular}
\end{table}

\subsection{Space radiation protection}
\label{sec:Transverse_SpaceRadiation}
An important safety priority for human spaceflight is the protection of astronauts from the harmful effects of the radiation environment in space \cite{Walsh2019, NASEMcancer, Freese2016, Chancellor2014}. The three main sources of radiation exposure are geomagnetically trapped particles (mostly protons and electrons \cite{Roederer2014}), solar energetic particles (mostly protons and light ions~\cite{Reames2021}), and \acrshort{gcr}s. The \acrshort{gcr} contribution is of primary importance for long-duration missions to the Moon, Mars, and other deep-space destinations, where little or no natural atmospheric and magnetospheric shielding is available \cite{Freese2016, Wilson2001}. The relevant part of the \acrshort{gcr} spectrum includes all nuclei up to Ni, with energies up to at least tens of \unit{GeV/n}, or even up to \unit{TeV/n} for some aspects. Nuclei of higher mass or energy are currently not considered in space radiation protection because of their much lower fluxes. This radiation field interacts with a spacecraft's hull and structures, and is significantly modified -- via electronic energy loss, fragmentation, and nuclear interactions~\cite{Zeitlin2016} -- by the time it reaches an astronaut inside the vessel~\cite{Zeitlin2021, Dobynde2020}. It is then modified even further throughout the astronaut's body \cite{Cucinotta2011}. The most important particles that contribute to the radiation environment (inside a spacecraft or habitat) are neutrons \cite{Horst2022, Heilbronn2015, Slaba2011}, protons \cite{DeWitt2024}, light ions (i.e., isotopes of H and He), and pions \cite{Vozenin2024}. All these particles are light in mass and are therefore scattered at large angles. Thus, fully three-dimensional transport codes, which use double-differential cross-sections in solid angle and energy as input, are required to assess how spacecraft shielding alters the primary \acrshort{cr} radiation field.

\subsubsection{Transport code disagreements}
\label{sec:xscode_spaceradiation}
\begin{figure}
    \centering
    \includegraphics[width=.5\linewidth]{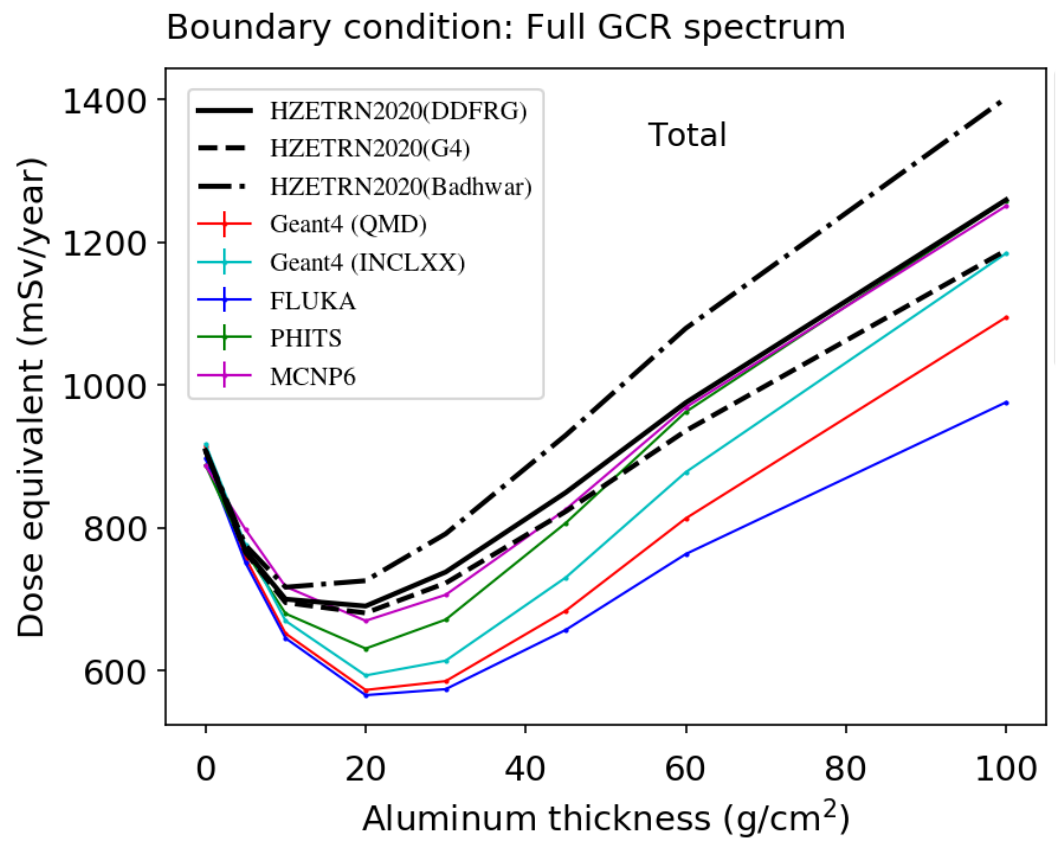}
    \caption{Simulated dose equivalent (total all particles) as a function of shielding depth for a detector exposed to the full \acrshort{gcr} spectrum and located between two slab shields, {\em each} of them having the thickness indicated on the horizontal axis. The results of the \texttt{Geant4} \acrshort{mc} code \cite{Agostinelli2003, Allison2016} with the \texttt{QMD} and \texttt{INCLXX} nuclear models are shown in red and cyan, respectively. Those of the \texttt{HZETRN} \cite{Wilson2014, Wilson2015, Wilson2015a, Wilson2016} transport code with three different nuclear models (\texttt{DDFRG} \cite{Norbury2021, Norbury2023}, Badhwar \cite{Blattnig2000}, and G4, where G4 refers to the \texttt{INCLXX} model from \texttt{Geant4}) are shown as solid and dashed black lines. The other evaluated \acrshort{mc} codes are \texttt{FLUKA} (blue) \cite{Battistoni2015, Ahdida2022}, \texttt{PHITS} (green) \cite{Sato2023, Iwamoto2021} and \texttt{MCNP6} (purple) \cite{Rising2023}. Updated version of Fig.~13 in Ref.~\cite{Slaba2017}, provided by Tony Slaba.}
    \label{spacerad:fig1}
\end{figure}

To date, engineers have designed most spacecraft with relatively thin shielding, with typical aerial thicknesses of about \unit[20]{g/cm$^2$} (compared to \unit[1000]{g/cm$^2$} and \unit[20]{g/cm$^2$} for Earth's and Mars' atmospheres, respectively). The \acrshort{iss} shielding, for example, shows considerable variation, from less than \unit[10]{g/cm$^2$} to about \unit[100]{g/cm$^2$} or more~\cite{Koontz2005}. Figure~\ref{spacerad:fig1} shows the results of a variety of transport-code calculations that assess the dose equivalent produced by the full \acrshort{gcr} spectrum, in a detector between two slab shields of varying thickness~\cite{Slaba2017}. As the shielding increases, the dose equivalent drops steadily and reaches a minimum around \unit[20]{g/cm$^2$}, beyond which it starts to rise again due to the increased production of secondary particles. The existence of this minimum shows that there is an optimal shield thickness for radiation protection, if only \acrshort{gcr} are taken into account \cite{Slaba2017}. Martian and lunar habitats for crewed long-duration missions will likely have shielding approaching or exceeding this optimal thickness \cite{Townsend2018, Dobynde2021} and may, in extreme cases, reach several hundreds of \unit{g/cm$^2$} \cite{Akisheva2021}. Good radiation protection criteria will thus need to rely heavily on accurate and reliable estimates of the radiation environment inside these thick shields. Unfortunately, the transport-code calculations presented in Fig.~\ref{spacerad:fig1} show disagreements on the order of 30\% or more at large shielding depths.

The key question is: what causes this large variation, and what can we do about it? The main reason for the disagreement between the codes are the different nuclear-reaction models they use, which is nicely illustrated by Fig.~\ref{spacerad:fig1}. For example, two different nuclear models, \texttt{QMD} and \texttt{INCLXX}, are used as input to the otherwise identically configured \texttt{Geant4} \acrshort{mc} framework \cite{Agostinelli2003, Allison2016}, showing variation of about 10\% at large shielding depth. Another example is the deterministic transport code \texttt{HZETRN}~\cite{Wilson2014, Wilson2015, Wilson2015a, Wilson2016} that was evaluated with three different nuclear models, producing even larger variations of about 40\%. The other \acrshort{mc} codes, \texttt{FLUKA}~\cite{Battistoni2015, Ahdida2022}, \texttt{PHITS}~\cite{Sato2023, Iwamoto2021} and \texttt{MCNP6}~\cite{Rising2023}, did not allow changes to their underlying nuclear models, though it can be argued that they are among the primary differences between the otherwise similar codes~\cite{Kim2012}. Clearly, the key to resolving the observed disagreements lies in our ability to significantly improve the nuclear-reaction models. What, then, is the obstacle to doing so? It is not the shortage of models or the lack of nuclear theorists working on model development. Rather, it is the lack of sufficient experimental cross-section data that would allow validating the available models.

\subsubsection{Nuclear data: availability and gaps}
\begin{figure}
    \includegraphics[width=0.32\textwidth]{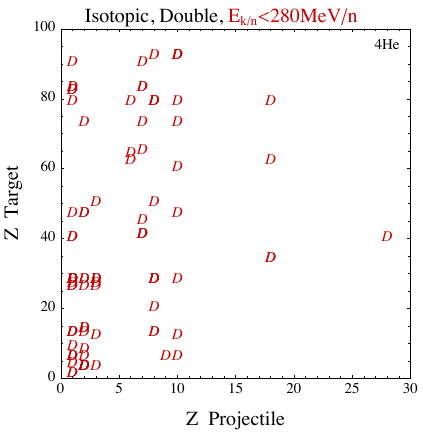}
    \includegraphics[width=0.32\textwidth]{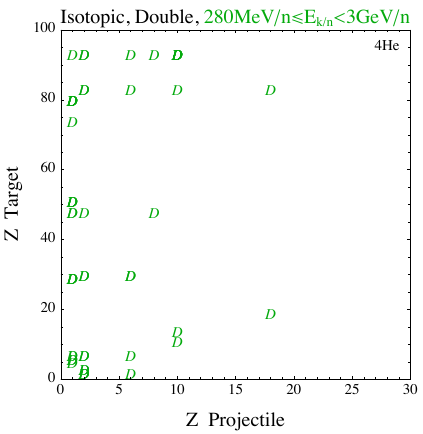}
    \includegraphics[width=0.32\textwidth]{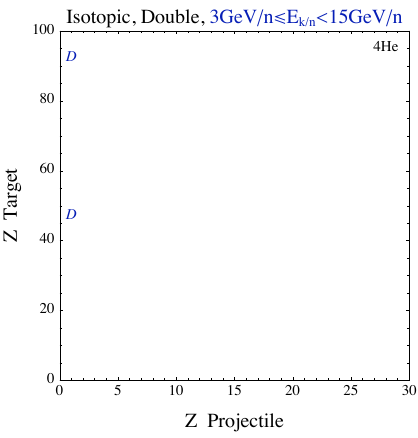}
    \caption{Example of available nuclear-reaction double-differential cross-section measurements (represented by the symbols “D'') for the production of $^4$He particles from nucleus--nucleus reactions with $E_{\rm k/n}<280$\,MeV/n (left), 280\,MeV/n\,$\leq E_{\rm k/n}<3$\,GeV/n (middle) and 3\,GeV/n\,$\leq E_{\rm k/n} <15$\,GeV/n (right). There are no measurements available for $E_{\rm k/n}>15$\,GeV/n. Figure reproduced from Ref.~\cite{Norbury2012}.}
    \label{spacerad:fig2}
\end{figure}

Norbury et al.~\cite{Norbury2012} performed an exhaustive survey of the availability of nuclear-reaction cross-section data for charged-nuclei production relevant to space radiation protection. An example, highlighting the availability of double-differential cross-sections for nucleus--nucleus reactions producing $^4$He particles, is shown in Fig.~\ref{spacerad:fig2}. The presence of the symbol “D'' on the plots indicates that experimental data are available for a given projectile--target combination, while the quality of the data is not ascertained. Below the pion-production threshold, \unit[280]{MeV/n}, there is an abundance of data available for a variety of targets and for projectile charges $Z \leq 10$. Above $Z = 10$, there is very little data. The same is true to a lesser extent for projectile energies, $E_{\rm k/n}$, between the pion threshold and \unit[3]{GeV/n}; for $\unit[3]{GeV/n} < E_{\rm k/n} < \unit[15]{GeV/n}$, there are only two data sets; for \unit[$E_{\rm k/n} > 15$]{GeV/n}, no data is available. Figure~\ref{spacerad:fig2} only shows the production of $^4$He particles, but Norbury et al.~\cite{Norbury2012} compiled measurement data for a broad range of nuclear fragments and types\footnote{Yield distributions were not included because they cannot be used without prior conversion into cross-sections.} of cross-sections (total, charge-changing, single-differential, etc.).

The most important reference for neutron production data is that by Nakamura and Heilbronn \cite{Nakamura2005}. They collected all the existing world data for both cross-sections and thick-target yields. Most data came from experiments at the \acrshort{riken} (\acrlong{riken}, Japan), \acrshort{himac} (\acrlong{himac}, Japan) and Bevalac (at \acrshort{lbnl}) accelerators. Subsequently, Satoh et al.~\cite{Satoh2011} and Itashiki et al.~\cite{Itashiki2016} reported more data, some of which overlaps with that listed in Ref.~\cite{Nakamura2005}, enabling comparisons between experiments. When one analyses all these data, some significant disagreements are found.
Bevalac data (\unit[337]{MeV/n})~\cite{Nakamura2005} show inconsistencies with \acrshort{himac} data, though the recent measurements by Satoh and Itashiki are broadly consistent with the latter. The low-energy (\unit[95]{MeV/n} and \unit[135]{MeV/n}) \acrshort{riken} data~\cite{Nakamura2005} also show some inconsistencies. Overall, the complete neutron data set needs a re-evaluation and extensive new experimental data to resolve disagreements between the different available measurements.

\begin{table}
    \small
    \setlength{\tabcolsep}{6pt}
    \centering
    \caption{Summary of highest-priority cross-section measurements for light-ion, neutron and pion production relevant to space radiation protection. Experiments should determine the multiplicities and energy spectra of the reaction products. If possible, neutron energies should be measured down to 1\,MeV.\vspace{1.5mm}}
    \label{tab:spacerad-measurements}
    \begin{tabular}{cccccc}
        \toprule
        \textbf{Projectile $i$} & \textbf{Targets $j$} & \textbf{Products $k$} & \textbf{Measurements} & \textbf{Projectile $\boldmath E_{\rm k/n}$} & \textbf{Precision}\\
        \midrule
        \multirow{2}{*}{He} & \multirow{2}{*}{p, C, Al, Fe} & \multirow{2}{*}{$^{1,2,3}$H, $^{3,4}$He, n, $\pi^\pm$} & \multirow{2}{*}{$\displaystyle\frac{\dd^2\sigma^{i+j\to k}}{\dd\Omega\dd E}$, $\sigma_{\rm inel}^{i+j}$} &  \multirow{2}{*}{0.1 to 50\,GeV/n} & (any data)\\
        & & & & & pref. $<20\%$ \\
        \midrule
        \multirow{2}{*}{O, Si, Fe} & \multirow{2}{*}{p, C, Al, Fe} & \multirow{2}{*}{$^{1,2,3}$H, $^{3,4}$He, n, $\pi^\pm$} & \multirow{2}{*}{$\displaystyle\frac{\dd^2\sigma^{i+j\to k}}{\dd\Omega\dd E}$, $\sigma_{\rm inel}^{i+j}$} & \multirow{2}{*}{0.1 to 50\,GeV/n} & (any data)\\
        & & & & & pref. $<20\%$ \\
        \bottomrule
    \end{tabular}
\end{table}

\subsubsection{Summary and wish list}
As summarised in Table~\ref{tab:spacerad-measurements}, the primary projectiles of interest for future measurements of light-ion production relevant to space radiation protection are Fe, Si, O and He. Energies of interest are between \unit[100]{MeV/n} and \unit[50]{GeV/n}, with an emphasis on data at higher energies, where there are significant data gaps. Other projectiles and energies above \unit[10]{GeV/n} are also of interest for testing models and transport codes, if such data is all that is available. A variety of targets are of interest, including the major constituent elements of the most commonly used aerospace materials and the human body (e.g., H, C, Al and Fe). To be useful as input to nuclear-reaction models and hence transport codes, measurements must determine double-differential as a function of the solid angle and energy, and total inelastic cross-sections $\sigma_{\rm inel}$ -- also denoted $\sigma_{\rm R}$, see discussion of Eq.~\eqref{eq:def_tot-inel-nuclei} --, and identify fragments and secondary particles over as broad an energy range as possible. For secondary particles, pions are of particular interest, for which production cross-sections should be measured with high precision. In principle, any data is helpful for improving nuclear models, though preferably measurements should reach a precision of 20\% or better.

There is also a variety of other measurement needs, the most important being double-differential cross-sections for neutron production, especially above \unit[1]{GeV/n}, where there is essentially no data, and for energies as low as \unit[1]{MeV}, where the biological damage from neutrons is the largest \cite{Stricklin2021}. Many measurements of yield distributions are already available in the literature (see, e.g., \cite{Lynch1987}), for which it would be very helpful if a methodology was developed to reliably convert them into cross-sections. Finally, extensive comparisons of the most important nuclear models used in transport codes are needed, including validations against currently available and future measurements.

\subsection{Hadrontherapy}
\label{sec:Transverse_Hadrontherapy}

The study of nuclear processes involved in the interaction of protons and heavier ions is crucial for hadrontherapy, as well as for space radiation protection. In fact, the energies and ions of interest for clinical applications overlap the ion types and energies composing the \acrshort{gcr} particle field. Previous works have demonstrated that ion fragmentation processes in clinical treatments are a source of uncertainties in the calculation of the relative biological effectiveness, which is commonly used to calculate the dose delivered to the patient via the \acrshort{tps} \cite{Luehr2012}.

\begin{figure}
    \centering
    \includegraphics[width=1.0\linewidth]{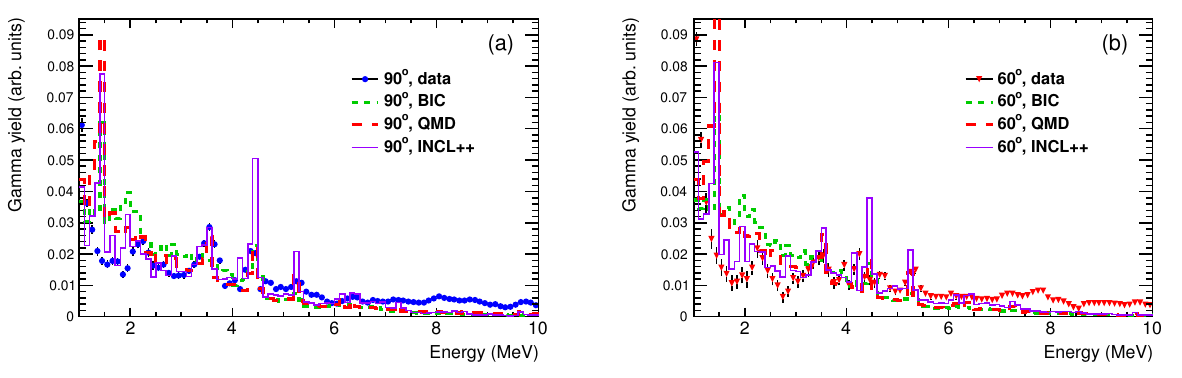}
    \caption{Measured and simulated energy spectra of prompt-gamma produced by 220\,MeV/n $^{12}$C interacting in a 20-cm thick target of PMMA, measured at 90 and 60$^o$. Reproduced from Ref.~\cite{Vanstalle2017}.}
    \label{hadrontherapy:fig1}
\end{figure}

It is also essential to correctly reproduce nuclear reactions during a particle-therapy treatment, as many dose monitoring techniques are based on the detection of secondary particles emitted by these reactions. For example, many research teams are developing monitoring systems based on prompt-gamma detection, these prompt-gamma being emitted during hadronic processes occurring in the patient \cite{Krimmer2015, Marcatili2020}. Other techniques suggest detecting the annihilation $\gamma$ from $^{11}$C, produced by the incoming beam interacting with the target~\cite{Parodi2002, Enghardt2004}. Finally, various teams have also suggested developing monitoring techniques based on the detection of secondary protons that are produced by ion fragmentation in the patient (only if $Z>1$) \cite{Finck2017, Bautista2021}. Most of these studies rely on \acrshort{mc} simulations, although it was demonstrated that important discrepancies exist between measured data and simulated output~\cite{Dudouet2014, Vanstalle2017}. An example of the differences in energy spectra simulated by different hadronic models of the \texttt{Geant4} \acrshort{mc} code (\texttt{BIC}, \texttt{QMD} and \texttt{INCL}), compared to measured data, is presented in Fig.~\ref{hadrontherapy:fig1}. The significant discrepancies, observed between the three models and the experimental results, mainly arise from the difficulties \acrshort{mc} simulations encounter in accurately reproducing the hadronic processes undergone by the incoming carbon ion.
Therefore, we must improve the accuracy of the nuclear cross-sections that govern particle-therapy treatment.

\subsubsection{Current status}
\begin{figure}
    \centering
    \includegraphics[width=1.0\linewidth]{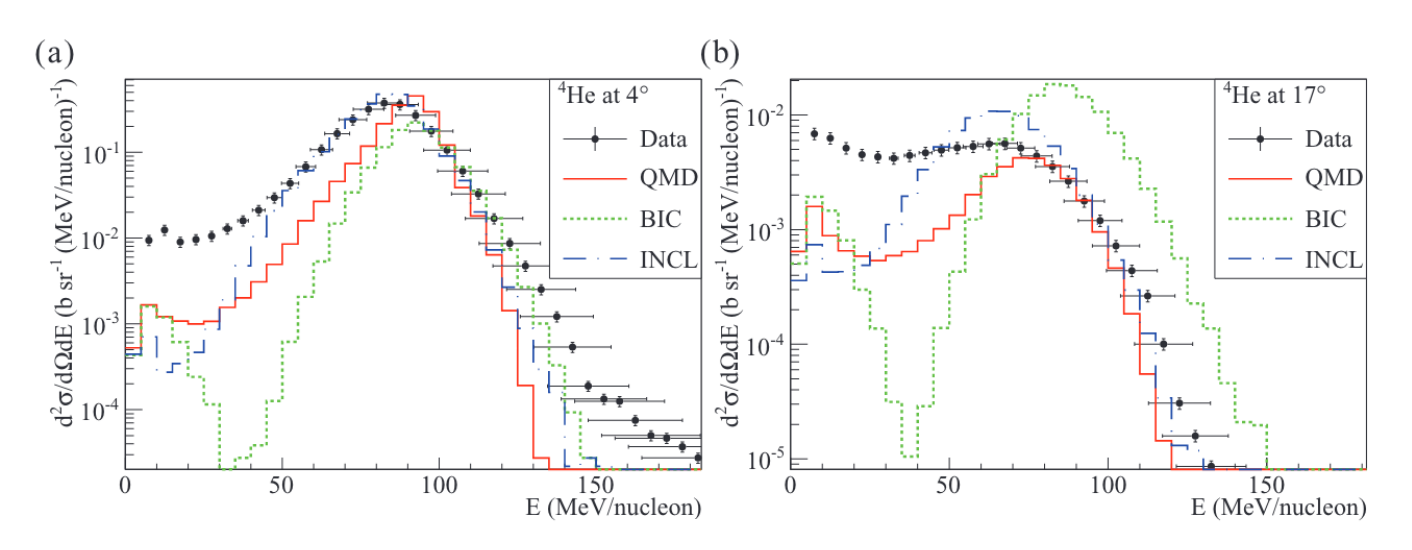}
    \caption{Double differential cross-sections of $^{4}$He produced by 95\,MeV/n $^{12}$C ion on carbon target, measured at 4 and 17$^o$. Reproduced from \cite{Dudouet2014}.}
    \label{hadrontherapy:fig2}
\end{figure}

Many studies have already been carried out to characterise the radiation fields of secondary particles produced by MeV to GeV ions \cite{Haettner2013, Dudouet2014, Horst2019, Divay2017, Mattei2020}. For example, the \acrshort{gsi} Biophysics Department made available a fragmentation cross-section database, providing useful and crucial data for \acrshort{gcr} field characterisation \cite{2021NJPh...23j1201L}. Measurements on thin and thick targets were also carried out. Several experiments were performed at the \acrshort{ganil} (\acrlong{ganil}, Caen, France) facility with 50 and 95\,MeV/n $^{12}$C ions, allowing to extract double-differential cross-sections on different targets (H, C, O, Al and $^{\textrm{nat}}$Ti). For example, the measured $\dd^2\sigma/(\dd\Omega \dd E)$ of $^4$He produced by $^{12}$C ions on carbon target at different angles can be seen in Fig.~\ref{hadrontherapy:fig2}, compared to several hadronic models provided by \texttt{Geant4}.
Different measurement strategies were investigated to characterise secondary particles, from \acrshort{tof} techniques with scintillating detectors to dose measurements, with thermoluminescent dosimeters-based \cite{Boscolo2020} or Bonner sphere spectroscopy for neutron measurements, or {\em tissue equivalent proportional counter} that can provide a direct measurement of the linear energy transfer.

Currently, the international \acrshort{foot} (\acrlong{foot}) collaboration intends to perform systematic measurements of differential cross-sections of secondary particles produced by radiation on tissue-equivalent targets
\cite{Biondi:2020rip}. The first results produced by the collaboration were measurements of elemental cross-sections in \cite{Toppi2022}, performed at the \acrshort{gsi} facility, from 400\,MeV/n $^{16}$O interacting with a carbon target. Another experiment at \acrshort{gsi} with the same ion allowed the first measured differential cross-sections by the collaboration, available in \cite{Ridolfi2024}. 

\subsubsection{Wish list}

Several studies have already been carried out, but there is still an important lack of data on double-differential cross-sections in the energy range of hadrontherapy (i.e., between 80 and 400\,MeV/n), as presented in Fig.~\ref{spacerad:fig2}. The needs of charge-changing, differential and double-differential cross-section measurements for ion therapy overlap some recommendations for space radiation protection (see previous section). The most important needs (reactions and precision) for the hadrontherapy community are summarised in Table~\ref{tab:hadrontherapy-measurements}.

For therapy, the most used ions are currently p and $^{12}$C, but a renewed interest has recently emerged for $^4$He and $^{16}$O \cite{Tommasino2015, Tessonnier2017}. Therefore, the priority in the hadrontherapy field is to measure double-differential cross-sections of $^4$He, $^{12}$C and $^{16}$O-induced reactions on targets of interest for clinical applications: mainly H, C, O and Ca. The measurements of hadronic reactions on H targets will allow the evaluation of target fragmentation through an inverse kinematic approach, as proposed by the \acrshort{foot} collaboration~\cite{Toppi2022}. Nuclear reactions on Al can also be of interest to take into account the activation of accelerator materials.
The precision required for the cross-sections measurements in hadrontherapy is based on compliance with requirements on delivered dose uncertainties in clinical practice. Indeed, the 62$^{\rm nd}$ report of the \acrshort{icru} (\acrlong{icru}) recommends a maximal variation around the delivered dose of $+7\%$ and $-5\%$~\cite{ICRU62}. The charge-changing cross-sections, owing to their large contribution to the target fragmentation, dominate the error budget of the delivered dose, and hence require the best precision. As secondary particles produced during hadrontherapy treatments are responsible for an additional dose delivered outside the tumour volume, the double-differential cross-sections accuracy will have a more important impact on the out-of-field dose. 

\begin{table}
    \small
    \setlength{\tabcolsep}{6pt}
    \centering
    \caption{Summary of highest-priority cross-section measurements and precision required for hadrontherapy. See text for the motivation.\vspace{1.2mm}}
    \label{tab:hadrontherapy-measurements}
    \begin{tabular}{cccccc}
        \toprule
        \textbf{Particle $i$} & \textbf{Targets $j$} & \textbf{Products $k$} & \textbf{Measurements} & \textbf{Projectile $\boldmath E_{\rm k/n}$} & \textbf{Precision}\\
        \midrule
        \multirow{4}{*}{$^4$He, $^{12}$C, $^{16}$O} & \multirow{4}{*}{H, C, O, Ca} & \multirow{4}{*}{All} & $\sigma^{i+j\to k}_{\Delta Z}$ (charge~changing) & \multirow{4}{*}{$80<E_{\rm k/n}<400$\,MeV/n} & $<2\%$\\
        & & & $\sigma^{i+j\to k}_{\Delta A}$ (mass~changing) & & $<5\%$\\
        & & & $\dd\sigma^{i+j\to k}/\dd E$ & & $<10\%$\\
        &&&$\dd^2\sigma^{i+j\to k}/(\dd\Omega\dd E)$ & & $<5\%$\\
        \bottomrule
    \end{tabular}
\end{table}

\subsection{Further synergies between astroparticle and high-energy physics}
\label{sec:Transverse_Other}

Despite not described in detail in this paper, other synergies between the astroparticle and high-energy physics communities exist, where experimental inputs on the cross-sections are also needed. A first example discusses how the current interpretation of the atmospheric shower data induced by \acrshort{uhecr}s is limited by the knowledge of the hadronic cross-sections (Sec.~\ref{sec:UHECR}). A second example explains how measurements of two-particle correlations is relevant to the understanding of the structure of neutron stars (Sec.~\ref{sec:femto}).

\subsubsection{Ultra-High-Energy Cosmic Rays}
\label{sec:UHECR}

The physics case of \acrshort{uhecr}s was discussed in Sec.~\ref{sec:physicscase_UHECRs}. We highlight below two interaction mechanisms for which better nuclear data are needed, and how they impact the measurement or interpretation of \acrshort{uhecr} data.

\paragraph{Hadronic interactions in the atmosphere}
Knowledge of hadronic cross-sections is fundamental to understand the physics of air showers, namely cascades of secondary particles generated when \acrshort{uhecr}s (see Fig.~\ref{fig:CRspectrum}) interact with Earth's atmosphere~\cite{Abreu2012PhRvL}. These interactions, dominated by hadronic processes, govern the development, energy distribution, and particle composition of air showers. Precise measurements of hadronic cross-sections are essential for interpreting air shower data, with profound implications for analysing the \acrshort{uhecr}s spectrum and composition, as observed by facilities like the Pierre Auger Observatory and the \acrshort{ta}.
At the heart of air shower physics lies the challenge of modelling hadronic interactions across an enormous energy range, often surpassing the energies achievable in terrestrial accelerators such as the \acrshort{lhc}. These high-energy interactions involve nuclei from primary \acrshort{cr}s colliding with atmospheric atoms, resulting in a complex cascade of secondary particles, including pions, kaons, and baryons. Measuring hadronic cross-sections provides critical constraints on the theoretical models used to predict particle multiplicities, energy spectra and angular distributions within the shower.
Accurate cross-section data enable more reliable extrapolations of hadronic interactions at ultra-high energies, where theoretical uncertainties and model dependencies become significant~\cite{Ulrich:2010rg}.
Reducing uncertainties in these measurements enhances our ability to distinguish between different \acrshort{cr} composition models and deepens our understanding of the origins and acceleration mechanisms of these particles.

\paragraph{Photo-disintegration and giant dipole resonances}
The observed \acrshort{uhecr}s exhibit a predominantly heavy composition, suggesting that photo-disintegration processes play a crucial role in modifying their nuclear species as they travel from their sources to Earth. 
The dominant contribution to photo-disintegration arises from the excitation of the giant dipole resonance, a collective vibration of protons and neutrons within the nucleus. This resonance typically occurs at photon energies above $\sim\!8$\,MeV (in the nucleus' rest frame) and leads to the emission of one or two nucleons. At higher energies, the quasi-deuteron process becomes significant, wherein the photon interacts with a nucleon pair, resulting in the ejection of nucleons or light fragments. For photon energies exceeding $\sim\!150$\,MeV, even more energetic processes, such as baryonic resonances, begin to dominate.
Given the average energy of today's photons from the \acrfull{cmb}, \acrshort{uhecr} nuclei with Lorentz factors exceeding a few $10^9$ experience interactions where \acrshort{cmb} photons can reach tens of MeV in their rest frame. In the sources of \acrshort{uhecr}s, intense photon fields from local structures (e.g., accretion discs) can even reach beyond 150\,MeV in the \acrshort{uhecr}'s rest frame and lead to photopion production, with observable predictions of coincident neutrino and gamma-ray emissions \cite{1997PhRvL..78.2292W}, besides also producing photodisintegrations of these \acrshort{cr}.
Therefore, photonuclear interactions contribute significantly to the evolution of the \acrshort{uhecr} composition during propagation ~\cite{AlvesBatista:2015jem, Boncioli:2023gbl} and in the sources~\cite{2018A&A...611A.101B}.

Despite extensive studies, photonuclear cross-section data remain incomplete and inconsistent. Systematic discrepancies persist among different experimental techniques at various accelerator facilities, and there is a lack of measurements for many nuclear species~\cite{Boncioli:2016lkt}. 
Several models have been developed to describe photonuclear reactions in \acrshort{uhecr} propagation. The Puget--Stecker--Bredekamp model historically provided a simplified approach by implementing a single decay chain per nucleus. However, more advanced simulations, such as those in \texttt{TALYS}~\cite{2023EPJA...59..131K}, allow for multiple decay chains and have been incorporated into modern propagation codes like \texttt{CRPropa}~\cite{Aerdker:2023tlu} and \texttt{SimProp}~\cite{Aloisio:2017iyh}. Nonetheless, discrepancies between model predictions and available cross-section data introduce systematic uncertainties in interpreting current \acrshort{uhecr} observations~\cite{Rossoni:2024ial}.
In the case of photomeson interactions, the limited data available have been gathered to produce models beyond the prevalent simplistic {\em nucleon superposition} approach \cite{2019JCAP...11..007M}, but the description is still lacking, and the required precision in the cross-sections and secondary yields have not been attained.

Addressing these experimental and theoretical challenges is a key motivation for ongoing efforts, such as the \acrshort{pandora} (\acrlong{pandora}) project, which aims to refine our understanding of photonuclear interactions relevant to \acrshort{uhecr} physics~\cite{Kido:2022ads, PANDORA:2022zky}.
Measurements of photomeson-related quantities are not currently under planning, but they could be carried out at \acrshort{cern}, where the necessary projectile energies are already available.

\subsubsection{Femtoscopy for neutron-star research}
\label{sec:femto}
Femtoscopy allows the study of residual strong interaction between hadrons, by measuring momentum correlations of hadron pairs produced in collisions at accelerator facilities. It is especially beneficial for the studies of more exotic particles, for which direct scattering experiments are limited or even impossible, due to their short-lived nature -- for instance, hadrons containing strange quark(s)~\cite{Fabbietti2020}. The residual strong interactions between nucleons and hyperons are of especial interest for astrophysics (as introduced in Sec.~\ref{sec:physicscase_femto}), as they are the necessary components to understand what are the constituents of neutron stars~\cite{Tolos:2020aln}. Whether and at which density hyperons appear in neutron stars depends on the equation of state of dense matter, and thus on the nucleon--hyperon interactions included in it. In the last decade, correlation functions were measured for p--$\Sigma$ \cite{ALICE:2019buq}, p--$\Xi$ hyperons \cite{ALICE2020} and p--$\Lambda$~\cite{STAR2019, ALICE:2021njx}, providing unprecedented precision data to constrain the relevant two-body strong interactions. Moreover, femtoscopy technique was recently extended to the three-body sector, allowing the study of three free hadron scattering process $3\to3$, which is not accessible with any other experimental setup~\cite{ALICE2022, Kievsky:2023maf}. Hyperon--nucleon--nucleon three-body force is one of the most crucial components for the proper description of the equation of state of dense matter~\cite{Hammer2013, Gerstung:2020ktv}.

The facilities and experiments capable of performing femtoscopy in high-energy collisions are shared laboratories for hadron and nuclear physics (e.g., \acrshort{alice}, \acrshort{amber}, \acrshort{lhcb}, NA61/\acrshort{shine}). They provide important input for different research questions in astrophysics. For more details, see, for instance, the recent \acrshort{jenaa}\,2024 workshop at \acrshort{cern}~\cite{JENAA2024}.

\begin{figure}
\centering
\includegraphics[width=0.7\linewidth]{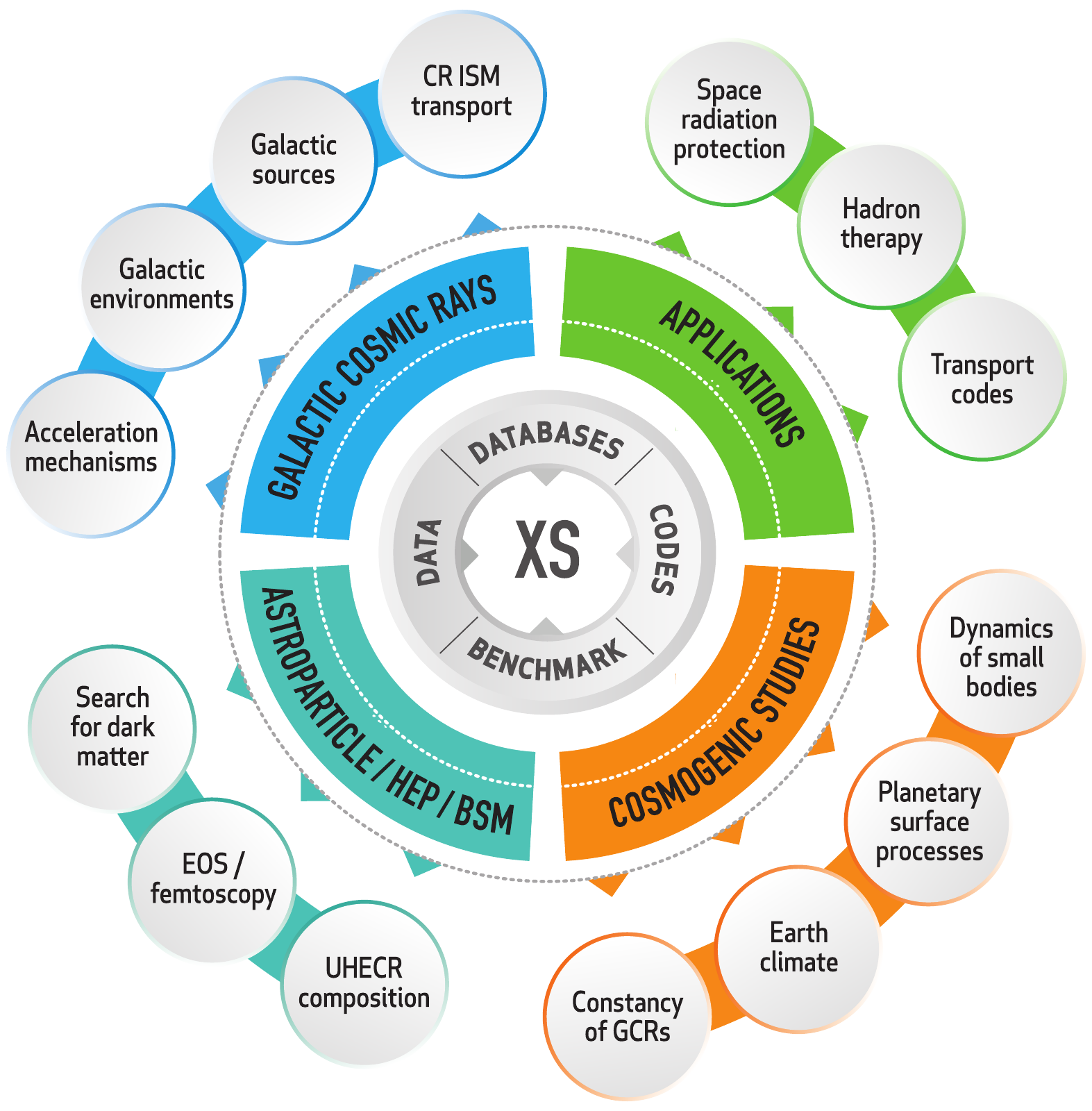}
\caption{Sketch illustrating the central role of cross-sections (denoted XS) for many physics cases discussed in this review. At the core of all studies lie the cross-sections. The first layer includes the measurements of cross-section \texttt{DATA} in diverse facilities (Sec.~\ref{sec:facilities}), ideally curated and made available in public \texttt{DATABASES}, included and fitted in public nuclear \texttt{CODES}, ideally regularly \texttt{BENCHMARKED} (against the data and one another). The second layer consists in the physics topics, some having been deeply discussed in this review, with high-precision era of \acrshort{cr} experiments (Sec.~\ref{sec:CRdata}) motivating \acrshort{gcr} studies (Sec.~\ref{sec:XSneeds}), but also others more briefly described, such as cosmogenic studies (Sec.~\ref{sec:Transverse_Cosmogenic}), space and medicine applications (Secs.~\ref{sec:Transverse_SpaceRadiation} and \ref{sec:Transverse_Hadrontherapy}), and other astroparticle/\acrshort{hep}/\acrshort{bsm} (\acrlong{bsm}) synergies (Secs.~\ref{sec:XSneeds} and~\ref{sec:Transverse_Other}). The last layer highlights some salient physics cases (Sec.~\ref{sec:physics_cases}) that can be advanced with high-precision cross-section measurements (and the use of nuclear codes, ideally for the less important unmeasured reactions only).}
\label{fig:conclusions}
\end{figure}

\section{Conclusions and long-term plans}

This paper is a call to start, support and develop long-term campaigns to measure a variety of cross-sections. High-precision cross-sections are mandatory to fully exploit recent high-precision \acrshort{gcr} measurements, and of pivotal importance to address the associated physics puzzles. In the last decade, the community has pursued many studies to provide actionable lists of cross-sections to measure, and has built successful synergies to perform some of these measurements.
The most salient results presented in this review can be summarised as follows:
\begin{itemize}
    \item {\em Physics and societal challenges.} The quest for difficult physics questions, relatively new (What is \acrshort{dm}?) or much older (What are the sources of \acrshort{gcr}s and the origin of life on terrestrial planets?), cannot go forward without new campaigns of high-precision cross-section measurements. Some of these cross-sections, and many others, are also needed for several societal and applied topics, ranging from human space exploration to cancer treatment. These synergies are summarised and highlighted in Fig.~\ref{fig:conclusions}.

    \item {\em Ongoing and future \acrshort{cr} experiments.} In the last decade, direct \acrshort{cr} experiments broke several barriers in terms of precision (percent level), energy (hundreds of TeV) and variety (anti-matter, leptons, nuclei) -- see Figs.~\ref{fig:CRdata_leptons} to~\ref{fig:CRdata_secprim}. The next decade will see even more results (see Fig.~\ref{fig:CRexperiments}), and despite a foggy horizon after 2030, the sub-percent-precision frontier is one of the many ambitious objectives of projects beyond 2040.

    \item {\em Cross-section needs for \acrshort{gcr}s.} The methodology to establish and rank the list of cross-sections to improve is sound, and we have identified the desired reactions. Moreover, forecasts have shown that these measurements are guaranteed to be a game changer for the field. We refer readers and experimentalists to our summary tables and plots of cross-section needs (reactions, energy range and precision): production cross-sections (for \acrshort{gcr} flux modelling) are presented in Tables~\ref{table:xs_ranked}, \ref{tab:ninter} and Fig.~\ref{fig:nucdata_precision} for nuclei, and in Table~\ref{tab:measurements} for antinuclei, positrons and $\gamma$ rays; inelastic, annihilating and non-annihilating cross-sections -- required for \acrshort{gcr} propagation studies or the analysis of \acrshort{cr} experiments -- are presented for nuclei and antinuclei in Table~\ref{tab:antinuclei-measurements} (and further detailed in Fig.~\ref{fig:XS_inel_data} for nuclei).

    \item {\em Other cross-section needs.} A large variety of reactions (projectiles, targets, and products) and cross-sections (total, simply or doubly differential) are also required in adjacent fields, where some overlap exists with the \acrshort{gcr} needs. Tables~\ref{tab:Cosmogenic-measurements}, \ref{tab:spacerad-measurements} and \ref{tab:hadrontherapy-measurements} summarise the highest-priority cross-section needs for cosmogenic studies, space-radiation protection and hadrontherapy. We do not provide wish lists for \acrshort{uhecr}s or neutron stars/femtoscopy physics, as these topics were presented mostly to illustrate further existing synergies between the astrophysics and \acrshort{hep} communities.

    \item {\em Key facilities and experiments.} Given the wide range of energies (hundreds of MeV to hundreds of TeV), projectiles (all nuclei, anti-matter, neutrons and muons), targets (H, He, C, O, etc.) and fragments to measure, not a single facility will provide all the needs. Luckily, facilities at \acrshort{ps}, \acrshort{sps} and \acrshort{lhc} at \acrshort{cern} (see Fig.~\ref{fig:LHC_schedule}) are already being taken advantage of, and several near-future nuclear physics facilities will have the capability to perform some desired measurements (see Table~\ref{tab:beams}). In addition, \acrshort{cr} experiments themselves have proven to be excellent detectors and complementary setups to measure cross-sections based on flight data. However, measurements of the cross-sections highlighted in this review remain marginally supported in current physics programs. We hope this paper will give more visibility to the current efforts and motivate experimentalists to join, in order to achieve at least the most urgent measurements in our wish lists.

    \item {\em Nuclear databases, transport codes and \acrshort{mc} event generators.} The number and energy coverage of the cross-sections needed is huge. On the one-hand, easy-to-use and up-to-date databases of nuclear data, possibly specific to physics sub-topics, would be a huge boost to facilitate their comparison and use. On the other-hand, general-purpose parametrisations and codes will always play a critical role for filling the gaps. Several efforts and initiative already exist to provide public and verified data, and to benchmark codes in the nuclear physics and \acrshort{hep} communities. Nevertheless, more synergies and coordinated efforts would benefit all communities. This possibly calls for a dedicated road map.
\end{itemize}

Looking at the past often gives a good vantage point to prepare for the future. In that respect, it is worth stressing that in the 1980s, similar synergies, though at a smaller scale, and efforts were carried for over 20~years to provide the cross-section data needed to interpret \acrshort{gcr} data of the time with a precision of tens of percent. These nuclear data and codes still have a legacy status. Today's and tomorrow's challenges are no less difficult than these past ones, but we are clearly embarked on the premises and promises of a long term programme that will secure legacy nuclear data for the next 20~years.

Besides pursuing the current data campaigns highlighted in this paper, there are straightforward and {\em relatively easy} directions to follow: on the modelling side, one need to provide more robust and comprehensive compilations of nuclear cross-sections, extend the wish list of desired cross-section to ultra-heavy \acrshort{gcr} species, quantify in more detail the impact and needs for other cross-section types (inelastic, non-annihilating, \dots), etc. There are also somehow {\em more difficult} or involved directions to follow: survey more closely the possibilities offered by current facilities and experiments, and prepare and plan for new opportunities and new detectors to make strong proposals for beam time for these cross-section measurements. On the \acrshort{cr} experiment side, the next generation of detectors will bring even more challenging precisions in terms of cross-sections needed to analyse and interpret their data. As these particle physics detectors already rely on beam tests at \acrshort{cern} (integration and validation), it could be interesting to think about, and possibly plan as well for, dedicated cross-section measurements campaigns at this early stage.

The clear, long-term, challenging but rewarding programme ahead of us, however, faces the difficulties of limited human resources and funding, in a future constrained by a necessary decrease of our footprint in terms of greenhouse gas emissions. Its advantage is that it relies on existing facilities, taking a priori a very small fraction of the physics programmes for which they were conceived. In this respect, given the broad and interdisciplinary questions, we hope that the road map provided in this review will convince deciding committees and the many agencies financing research world-wide to gather support for this endeavour.

\section*{Contribution statements}
F.~Donato, S.~Mariani and D.~Maurin proposed the project and coordinated the discussions during \acrshort{xscrc}\,2024~\cite{XSCRC2024}, then supervised the preparation of the article.

L.~Audouin, E.~Berti, P.~Coppin, M.~Di Mauro, P.~von~Doetinchem, F.~Donato, C.~Evoli, Y.~Génolini, P.~Ghosh, I.~Leya, M.~J.~Losekamm, S.~Mariani, D.~Maurin, J.~W.~Norbury, L.~Orusa, M.~Paniccia, T.~Poeschl, P.~D.~Serpico, A.~Tykhonov, M.~Unger, M.~Vanstalle and M.-J.~Zhao wrote the article, also contributing to its organisation and content.

The participants of \acrshort{xscrc}\,2024~\cite{XSCRC2024} who gave a talk or contributed to the discussions at the workshop were invited to participate in the paper: D.~Boncioli, M.~Chiosso, D.~Giordano, D.~M.~Gomez~Coral, G.~Graziani, C.~Lucarelli, P.~Maestro, M.~Mahlein, J.~Ocampo-Peleteiro, A.~Oliva, T.~Pierog and L.~Šerkšnytė, provided comments, additional text and references.

All authors read and commented on the manuscript to provide the final version.

\section*{Acknowledgements}
We thank our colleagues of the \acrshort{xscrc} series for lively discussions over the years, which helped to crystallise and define the cross-section needs for various physics cases, leading to new cross-sections measurements, and many successful synergies between our communities. We thank \acrshort{cern}, and in particular the Theoretical Physics Department, for hosting and supporting all the editions of the \acrshort{xscrc} workshop.
We thank R.~Delorme and V.~Tatischeff for their helpful suggestions regarding nuclear and medical physics experts for the \acrshort{xscrc}\,2024 workshop.

F.~Donato is supported by the 
Research grant {\sl The Dark Universe: A Synergic Multimessenger Approach}, No.~2017X7X85K, funded by the {\sc Miur}.
P.~Coppin and A.~Tykhonov acknowledge the support of the European Research Council (ERC) under the European Union’s Horizon~2020 research and innovation programme (Grant No.~851103). P.~Coppin is also supported by the Swiss National Science Foundation (SNSF).
Philip von Doetinchem knowledges the support of the National Science Foundation grant~PHY-2411633.
F.~Donato, C.~Evoli and M.~Di~Mauro acknowledge support from the research grant {\sl TAsP (Theoretical Astroparticle Physics)} funded by Istituto Nazionale di Fisica Nucleare (INFN).
M.D.M.~acknowledges support from the Italian Ministry of University and Research (MUR), PRIN 2022 ``EXSKALIBUR – Euclid-Cross-SKA: Likelihood Inference Building for Universe’s Research'', Grant No.~20222BBYB9, CUP I53D23000610~0006, and from the European Union -- Next Generation EU.
Y.~Génolini et P.~D.~Serpico acknowledge support by the Université Savoie Mont Blanc via the research grant NoBaRaCo.
P.~Ghosh was supported by the Pioneers mission, the Trans-Iron Galactic Element Recorder for the International Space Station, TIGERISS, an Exceptional Nucleosynthesis Pioneer.
The work of D.~M.~Gomez~Coral is supported by UNAM-PAPIIT IA101624 and SECIHTI under grant CBF2023-2024-118.
M.~Mahlein is supported by the European Research Council (ERC) under the European Union’s Horizon 2020 research and innovation programme (Grant Agreement No~950692) and the BMBF 05P24W04 ALICE.
The work of I.~Leya is funded by the Swiss National Science Foundation (200020\_219357, 200020\_196955).
The work of M.J.~Losekamm is funded by the Deutsche Forschungsgemeinschaft (DFG, German Research Foundation) via Germany's Excellence Initiative -- EXC-2094 -- 390783311.
The work of D.~Maurin was supported by the {\em Action thématique Phénomènes extrêmes et Multi-messagers (AT-PEM)} of CNRS/INSU PN Astro with INP and IN2P3, co-funded by CEA and CNES, and by the INTERCOS project funded by IN2P3.
J.W.~Norbury was supported by the RadWorks project within Exploration Capabilities of the Mars Campaign Development Division in the Exploration Systems Development Mission Directorate of NASA, United States.
J.~Ocampo-Peleteiro and A.~Oliva are supported by INFN and ASI under ASI-INFN Agreements No.~2019-19-HH.0, No.~2021-43-HH.0, and their amendments.
L.~Orusa acknowledges the support of the Multimessenger Plasma Physics Center (MPPC), NSF grants~PHY2206607.
M.~Unger acknowledges the support of the German Research Foundation DFG (Project No.~426579465).
M.-J.~Zhao acknowledges the support from the National Natural Science Foundation of China under Grants No.~12175248 and No.~12342502.

\addcontentsline{toc}{section}{Acronyms}
\printglossary[type=\acronymtype, title=Acronyms, toctitle=Acronyms]

\addcontentsline{toc}{section}{References}
\bibliographystyle{elsarticle-num-names}
\bibliography{bibliosorted}

\end{document}